\documentclass[a4paper,11pt]{article}
\pdfoutput=1
\usepackage{jheppub}
\usepackage[utf8]{inputenc}
\usepackage{url}

\usepackage{pdflscape}
\usepackage{tikz}
\usetikzlibrary{positioning}
\usetikzlibrary{arrows}
\usetikzlibrary{decorations.pathreplacing}
\usetikzlibrary{shapes.geometric}
\usetikzlibrary{calc}

\usetikzlibrary{decorations.markings}
\usetikzlibrary{arrows.meta}
\usetikzlibrary{shapes}
\usetikzlibrary{chains}
\usetikzlibrary{arrows,fit,decorations.pathreplacing, shapes.misc,decorations.pathmorphing}
\usepackage{refcount}
\usepackage{multirow}

\usepackage{paralist}

\newcommand{\be}{\begin{equation}}
\newcommand{\ee}{\end{equation}}
\def\Tr{\mathop{\mathrm{Tr}}\nolimits}

\tikzset{bd/.style={circle, draw=black, inner sep=0pt, fill=black, minimum size=2mm}}
\tikzset{wd/.style={circle, draw=black, inner sep=0pt, fill=white, minimum size=2mm}}
\tikzstyle{ligne}=[draw, thick] 
\tikzset{D7/.style={circle, draw=black, inner sep=0pt, fill=white, minimum size=3mm}}

\newcommand{\ess}[3]{\mathcal{S}_{#1,#2}^{(#3)}}
\newcommand{\tee}[3]{\mathcal{T}_{#1,#2}^{(#3)}}
\newcommand{\essdg}[3]{\mathring{\mathcal{S}}_{#1,#2}^{(#3)}}
\newcommand{\teedg}[3]{\mathring{\mathcal{T}}_{#1,#2}^{(#3)}}

\usetikzlibrary{shapes.misc}
\tikzset{gauge1/.style={draw=none,minimum size=0.6cm,fill=white,circle, draw}}
\tikzset{gauge3/.style={draw=none,minimum size=0.4cm,fill=white,circle, draw}}
\tikzset{crosses/.style={cross out, draw=black, minimum size=0.3cm, inner sep=0pt, outer sep=0pt},
cross/.default={1pt}}
\tikzset{blank/.style={draw=none,minimum size=0.4cm,fill=none,circle, draw}}
\tikzset{flavour2/.style={draw=none,minimum size=0.5cm,fill=white,regular polygon sides=4,draw}}
\tikzset{flavourBlue/.style={draw=none,minimum size=0.4cm,fill=blue,regular polygon sides=4,draw}}
\tikzset{flavourRed/.style={draw=none,minimum size=0.4cm,fill=red,regular polygon sides=4,draw}}
\tikzset{none/.style={draw=none}}
\tikzset{redgauge/.style={draw=none,minimum size=0.4cm,fill=red,circle, draw}}
\tikzset{miniU/.style={draw=none,minimum size=0.1cm,fill=red,circle, draw}}
\tikzset{smallgauge1/.style={draw=none,minimum size=0.1cm,fill=white,circle, draw}}
\tikzset{miniBlue/.style={draw=none,minimum size=0.1cm,fill=blue,circle, draw}}
\tikzset{gauge2/.style={draw=none,minimum size=0.35mm,fill=red,circle, draw}}
\tikzset{bluegauge/.style={draw=none,minimum size=0.4cm,fill=blue,circle, draw}}
\tikzset{flavour1/.style={draw=none,minimum size=0.35mm,fill=blue, regular polygon,regular polygon sides=4,draw}}
\tikzset{flavour0/.style={draw=none,minimum size=0.35mm,fill=white, regular polygon,regular polygon sides=4,draw}}
\tikzset{smalldot/.style={draw=none,minimum size=0.1mm,fill=black, circle,draw}}
\tikzset{dotsize/.style={circle,fill,inner sep=1.5pt,draw}}
\tikzset{doubleguys/.style={double, double distance = 3pt}}
\tikzset{tripleguys/.style={triple}}
\tikzset{new edge style 1/.style={dashed}}
\tikzset{thickline/.style={line width=0.06cm}}
\tikzset{darke/.style={line width=0.3mm,black}}
\tikzset{brace/.style={decorate,decoration={brace,amplitude=10pt}}}
\pgfdeclarelayer{edgelayer}
\pgfdeclarelayer{nodelayer}
\usetikzlibrary{decorations.pathreplacing}
\pgfsetlayers{edgelayer,nodelayer,main}

\usepackage{booktabs}
\usepackage{todonotes}

\usepackage{enumitem}

\usepackage[font={small},labelfont=bf]{caption}
\usepackage{subcaption}
\usepackage{adjustbox}
\tikzset{hasse/.style={circle, fill,inner sep=2pt}}
\tikzset{gauge/.style={inner sep=1mm,draw=none,fill=white,minimum size=2mm,circle, draw}}
\tikzset{flavour/.style={draw=none,minimum size=0.3mm,fill=white, regular polygon,regular polygon sides=4,draw}}
\tikzset{bd/.style={circle, draw=black, inner sep=0pt, fill=black, minimum size=2mm}}
\tikzset{gd/.style={circle, draw=green, inner sep=0pt, fill=green, minimum size=2mm}}
\tikzset{gauge3/.style={draw=none,minimum size=0.35mm,fill=white,circle, draw}}
\tikzset{none/.style={draw=none}}
\tikzset{flavor1/.style={draw=none,minimum size=0.35mm,fill=white, regular polygon,regular polygon sides=4,draw}}
\tikzset{flavour2/.style={draw=none,minimum size=0.35mm,fill=white, regular polygon,regular polygon sides=4,draw}}
\tikzset{blankflavor/.style={draw=none,minimum size=0.5mm,fill=none, regular polygon,regular polygon sides=4,draw}}
\pgfdeclarelayer{edgelayer}
\pgfdeclarelayer{nodelayer}
\pgfsetlayers{edgelayer,nodelayer,main}
\tikzset{brace/.style={decorate,decoration={brace,amplitude=10pt}}}
\DeclareFontFamily{U}{rcjhbltx}{}
\DeclareFontShape{U}{rcjhbltx}{m}{n}{<->rcjhbltx}{}
\DeclareSymbolFont{hebrewletters}{U}{rcjhbltx}{m}{n}

\allowdisplaybreaks

\let\aleph\relax\let\beth\relax
\let\gimel\relax\let\daleth\relax

\DeclareMathSymbol{\aleph}{\mathord}{hebrewletters}{39}
\DeclareMathSymbol{\beth}{\mathord}{hebrewletters}{98}\let\bet\beth
\DeclareMathSymbol{\gimel}{\mathord}{hebrewletters}{103}
\DeclareMathSymbol{\daleth}{\mathord}{hebrewletters}{100}\let\dalet\daleth

\DeclareMathSymbol{\lamed}{\mathord}{hebrewletters}{108}
\DeclareMathSymbol{\mem}{\mathord}{hebrewletters}{109}\let\mim\mem
\DeclareMathSymbol{\ayin}{\mathord}{hebrewletters}{96}
\DeclareMathSymbol{\tsadi}{\mathord}{hebrewletters}{118}
\DeclareMathSymbol{\qof}{\mathord}{hebrewletters}{114}
\DeclareMathSymbol{\shin}{\mathord}{hebrewletters}{152}
\DeclareMathSymbol{\thet}{\mathord}{hebrewletters}{84}

\preprint{Imperial/TP/20/AH/08}

\title{S-fold magnetic quivers}

\author[\lamed]{Antoine Bourget,}
\author[\aleph]{Simone Giacomelli,}
\author[\lamed]{Julius F. Grimminger,}
\author[\lamed]{Amihay Hanany,}
\author[\beth]{Marcus Sperling,}
\author[\lamed]{and Zhenghao Zhong\,}
\affiliation[\lamed]{Theoretical Physics Group, The Blackett Laboratory, Imperial College London, Prince Consort Road
London, SW7 2AZ, UK}
\affiliation[\aleph]{Mathematical Institute, University of Oxford, Woodstock Road, Oxford, OX2 6GG, UK}
\affiliation[\beth]{Yau Mathematical Sciences Center, Tsinghua University, Haidian District, Beijing, 100084, China}
\emailAdd{a.bourget@imperial.ac.uk}
\emailAdd{simone.giacomelli@maths.ox.ac.uk}
\emailAdd{julius.grimminger17@imperial.ac.uk}
\emailAdd{a.hanany@imperial.ac.uk}
\emailAdd{marcus.sperling@univie.ac.at}
\emailAdd{zhenghao.zhong14@imperial.ac.uk}

\abstract{Magnetic quivers and Hasse diagrams for Higgs branches of rank $r$ 4d $\mathcal{N}=2$ SCFTs arising from $\mathbb{Z}_{\ell}$ $\mathcal{S}$-fold constructions are discussed. The magnetic quivers are derived using three different methods: 1) Using clues like dimension, global symmetry, and the folding parameter $\ell$ to guess the magnetic quiver. 2) From 6d $\mathcal{N}=(1,0)$ SCFTs as UV completions of 5d marginal theories, and specific FI deformations on their magnetic quiver, which is further folded by $\mathbb{Z}_{\ell}$. 3) From T-duality of Type IIA brane systems of 6d $\mathcal{N}=(1,0)$ SCFTs and explicit mass deformation of the resulting brane web followed by $\mathbb{Z}_{\ell}$ folding. A choice of the ungauging scheme, either on a long node or on a short node, yields two different moduli spaces related by an orbifold action, thus suggesting a larger set of SCFTs in four dimensions than previously expected.}

\begin{document}

\maketitle

\section{Introduction}

This paper aims to study the Higgs branch of 4d $\mathcal{N}=2$ superconformal field theories (SCFTs) defined as the world-volume theories on D$3$ branes probing certain Type IIB backgrounds called $\mathcal{S}$-folds \cite{Garcia-Etxebarria:2015wns,Aharony:2016kai,Apruzzi:2020pmv,Giacomelli:2020jel,Heckman:2020svr,Giacomelli:2020gee}. At rank $1$ these theories are well known; they were constructed through compactifications of 6d $\mathcal{N}=(1,0)$ theories in \cite{Xie:2014pua,Zafrir:2016wkk,Ohmori:2018ona}, their Coulomb branches were studied in detail \cite{Argyres:2015ffa,Argyres:2015gha,Argyres:2016xua,Argyres:2016xmc,Argyres:2016yzz,Caorsi:2019vex}, and their Higgs branches were recently studied through \emph{magnetic quivers} in \cite{Bourget:2020asf}. The concept of magnetic quivers proves useful to study Higgs branches of theories, both Lagrangian and non-Lagrangian \cite{Hanany:1996ie,DelZotto:2014kka,Cremonesi:2015lsa,Ferlito:2017xdq,Mekareeya:2017jgc,Hanany:2018vph,Hanany:2018uhm,Cabrera:2018jxt,Cabrera:2019izd,Cabrera:2019dob,Bourget:2019rtl,Bourget:2020gzi,Bourget:2020xdz,Eckhard:2020jyr,Closset:2020scj,Akhond:2020vhc,vanBeest:2020kou}. 
A convenient way to understand a hyper-K\"ahler moduli space, like the Higgs branch, of a quantum field theory with 8 supercharges is through its Hasse diagram \cite{Cabrera:2016vvv,Heckman:2016ssk,Cabrera:2017njm,Rogers:2018dez,Heckman:2018jxk,Rogers:2019pqe,Hassler:2019eso,Bourget:2019aer, Grimminger:2020dmg}. More specifically, Hasse diagrams allow one to study the following moduli spaces: Higgs branches of theories in dimension $3{-}6$; Coulomb branches of 3d theories; and even the full moduli space in 3d, which contains mixed branches. Hasse diagrams were also used to study the Coulomb branch and the full moduli space of 4d $\mathcal{N}=2$ theories in \cite{Argyres:2020wmq}. A feature of all magnetic quivers coming from $\mathcal{S}$-folds studied in the following is the appearance of non-simply laced edges, where the order of the non-simply laced edges corresponds to the order of the $\mathcal{S}$-fold (the folding parameter). The 3d $\mathcal{N}=4$ Coulomb branches of non-simply laced quivers were studied in \cite{Cremonesi:2014xha,Dey:2016qqp,Hanany:2020jzl,Bourget:2020bxh}. The non-simply laced quivers can be obtained from simply laced ones through an operation called \emph{folding} \cite{Dey:2014tka,Kimura:2017hez,Nakajima:2019olw}.

\subsection{\texorpdfstring{Review of $\mathcal{N}=2$ $\mathcal{S}$-fold theories}{Review of N=2 S-fold theories}}
\label{subsectionReview}
For convenience of the reader, we recall the definition of $\mathcal{S}$-fold theories in the following.
\paragraph{7-branes and $\mathcal{S}$-folds in F-theory.}
The  $\mathcal{N}=2$ $\mathcal{S}$-fold theories are defined as the world-volume theories on $r$ D3 branes in Type IIB probing backgrounds which generalize the $\mathcal{N}=3$ $\mathcal{S}$-folds constructed in \cite{Garcia-Etxebarria:2015wns,Aharony:2016kai}. The distinctive feature of $\mathcal{S}$-fold geometries is the presence of a quotient involving a discrete $\mathbb{Z}_{\ell}$ subgroup of the $SL(2,\mathbb{Z})$ S-duality group of Type IIB string theory, generalizing the standard orientifold projection which corresponds to the case $\ell=2$. For $\ell>2$,  $\mathbb{Z}_{\ell}$ is a symmetry of the theory only for specific nonzero values of the axio-dilaton. The construction is therefore non-perturbative since the quotient is allowed only when the string coupling has the required order one value.

As was shown in \cite{Apruzzi:2020pmv}, more general $\mathcal{N}=2$ $\mathcal{S}$-folds can be defined by combining the $\mathcal{N}=3$ $\mathcal{S}$-folds together with 7-branes characterized by a constant axio-dilaton \cite{Dasgupta:1996ij}. The latter are specified by their deficit angle $\Delta_7$. The angular coordinate around the 7-branes has periodicity $\frac{2\pi}{\Delta_7}$, where $\Delta_7$ and the gauge algebra $G$ supported on the 7-branes are as follows: 
\begin{center}
\begin{tabular}{c|cccccccc}
\toprule
$G$ & $\emptyset$ & $A_1$ & $A_2$ & $D_4$ & $E_6$ & $E_7$ & $E_8$ \\
\midrule
$\Delta_7$ & $\frac{6}{5}$ & $\frac{4}{3}$ & $\frac{3}{2}$ & $2$ & $3$ & $4$ &  $6$ \\
\bottomrule 
\end{tabular}
\end{center}
The idea is to consider a $\mathbb{Z}_{\ell}$ quotient of the Kodaira singularity which describes the 7-branes in F-theory. In Type IIB this is implemented by performing a $\mathbb{Z}_{\ell}$ quotient of the plane transverse to the 7-branes, accompanied by a $\mathbb{Z}_{\ell\Delta_7}\subset SL(2,\mathbb{Z})$ quotient to preserve supersymmetry. Furthermore, we take a $\mathbb{Z}_{\ell}$ quotient of the $\mathbb{C}^2$ along the 7-branes world-volume but transverse to the D3 branes. The construction can be thought of as a generalization of the $\mathcal{N}=3$ $\mathcal{S}$-fold since it precisely reduces to the more supersymmetric background  in the absence of 7-branes (namely $\Delta_7=1$). 

Requiring $\mathbb{Z}_{\ell\Delta_7}$ to be a subgroup of $SL(2,\mathbb{Z})$, we find that 
\be\label{quotient} \ell\Delta_7=1,2,3,4\;\;\text{or}\;\;6\ee 
and we should further impose the compatibility between the 7-branes (which freezes the axio-dilaton at a specific value) and the  $\mathbb{Z}_{\ell\Delta_7}$ quotient. We then easily conclude that\footnote{In principle also $\ell=5,6$ could be considered but we concentrate on the $\ell\leq4$ case in this paper since the pattern of Higgsings and RG flows, which is the main focus of this work, is much richer.}
\begin{compactitem}
\item For $\ell=2$ the allowed solutions are $\Delta_7=\frac{3}{2}$, $2$, and $3$. 
\item For $\ell=3$ we can have $\Delta_7=\frac{4}{3}$ and $2$. 
\item For $\ell=4$ only the 7-branes of type $A_2$, namely $\Delta_7=\frac{3}{2}$ is allowed.
\end{compactitem}
The models with $\ell=2,3,4$ come in two variants, depending on whether we include discrete flux for the B-field in Type IIB or not. The models with discrete flux, the $\mathcal{S}^{(r)}_{G,\ell}$ theories, have been constructed in \cite{Apruzzi:2020pmv}. The rank of the theory is $r$, the number of probe D3 branes, and for $r=1$ they coincide with the rank-1 models with non-trivial \emph{Enhanced Coulomb Branch} (ECB). In \cite{Giacomelli:2020jel} it was argued that by moving along the ECB, the $\mathcal{S}^{(r)}_{G,\ell}$ theories can be higgsed to another family of theories of the same rank, called $\mathcal{T}^{(r)}_{G,\ell}$. For $\ell=2$, this Higgsing can be described by activating a minimal nilpotent vacuum expectation value (VEV) for the moment map of the global symmetry of $\mathcal{S}^{(r)}_{G,2}$ theories, specifically the global symmetry factor arising from the 7-brane of type $G$. It turns out that the  $\mathcal{T}^{(r)}_{G,\ell}$ models describe the theory on the world-volume of $r$ D3 branes probing $\mathcal{N}=2$ $\mathcal{S}$-folds without discrete flux \cite{Heckman:2020svr,Giacomelli:2020gee}. 

The presence of the discrete flux has indeed deep effects on the 4d SCFTs and by turning it on, both the global symmetry and the dimension of the Higgs Branch of the theory are affected, as discussed in detail below. Let us briefly discuss some basic properties of the Higgs branch. Indeed, we can exploit the geometric setup described above to understand part of the moduli space of the theory simply by moving D3 branes away from the $\mathbb{Z}_{\ell}$ fixed point. Moving them along the plane transverse to the 7-branes corresponds to probing the Coulomb branch, whereas moving them along the 7-branes describes a motion onto the Higgs branch. 

From the geometric setup we can evaluate the dimension of the Higgs Branch. By moving the $r$ D3 branes at generic points along the 7-brane, we find a low-energy effective theory given by $r$ copies of a rank-1 theory which can be identified, as pointed out in  \cite{Apruzzi:2020pmv}, with the one $G$ instanton theory, $\mathcal{I}^{(1)}_G$: the D3 brane does not `see' anymore the action of $\mathbb{Z}_{\ell}$ and only sees the underlying 7-branes of type $G$. Since the dimension of the Higgs branch of $\mathcal{I}^{(1)}_G$ (including the center of mass mode) is $6\Delta_7-6$ we conclude that the Higgs branch has at least dimension $6r(\Delta_7-1)$. This is the correct result for $\mathcal{T}^{(r)}_{G,\ell}$, namely without discrete flux, whereas we should add an extra contribution for the $\mathcal{S}^{(r)}_{G,\ell}$ theories. This is due to the fact that $\mathcal{N}=2$ $\mathcal{S}$-folds with discrete flux provide extra massless modes localized at the $\mathcal{S}$-fold singularity which contribute to the dimension of the Higgs branch an $r$ independent contribution of $\ell(\Delta_7-1)$. This can be computed e.g.\ by comparing with the ECB dimension of known rank-1 theories. Hence, the dimension of the Higgs branch of the $\mathcal{S}^{(r)}_{G,\ell}$ theories is $(6r+\ell)(\Delta_7-1)$. 

The main properties of $\mathcal{S}$-fold theories used below together with the corresponding magnetic quivers are provided in Tables \ref{resulttable} and \ref{resulttableT}.

\begin{table}[t]
\centering
\begin{adjustbox}{center}
	\begin{tabular}{cccccc}
\toprule
		\multirow{2}{*}{SCFT}  & \multirow{2}{*}{Magnetic quiver} & \multicolumn{2}{c}{Global Symmetry}  & \multirow{2}{*}{$(\ell , \Delta_7)$}  &  \multirow{2}{*}{Dimension}  \\ 
		 & & $r>1$ & $r=1$ & & \\
\midrule
       $\ess{E_6}{2}{r}$ &		
    \begin{tabular}{c}
    \scalebox{0.70}{\begin{tikzpicture}
	\begin{pgfonlayer}{nodelayer}
		\node [style=gauge3] (2) at (-1, 0) {};
		\node [style=none] (6) at (-1, -0.5) {$1{+}3r$};
		\node [style=gauge3] (12) at (-2, 0) {};
		\node [style=none] (15) at (-2, -0.5) {$1{+}2r$};
		\node [style=gauge3] (24) at (0.2, 0) {};
		\node [style=none] (29) at (0.2, -0.5) {$1{+}4r$};
		\node [style=none] (30) at (-0.85, 0.075) {};
		\node [style=none] (31) at (0.2, 0.075) {};
		\node [style=none] (32) at (-0.85, -0.075) {};
		\node [style=none] (33) at (0.2, -0.075) {};
		\node [style=none] (34) at (-0.55, 0) {};
		\node [style=none] (35) at (-0.175, 0.375) {};
		\node [style=none] (36) at (-0.175, -0.375) {};
		\node [style=gauge3] (37) at (0.95, 0) {};
		\node [style=none] (38) at (0.95, -0.5) {$2r$};
		\node [style=gauge3] (39) at (-3, 0) {};
		\node [style=gauge3] (40) at (-4, 0) {};
		\node [style=none] (41) at (-4, -0.5) {1};
		\node [style=none] (42) at (-3, -0.5) {$1{+}r$};
	\end{pgfonlayer}
	\begin{pgfonlayer}{edgelayer}
		\draw (30.center) to (31.center);
		\draw (33.center) to (32.center);
		\draw (35.center) to (34.center);
		\draw (34.center) to (36.center);
		\draw (37) to (24);
		\draw (12) to (2);
		\draw (40) to (39);
		\draw (39) to (12);
	\end{pgfonlayer}
\end{tikzpicture}}
    \end{tabular} & $C_4 A_1$  &$C_5$ & $(2, 3)$ & $12r+4$	\\ 
          $\ess{D_4}{2}{r}$ &		
    \begin{tabular}{c}
    \scalebox{0.70}{\begin{tikzpicture}
	\begin{pgfonlayer}{nodelayer}
		\node [style=gauge3] (2) at (-1, 0) {};
		\node [style=none] (6) at (-1, -0.5) {$1{+}r$};
		\node [style=gauge3] (12) at (-2, 0) {};
		\node [style=none] (15) at (-2, -0.5) {1};
		\node [style=gauge3] (24) at (0.25, 0) {};
		\node [style=none] (29) at (0.25, -0.5) {$1{+}2r$};
		\node [style=none] (30) at (-1, 0.075) {};
		\node [style=none] (31) at (0.25, 0.075) {};
		\node [style=none] (32) at (-1, -0.075) {};
		\node [style=none] (33) at (0.25, -0.075) {};
		\node [style=none] (34) at (-0.5, 0) {};
		\node [style=none] (35) at (-0.125, 0.375) {};
		\node [style=none] (36) at (-0.125, -0.375) {};
		\node [style=gauge3] (37) at (1, 0) {};
		\node [style=none] (38) at (1, -0.5) {$2r$};
		\node [style=gauge3] (39) at (2.25, 0) {};
		\node [style=none] (40) at (1, 0.075) {};
		\node [style=none] (41) at (2.25, 0.075) {};
		\node [style=none] (42) at (1, -0.075) {};
		\node [style=none] (43) at (2.25, -0.075) {};
		\node [style=none] (44) at (1.75, 0) {};
		\node [style=none] (45) at (1.375, 0.375) {};
		\node [style=none] (46) at (1.375, -0.375) {};
		\node [style=none] (49) at (2.25, -0.5) {$r$};
	\end{pgfonlayer}
	\begin{pgfonlayer}{edgelayer}
		\draw (30.center) to (31.center);
		\draw (33.center) to (32.center);
		\draw (35.center) to (34.center);
		\draw (34.center) to (36.center);
		\draw (37) to (24);
		\draw (40.center) to (41.center);
		\draw (43.center) to (42.center);
		\draw (45.center) to (44.center);
		\draw (44.center) to (46.center);
		\draw (12) to (2);
	\end{pgfonlayer}
\end{tikzpicture}
}
    \end{tabular} & $C_2 A_1 A_1$ & $C_3  A_1$ & $(2, 2)$ &  $6r+2$	\\ 
      $\ess{A_2}{2}{r}$ &		
    \begin{tabular}{c}
    \scalebox{0.70}{\begin{tikzpicture}
	\begin{pgfonlayer}{nodelayer}
		\node [style=gauge3] (2) at (-1, 0) {};
		\node [style=none] (6) at (-1, -0.5) {1};
		\node [style=gauge3] (24) at (0.25, 0) {};
		\node [style=none] (29) at (0.25, -0.5) {$1{+}r$};
		\node [style=none] (30) at (-1, 0.075) {};
		\node [style=none] (31) at (0.25, 0.075) {};
		\node [style=none] (32) at (-1, -0.075) {};
		\node [style=none] (33) at (0.25, -0.075) {};
		\node [style=none] (34) at (-0.5, 0) {};
		\node [style=none] (35) at (-0.125, 0.375) {};
		\node [style=none] (36) at (-0.125, -0.375) {};
		\node [style=gauge3] (37) at (1.25, 0.5) {};
		\node [style=gauge3] (38) at (1.25, -0.5) {};
		\node [style=none] (39) at (1.75, 0.5) {$r$};
		\node [style=none] (40) at (1.75, -0.5) {$r$};
	\end{pgfonlayer}
	\begin{pgfonlayer}{edgelayer}
		\draw (30.center) to (31.center);
		\draw (33.center) to (32.center);
		\draw (35.center) to (34.center);
		\draw (34.center) to (36.center);
		\draw (37) to (33.center);
		\draw (38) to (31.center);
		\draw (37) to (38);
	\end{pgfonlayer}
\end{tikzpicture}}
    \end{tabular} & $C_1 A_1 U_1$ & $C_2  U_1$ &  $(2, \frac{3}{2})$ & 	$3r+1$ \\ 
    $\ess{D_4}{3}{r}$ &		
    \begin{tabular}{c}
    \scalebox{0.70}{
\begin{tikzpicture}
	\begin{pgfonlayer}{nodelayer}
		\node [style=gauge3] (2) at (-1, 0) {};
		\node [style=none] (6) at (-1, -0.5) {$1{+}2r$};
		\node [style=gauge3] (24) at (0.5, 0) {};
		\node [style=none] (29) at (0.5, -0.5) {$1{+}3r$};
		\node [style=none] (30) at (-1, 0.15) {};
		\node [style=none] (31) at (0.5, 0.15) {};
		\node [style=none] (32) at (-1, -0.15) {};
		\node [style=none] (33) at (0.5, -0.15) {};
		\node [style=none] (34) at (-0.5, 0) {};
		\node [style=none] (35) at (0, 0.5) {};
		\node [style=none] (36) at (0, -0.5) {};
		\node [style=gauge3] (37) at (-2, 0) {};
		\node [style=gauge3] (38) at (-3, 0) {};
		\node [style=none] (39) at (-2, -0.5) {$1{+}r$};
		\node [style=none] (40) at (-3, -0.5) {$1$};
	\end{pgfonlayer}
	\begin{pgfonlayer}{edgelayer}
		\draw (2) to (24);
		\draw (30.center) to (31.center);
		\draw (33.center) to (32.center);
		\draw (35.center) to (34.center);
		\draw (34.center) to (36.center);
		\draw (37) to (2);
		\draw (38) to (37);
	\end{pgfonlayer}
\end{tikzpicture}
}
    \end{tabular} & $A_2 U_1$ & $A_3$ & $(3, 2)$ & $6r+3$	\\ 
      $\ess{A_1}{3}{r}$ &		
    \begin{tabular}{c}
    \scalebox{0.70}{\begin{tikzpicture}
	\begin{pgfonlayer}{nodelayer}
		\node [style=gauge3] (2) at (-1, 0) {};
		\node [style=none] (6) at (-1, -0.5) {1};
		\node [style=gauge3] (24) at (0.5, 0) {};
		\node [style=none] (29) at (0.5, -0.5) {$1{+}r$};
		\node [style=none] (30) at (-1, 0.15) {};
		\node [style=none] (31) at (0.5, 0.15) {};
		\node [style=none] (32) at (-1, -0.15) {};
		\node [style=none] (33) at (0.5, -0.15) {};
		\node [style=none] (34) at (-0.5, 0) {};
		\node [style=none] (35) at (0, 0.5) {};
		\node [style=none] (36) at (0, -0.5) {};
		\node [style=gauge3] (37) at (1.5, 0) {};
		\node [style=none] (38) at (0.5, 0.075) {};
		\node [style=none] (39) at (0.5, -0.1) {};
		\node [style=none] (40) at (1.5, 0.075) {};
		\node [style=none] (41) at (1.5, -0.1) {};
		\node [style=none] (42) at (1.5, -0.5) {$r$};
	\end{pgfonlayer}
	\begin{pgfonlayer}{edgelayer}
		\draw (2) to (24);
		\draw (30.center) to (31.center);
		\draw (33.center) to (32.center);
		\draw (35.center) to (34.center);
		\draw (34.center) to (36.center);
		\draw (40.center) to (38.center);
		\draw (39.center) to (41.center);
	\end{pgfonlayer}
\end{tikzpicture}}
    \end{tabular} & $U_1U_1$& $A_1  U_1$ & $(3, \frac{4}{3})$ & $2r+1$\\ 
        $\ess{A_2}{4}{r}$ &		
    \begin{tabular}{c}
    \scalebox{0.70}{
\begin{tikzpicture}
	\begin{pgfonlayer}{nodelayer}
		\node [style=gauge3] (2) at (-1, 0) {};
		\node [style=none] (6) at (-1, -0.5) {$1{+}r$};
		\node [style=gauge3] (24) at (0.5, 0) {};
		\node [style=none] (29) at (0.5, -0.5) {$1{+}2r$};
		\node [style=none] (30) at (-1, 0.05) {};
		\node [style=none] (31) at (0.5, 0.05) {};
		\node [style=none] (32) at (-1, -0.075) {};
		\node [style=none] (33) at (0.5, -0.075) {};
		\node [style=none] (34) at (-0.5, 0) {};
		\node [style=none] (35) at (0, 0.5) {};
		\node [style=none] (36) at (0, -0.5) {};
		\node [style=none] (37) at (-1, 0.15) {};
		\node [style=none] (38) at (0.5, 0.15) {};
		\node [style=none] (39) at (-1, -0.175) {};
		\node [style=none] (40) at (0.5, -0.175) {};
		\node [style=gauge3] (41) at (-2, 0) {};
		\node [style=none] (42) at (-2, -0.5) {1};
	\end{pgfonlayer}
	\begin{pgfonlayer}{edgelayer}
		\draw (30.center) to (31.center);
		\draw (33.center) to (32.center);
		\draw (35.center) to (34.center);
		\draw (34.center) to (36.center);
		\draw (37.center) to (38.center);
		\draw (39.center) to (40.center);
		\draw (41) to (2);
	\end{pgfonlayer}
\end{tikzpicture}}
    \end{tabular} & $A_1 U_1$& $A_2$ & $(4, \frac{3}{2})$ & $3r+2$ 	\\ 
    \bottomrule
	\end{tabular}
\end{adjustbox}
\caption{The magnetic quivers of 4d $\mathcal{N}=2$ rank $r$ $\mathcal{S}$-fold theories. The Higgs branch dimension of $\ess{G}{\ell}{r}$ is $(6r + \ell)(\Delta_7 -1)$, which matches the Coulomb branch dimension of the magnetic quiver. The folding parameter $\ell$ also indicates the multiplicity of the non-simply laced edge. The global symmetry of the magnetic quiver displays the expected enhancement for $r=1$ \cite{Giacomelli:2020jel}. Recall that a $\mathrm{U}(1)$ is ungauged on a long node for all the quivers. }
\label{resulttable}
\end{table}

\begin{table}[t]
\centering
\begin{adjustbox}{center}
	\begin{tabular}{cccccc}
\toprule
		\multirow{2}{*}{SCFT}  &  \multirow{2}{*}{Magnetic quiver}  & \multicolumn{3}{c}{Global Symmetry}   &  \multirow{2}{*}{Dimension}  \\ 
		& & $r>2$  & $r=2$ & $r=1$   & \\
\midrule
       $\tee{E_6}{2}{r}$ &		
    \begin{tabular}{c}
    \scalebox{0.70}{\begin{tikzpicture}
	\begin{pgfonlayer}{nodelayer}
		\node [style=gauge3] (2) at (-1, 0) {};
		\node [style=none] (6) at (-1, -0.5) {$3r$};
		\node [style=gauge3] (12) at (-2, 0) {};
		\node [style=none] (15) at (-2, -0.5) {$2r$};
		\node [style=gauge3] (24) at (0.2, 0) {};
		\node [style=none] (29) at (0.2, -0.5) {$4r$};
		\node [style=none] (30) at (-0.85, 0.075) {};
		\node [style=none] (31) at (0.2, 0.075) {};
		\node [style=none] (32) at (-0.85, -0.075) {};
		\node [style=none] (33) at (0.2, -0.075) {};
		\node [style=none] (34) at (-0.55, 0) {};
		\node [style=none] (35) at (-0.175, 0.375) {};
		\node [style=none] (36) at (-0.175, -0.375) {};
		\node [style=gauge3] (37) at (0.95, 0) {};
		\node [style=none] (38) at (0.95, -0.5) {$2r$};
		\node [style=gauge3] (39) at (-3, 0) {};
		\node [style=gauge3] (40) at (-4, 0) {};
		\node [style=none] (41) at (-4, -0.5) {1};
		\node [style=none] (42) at (-3, -0.5) {$r$};
	\end{pgfonlayer}
	\begin{pgfonlayer}{edgelayer}
		\draw (30.center) to (31.center);
		\draw (33.center) to (32.center);
		\draw (35.center) to (34.center);
		\draw (34.center) to (36.center);
		\draw (37) to (24);
		\draw (12) to (2);
		\draw (40) to (39);
		\draw (39) to (12);
	\end{pgfonlayer}
\end{tikzpicture}}
    \end{tabular} & $F_4 A_1$ & $F_4 A_1 A_1$ & $E_6A_1$ & $12r$\\ 
          $\tee{D_4}{2}{r}$ &		
    \begin{tabular}{c}
    \scalebox{0.70}{\begin{tikzpicture}
	\begin{pgfonlayer}{nodelayer}
		\node [style=gauge3] (2) at (-1, 0) {};
		\node [style=none] (6) at (-1, -0.5) {$r$};
		\node [style=gauge3] (12) at (-2, 0) {};
		\node [style=none] (15) at (-2, -0.5) {1};
		\node [style=gauge3] (24) at (0.25, 0) {};
		\node [style=none] (29) at (0.25, -0.5) {$2r$};
		\node [style=none] (30) at (-1, 0.075) {};
		\node [style=none] (31) at (0.25, 0.075) {};
		\node [style=none] (32) at (-1, -0.075) {};
		\node [style=none] (33) at (0.25, -0.075) {};
		\node [style=none] (34) at (-0.5, 0) {};
		\node [style=none] (35) at (-0.125, 0.375) {};
		\node [style=none] (36) at (-0.125, -0.375) {};
		\node [style=gauge3] (37) at (1, 0) {};
		\node [style=none] (38) at (1, -0.5) {$2r$};
		\node [style=gauge3] (39) at (2.25, 0) {};
		\node [style=none] (40) at (1, 0.075) {};
		\node [style=none] (41) at (2.25, 0.075) {};
		\node [style=none] (42) at (1, -0.075) {};
		\node [style=none] (43) at (2.25, -0.075) {};
		\node [style=none] (44) at (1.75, 0) {};
		\node [style=none] (45) at (1.375, 0.375) {};
		\node [style=none] (46) at (1.375, -0.375) {};
		\node [style=none] (49) at (2.25, -0.5) {$r$};
	\end{pgfonlayer}
	\begin{pgfonlayer}{edgelayer}
		\draw (30.center) to (31.center);
		\draw (33.center) to (32.center);
		\draw (35.center) to (34.center);
		\draw (34.center) to (36.center);
		\draw (37) to (24);
		\draw (40.center) to (41.center);
		\draw (43.center) to (42.center);
		\draw (45.center) to (44.center);
		\draw (44.center) to (46.center);
		\draw (12) to (2);
	\end{pgfonlayer}
\end{tikzpicture}
}
    \end{tabular} & $B_3 A_1$ & $B_3 A_1 A_1$ & $D_4A_1$ &  $6r$	\\ 
      $\tee{A_2}{2}{r}$ &		
    \begin{tabular}{c}
    \scalebox{0.70}{\begin{tikzpicture}
	\begin{pgfonlayer}{nodelayer}
		\node [style=gauge3] (2) at (-1, 0) {};
		\node [style=none] (6) at (-1, -0.5) {1};
		\node [style=gauge3] (24) at (0.25, 0) {};
		\node [style=none] (29) at (0.25, -0.5) {$r$};
		\node [style=none] (30) at (-1, 0.075) {};
		\node [style=none] (31) at (0.25, 0.075) {};
		\node [style=none] (32) at (-1, -0.075) {};
		\node [style=none] (33) at (0.25, -0.075) {};
		\node [style=none] (34) at (-0.5, 0) {};
		\node [style=none] (35) at (-0.125, 0.375) {};
		\node [style=none] (36) at (-0.125, -0.375) {};
		\node [style=gauge3] (37) at (1.25, 0.5) {};
		\node [style=gauge3] (38) at (1.25, -0.5) {};
		\node [style=none] (39) at (1.75, 0.5) {$r$};
		\node [style=none] (40) at (1.75, -0.5) {$r$};
	\end{pgfonlayer}
	\begin{pgfonlayer}{edgelayer}
		\draw (30.center) to (31.center);
		\draw (33.center) to (32.center);
		\draw (35.center) to (34.center);
		\draw (34.center) to (36.center);
		\draw (37) to (33.center);
		\draw (38) to (31.center);
		\draw (37) to (38);
	\end{pgfonlayer}
\end{tikzpicture}}
    \end{tabular} & $A_2 A_1$ & $A_2 A_1 A_1$ & $A_2A_1$ & $3r$ \\ 
    $\tee{D_4}{3}{r}$ &		
    \begin{tabular}{c}
    \scalebox{0.70}{
\begin{tikzpicture}
	\begin{pgfonlayer}{nodelayer}
		\node [style=gauge3] (2) at (-1, 0) {};
		\node [style=none] (6) at (-1, -0.5) {$2r$};
		\node [style=gauge3] (24) at (0.5, 0) {};
		\node [style=none] (29) at (0.5, -0.5) {$3r$};
		\node [style=none] (30) at (-1, 0.15) {};
		\node [style=none] (31) at (0.5, 0.15) {};
		\node [style=none] (32) at (-1, -0.15) {};
		\node [style=none] (33) at (0.5, -0.15) {};
		\node [style=none] (34) at (-0.5, 0) {};
		\node [style=none] (35) at (0, 0.5) {};
		\node [style=none] (36) at (0, -0.5) {};
		\node [style=gauge3] (37) at (-2, 0) {};
		\node [style=gauge3] (38) at (-3, 0) {};
		\node [style=none] (39) at (-2, -0.5) {$r$};
		\node [style=none] (40) at (-3, -0.5) {1};
	\end{pgfonlayer}
	\begin{pgfonlayer}{edgelayer}
		\draw (2) to (24);
		\draw (30.center) to (31.center);
		\draw (33.center) to (32.center);
		\draw (35.center) to (34.center);
		\draw (34.center) to (36.center);
		\draw (37) to (2);
		\draw (38) to (37);
	\end{pgfonlayer}
\end{tikzpicture}
}
    \end{tabular} & $G_2 U_1$ & $G_2 A_1$ & $D_4A_1$ & $6r$	\\ 
      $\tee{A_1}{3}{r}$ &		
    \begin{tabular}{c}
    \scalebox{0.70}{\begin{tikzpicture}
	\begin{pgfonlayer}{nodelayer}
		\node [style=gauge3] (2) at (-1, 0) {};
		\node [style=none] (6) at (-1, -0.5) {1};
		\node [style=gauge3] (24) at (0.5, 0) {};
		\node [style=none] (29) at (0.5, -0.5) {$r$};
		\node [style=none] (30) at (-1, 0.15) {};
		\node [style=none] (31) at (0.5, 0.15) {};
		\node [style=none] (32) at (-1, -0.15) {};
		\node [style=none] (33) at (0.5, -0.15) {};
		\node [style=none] (34) at (-0.5, 0) {};
		\node [style=none] (35) at (0, 0.5) {};
		\node [style=none] (36) at (0, -0.5) {};
		\node [style=gauge3] (37) at (1.5, 0) {};
		\node [style=none] (38) at (0.5, 0.075) {};
		\node [style=none] (39) at (0.5, -0.1) {};
		\node [style=none] (40) at (1.5, 0.075) {};
		\node [style=none] (41) at (1.5, -0.1) {};
		\node [style=none] (42) at (1.5, -0.5) {$r$};
	\end{pgfonlayer}
	\begin{pgfonlayer}{edgelayer}
		\draw (2) to (24);
		\draw (30.center) to (31.center);
		\draw (33.center) to (32.center);
		\draw (35.center) to (34.center);
		\draw (34.center) to (36.center);
		\draw (40.center) to (38.center);
		\draw (39.center) to (41.center);
	\end{pgfonlayer}
\end{tikzpicture}}
    \end{tabular} & $A_1 U_1$ & $A_1 A_1$  & $A_1 A_1$  & $2r$\\ 
        $\tee{A_2}{4}{r}$ &		
    \begin{tabular}{c}
    \scalebox{0.70}{
\begin{tikzpicture}
	\begin{pgfonlayer}{nodelayer}
		\node [style=gauge3] (2) at (-1, 0) {};
		\node [style=none] (6) at (-1, -0.5) {$r$};
		\node [style=gauge3] (24) at (0.5, 0) {};
		\node [style=none] (29) at (0.5, -0.5) {$2r$};
		\node [style=none] (30) at (-1, 0.05) {};
		\node [style=none] (31) at (0.5, 0.05) {};
		\node [style=none] (32) at (-1, -0.075) {};
		\node [style=none] (33) at (0.5, -0.075) {};
		\node [style=none] (34) at (-0.5, 0) {};
		\node [style=none] (35) at (0, 0.5) {};
		\node [style=none] (36) at (0, -0.5) {};
		\node [style=none] (37) at (-1, 0.15) {};
		\node [style=none] (38) at (0.5, 0.15) {};
		\node [style=none] (39) at (-1, -0.175) {};
		\node [style=none] (40) at (0.5, -0.175) {};
		\node [style=gauge3] (41) at (-2, 0) {};
		\node [style=none] (42) at (-2, -0.5) {1};
	\end{pgfonlayer}
	\begin{pgfonlayer}{edgelayer}
		\draw (30.center) to (31.center);
		\draw (33.center) to (32.center);
		\draw (35.center) to (34.center);
		\draw (34.center) to (36.center);
		\draw (37.center) to (38.center);
		\draw (39.center) to (40.center);
		\draw (41) to (2);
	\end{pgfonlayer}
\end{tikzpicture}}
    \end{tabular} & $A_1 U_1$ & $A_1 A_1$ & $A_2 A_1$  & $3r$ \\ 
    \bottomrule
	\end{tabular}
\end{adjustbox}
\caption{Proposal for the magnetic quivers of the $\tee{G}{\ell}{r}$ theories with $r\geq1$. The Higgs branch dimension of $\tee{G}{\ell}{r}$ is reproduced by the Coulomb branch dimension of the magnetic quiver. The folding parameter $\ell$ also indicates the multiplicity of the non-simply laced edge. In addition, the magnetic quivers display the expected symmetry enhancement for $r=2$ \cite{Giacomelli:2020jel}. For $r=1$ the magnetic quivers describe the moduli space of one $G$ instanton on $\mathbb{H}$, naturally including the center of mass, $\tee{G}{\ell}{1}=\mathcal{I}_{G}^{(1)}\otimes\mathbb{H}$, and the global symmetry is further enhanced. Recall that a $\mathrm{U}(1)$ is ungauged on a long node for all the quivers. }
\label{resulttableT}
\end{table}

\paragraph{$\mathcal{S}$-fold SCFTs from six dimensions.} 
Both $\mathcal{S}^{(r)}_{G,\ell}$ and $\mathcal{T}^{(r)}_{G,\ell}$ theories (with $\ell\Delta_7=6$) can alternatively be defined in terms of twisted compactification of 6d $\mathcal{N}=(1,0)$ theories \cite{Giacomelli:2020jel}. We exploit this fact in the derivation of magnetic quivers below, so we review this in detail. The relevant six-dimensional theories can be uniformly described as the world-volume theories of $r$ M5 branes inside the M9 wall near a $\mathbb{C}^2/\mathbb{Z}_{\ell}$ singularity. An M5 brane approaching the M9 plane yields the theory of the small $E_8$ instantons, \cite{Witten:1995gx,Ganor:1996mu,Seiberg:1996vs}. The inclusion of an orbifold into this setting has been studied in some detail, see for instance \cite{Aspinwall:1997ye,DelZotto:2014hpa,Heckman:2015bfa,Zafrir:2015rga,Ohmori:2015tka,Hayashi:2015zka,Mekareeya:2017jgc}.
The six-dimensional theories can be characterized in terms of the effective low-energy Lagrangian theory at a generic point of their tensor branch and we adopt this point of view in the present paper. 

The models relevant for constructing the $\mathcal{S}^{(r)}_{G,\ell}$ theories are described by the effective Lagrangian 
\begin{align}
 \raisebox{-.5\height}{
 	\begin{tikzpicture}
	\tikzstyle{gauge} = [circle, draw,inner sep=3pt];
	\tikzstyle{flavour} = [regular polygon,regular polygon sides=4,inner 
sep=3pt, draw];
	\node (g1) [gauge,label={[rotate=-45]below right:{$\scriptstyle{ \mathrm{SU}(\ell) }$}}] {};
	\node (g2) [right of=g1] {$\ldots$};
	\node (g3) [gauge,right of=g2,label={[rotate=-45]below right:{$\scriptstyle{ \mathrm{SU}(\ell) }$}}] {};
	\node (g4) [gauge,right of=g3,label={[rotate=-45]below right:{$\scriptstyle{ \mathrm{SU}(\ell) }$}}] {};
	\node (g5) [gauge,right of=g4,label={[rotate=-45]below right:{$\scriptstyle{ \mathrm{SU}(\ell) }$}}] {};
	\node (f1) [flavour0,left of=g1,label=below:{$\scriptstyle{ \mathrm{SU}(\ell) }$}] {};
	\node (f5) [flavour0,above of=g5,label=above:{$\scriptstyle{ 8 }$}] {};
	\node (fAS) [flavour0,right of=g5,label=below:{$\scriptstyle{1}$}] {};
	\draw [line join=round,decorate, decoration={zigzag, segment length=4,amplitude=.9,post=lineto,post length=2pt}]  (g5)--(fAS);
	\draw (4.5,0.25) node {$\scriptscriptstyle{\Lambda^2}$};
	\draw (g1)--(g2) (g2)--(g3) (g3)--(g4) (g1)--(f1) (g4)--(g5) (g5)--(f5);
    \draw[decoration={brace,mirror,raise=20pt},decorate,thick]
  (-0.25,-0.6) -- node[below=20pt] {$r-1$} (3.75,-0.6);
	\end{tikzpicture}
	} \quad + \quad  \text{$r$ tensors}\; ,
	\label{eq:6d_theory_ess}
\end{align}
 where we have $\ell$ fundamental hypermultiplets on the left and eight of them on the right. The wiggly line on the right denotes a single hypermultiplet in the rank-two anti-symmetric representation $\Lambda^2$. This representation can be ignored for $\ell=2$, it provides an extra fundamental on the right for $\ell=3$, and carries an $\mathrm{SU}(2)$ global symmetry for $\ell=4$ (since it is equivalent to the $\bf{6}$ of $\mathrm{SO}(6)$). The low-energy effective description \eqref{eq:6d_theory_ess} can be conveniently derived from brane constructions in Type IIA superstring theory \cite{Brunner:1997gf,Hanany:1997gh,Cabrera:2019izd} (see also \cite{Gorbatov:2001pw}) involving D$6$ and NS$5$ branes as well as $8$ D$8$ branes on top of an O$8^-$ orientifold.

In order to construct the $\mathcal{T}^{(r)}_{G,\ell}$ theories the starting point in six dimensions is given by the following family of theories \cite{Giacomelli:2020jel}:
\begin{align}
 \raisebox{-.5\height}{
 	\begin{tikzpicture}
	\tikzstyle{gauge} = [circle, draw,inner sep=3pt];
	\tikzstyle{flavour} = [regular polygon,regular polygon sides=4,inner 
sep=3pt, draw];
	\node (g1) [gauge,label={[rotate=-45]below right:{$\scriptstyle{ \mathrm{SU}(\ell) }$}}] {};
	\node (g2) [right of=g1] {$\ldots$};
	\node (g3) [gauge,right of=g2,label={[rotate=-45]below right:{$\scriptstyle{ \mathrm{SU}(\ell) }$}}] {};
	\node (g4) [gauge,right of=g3,label={[rotate=-45]below right:{$\scriptstyle{ \mathrm{SU}(\ell) }$}}] {};
	\node (f1) [flavour,left of=g1,label=below:{$\scriptstyle{ \mathrm{SU}(\ell) }$}] {};
	\node (f5) [flavour,right of=g4,label=below:{$\scriptstyle{ \mathrm{SU}(\ell) }$}] {};
	\draw (g1)--(g2) (g2)--(g3) (g3)--(g4) (g1)--(f1) (g4)--(f5);
    \draw[decoration={brace,mirror,raise=20pt},decorate,thick]
  (-0.25,-0.6) -- node[below=20pt] {$r-1$} (3.75,-0.6);
	\end{tikzpicture}
	} 
	\quad + \quad
    \text{$r$ tensors} 
	\; ,
\end{align} 
where we have $\ell$ flavors on each side of the quiver. In terms of brane systems in Type IIA, these theories are obtained by choosing different boundary conditions of the D6 on the 8 D8 branes. Most obviously, the $\ell$ D6 branes end on the first $\ell$ D8 branes next to the orientifold 8-plane, see \cite{Cabrera:2019izd} and Section \ref{sec:6d_magQuiv} below. Put differently, a different embedding $\mathbb{Z}_{\ell} \hookrightarrow E_8$ is chosen. 

Once we have these theories at hand, we can engineer $\mathcal{S}$-fold theories by compactifying them on $T^2$ with almost commuting holonomies along the cycles of the torus following the procedure described in \cite{Ohmori:2018ona}.

\subsection{Organization of the paper}
\begin{figure}[t]
    \centering
   \begin{tikzpicture}[scale=0.8, every node/.style={scale=0.8}]
\node[text width=2cm,align=center,draw](1) at (0,4) {6d SCFT};
\node[text width=2cm,align=center,draw](2) at (5,4) {6d MQ};
\node[text width=2cm,align=center,draw](3) at (0,0) {5d SCFTs};
\node[text width=2cm,align=center,draw](4) at (5,0) {5d MQs};
\node[text width=2cm,align=center,draw](5) at (5,-4) {4d MQs};
\draw[<-] (2) -- (1);
\draw[<-] (4) -- (3);
\draw[<-] (3) -- (1);
\draw[<-] (4) -- (2);
\draw[<-] (5) -- (4);
\node[text width=4cm,align=center] at (-2,2) {$S^1$ compactification and mass deformation};
\node[text width=4cm,align=center] at (7,-2) {$\mathbb{Z}_{\ell}$ twisted $S^1$ compactification};
\node[text width=5cm,align=center] at (2.5,5) {M/IIA construction};
\node[text width=3cm,align=center] at (2.5,-.5) {IIB brane web};
\node[text width=3cm,align=center] at (7,2) {FI deformations};
\draw[dashed,gray] (1) to [out=-10,in=100,looseness=2] (4);
\draw[dashed,gray] (1) to [out=-80,in=170,looseness=2] (4);
\node[text width=3cm,align=center] at (3.5,3) {Section \ref{SectionFI}};
\node[text width=3cm,align=center] at (1.5,1) {Section \ref{sectionBW}};
\end{tikzpicture} 
    \caption{Summary of the derivations of the 4d magnetic quivers discussed in Sections \ref{SectionFI} and \ref{sectionBW}. }
    \label{figSummary}
\end{figure}
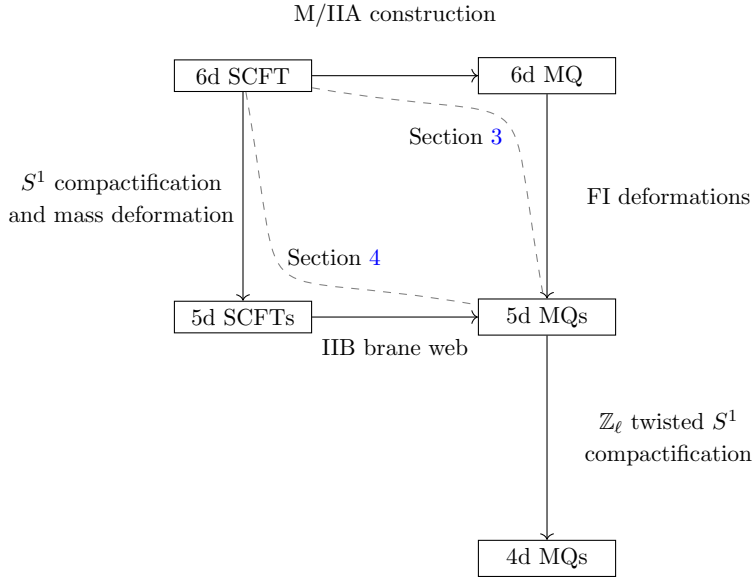

This article is organized as follows. The next three sections offer three independent derivations of the quivers of Tables \ref{resulttable} and \ref{resulttableT}. In Section \ref{sectionMQ}, the quivers are inferred from the field theoretic properties of the $\ess{G}{\ell}{r}$ and $\tee{G}{\ell}{r}$ theories in conjunction with quiver technology. We also provide the magnetic quivers for the $\essdg{G}{\ell}{r}$ and $\teedg{G}{\ell}{r}$ theories. 
Thereafter, in Sections \ref{SectionFI} and \ref{sectionBW}, we provide two constructions based on the twisted torus compactification of 6d SCFTs. The approaches in those sections are summarized in Figure \ref{figSummary}. 
Finally, in Section \ref{sec:Hasse} we use the magnetic quivers to derive part of the Hasse diagrams for the Higgs branches of the $\ess{G}{\ell}{r}$, $\tee{G}{\ell}{r}$, $\essdg{G}{\ell}{r}$ and $\teedg{G}{\ell}{r}$ theories, emphasizing transitions between them. 
The appendices contain several new results about quiver subtraction, derivation of magnetic quivers from brane webs and the Hilbert series for instantons on orbifolds. 

\subsubsection*{Conventions on quivers}
Quivers are represented as graphs where a Lie group is associated to each vertex. We adopt the following conventions throughout the paper: 
\begin{compactitem}
\item For electric quivers (quivers representing gauge theories in 5d or 6d) the names of the groups are written in full (e.g. $\mathrm{SU}(n)$, $\mathrm{USp}(2n)$, etc.). 
\item For magnetic quivers, nodes are simply labeled by an integer $n$ which represents a group $\mathrm{U}(n)$. 
\item \emph{All the non-simply laced quivers should be interpreted as having an ungauged $\mathrm{U}(1)$ on the long side. } 
\end{compactitem}

\section{Magnetic quivers}
\label{sectionMQ}

To begin with, the magnetic quivers for the 4d $\ess{G}{\ell}{r}$ and $\tee{G}{\ell}{r}$ theories are proposed based on their known properties, like global symmetry or Higgs branch dimension. 
\subsection{\texorpdfstring{$\ess{G}{\ell}{r}$ theories}{S-theories}}
In this section, we use a similar approach as \cite{Bourget:2020asf} in deriving the magnetic quivers of the  $\ess{G}{\ell}{r}$ theories. This is done by constructing quivers with unitary gauge groups using the following information \cite[Tab.\ 1]{Giacomelli:2020jel}: the global symmetry, Coulomb branch dimension, and multiplicity of the non-simply laced edge. Furthermore, the limiting cases $\ess{G}{\ell}{1}$ reduce to the magnetic quivers of 4d $\mathcal{N}=2$ rank 1 SCFTs given in \cite{Bourget:2020asf}. This restricts the form of the magnetic quiver for higher $r$.  Following these constraints, we find the magnetic quivers in Table \ref{resulttable} for general $r$. Let us discuss how the magnetic quivers satisfy the expected properties. 
\paragraph{Global symmetry.}
The Higgs branch global symmetry of $\ess{G}{\ell}{r}$ is given in \cite[Tab.\ 1]{Giacomelli:2020jel} for general $r$. In terms of the magnetic quivers, this is the Coulomb branch global symmetry. It is well known that the Coulomb branch global symmetry of quivers with only unitary gauge groups can be read off from the subset of gauge nodes that are balanced (with some exceptions that are discussed below). For a unitary node, the balance $e$ equals the number of hypermultiplets connected to that node minus twice the rank of the node. A node is balanced if $e=0$, otherwise it is over (or under) balanced if $e >0$ ($e<0$).
Connected balanced nodes form a Dynkin diagram of a semi-simple Lie algebra $\mathfrak{g}_i$. The candidate algebra of the global symmetry group is then determined by the product $\mathfrak{g}_{\mathrm{global}}=\prod_i \mathfrak{g}_i$ of the algebras associated to all each connected, balanced subset of nodes. 
However, it may be the case that studying the subset of balanced nodes only allows one to deduce a sub-algebra of the full global symmetry realized on the moduli space.
This is especially the case when one has non-simply laced edges in the quiver where extra enhancements could be expected.  Among the cases considered, the following magnetic quivers of Table \ref{resulttable} display enhancement beyond the algebra from the balanced nodes:
\begin{compactitem}
\item For $\ess{E_6}{2}{r}$ with $r>1$, the balanced nodes imply a $C_4 U_1$ global symmetry algebra. However, the overbalanced $\mathrm{U}(2r)$ and $\mathrm{U}(1)$ gauge nodes actually enhanced the $U_1$ to an $A_1$; hence, the $C_4A_1$ global symmetry. 
\item For $\ess{D_4}{2}{r}$ with $r>1$, the balance analysis shows a $C_2 A_1 U_1$ symmetry. Again, the overbalanced $\mathrm{U}(2r)$ and $\mathrm{U}(1)$ gauge nodes enhance the $U_1$ to an $A_1$; thus, the symmetry algebra becomes $C_2 A_1 A_1$.
\item For $\ess{A_2}{2}{r}$ with $r>1$, the subset of balanced nodes indicate a $C_1 U_1 U_1$ global symmetry. Here, the two overbalanced $\mathrm{U}(r)$ nodes and one overbalanced $\mathrm{U}(1)$ gauge nodes enhance one of the $U_1$ to an $A_1$; such that the symmetry becomes $C_1 U_1 A_1$.
\end{compactitem}%
The global symmetry of the proposed magnetic quivers in Table \ref{resulttable} is verified by computing its refined Coulomb branch Hilbert series and checking the coefficient of the $t^2$ term in the perturbative expansion which gives the character of the adjoint representation of the global symmetry group.

\paragraph{Dimension.}
The quaternionic Higgs branch dimension is given by 
\begin{align}
 \dim_{\mathbb{H}}\mathcal{H}\left( \ess{G}{\ell}{r} \right)=(6r+\ell)(\Delta_7-1) = h^{\vee}_{G}\left(r+\frac{\ell}{6}\right)  
\end{align}
where $\Delta_7=(h^{\vee}_{G}+6)/6$ is the deficit angle of the 7-branes \cite[Tab.\ 1]{Giacomelli:2020jel}. In the magnetic quivers, we expect to match $\dim_{\mathbb{H}}\mathcal{H}(\ess{G}{\ell}{r})$ with the 3d $\mathcal{N}=4$ Coulomb branch dimension, which is given by 
\begin{align}
 \dim_{\mathbb{H}}\mathcal{H} \left( \ess{G}{\ell}{r} \right)
 = \dim_{\mathbb{H}}\mathcal{C} \left(\text{Magnetic Quiver} \right)=\sum_{i=1}^k\mathrm{rank}(\mathrm{U}(n_i))-1   
\end{align}
and the right hand side is the sum of the ranks of the unitary gauge groups (minus one due to decoupling an overall $\mathrm{U}(1)$). The dimension of each family of $\ess{G}{\ell}{r}$ is given in  Table \ref{resulttable},  which agrees with the Coulomb branch dimension of the magnetic quivers. 

\paragraph{Multiplicity of non-simply laced edge.}
The $\ess{G}{\ell}{r}$ theories are   $\mathbb{Z}_\ell$ $\mathcal{S}$-folds, which translates to a non-simply laced edge with multiplicity $\ell$ in the magnetic quiver.

Once the magnetic quivers are proposed, we compute the Coulomb branch Hilbert series of the magnetic quivers using the monopole formula \cite{Cremonesi:2013lqa}. From the Hilbert series one can extract information of the chiral ring such as the generators and relations by further computing the plethystic logarithm (PL) \cite{Benvenuti:2006qr}. Generators correspond to positive contributions, whereas the relations between them are given by negative contributions. Thus, by studying the first few orders in the perturbative series, one is able to identify the generators and lower order relations. Higher order terms are often obscured by higher syzygies such that extracting higher order relations or generators is more laborious. The unrefined Hilbert series and its plethystic logarithm for the $r=2$ cases are given in Table \ref{HStable}. We provide the refined PL up to $t^4$ which includes how the generators and relations form irreducible representations of the global symmetry group. The results are presented in Table \ref{refinedPL}.

For generators up to order $t^4$, their presence and transformation properties are consistent with the predictions in \cite{Apruzzi:2020pmv,Giacomelli:2020jel}. This further offers a non-trivial consistency check that the proposed magnetic quivers are compatible with the properties of $\ess{G}{\ell}{r}$ theories. However, as can be seen in the unrefined PL in Table \ref{HStable}, there exists higher order generators which will take some effort to identify. We leave this for future work.

\begin{table}[t]
\small
\centering
\begin{adjustbox}{center}
\scalebox{0.8}{
	\begin{tabular}{ccc}
\toprule
		SCFT  & Hilbert Series & PL[HS] \\ 
\midrule
       $\ess{E_6}{2}{2}$ &	 $1+39t^2+108t^3+989t^4+4540t^5+O(t^6)$ &$39t^2 +108t^3 +209t^4 +328t^5 +O[t^6]$	\\ \hline
          $\ess{D_4}{2}{2}$ &	 $\dfrac
	{\scriptsize \left(\begin{array}{c}1+t+11 t^2+42 t^3+159 t^4+551 t^5+1829 t^6+5584 t^7+16155 t^8+43955 t^9+113250 t^{10}\\+276689
   t^{11}+643191 t^{12}+1424394 t^{13}+3012854 t^{14}+6095898 t^{15}+11818736 t^{16}\\+21987612
   t^{17}+39306452 t^{18}+67596644 t^{19}+111957432 t^{20}+178757670 t^{21}+275393063 t^{22}\\+409695251
   t^{23}+588987515 t^{24}+818774420 t^{25}+1101255657 t^{26}+1433825259 t^{27}\\+1807938114
   t^{28}+2208595573 t^{29}+2614801861 t^{30}+3001022474 t^{31}+3339685060 t^{32}\\+3604301588
   t^{33}+3772852500 t^{34}+3830735670 t^{35} +\dots \mathrm{palindrome} +\dots + t^{70}
 \end{array}\right)} {(1-t)^{-1}\left(1-t^2\right)^6 \left(1-t^3\right)^{11} \left(1-t^4\right)^6 \left(1-t^5\right)^6}$ & 
 $\begin{array}{c} 16 t^2+42 t^3+68 t^4+88 t^5-58 t^6-692 t^7\\-2429 t^8-4572 t^9-204 t^{10}+O\left(t^{11}\right)\end{array}$ 

 \\ 
    \midrule
    $\ess{A_2}{2}{2}$ &		 $\dfrac
	{\scriptsize \left(\begin{array}{c}1+2 t+6 t^2+19 t^3+55 t^4+133 t^5+303 t^6+637 t^7+1258 t^8+2312 t^9+3986 t^{10}\\+6422 t^{11}+9754
   t^{12}+13947 t^{13}+18841 t^{14}+24047 t^{15}+29059 t^{16}+33215 t^{17}+35995 t^{18}+36958
   t^{19}\\+\dots \mathrm{palindrome}+\dots+t^{38}
 \end{array}\right)} {(1-t)^{-2}\left(1-t^2\right)^4 \left(1-t^3\right)^5 \left(1-t^4\right)^4 \left(1-t^5\right)^3} $ &
 
 $\begin{array}{c} 7 t^2+14 t^3+21 t^4+18 t^5-14 t^6-88 t^7\\ -211 t^8-236 t^9+157 t^{10}+O\left(t^{11}\right)\end{array}$\\ 
    \midrule
     $\ess{D_4}{3}{2}$ &		 $1+9t^2+28t^3+92t^4+338t^5+O(t^6)$	& $9 t^2+28 t^3+47 t^4+86 t^5+O\left(t^6\right)$\\ 
    \midrule
      $\ess{A_1}{3}{2}$&	$\dfrac
	{\scriptsize \left(\begin{array}{c}1-t+2 t^2+t^3+6 t^4+6 t^5+13 t^6+18 t^7+32 t^8+39 t^9+59 t^{10}+70 t^{11}\\+93 t^{12}+101 t^{13}+120
   t^{14}+120 t^{15}+133 t^{16}+120 t^{17}+120 t^{18}+101 t^{19}\\+93 t^{20}+70 t^{21}+59 t^{22}+39
   t^{23}+32 t^{24}+18 t^{25}+13 t^{26}+6 t^{27}+6 t^{28}+t^{29}+2 t^{30}-t^{31}+t^{32}
 \end{array}\right)} {(1-t) \left(1-t^3\right)^3 \left(1-t^4\right)^2 \left(1-t^5\right) \left(1-t^6\right)^2 \left(1-t^7\right)} $ & $\begin{array}{c} 2t^2+6 t^3+8 t^4+10 t^5+8 t^6+2 t^7\\-17 t^8-42 t^9-72 t^{10}+O\left(t^{11}\right)\end{array}$\\  
    \midrule
        $\ess{A_2}{4}{2}$ &	$\dfrac
	{\scriptsize \left(\begin{array}{c}1 - 2 t + 2 t^2 + 11 t^4 - 12 t^5 + 30 t^6 - 10 t^7 + 94 t^8 - 
 30 t^9 + 210 t^{10} - 8 t^{11}\\ + 493 t^{12} + 16 t^{13} + 908 t^{14} + 
 166 t^{15} + 1633 t^{16} + 356 t^{17} + 2507 t^{18} + 714 t^{19} + 3579 t^{20}\\ + 
 1074 t^{21} + 4520 t^{22} + 1422 t^{23} + 5293 t^{24} + 1626 t^{25} + 5486 t^{26}+\dots\mathrm{palindrome}+\dots+t^{52}
 \end{array}\right)} {(1-t)^2 \left(1-t^2\right)^3 \left(1-t^3\right)^2 \left(1-t^4\right)^3 \left(1-t^6\right)^3 \left(1-t^8\right)^3} $ &	$\begin{array}{c} 4 t^2+4 t^3+16 t^4+12 t^5+30 t^6+24 t^7\\+22 t^8-16 t^9-138 t^{10}+O\left(t^{11}\right)\end{array}$\\ 
    \bottomrule
	\end{tabular}}
\end{adjustbox}
\caption{Coulomb branch Hilbert series and plethystic logarithm for the magnetic quivers of $\ess{G}{\ell}{r=2}$ theories in Table \ref{resulttable}. The unrefined monopole formula can be evaluated exactly for all but two cases. The unrefined PL confirms the dimension of the global symmetry. }
\label{HStable}
\end{table}
\begin{table}[t]
\small
\centering
\begin{adjustbox}{center}
\scalebox{1}{
	\begin{tabular}{ccc}
\toprule
		SCFT   & Refined PL[HS] \\ 
\midrule
       $\ess{E_6}{2}{2}$ &	 \begin{tabular}{c}
        \parbox{13.75cm}{$t^2:[2]_{A_1}[0000]_{C_4}+[0]_{A_1}[2000]_{C_4}\\ t^3:[1]_{A_1}[0001]_{C_4}+[2]_{A_1}[1000]_{C_4}\\ t^4:[4]_{A_1}[0000]_{C_4}+[1]_{A_1}[0010]_{C_4}+[2]_{A_1}[2000]_{C_4}$}
    \end{tabular} 	\\  
    \midrule
          $\ess{D_4}{2}{2}$   &	\begin{tabular}{c}
        \parbox{13.75cm}{$t^2:[2]_{A_1}[0]_{A_1}[00]_{C_2}+[0]_{A_1}[2]_{A_1}[00]_{C_2}+[0]_{A_1}[0]_{A_1}[20]_{C_2}\\ t^3:[1]_{A_1}[2]_{A_1}[01]_{C_2}+[2]_{A_1}[0]_{A_1}[10]_{C_2}\\ t^4:[4]_{A_1}[0]_{A_1}[00]_{C_2}+[2]_{A_1}[2]_{A_1}[00]_{C_2}+[4]_{A_1}[0]_{A_1}[00]_{C_2}+[2]_{A_1}[2]_{A_1}[10]_{C_2}+[2]_{A_1}[0]_{A_1}[20]_{C_2}$}
    \end{tabular} 
 \\ 
    \midrule
    $\ess{A_2}{2}{2}$ &	\begin{tabular}{c}
        \parbox{13.75cm}{$t^2:[2]_{A_1}[0]_{C_1}+[0]_{A_1}[2]_{C_1}+[0]_{A_1}[0]_{C_1}\\ t^3:(q+1/q)[1]_{A_1}[1]_{C_1}+[2]_{A_1}[1]_{C_1}\\ t^4:(q+1/q)[1]_{A_1}[0]_{C_1}+[2]_{A_1}[0]_{C_1}+[4]_{A_1}[0]_{C_1} +[2]_{A_1}[2]_{C_1}$}
    \end{tabular} 	\\ 
    \midrule
     $\ess{D_4}{3}{2}$ &\begin{tabular}{c}
        \parbox{13.75cm}{$t^2:[00]_{A_2}+[11]_{A_2}\\ t^3:q^9[00]_{A_2}+q^7[01]_{A_2}+q^3[03]_{A_2}+q^{-3}[30]_{A_2}+q^{-7}[10]_{A_2}+q^{-9}[00]_{A_2}\\ t^4:q^6[30]_{A_2}+q^4[20]_{A_2}+q^2[10]_{A_2}+[00]_{A_2}+[11]_{A_2}+q^{-2}[01]_{A_2}+q^{-4}[02]_{A_2}+q^{-6}[03]_{A_2}$}
    \end{tabular} 	\\ 
    \midrule
      $\ess{A_1}{3}{2}$&\begin{tabular}{c}
        \parbox{13.75cm}{$t^2:2 \\ t^3:b^2 q+\frac{1}{b^2 q}+\frac{b}{q}+\frac{q}{b}+q+\frac{1}{q}\\ t^4:b^3+\frac{1}{b^3}+b^2+\frac{1}{b^2}+b+\frac{1}{b}+2$}
    \end{tabular}  \\  
    \midrule
        $\ess{A_2}{4}{2}$ &\begin{tabular}{c}
        \parbox{13.75cm}{$t^2:[0]_{A_1}+[2]_{A_1}\\ t^3:(q^5+q^{-5})[1]_{A_1}\\ t^4:(1+q^8+q^{-8})[0]_{A_1}+[2]_{A_1}+(q^4+q^{-4})[4]_{A_1}$}
    \end{tabular} \\ 
    \bottomrule
	\end{tabular}}
\end{adjustbox}
\caption{Refined plethystic logarithm of the Hilbert series for the magnetic quivers of the  $\ess{G}{\ell}{r=2}$ theories in Table \ref{resulttable}. In abuse of notation, $[\ldots]_G$ denotes the $G$-character of a representation with Dynkin labels $[\ldots]$. Moreover, $q$ and $b$ label $\mathrm{U}(1)$ charges. }
\label{refinedPL}
\end{table}

\subsection{\texorpdfstring{$\tee{G}{\ell}{r}$ theories}{T-theories}}
One can proceed for the $\tee{G}{\ell}{r}$ theories in an analogous fashion as for the $\ess{G}{\ell}{r}$ theories. Using properties like global symmetry, Higgs branch dimensions, and the multiplicity of the non-simply laced edge, the form of the candidate magnetic quivers is highly constrained. This procedure results in the quivers summarized in Table \ref{resulttableT}. 

The proposal can be checked via Hilbert series techniques, as displayed in Table \ref{HStableT} and \ref{refinedPLT}. Wherever possible, the computational results are compared with the findings of \cite{Giacomelli:2020jel}. 
An important consistency check is to recover the global symmetry and the expected enhanced global symmetry for the $r=2$ case. This can be explicitly verified by the monopole formula and the analysis of the magnetic quivers. As with the $\ess{G}{\ell}{r}$ theories, the generators up to $t^4$ are given in Table \ref{refinedPLT} along with their transformation properties under the global symmetry and are consistent with the results in \cite{Giacomelli:2020jel}. As above, the presence of higher order generators can be seen from the unrefined PL. We leave the identification of these higher order generators for future work. 

As detailed in \cite{Giacomelli:2020jel}, a $\tee{G}{\ell}{r}$ theory can be obtained from the $\ess{G}{\ell}{r}$ theory via a Higgsing, which is further elaborated in Section \ref{sec:Hasse}.

\begin{table}[t]
\small
\centering
\begin{adjustbox}{center}
\scalebox{0.8}{
\begin{tabular}{ccc}
\toprule
		SCFT  &Hilbert Series & Plethystic Logarithm \\ 
\midrule
       $\tee{E_6}{2}{2}$ &$1 + 58t^2 + 104t^3 + 1944t^4 +5928t^5 + O(t^6)$ &$58 t^2+104 t^3+233 t^4-104 t^5+O\left(t^6\right)$\\ 
       \midrule
          $\tee{D_4}{2}{2}$ &$\frac{ \scriptsize \left(\begin{array}{c}1+3 t+23 t^2+82 t^3+406 t^4+1324 t^5+4798 t^6+13922
   t^7+41141 t^8+104949 t^9\\+262807 t^{10}+591960
   t^{11}+1290426 t^{12}+2586054 t^{13}+4984814
   t^{14}+8950824 t^{15}\\+15429604 t^{16}+24985684
   t^{17}+38823750 t^{18}+56981282 t^{19}+80270253
   t^{20}\\+107209505 t^{21}+137495343 t^{22}+167592304
   t^{23}+196232551 t^{24}+218705369 t^{25}\\+234220563
   t^{26}+238929060 t^{27}+\dots\mathrm{palindromic}\dots+t^{54}\end{array}\right)}{(1-t)^{-3} (1-t^2)^{10} (1-t^3)^7 (1-t^4)^{10} }$& $\begin{array}{c}27 t^2+28 t^3+83 t^4-28 t^5-479 t^6\\-1176 t^7-1466 t^8+6692
   t^9+35742 t^{10} +O\left(t^{11}\right)\end{array}$	\\ 
    \midrule
   $\tee{A_2}{2}{2}$ &$\frac{1+8 t^2+60 t^4+211 t^6+569 t^8+933 t^{10}+1164
   t^{12}+933 t^{14}+569 t^{16}+211 t^{18}+60 t^{20}+8
   t^{22}+t^{24}}{ (1-t^2)^6 (1-t^4)^6}$ &$\begin{array}{c} 14 t^2+30 t^4-101 t^6\\-117 t^8+1857
   t^{10}+O\left(t^{11}\right)\end{array}$  \\ 
    \midrule
    $\tee{D_4}{3}{2}$ &$1+17t^2 +32t^3 +211t^4+614t^5+2383t^6+7238t^7+O\left(t^{8}\right)$ & $\begin{array}{c} 17 t^2+32 t^3+58 t^4+70 t^5-100 t^6\\-7942 t^7+O\left(t^8\right)\end{array}$\\ 
    \midrule

      $\tee{A_1}{3}{2}$ & $\frac{\scriptsize \left(\begin{array}{c}1+t+5 t^2+7 t^3+18 t^4+33 t^5+56 t^6+84 t^7+115
   t^8+146 t^9+180 t^{10}\\+199 t^{11}+212 t^{12}+\dots \mathrm{palindromic}\dots+t^{24}\end{array}\right)}{(1-t)^{-1} (1-t^2)^2 (1-t^3)^2 (1-t^4)^2 (1-t^5)^3 }$ &$\begin{array}{c}6 t^2+4 t^3+3 t^4+10 t^5-4 t^6-22 t^7\\-25 t^8-2 t^9+25
   t^{10}+O\left(t^{11}\right)\end{array}$	\\ \hline
        $\tee{A_2}{4}{2}$ &	$\frac{\scriptsize \left(\begin{array}{c}1-2 t+4 t^2+9 t^4+4 t^5+26 t^6+6 t^7+72 t^8+30 t^9\\+94
   t^{10}+68 t^{11}+138 t^{12}+70 t^{13}+172 t^{14}+\mathrm{palindromic}+t^{28}\end{array}\right)}{(1-t)^2(1-t^2)^3 (1-t^3)^2 (1-t^4)^2 (1-t^6)^3}$ &$\begin{array}{c} 6 t^2+8 t^3+14 t^4+16 t^5+12 t^6-32 t^7\\-107 t^8-184 t^9-188
   t^{10}+O\left(t^{11}\right)\end{array}$\\ 
    \bottomrule
	\end{tabular}}
\end{adjustbox}
\caption{Coulomb branch Hilbert series and plethystic logarithm for the magnetic quivers of the $\tee{G}{\ell}{r=2}$ theories of Table \ref{resulttableT}. The exact unrefined monopole formula can be computed in all but two cases. The unrefined PL confirms the enhanced global symmetry for $r=2$.}
\label{HStableT}
\end{table}

\begin{table}[t]
\centering
\begin{adjustbox}{center}
	\begin{tabular}{ccc}
\toprule
		SCFT   & Refined PL[HS] \\ 
\midrule
       $\tee{E_6}{2}{2}$ &	 \begin{tabular}{c}
   \parbox{11cm}{$t^2:[2]_{A_1}[0]_{A_1}[0000]_{F_4}+[0]_{A_1}[2]_{A_1}[0000]_{F_4}+[0]_{A_1}[0]_{A_1}[1000]_{F_4}\\ t^3:[1]_{A_1}[0]_{A_1}[0001]_{F_4}+[0]_{A_1}[1]_{A_1}[0001]\\ t^4:[1]_{A_1}[1]_{A_1}[1000]_{F_4}+[0]_{A_1}[0]_{A_1}[0001]_{F_4}-[0]_{A_1}[0]_{A_1}[0000]_{F_4}$}
    \end{tabular} 	\\ 
    \midrule
          $\tee{D_4}{2}{2}$   &	\begin{tabular}{c}
        \parbox{11cm}{$t^2:[2]_{A_1}[0]_{A_1}[000]_{B_3}+[0]_{A_1}[2]_{A_1}[000]+[0]_{A_1}[0]_{A_1}[010]_{B_3}\\ t^3:[1]_{A_1}[0]_{A_1}[100]_{B_3}+[0]_{A_1}[1]_{A_1}[100]_{B_3}\\ t^4:[1]_{A_1}[1]_{A_1}[010]_{B_3}-[0]_{A_1}[0]_{A_1}[000]_{B_3}$}
    \end{tabular} 
 \\ 
    \midrule
    $\tee{A_2}{2}{2}$ &	\begin{tabular}{c}
           \parbox{11cm}{$t^2:[2]_{A_1}[0]_{A_1}[00]_{A_2}+[0]_{A_1}[2]_{A_1}[00]_{A_2}+[0]_{A_1}[0]_{A_1}[11]_{A_2}\\ t^4:[2]_{A_1}[2]_{A_1}[11]_{A_2}-2[0]_{A_1}[0]_{A_1}[00]_{A_2}$}
    \end{tabular} 	\\ 
    \midrule
     $\tee{D_4}{3}{2}$ &\begin{tabular}{c}
          \parbox{11cm}{$t^2:[2]_{A_1}[00]_{G_2}+[0]_{A_1}[01]_{G_2}\\ t^3:[3]_{A_1}[00]_{G_2}+2[2]_{A_1}[10]_{G_2}\\ t^4:3[0]_{A_1}[00]_{G_2}+[0]_{A_1}[01]_{G_2}+2[2]_{A_1}[10]_{G_2}-[0]_{A_1}[00]_{G_2}$}
    \end{tabular} 	\\ 
    \midrule
      $\tee{A_1}{3}{2}$&\begin{tabular}{c}
         \parbox{11cm}{$t^2:[2]_{A_1}[0]_{A_1}+[0]_{A_1}[2]_{A_1}\\ t^3:[3]_{A_1}[0]_{A_1}\\ t^4:[0]_{A_1}[2]_{A_1}$}
    \end{tabular}  \\  
    \midrule
        $\tee{A_2}{4}{2}$ &\begin{tabular}{c}
         \parbox{11cm}{$t^2:[2]_{A_1}[0]_{A_1}+[0]_{A_1}[2]_{A_1}\\ t^3:2[1]_{A_1}[1]_{A_1}\\ t^4:[2]_{A_1}[0]_{A_1}+2[0]_{A_1}[2]_{A_1}+[0]_{A_1}[4]_{A_1}$}
    \end{tabular} \\ 
    \bottomrule
	\end{tabular}
\end{adjustbox}
\caption{Refined plethystic logarithm of the Hilbert series for the magnetic quivers of the  $\tee{G}{\ell}{r=2}$ theories in Table \ref{resulttableT}. In abuse of notation, $[\ldots]_G$ denotes the $G$-character of a representation with Dynkin labels $[\ldots]$.}
\label{refinedPLT}
\end{table}

\subsection{\texorpdfstring{Instantons on $\mathbb{C}^2/\mathbb{Z}_{\ell}$ --- The $\essdg{G}{\ell}{r}$ and $\teedg{G}{\ell}{r}$ theories}{Instantons on C2 orbifold -- The cSGlr and cTGlr theories}}\label{shortside}

In deriving the quivers in Section \ref{SectionFI} and in Section \ref{sectionBW} we only use the definition of $\mathcal{S}$-fold theories as twisted compactifications on $T^2$ of certain $\mathcal{N}=(1,0)$ theories in six dimensions and do not make use of the F-theory realization which is reviewed in Section \ref{subsectionReview}. A natural expectation from the F-theory setup is that the Higgs branch of $\mathcal{S}$-fold theories can be interpreted as the moduli space of $r$ $G$ instantons on $\mathbb{C}^2/\mathbb{Z}_{\ell}$ since we have 7-branes probed by a stack of $r$ D3 branes. As explained below, this correlates with the existence of discretely-gauged versions of $\mathcal{S}$-fold theories and we find that this can be understood at the level of magnetic quivers by changing the ungauging scheme \cite{Hanany:2020jzl}. 

\subsubsection*{The $\tee{G}{\ell}{r}$ and $\teedg{G}{\ell}{r}$ theories}
Let us start by considering the case of trivial flux in F-theory, namely $\mathcal{T}^{(r)}_{G,\ell}$ theories. 
For $r=1$ their Higgs branches are the full moduli space of one $G$-instanton on $\mathbb{C}^2$, which can be written as a product $\widetilde{\mathcal{M}}_\text{1-inst}(G)\times \mathbb{C}^2$. The $\mathbb{C}^2$ factor is the center of mass of the instanton, and $\widetilde{\mathcal{M}}_\text{1-inst}(G)$ can be identified with the closure of the minimal nilpotent orbit of the Lie algebra of $G$.  
As was pointed out in \cite{Argyres:2016yzz}, these models, for any $r$, have a global $\mathbb{Z}_{\ell}$ symmetry (embedded in the $\mathrm{U}(1)_r$ symmetry) acting on the Coulomb branch. We can also understand this symmetry starting from six dimensions: For $r=1$ we are compactifying on $T^2$ the rank-1 E-string theory and the Coulomb branch of the resulting four-dimensional model is an $\ell$-fold cover of the Coulomb branch of the $E_8$ SCFT. More generally (see \cite{Ohmori:2018ona}), for arbitrary $r$ the parent six-dimensional theory includes a tensor multiplet which is not paired with any vector multiplet and in such a situation, upon torus compactification with $\mathbb{Z}_{\ell}$ holonomies, we find a four-dimensional theory with a discrete global $\mathbb{Z}_{\ell}$ symmetry. 

Therefore the six-dimensional picture, which is the only input we use in the derivation of the quivers in Table \ref{resulttableT}, provides us with a family of theories with a discrete global $\mathbb{Z}_{\ell}$ symmetry and we may consider gauging it. The expectation is that the Higgs branch of the gauged theory (called $\teedg{G}{\ell}{r}$, where the circle symbolizes the cyclic gauging) is a $\mathbb{Z}_{\ell}$ quotient of the Higgs branch of the parent theory $\tee{G}{\ell}{r}$. Since the associated magnetic quivers are non simply-laced, we know there is a natural way to implement such a quotient -- the ungauging scheme \cite{Hanany:2020jzl}:
Recall that for the quivers in Table \ref{resulttableT}, the ungauging is on a long node.
Furthermore, using the Crawley-Boevey trick \cite{crawley-boevey}, the ungauging on the short node can be made into an ungauging of a long node by replacing the flavor with a suitable long node. 
In more detail, the additional long node has to have the same length as the original long node in the quiver.
Hence, one can ungauge on a short node of Table \ref{resulttableT}, which results in a new family of quivers, as given in Table \ref{shortR}. 
 
For instance, the magnetic quiver for $\tee{D_4}{2}{2}$, which is 
\begin{equation}
   \tee{D_4}{2}{2} \, : \qquad    \raisebox{-.5\height}{ \scalebox{0.70}{\begin{tikzpicture}
	\begin{pgfonlayer}{nodelayer}
		\node [style=gauge3] (2) at (-1, 0) {};
		\node [style=none] (6) at (-1, -0.5) {$r$};
		\node [style=gauge3] (12) at (-2, 0) {};
		\node [style=none] (15) at (-2, -0.5) {1};
		\node [style=gauge3] (24) at (0.25, 0) {};
		\node [style=none] (29) at (0.25, -0.5) {$2r$};
		\node [style=none] (30) at (-1, 0.075) {};
		\node [style=none] (31) at (0.25, 0.075) {};
		\node [style=none] (32) at (-1, -0.075) {};
		\node [style=none] (33) at (0.25, -0.075) {};
		\node [style=none] (34) at (-0.5, 0) {};
		\node [style=none] (35) at (-0.125, 0.375) {};
		\node [style=none] (36) at (-0.125, -0.375) {};
		\node [style=gauge3] (37) at (1, 0) {};
		\node [style=none] (38) at (1, -0.5) {$2r$};
		\node [style=gauge3] (39) at (2.25, 0) {};
		\node [style=none] (40) at (1, 0.075) {};
		\node [style=none] (41) at (2.25, 0.075) {};
		\node [style=none] (42) at (1, -0.075) {};
		\node [style=none] (43) at (2.25, -0.075) {};
		\node [style=none] (44) at (1.75, 0) {};
		\node [style=none] (45) at (1.375, 0.375) {};
		\node [style=none] (46) at (1.375, -0.375) {};
		\node [style=none] (49) at (2.25, -0.5) {$r$};
	\end{pgfonlayer}
	\begin{pgfonlayer}{edgelayer}
		\draw (30.center) to (31.center);
		\draw (33.center) to (32.center);
		\draw (35.center) to (34.center);
		\draw (34.center) to (36.center);
		\draw (37) to (24);
		\draw (40.center) to (41.center);
		\draw (43.center) to (42.center);
		\draw (45.center) to (44.center);
		\draw (44.center) to (46.center);
		\draw (12) to (2);
	\end{pgfonlayer}
\end{tikzpicture}}} \quad =  \quad  \raisebox{-.5\height}{\scalebox{0.70}{\begin{tikzpicture}
	\begin{pgfonlayer}{nodelayer}
		\node [style=gauge3] (2) at (-1, 0) {};
		\node [style=none] (6) at (-1, -0.5) {$r$};
		\node [style=gauge3] (12) at (-2, 0) {};
		\node [style=none] (15) at (-2, -0.5) {1};
		\node [style=gauge3] (24) at (0.25, 0) {};
		\node [style=none] (29) at (0.25, -0.5) {$2r$};
		\node [style=none] (30) at (-1, 0.075) {};
		\node [style=none] (31) at (0.25, 0.075) {};
		\node [style=none] (32) at (-1, -0.075) {};
		\node [style=none] (33) at (0.25, -0.075) {};
		\node [style=none] (34) at (-0.5, 0) {};
		\node [style=none] (35) at (-0.125, 0.375) {};
		\node [style=none] (36) at (-0.125, -0.375) {};
		\node [style=gauge3] (37) at (1, 0) {};
		\node [style=none] (38) at (1, -0.5) {$2r$};
		\node [style=gauge3] (39) at (2.25, 0) {};
		\node [style=none] (40) at (1, 0.075) {};
		\node [style=none] (41) at (2.25, 0.075) {};
		\node [style=none] (42) at (1, -0.075) {};
		\node [style=none] (43) at (2.25, -0.075) {};
		\node [style=none] (44) at (1.75, 0) {};
		\node [style=none] (45) at (1.375, 0.375) {};
		\node [style=none] (46) at (1.375, -0.375) {};
		\node [style=none] (49) at (2.25, -0.5) {$r$};
		\node [style=blankflavor] (50) at (0.25, 0) {};
	\end{pgfonlayer}
	\begin{pgfonlayer}{edgelayer}
		\draw (30.center) to (31.center);
		\draw (33.center) to (32.center);
		\draw (35.center) to (34.center);
		\draw (34.center) to (36.center);
		\draw (37) to (24);
		\draw (40.center) to (41.center);
		\draw (43.center) to (42.center);
		\draw (45.center) to (44.center);
		\draw (44.center) to (46.center);
		\draw (12) to (2);
	\end{pgfonlayer}
\end{tikzpicture}}}
\end{equation}
becomes when ungauged on the short node 
\begin{equation}
       \teedg{D_4}{2}{2} \, : \qquad \raisebox{-.5\height}{ \scalebox{0.70}{\begin{tikzpicture}
	\begin{pgfonlayer}{nodelayer}
		\node [style=gauge3] (2) at (-1, 0) {};
		\node [style=none] (6) at (-1, -0.5) {$r$};
		\node [style=none] (12) at (-2, 0) {};
		\node [style=none] (15) at (-2, -0.5) {1};
		\node [style=gauge3] (24) at (0.25, 0) {};
		\node [style=none] (29) at (0.25, -0.5) {$2r$};
		\node [style=none] (30) at (-1, 0.075) {};
		\node [style=none] (31) at (0.25, 0.075) {};
		\node [style=none] (32) at (-1, -0.075) {};
		\node [style=none] (33) at (0.25, -0.075) {};
		\node [style=none] (34) at (-0.5, 0) {};
		\node [style=none] (35) at (-0.125, 0.375) {};
		\node [style=none] (36) at (-0.125, -0.375) {};
		\node [style=gauge3] (37) at (1, 0) {};
		\node [style=none] (38) at (1, -0.5) {$2r$};
		\node [style=gauge3] (39) at (2.25, 0) {};
		\node [style=none] (40) at (1, 0.075) {};
		\node [style=none] (41) at (2.25, 0.075) {};
		\node [style=none] (42) at (1, -0.075) {};
		\node [style=none] (43) at (2.25, -0.075) {};
		\node [style=none] (44) at (1.75, 0) {};
		\node [style=none] (45) at (1.375, 0.375) {};
		\node [style=none] (46) at (1.375, -0.375) {};
		\node [style=none] (49) at (2.25, -0.5) {$r$};
		\node [style=blankflavor] (50) at (-2, 0) {};
	\end{pgfonlayer}
	\begin{pgfonlayer}{edgelayer}
		\draw (30.center) to (31.center);
		\draw (33.center) to (32.center);
		\draw (35.center) to (34.center);
		\draw (34.center) to (36.center);
		\draw (37) to (24);
		\draw (40.center) to (41.center);
		\draw (43.center) to (42.center);
		\draw (45.center) to (44.center);
		\draw (44.center) to (46.center);
		\draw (50) to (2);
	\end{pgfonlayer}
\end{tikzpicture}}} \quad = \quad     \raisebox{-.5\height}{\scalebox{0.70}{
\begin{tikzpicture}
	\begin{pgfonlayer}{nodelayer}
		\node [style=gauge3] (2) at (-1, 0) {};
		\node [style=none] (6) at (-1, -0.5) {$r$};
		\node [style=gauge3] (24) at (0.25, 0) {};
		\node [style=none] (29) at (0.25, -0.5) {$2r$};
		\node [style=none] (30) at (-1, 0.075) {};
		\node [style=none] (31) at (0.25, 0.075) {};
		\node [style=none] (32) at (-1, -0.075) {};
		\node [style=none] (33) at (0.25, -0.075) {};
		\node [style=none] (34) at (-0.5, 0) {};
		\node [style=none] (35) at (-0.125, 0.375) {};
		\node [style=none] (36) at (-0.125, -0.375) {};
		\node [style=gauge3] (37) at (1, 0) {};
		\node [style=none] (38) at (1, -0.5) {$2r$};
		\node [style=gauge3] (39) at (2.25, 0) {};
		\node [style=none] (40) at (1, 0.075) {};
		\node [style=none] (41) at (2.25, 0.075) {};
		\node [style=none] (42) at (1, -0.075) {};
		\node [style=none] (43) at (2.25, -0.075) {};
		\node [style=none] (44) at (1.75, 0) {};
		\node [style=none] (45) at (1.375, 0.375) {};
		\node [style=none] (46) at (1.375, -0.375) {};
		\node [style=none] (49) at (2.25, -0.5) {$r$};
		\node [style=gauge3] (50) at (-1, 0) {};
		\node [style=none] (51) at (-2.25, -0.5) {1};
		\node [style=none] (52) at (-1, 0.075) {};
		\node [style=none] (53) at (-2.05, 0.075) {};
		\node [style=none] (54) at (-1, -0.075) {};
		\node [style=none] (55) at (-2.05, -0.075) {};
		\node [style=none] (56) at (-1.55, 0) {};
		\node [style=none] (57) at (-1.925, 0.375) {};
		\node [style=none] (58) at (-1.925, -0.375) {};
		\node [style=none] (59) at (-2.25, 0) {};
		\node [style=flavour2] (60) at (-2.25, 0) {};
	\end{pgfonlayer}
	\begin{pgfonlayer}{edgelayer}
		\draw (30.center) to (31.center);
		\draw (33.center) to (32.center);
		\draw (35.center) to (34.center);
		\draw (34.center) to (36.center);
		\draw (37) to (24);
		\draw (40.center) to (41.center);
		\draw (43.center) to (42.center);
		\draw (45.center) to (44.center);
		\draw (44.center) to (46.center);
		\draw (52.center) to (53.center);
		\draw (55.center) to (54.center);
		\draw (57.center) to (56.center);
		\draw (56.center) to (58.center);
	\end{pgfonlayer}
\end{tikzpicture}}}  \quad  =  \quad   \raisebox{-.5\height}{\scalebox{0.70}{
\begin{tikzpicture}
	\begin{pgfonlayer}{nodelayer}
		\node [style=gauge3] (2) at (-1, 0) {};
		\node [style=none] (6) at (-1, -0.5) {$r$};
		\node [style=gauge3] (24) at (0.25, 0) {};
		\node [style=none] (29) at (0.25, -0.5) {$2r$};
		\node [style=none] (30) at (-1, 0.075) {};
		\node [style=none] (31) at (0.25, 0.075) {};
		\node [style=none] (32) at (-1, -0.075) {};
		\node [style=none] (33) at (0.25, -0.075) {};
		\node [style=none] (34) at (-0.5, 0) {};
		\node [style=none] (35) at (-0.125, 0.375) {};
		\node [style=none] (36) at (-0.125, -0.375) {};
		\node [style=gauge3] (37) at (1, 0) {};
		\node [style=none] (38) at (1, -0.5) {$2r$};
		\node [style=gauge3] (39) at (2.25, 0) {};
		\node [style=none] (40) at (1, 0.075) {};
		\node [style=none] (41) at (2.25, 0.075) {};
		\node [style=none] (42) at (1, -0.075) {};
		\node [style=none] (43) at (2.25, -0.075) {};
		\node [style=none] (44) at (1.75, 0) {};
		\node [style=none] (45) at (1.375, 0.375) {};
		\node [style=none] (46) at (1.375, -0.375) {};
		\node [style=none] (49) at (2.25, -0.5) {$r$};
		\node [style=gauge3] (50) at (-1, 0) {};
		\node [style=none] (51) at (-2.25, -0.5) {1};
		\node [style=none] (52) at (-1, 0.075) {};
		\node [style=none] (53) at (-2.3, 0.075) {};
		\node [style=none] (54) at (-1, -0.075) {};
		\node [style=none] (55) at (-2.3, -0.075) {};
		\node [style=none] (56) at (-1.55, 0) {};
		\node [style=none] (57) at (-1.925, 0.375) {};
		\node [style=none] (58) at (-1.925, -0.375) {};
		\node [style=gauge3] (59) at (-2.25, 0) {};
	\end{pgfonlayer}
	\begin{pgfonlayer}{edgelayer}
		\draw (30.center) to (31.center);
		\draw (33.center) to (32.center);
		\draw (35.center) to (34.center);
		\draw (34.center) to (36.center);
		\draw (37) to (24);
		\draw (40.center) to (41.center);
		\draw (43.center) to (42.center);
		\draw (45.center) to (44.center);
		\draw (44.center) to (46.center);
		\draw (52.center) to (53.center);
		\draw (55.center) to (54.center);
		\draw (57.center) to (56.center);
		\draw (56.center) to (58.center);
	\end{pgfonlayer}
\end{tikzpicture}}}
\end{equation}
Note that changing the ungauging scheme from long to short for the $\ess{A_2}{2}{r}$, $\ess{A_1}{3}{r}$, $\tee{A_2}{2}{r}$ and $\tee{A_1}{3}{r}$ cases results in a simply laced quiver, where the non-simply laced edge of order $\ell$ turns into $\ell$ edges.\\

Our claim is that this operation provides the Higgs branch of the gauged version of the underlying 4d theory and that this moduli space is naturally interpreted as the moduli space of $r$ $G$ instantons on $\mathbb{C}^2/\mathbb{Z}_{\ell}$, with suitable boundary conditions, as required by the F-theory picture. See \cite{Cherkis:2008ip,Witten:2009xu,Mekareeya:2015bla} for the rank-preserving cases. We now provide some consistency checks of our claim. This at the same time confirms the validity of the F-theory realization and of our magnetic quivers. Notice that the idea of ungauging on the short node applies equally well to $\ess{G}{\ell}{r}$ theories, leading to the quivers in Table \ref{shortS}. Below we discuss these as well.

From the global symmetry of the theory (for example $F_4$ in the case $G=E_6$), we know that in general the holonomy at infinity for the gauge field involves an outer automorphism of the group $G$ leading to a complicated instanton moduli space about which little is known. This happens for all the models considered in this paper with the exception of two cases: $\ell=2$, $G=A_2$ and $\ell=3$, $G=A_1$ \footnote{This is suggested by the fact that the global symmetry of the theory in those cases includes the whole $G$. See also \cite[Sec.\ 2]{Giacomelli:2020gee} for a detailed discussion.}. We therefore restrict to the cases which do not involve any outer automorphisms, since for these we should be able to match against results available in the literature.  

\paragraph{The case $\ell=2$ and $G=A_2$.} 
The F-theory realization implies that the Higgs branch of these theories should be identified with the moduli space of $r$ $\mathrm{SU}(3)$ instantons on $\mathbb{C}^2/\mathbb{Z}_2$. Furthermore, the global symmetry of the $\mathcal{T}^{(r)}_{A_2,2}$ theories suggests that the holonomy at infinity is trivial. This is nicely confirmed by the magnetic quiver in Table \ref{resulttableT}: If we ungauge on the short node of the magnetic quiver of $\mathcal{T}^{(r)}_{A_2,2}$ theories (there is only one way to do it), it reduces to an ordinary unitary quiver and we can easily see that the resulting Coulomb branch is identical to that of the quiver
\be\label{shortA2} 
\raisebox{-.5\height}{
\begin{tikzpicture}
\tikzstyle{gauge} = [circle, draw,inner sep=3pt];
	\tikzstyle{flavour} = [regular polygon,regular polygon sides=4,inner 
sep=3pt, draw];
\node (C) [gauge,label=right:{$\scriptstyle{r}$}] at (1,.66) {};
\node (D) [gauge,label=right:{$\scriptstyle{r}$}] at (1,-.66) {};
\node (A) [gauge,label=below:{$\scriptstyle{r}$}] at (0,0) {};
\node (B) [flavour,left of=A,label=left:{$\scriptstyle{2}$}] {};
\draw (B)--(A)--(C)--(D)--(A);
\end{tikzpicture} 
}
\ee 
which is indeed well known to describe $r$ $\mathrm{SU}(3)$ instantons on $\mathbb{C}^2/\mathbb{Z}_2$ with $\mathrm{SU}(3)$-preserving holonomy \cite{Dey:2013fea,Mekareeya:2015bla}. 

\paragraph{The case $\ell=3$ and $G=A_1$.} From F-theory we expect to find in this case the moduli space of $r$ $\mathrm{SU}(2)$ instantons on $\mathbb{C}^2/\mathbb{Z}_3$ and again, by ungauging on the short node in the only possible way the magnetic quiver of the $\mathcal{T}^{(r)}_{A_1,3}$ theories, we find the Coulomb branch of the quiver (see Table \ref{shortR})
\be\label{shortA1}
\raisebox{-.5\height}{
\begin{tikzpicture}
\tikzstyle{gauge} = [circle, draw,inner sep=3pt];
	\tikzstyle{flavour} = [regular polygon,regular polygon sides=4,inner 
sep=3pt, draw];
\node (C) [gauge,label=below:{$\scriptstyle{r}$}] at (1.5,0) {};
\node (A) [gauge,label=below:{$\scriptstyle{r}$}] at (0,0) {};
\node (B) [flavour,left of=A,label=left:{$\scriptstyle{3}$}] {};
\draw (B)--(A); 
\path[every node/.style={font=\sffamily\small,
  		fill=white,inner sep=1pt}]
(A) edge [loop, out=45, in=135, looseness=1.2] (C);
\path[every node/.style={font=\sffamily\small,
  		fill=white,inner sep=1pt}]
(C) edge [loop, out=225, in=315, looseness=1.2] (A);
\end{tikzpicture} 
}
\ee
which again is perfectly consistent with the expectation from F-theory. We therefore see that the six-dimensional construction naturally leads us to the ungauged models, which come equipped with their ungauging prescription. Changing the ungauging prescription instead gives us the discretely gauged theories, whose moduli space reproduce more manifestly the properties of the underlying F-theory background.

\subsubsection*{The $\ess{G}{\ell}{r}$ and $\essdg{G}{\ell}{r}$ theories}

We can repeat the analysis for $\ess{G}{\ell}{r}$ theories and consider the moduli spaces we get by ungauging on the short node of the magnetic quivers in Table \ref{resulttable}, thus obtaining the so called the $\essdg{G}{\ell}{r}$ theories. The expectation from F-theory is that we again find moduli spaces of instantons but with a different choice of holonomy at infinity. 

\paragraph{The case $\ell=2$ and $G=A_2$.}
For $\ell=2$ and $G=A_2$ we find the following description of the Higgs branch
\begin{equation}
\label{instantonsell2A2}
    \mathcal{H}^{\mathrm{4d}} \left( \essdg{A_2}{2}{r} \right) = \mathcal{C}^{\mathrm{3d}} \left( \raisebox{-.5\height}{
\begin{tikzpicture}
\tikzstyle{gauge} = [circle, draw,inner sep=3pt];
	\tikzstyle{flavour} = [regular polygon,regular polygon sides=4,inner 
sep=3pt, draw];
\node (C) [gauge,label=right:{$\scriptstyle{r}$}] at (1,.66) {};
\node (D) [gauge,label=right:{$\scriptstyle{r}$}] at (1,-.66) {};
\node (A) [gauge,label=below:{$\scriptstyle{r{+}1}$}] at (0,0) {};
\node (B) [flavour,left of=A,label=left:{$\scriptstyle{2}$}] {};
\draw (B)--(A)--(C)--(D)--(A);
\end{tikzpicture} 
} \right) = \mathcal{H}^{\mathrm{class}} \left( \raisebox{-.5\height}{
\begin{tikzpicture}
\tikzstyle{gauge} = [circle, draw,inner sep=3pt];
	\tikzstyle{flavour} = [regular polygon,regular polygon sides=4,inner sep=3pt, draw];
\node (C) [gauge,label={[rotate=-45]below right:{$\scriptstyle{\mathrm{U}(r)}$}}] at (1.5,0) {};
\node (A) [gauge,label={[rotate=-45]below right:{$\scriptstyle{\mathrm{U}(r{+}1)}$}}] at (0,0) {};
\node (B) [flavour,left of=A,label=left:{$\scriptstyle{2}$}] {};
\node (D) [flavour,right of=C,label=right:{$\scriptstyle{1}$}] {};
\draw (B)--(A);
\draw (C)--(D);
\path[every node/.style={font=\sffamily\small,
  		fill=white,inner sep=1pt}]
(A) edge [loop, out=45, in=135, looseness=1.2] (C);
\path[every node/.style={font=\sffamily\small,
  		fill=white,inner sep=1pt}]
(C) edge [loop, out=225, in=315, looseness=1.2] (A);
\end{tikzpicture} 
} \right)  \, . 
\end{equation}
The second equality can be derived using 3d mirror symmetry. From these quivers we see that we are dealing with a moduli space of $\mathrm{SU}(3)$ instantons on $\mathbb{C}^2/\mathbb{Z}_2$, but this time with a nontrivial first Chern class. From the right hand side of \eqref{instantonsell2A2}, it appears that the holonomy at infinity preserves an $\mathrm{SU}(2)\times \mathrm{U}(1)$ subgroup of $\mathrm{SU}(3)$, in agreement with the global symmetry of the $\mathcal{S}^{(r)}_{A_2,2}$ theories. 

\paragraph{The case $\ell=3$ and $G=A_1$.}  
In the $\ell=3$ and $G=A_1$ case we find instead 
\begin{equation}
\label{instantonsell3A1}
    \mathcal{H}^{\mathrm{4d}} \left( \essdg{A_1}{3}{r} \right) = \mathcal{C}^{\mathrm{3d}} \left( \raisebox{-.5\height}{
\begin{tikzpicture}
\tikzstyle{gauge} = [circle, draw,inner sep=3pt];
	\tikzstyle{flavour} = [regular polygon,regular polygon sides=4,inner 
sep=3pt, draw];
\node (C) [gauge,label=below:{$\scriptstyle{r}$}] at (1.5,0) {};
\node (A) [gauge,label=below:{$\scriptstyle{r{+}1}$}] at (0,0) {};
\node (B) [flavour,left of=A,label=left:{$\scriptstyle{3}$}] {};
\draw (B)--(A); 
\path[every node/.style={font=\sffamily\small,
  		fill=white,inner sep=1pt}]
(A) edge [loop, out=45, in=135, looseness=1.2] (C);
\path[every node/.style={font=\sffamily\small,
  		fill=white,inner sep=1pt}]
(C) edge [loop, out=225, in=315, looseness=1.2] (A);
\end{tikzpicture} 
} \right) = \mathcal{H}^{\mathrm{class}} \left(\raisebox{-.5\height}{
\begin{tikzpicture}
\tikzstyle{gauge} = [circle, draw,inner sep=3pt];
	\tikzstyle{flavour} = [regular polygon,regular polygon sides=4,inner 
sep=3pt, draw];
\node (C) [gauge,label=above:{$\scriptstyle{\mathrm{U}(r{+}1)}$}] at (2,0) {};
\node (D) [gauge,label=below:{$\scriptstyle{\mathrm{U}(r)}$}] at (1,-1.5) {};
\node (A) [gauge,label=above:{$\scriptstyle{\mathrm{U}(r{+}1)}$}] at (0,0) {};
\node (B) [flavour,left of=A,label=left:{$\scriptstyle{1}$}] {};
\node (E) [flavour,right of=C,label=right:{$\scriptstyle{1}$}] {};
\draw (B)--(A)--(C)--(D)--(A);
\draw (C)--(E);
\end{tikzpicture} 
} \right)  \, . 
\end{equation}
Again we find agreement with the global symmetry of $\mathcal{S}^{(r)}_{A_1,3}$ theories: the holonomy breaks the $\mathrm{SU}(2)$ gauge group to $\mathrm{U}(1)$. 

Let us also notice that from the electric quivers in \eqref{instantonsell2A2} and \eqref{instantonsell3A1} we can easily determine the instanton number, i.e.\ the D3 charge in the F-theory setup, using the formula given in \cite{Dey:2013fea}: 
\be\label{charge}\text{D3 charge} = \frac{1}{\ell}\sum_in_i,\ee 
where the sum is over the gauge nodes of the electric quiver and $n_i$ denotes the corresponding rank. Using \eqref{charge} from \eqref{instantonsell2A2} we find $r+\frac{1}{2}$ and from \eqref{instantonsell3A1} we get $r+\frac{2}{3}$, which in both cases is equal to $r+\frac{\ell-1}{\ell}$ as it should be since we have $r$ D3 branes and the contribution from the discrete flux is known to be $\frac{\ell-1}{\ell}$ \cite{Apruzzi:2020pmv}.

\paragraph{Other cases.}
Armed with this intuition that ungauging on the short node of the magnetic quivers in Table \ref{resulttable} and in Table \ref{resulttableT} provides us with the moduli spaces of instantons on $\mathbb{C}^2/\mathbb{Z}_{\ell}$, we can propose the Hilbert series of these spaces, even for the cases involving an outer automorphism twist. We present our findings for $r=1,2$ in Appendix \ref{shortHS}.
 
The connection between the quivers in Table \ref{shortS} and Table \ref{shortR} with the moduli space of instantons, as required by the F-theory picture with D3 branes probing 7-branes, strongly suggests a four dimensional interpretation of the ungauging on the short node, namely the existence of the four-dimensional theories $\essdg{G}{\ell}{r}$ and $\teedg{G}{\ell}{r}$ as discretely gauged versions of $\ess{G}{\ell}{r}$ and $\tee{G}{\ell}{r}$ respectively. As we have seen, in the case of $\tee{G}{\ell}{r}$ models also the construction from six-dimensions predicts the existence of a discrete $\mathbb{Z}_{\ell}$ symmetry acting on the Coulomb branch of the theory. In the case of $\ess{G}{\ell}{r}$ models the presence of a $\mathbb{Z}_{\ell}$ symmetry is less obvious, since it does not act on the Coulomb branch and therefore does not arise in the same way from the six-dimensional construction. Given the uniform treatment at the level of magnetic quivers we expect this symmetry to exist as a global symmetry and hence another theory where this symmetry is gauged. It would be important to understand better how this symmetry acts on the theory.

\begin{table}[t]
\centering
\begin{adjustbox}{center}
	\begin{tabular}{cccc}
\toprule
		SCFT & Magnetic quiver & Global Symmetry   &  Dimension  
 \\
\midrule
       $\essdg{E_6}{2}{r}$ &		
    \begin{tabular}{c}
    \scalebox{0.70}{\begin{tikzpicture}
	\begin{pgfonlayer}{nodelayer}
		\node [style=gauge3] (2) at (-1, 0) {};
		\node [style=none] (6) at (-1, -0.5) {$1{+}3r$};
		\node [style=gauge3] (12) at (-2, 0) {};
		\node [style=none] (15) at (-2, -0.5) {$1{+}2r$};
		\node [style=gauge3] (24) at (0.2, 0) {};
		\node [style=none] (29) at (0.2, -0.5) {$1{+}4r$};
		\node [style=none] (30) at (-0.85, 0.075) {};
		\node [style=none] (31) at (0.2, 0.075) {};
		\node [style=none] (32) at (-0.85, -0.075) {};
		\node [style=none] (33) at (0.2, -0.075) {};
		\node [style=none] (34) at (-0.55, 0) {};
		\node [style=none] (35) at (-0.175, 0.375) {};
		\node [style=none] (36) at (-0.175, -0.375) {};
		\node [style=gauge3] (37) at (0.95, 0) {};
		\node [style=none] (38) at (0.95, -0.5) {$2r$};
		\node [style=gauge3] (39) at (-3, 0) {};
		\node [style=none] (42) at (-3, -0.5) {$1{+}r$};
		\node [style=none] (43) at (-4.25, -0.5) {1};
		\node [style=none] (44) at (-3, 0.075) {};
		\node [style=none] (45) at (-4.3, 0.075) {};
		\node [style=none] (46) at (-3, -0.075) {};
		\node [style=none] (47) at (-4.3, -0.075) {};
		\node [style=none] (48) at (-3.55, 0) {};
		\node [style=none] (49) at (-3.925, 0.375) {};
		\node [style=none] (50) at (-3.925, -0.375) {};
		\node [style=gauge3] (51) at (-4.25, 0) {};
	\end{pgfonlayer}
	\begin{pgfonlayer}{edgelayer}
		\draw (30.center) to (31.center);
		\draw (33.center) to (32.center);
		\draw (35.center) to (34.center);
		\draw (34.center) to (36.center);
		\draw (37) to (24);
		\draw (12) to (2);
		\draw (39) to (12);
		\draw (44.center) to (45.center);
		\draw (47.center) to (46.center);
		\draw (49.center) to (48.center);
		\draw (48.center) to (50.center);
	\end{pgfonlayer}
\end{tikzpicture}}
    \end{tabular} & $C_4 A_1$ & $12r+4$	\\ 
          $\essdg{D_4}{2}{r}$ &		
    \begin{tabular}{c}
    \scalebox{0.70}{\begin{tikzpicture}
	\begin{pgfonlayer}{nodelayer}
		\node [style=gauge3] (2) at (-1, 0) {};
		\node [style=none] (6) at (-1, -0.5) {$1{+}r$};
		\node [style=gauge3] (24) at (0.25, 0) {};
		\node [style=none] (29) at (0.25, -0.5) {$1{+}2r$};
		\node [style=none] (30) at (-1, 0.075) {};
		\node [style=none] (31) at (0.25, 0.075) {};
		\node [style=none] (32) at (-1, -0.075) {};
		\node [style=none] (33) at (0.25, -0.075) {};
		\node [style=none] (34) at (-0.5, 0) {};
		\node [style=none] (35) at (-0.125, 0.375) {};
		\node [style=none] (36) at (-0.125, -0.375) {};
		\node [style=gauge3] (37) at (1, 0) {};
		\node [style=none] (38) at (1, -0.5) {$2r$};
		\node [style=gauge3] (39) at (2.25, 0) {};
		\node [style=none] (40) at (1, 0.075) {};
		\node [style=none] (41) at (2.25, 0.075) {};
		\node [style=none] (42) at (1, -0.075) {};
		\node [style=none] (43) at (2.25, -0.075) {};
		\node [style=none] (44) at (1.75, 0) {};
		\node [style=none] (45) at (1.375, 0.375) {};
		\node [style=none] (46) at (1.375, -0.375) {};
		\node [style=none] (49) at (2.25, -0.5) {$r$};
		\node [style=none] (50) at (-2.2, -0.5) {1};
		\node [style=none] (51) at (-0.95, 0.075) {};
		\node [style=none] (52) at (-2.25, 0.075) {};
		\node [style=none] (53) at (-0.95, -0.075) {};
		\node [style=none] (54) at (-2.25, -0.075) {};
		\node [style=none] (55) at (-1.5, 0) {};
		\node [style=none] (56) at (-1.875, 0.375) {};
		\node [style=none] (57) at (-1.875, -0.375) {};
		\node [style=gauge3] (58) at (-2.2, 0) {};
	\end{pgfonlayer}
	\begin{pgfonlayer}{edgelayer}
		\draw (30.center) to (31.center);
		\draw (33.center) to (32.center);
		\draw (35.center) to (34.center);
		\draw (34.center) to (36.center);
		\draw (37) to (24);
		\draw (40.center) to (41.center);
		\draw (43.center) to (42.center);
		\draw (45.center) to (44.center);
		\draw (44.center) to (46.center);
		\draw (51.center) to (52.center);
		\draw (54.center) to (53.center);
		\draw (56.center) to (55.center);
		\draw (55.center) to (57.center);
	\end{pgfonlayer}
\end{tikzpicture}
}
    \end{tabular} & $C_2 A_1 A_1$ &  $6r+2$	\\ 
      $\essdg{A_2}{2}{r}$ &		
    \begin{tabular}{c}
    \scalebox{0.70}{\begin{tikzpicture}
	\begin{pgfonlayer}{nodelayer}
		\node [style=gauge3] (24) at (0.25, 0) {};
		\node [style=none] (29) at (0.25, -0.5) {$1{+}r$};
		\node [style=none] (33) at (0.25, 0) {};
		\node [style=gauge3] (37) at (1.25, 0.5) {};
		\node [style=gauge3] (38) at (1.25, -0.5) {};
		\node [style=none] (39) at (1.75, 0.5) {$r$};
		\node [style=none] (40) at (1.75, -0.5) {$r$};
		\node [style=none] (41) at (-0.95, -0.5) {1};
		\node [style=none] (42) at (0.3, 0.075) {};
		\node [style=none] (43) at (-1, 0.075) {};
		\node [style=none] (44) at (0.3, -0.075) {};
		\node [style=none] (45) at (-1, -0.075) {};
		\node [style=gauge3] (49) at (-0.95, 0) {};
	\end{pgfonlayer}
	\begin{pgfonlayer}{edgelayer}
		\draw (37) to (33.center);
		\draw (37) to (38);
		\draw (33.center) to (38);
		\draw (42.center) to (43.center);
		\draw (45.center) to (44.center);
	\end{pgfonlayer}
\end{tikzpicture}}
    \end{tabular} & $C_1 A_1 U_1$  & 	$3r+1$ \\ 
    $\essdg{D_4}{3}{r}$ &		
    \begin{tabular}{c}
    \scalebox{0.70}{
\begin{tikzpicture}
	\begin{pgfonlayer}{nodelayer}
		\node [style=gauge3] (2) at (-1, 0) {};
		\node [style=none] (6) at (-1, -0.5) {$1{+}2r$};
		\node [style=gauge3] (24) at (0.5, 0) {};
		\node [style=none] (29) at (0.5, -0.5) {$1{+}3r$};
		\node [style=none] (30) at (-1, 0.15) {};
		\node [style=none] (31) at (0.5, 0.15) {};
		\node [style=none] (32) at (-1, -0.15) {};
		\node [style=none] (33) at (0.5, -0.15) {};
		\node [style=none] (34) at (-0.5, 0) {};
		\node [style=none] (35) at (0, 0.5) {};
		\node [style=none] (36) at (0, -0.5) {};
		\node [style=gauge3] (37) at (-2, 0) {};
		\node [style=none] (39) at (-2, -0.5) {$1{+}r$};
		\node [style=none] (41) at (-3.2, -0.5) {1};
		\node [style=none] (42) at (-1.95, 0.175) {};
		\node [style=none] (43) at (-3.25, 0.175) {};
		\node [style=none] (44) at (-1.95, -0.175) {};
		\node [style=none] (45) at (-3.25, -0.175) {};
		\node [style=none] (46) at (-2.5, 0) {};
		\node [style=none] (47) at (-2.875, 0.375) {};
		\node [style=none] (48) at (-2.875, -0.375) {};
		\node [style=gauge3] (49) at (-3.2, 0) {};
		\node [style=none] (50) at (-2, 0) {};
	\end{pgfonlayer}
	\begin{pgfonlayer}{edgelayer}
		\draw (2) to (24);
		\draw (30.center) to (31.center);
		\draw (33.center) to (32.center);
		\draw (35.center) to (34.center);
		\draw (34.center) to (36.center);
		\draw (37) to (2);
		\draw (42.center) to (43.center);
		\draw (45.center) to (44.center);
		\draw (47.center) to (46.center);
		\draw (46.center) to (48.center);
		\draw (49) to (50.center);
	\end{pgfonlayer}
\end{tikzpicture}
}
    \end{tabular} & $A_2 U_1$ & $6r+3$	\\ 
      $\essdg{A_1}{3}{r}$ &		
    \begin{tabular}{c}
    \scalebox{0.70}{\begin{tikzpicture}
	\begin{pgfonlayer}{nodelayer}
		\node [style=gauge3] (24) at (0.5, 0) {};
		\node [style=none] (29) at (0.5, -0.5) {$1{+}r$};
		\node [style=none] (31) at (0.5, 0.15) {};
		\node [style=none] (33) at (0.5, -0.15) {};
		\node [style=gauge3] (37) at (1.5, 0) {};
		\node [style=none] (38) at (0.5, 0.075) {};
		\node [style=none] (39) at (0.5, -0.1) {};
		\node [style=none] (40) at (1.5, 0.075) {};
		\node [style=none] (41) at (1.5, -0.1) {};
		\node [style=none] (42) at (1.5, -0.5) {$r$};
		\node [style=none] (43) at (-0.7, -0.5) {1};
		\node [style=none] (44) at (0.55, 0.175) {};
		\node [style=none] (45) at (-0.75, 0.175) {};
		\node [style=none] (46) at (0.55, -0.175) {};
		\node [style=none] (47) at (-0.75, -0.175) {};
		\node [style=gauge3] (51) at (-0.7, 0) {};
		\node [style=none] (52) at (0.5, 0) {};
	\end{pgfonlayer}
	\begin{pgfonlayer}{edgelayer}
		\draw (40.center) to (38.center);
		\draw (39.center) to (41.center);
		\draw (44.center) to (45.center);
		\draw (47.center) to (46.center);
		\draw (51) to (52.center);
	\end{pgfonlayer}
\end{tikzpicture}}
    \end{tabular} & $U_1U_1$ & $2r+1$\\ 
        $\essdg{A_2}{4}{r}$ &		
    \begin{tabular}{c}
    \scalebox{0.70}{
\begin{tikzpicture}
	\begin{pgfonlayer}{nodelayer}
		\node [style=gauge3] (2) at (-1, 0) {};
		\node [style=none] (6) at (-1, -0.5) {$1{+}r$};
		\node [style=gauge3] (24) at (0.5, 0) {};
		\node [style=none] (29) at (0.5, -0.5) {$1{+}2r$};
		\node [style=none] (30) at (-1, 0.05) {};
		\node [style=none] (31) at (0.5, 0.05) {};
		\node [style=none] (32) at (-1, -0.075) {};
		\node [style=none] (33) at (0.5, -0.075) {};
		\node [style=none] (34) at (-0.5, 0) {};
		\node [style=none] (35) at (0, 0.5) {};
		\node [style=none] (36) at (0, -0.5) {};
		\node [style=none] (37) at (-1, 0.15) {};
		\node [style=none] (38) at (0.5, 0.15) {};
		\node [style=none] (39) at (-1, -0.175) {};
		\node [style=none] (40) at (0.5, -0.175) {};
		\node [style=gauge3] (42) at (-2.5, 0) {};
		\node [style=none] (43) at (-2.5, -0.5) {$1$};
		\node [style=none] (44) at (-1, 0.05) {};
		\node [style=none] (45) at (-2.5, 0.05) {};
		\node [style=none] (46) at (-1, -0.075) {};
		\node [style=none] (47) at (-2.5, -0.075) {};
		\node [style=none] (48) at (-1.5, 0) {};
		\node [style=none] (49) at (-2, 0.5) {};
		\node [style=none] (50) at (-2, -0.5) {};
		\node [style=none] (51) at (-1, 0.15) {};
		\node [style=none] (52) at (-2.5, 0.15) {};
		\node [style=none] (53) at (-1, -0.175) {};
		\node [style=none] (54) at (-2.5, -0.175) {};
	\end{pgfonlayer}
	\begin{pgfonlayer}{edgelayer}
		\draw (30.center) to (31.center);
		\draw (33.center) to (32.center);
		\draw (35.center) to (34.center);
		\draw (34.center) to (36.center);
		\draw (37.center) to (38.center);
		\draw (39.center) to (40.center);
		\draw (44.center) to (45.center);
		\draw (47.center) to (46.center);
		\draw (49.center) to (48.center);
		\draw (48.center) to (50.center);
		\draw (51.center) to (52.center);
		\draw (53.center) to (54.center);
	\end{pgfonlayer}
\end{tikzpicture}}
    \end{tabular} & $A_1 U_1$ & $3r+2$ 	\\ 
    \bottomrule
	\end{tabular}
\end{adjustbox}
\caption{The $\essdg{G}{\ell}{r}$ theories with their magnetic quivers. The global symmetry is independent of $r$. The dimension of the Higgs branch of $\essdg{G}{\ell}{r}$ is equal to the dimension of the Higgs branch of $\ess{G}{\ell}{r}$. Recall that a $\mathrm{U}(1)$ is ungauged on a long node of all the quivers. }
\label{shortS}
\end{table}

\begin{table}[t]
\centering
\begin{adjustbox}{center}
	\begin{tabular}{cccc}
\toprule
		SCFT & Magnetic quiver & Global Symmetry  &  Dimension \\  
\midrule
       $\teedg{E_6}{2}{r}$ &		
    \begin{tabular}{c}
    \scalebox{0.70}{\begin{tikzpicture}
	\begin{pgfonlayer}{nodelayer}
		\node [style=gauge3] (2) at (-1, 0) {};
		\node [style=none] (6) at (-1, -0.5) {$3r$};
		\node [style=gauge3] (12) at (-2, 0) {};
		\node [style=none] (15) at (-2, -0.5) {$2r$};
		\node [style=gauge3] (24) at (0.2, 0) {};
		\node [style=none] (29) at (0.2, -0.5) {$4r$};
		\node [style=none] (30) at (-0.85, 0.075) {};
		\node [style=none] (31) at (0.2, 0.075) {};
		\node [style=none] (32) at (-0.85, -0.075) {};
		\node [style=none] (33) at (0.2, -0.075) {};
		\node [style=none] (34) at (-0.55, 0) {};
		\node [style=none] (35) at (-0.175, 0.375) {};
		\node [style=none] (36) at (-0.175, -0.375) {};
		\node [style=gauge3] (37) at (0.95, 0) {};
		\node [style=none] (38) at (0.95, -0.5) {$2r$};
		\node [style=gauge3] (39) at (-3, 0) {};
		\node [style=none] (41) at (-4.25, -0.5) {1};
		\node [style=none] (42) at (-3, -0.5) {$r$};
		\node [style=none] (44) at (-3, 0.075) {};
		\node [style=none] (45) at (-4.3, 0.075) {};
		\node [style=none] (46) at (-3, -0.075) {};
		\node [style=none] (47) at (-4.3, -0.075) {};
		\node [style=none] (48) at (-3.55, 0) {};
		\node [style=none] (49) at (-3.925, 0.375) {};
		\node [style=none] (50) at (-3.925, -0.375) {};
		\node [style=gauge3] (51) at (-4.25, 0) {};
	\end{pgfonlayer}
	\begin{pgfonlayer}{edgelayer}
		\draw (30.center) to (31.center);
		\draw (33.center) to (32.center);
		\draw (35.center) to (34.center);
		\draw (34.center) to (36.center);
		\draw (37) to (24);
		\draw (12) to (2);
		\draw (39) to (12);
		\draw (44.center) to (45.center);
		\draw (47.center) to (46.center);
		\draw (49.center) to (48.center);
		\draw (48.center) to (50.center);
	\end{pgfonlayer}
\end{tikzpicture}}
    \end{tabular} & $F_4 A_1$ & $12r$\\ 
          $\teedg{D_4}{2}{r}$ &		
    \begin{tabular}{c}
    \scalebox{0.70}{\begin{tikzpicture}
	\begin{pgfonlayer}{nodelayer}
		\node [style=gauge3] (2) at (-1, 0) {};
		\node [style=none] (6) at (-1, -0.5) {$r$};
		\node [style=gauge3] (24) at (0.25, 0) {};
		\node [style=none] (29) at (0.25, -0.5) {$2r$};
		\node [style=none] (30) at (-1, 0.075) {};
		\node [style=none] (31) at (0.25, 0.075) {};
		\node [style=none] (32) at (-1, -0.075) {};
		\node [style=none] (33) at (0.25, -0.075) {};
		\node [style=none] (34) at (-0.5, 0) {};
		\node [style=none] (35) at (-0.125, 0.375) {};
		\node [style=none] (36) at (-0.125, -0.375) {};
		\node [style=gauge3] (37) at (1, 0) {};
		\node [style=none] (38) at (1, -0.5) {$2r$};
		\node [style=gauge3] (39) at (2.25, 0) {};
		\node [style=none] (40) at (1, 0.075) {};
		\node [style=none] (41) at (2.25, 0.075) {};
		\node [style=none] (42) at (1, -0.075) {};
		\node [style=none] (43) at (2.25, -0.075) {};
		\node [style=none] (44) at (1.75, 0) {};
		\node [style=none] (45) at (1.375, 0.375) {};
		\node [style=none] (46) at (1.375, -0.375) {};
		\node [style=none] (49) at (2.25, -0.5) {$r$};
		\node [style=gauge3] (50) at (-1, 0) {};
		\node [style=none] (51) at (-2.25, -0.5) {1};
		\node [style=none] (52) at (-1, 0.075) {};
		\node [style=none] (53) at (-2.3, 0.075) {};
		\node [style=none] (54) at (-1, -0.075) {};
		\node [style=none] (55) at (-2.3, -0.075) {};
		\node [style=none] (56) at (-1.55, 0) {};
		\node [style=none] (57) at (-1.925, 0.375) {};
		\node [style=none] (58) at (-1.925, -0.375) {};
		\node [style=gauge3] (59) at (-2.25, 0) {};
	\end{pgfonlayer}
	\begin{pgfonlayer}{edgelayer}
		\draw (30.center) to (31.center);
		\draw (33.center) to (32.center);
		\draw (35.center) to (34.center);
		\draw (34.center) to (36.center);
		\draw (37) to (24);
		\draw (40.center) to (41.center);
		\draw (43.center) to (42.center);
		\draw (45.center) to (44.center);
		\draw (44.center) to (46.center);
		\draw (52.center) to (53.center);
		\draw (55.center) to (54.center);
		\draw (57.center) to (56.center);
		\draw (56.center) to (58.center);
	\end{pgfonlayer}
\end{tikzpicture}
}
    \end{tabular} & $B_3 A_1$ &  $6r$	\\ 
      $\teedg{A_2}{2}{r}$ &		
    \begin{tabular}{c}
    \scalebox{0.70}{\begin{tikzpicture}
	\begin{pgfonlayer}{nodelayer}
		\node [style=gauge3] (24) at (0.25, 0) {};
		\node [style=none] (29) at (0.25, -0.5) {$r$};
		\node [style=none] (31) at (0.25, 0.075) {};
		\node [style=none] (33) at (0.25, -0.075) {};
		\node [style=gauge3] (37) at (1.25, 0.5) {};
		\node [style=gauge3] (38) at (1.25, -0.5) {};
		\node [style=none] (39) at (1.75, 0.5) {$r$};
		\node [style=none] (40) at (1.75, -0.5) {$r$};
		\node [style=none] (43) at (-0.95, -0.5) {1};
		\node [style=none] (44) at (0.3, 0.075) {};
		\node [style=none] (45) at (-1, 0.075) {};
		\node [style=none] (46) at (0.3, -0.075) {};
		\node [style=none] (47) at (-1, -0.075) {};
		\node [style=gauge3] (51) at (-0.95, 0) {};
	\end{pgfonlayer}
	\begin{pgfonlayer}{edgelayer}
		\draw (37) to (33.center);
		\draw (38) to (31.center);
		\draw (37) to (38);
		\draw (44.center) to (45.center);
		\draw (47.center) to (46.center);
	\end{pgfonlayer}
\end{tikzpicture}}
    \end{tabular} & $A_2 A_1$ &  $3r$ \\ 
    $\teedg{D_4}{3}{r}$ &		
    \begin{tabular}{c}
    \scalebox{0.70}{
\begin{tikzpicture}
	\begin{pgfonlayer}{nodelayer}
		\node [style=gauge3] (2) at (-1, 0) {};
		\node [style=none] (6) at (-1, -0.5) {$2r$};
		\node [style=gauge3] (24) at (0.5, 0) {};
		\node [style=none] (29) at (0.5, -0.5) {$3r$};
		\node [style=none] (30) at (-1, 0.15) {};
		\node [style=none] (31) at (0.5, 0.15) {};
		\node [style=none] (32) at (-1, -0.15) {};
		\node [style=none] (33) at (0.5, -0.15) {};
		\node [style=none] (34) at (-0.5, 0) {};
		\node [style=none] (35) at (0, 0.5) {};
		\node [style=none] (36) at (0, -0.5) {};
		\node [style=gauge3] (37) at (-2, 0) {};
		\node [style=none] (39) at (-2, -0.5) {$r$};
		\node [style=none] (42) at (-1.95, 0.075) {};
		\node [style=none] (44) at (-1.95, -0.075) {};
		\node [style=none] (50) at (-3.2, -0.5) {1};
		\node [style=none] (51) at (-1.95, 0.175) {};
		\node [style=none] (52) at (-3.25, 0.175) {};
		\node [style=none] (53) at (-1.95, -0.175) {};
		\node [style=none] (54) at (-3.25, -0.175) {};
		\node [style=none] (55) at (-2.5, 0) {};
		\node [style=none] (56) at (-2.875, 0.375) {};
		\node [style=none] (57) at (-2.875, -0.375) {};
		\node [style=gauge3] (58) at (-3.2, 0) {};
		\node [style=none] (59) at (-2, 0) {};
	\end{pgfonlayer}
	\begin{pgfonlayer}{edgelayer}
		\draw (2) to (24);
		\draw (30.center) to (31.center);
		\draw (33.center) to (32.center);
		\draw (35.center) to (34.center);
		\draw (34.center) to (36.center);
		\draw (37) to (2);
		\draw (51.center) to (52.center);
		\draw (54.center) to (53.center);
		\draw (56.center) to (55.center);
		\draw (55.center) to (57.center);
		\draw (58) to (59.center);
	\end{pgfonlayer}
\end{tikzpicture}
}
    \end{tabular} & $G_2 U_1$  & $6r$	\\ 
      $\teedg{A_1}{3}{r}$ &		
    \begin{tabular}{c}
    \scalebox{0.70}{\begin{tikzpicture}
	\begin{pgfonlayer}{nodelayer}
		\node [style=gauge3] (24) at (0.5, 0) {};
		\node [style=none] (29) at (0.5, -0.5) {$r$};
		\node [style=none] (31) at (0.5, 0.15) {};
		\node [style=none] (33) at (0.5, -0.15) {};
		\node [style=gauge3] (37) at (1.5, 0) {};
		\node [style=none] (38) at (0.5, 0.075) {};
		\node [style=none] (39) at (0.5, -0.1) {};
		\node [style=none] (40) at (1.5, 0.075) {};
		\node [style=none] (41) at (1.5, -0.1) {};
		\node [style=none] (42) at (1.5, -0.5) {$r$};
		\node [style=none] (45) at (-0.7, -0.5) {1};
		\node [style=none] (46) at (0.55, 0.175) {};
		\node [style=none] (47) at (-0.75, 0.175) {};
		\node [style=none] (48) at (0.55, -0.175) {};
		\node [style=none] (49) at (-0.75, -0.175) {};
		\node [style=gauge3] (53) at (-0.7, 0) {};
		\node [style=none] (54) at (0.5, 0) {};
	\end{pgfonlayer}
	\begin{pgfonlayer}{edgelayer}
		\draw (40.center) to (38.center);
		\draw (39.center) to (41.center);
		\draw (46.center) to (47.center);
		\draw (49.center) to (48.center);
		\draw (53) to (54.center);
	\end{pgfonlayer}
\end{tikzpicture}}
    \end{tabular} & $A_1 U_1$  & $2r$\\ 
        $\teedg{A_2}{4}{r}$ &		
    \begin{tabular}{c}
    \scalebox{0.70}{
\begin{tikzpicture}
	\begin{pgfonlayer}{nodelayer}
		\node [style=gauge3] (2) at (-1, 0) {};
		\node [style=gauge3] (24) at (0.5, 0) {};
		\node [style=none] (29) at (0.5, -0.5) {$2r$};
		\node [style=none] (30) at (-1, 0.05) {};
		\node [style=none] (31) at (0.5, 0.05) {};
		\node [style=none] (32) at (-1, -0.075) {};
		\node [style=none] (33) at (0.5, -0.075) {};
		\node [style=none] (34) at (-0.5, 0) {};
		\node [style=none] (35) at (0, 0.5) {};
		\node [style=none] (36) at (0, -0.5) {};
		\node [style=none] (37) at (-1, 0.15) {};
		\node [style=none] (38) at (0.5, 0.15) {};
		\node [style=none] (39) at (-1, -0.175) {};
		\node [style=none] (40) at (0.5, -0.175) {};
		\node [style=none] (43) at (-1, -0.5) {$r$};
		\node [style=gauge3] (44) at (-2.5, 0) {};
		\node [style=none] (45) at (-2.5, -0.5) {$1$};
		\node [style=none] (46) at (-1, 0.05) {};
		\node [style=none] (47) at (-2.5, 0.05) {};
		\node [style=none] (48) at (-1, -0.075) {};
		\node [style=none] (49) at (-2.5, -0.075) {};
		\node [style=none] (50) at (-1.5, 0) {};
		\node [style=none] (51) at (-2, 0.5) {};
		\node [style=none] (52) at (-2, -0.5) {};
		\node [style=none] (53) at (-1, 0.15) {};
		\node [style=none] (54) at (-2.5, 0.15) {};
		\node [style=none] (55) at (-1, -0.175) {};
		\node [style=none] (56) at (-2.5, -0.175) {};
	\end{pgfonlayer}
	\begin{pgfonlayer}{edgelayer}
		\draw (30.center) to (31.center);
		\draw (33.center) to (32.center);
		\draw (35.center) to (34.center);
		\draw (34.center) to (36.center);
		\draw (37.center) to (38.center);
		\draw (39.center) to (40.center);
		\draw (46.center) to (47.center);
		\draw (49.center) to (48.center);
		\draw (51.center) to (50.center);
		\draw (50.center) to (52.center);
		\draw (53.center) to (54.center);
		\draw (55.center) to (56.center);
	\end{pgfonlayer}
\end{tikzpicture}}
    \end{tabular} & $A_1 U_1$ & $3r$ \\ 
    \bottomrule
	\end{tabular}
\end{adjustbox}
\caption{The $\teedg{G}{\ell}{r}$ theories with their magnetic quivers. The global symmetry is independent of $r$. The dimension of the Higgs branch is the same as for the corresponding $\tee{G}{\ell}{r}$ theory. Recall that a $\mathrm{U}(1)$ is ungauged on a long node of all the quivers. }
\label{shortR}
\end{table}

\section{6d magnetic quivers, FI deformations and folding}
\label{SectionFI}

In this section we derive the magnetic quivers from 5d. The idea is to start from the realization of $\mathcal{S}$-fold theories via compactification of 6d theories on $T^2$ with almost commuting holonomies. We can first compactify on a circle down to 5d, incorporating the corresponding holonomy with a mass deformation and then we perform a $\mathbb{Z}_{\ell}$ folding of the 5d magnetic quiver to extract the $\mathcal{S}$-fold magnetic quiver.

\subsection{Magnetic quivers for 6d theories}
\label{sec:6d_magQuiv}
To begin with, recall the construction of magnetic quivers for 6d $\mathcal{N}=(1,0)$ theories of \cite{Cabrera:2019izd}. The relevant 6d theories can be constructed via $r$ M5 branes transverse to a ${\mathbb{C}^2\slash\mathbb{Z}_{\ell}}$ singularity near an M9 plane. In the dual Type IIA set-up, the choice of boundary conditions of the $\ell$ D6 branes on the $8$ D8 branes that originate from the M9 plane break the exceptional $E_8$ symmetry. As detailed in \cite[Sec.\ 3.1]{Cabrera:2019izd}, the boundary conditions or, equivalently, the different embeddings $\mathbb{Z}_{\ell}\hookrightarrow E_8$ are conveniently encoded in Kac labels \cite{Kac:1994}, see Figure \ref{fig:Kac}. The Kac labels for the 6d origins of the  $\ess{G}{\ell}{r}$ and  $\tee{G}{\ell}{r}$ theories are summarized in Table \ref{fig:cases}. 
\begin{figure}[t]
\centering
    \begin{subfigure}{0.65\textwidth}
        \raisebox{-.5\height}{
 	\begin{tikzpicture}
	\tikzstyle{gauge} = [circle, draw,inner sep=3pt];
	\tikzstyle{flavour} = [regular polygon,regular polygon sides=4,inner sep=3pt, draw];
	\node (g1) [gauge, label=below:{$\scriptstyle{m^{\phantom{\prime}}_1}$}] {$\scriptscriptstyle{1}$};
	\node (g2) [gauge,right of=g1, label=below:{$\scriptstyle{m^{\phantom{\prime}}_2}$}] {$\scriptscriptstyle{2}$};
	\node (g3) [gauge,right of=g2, label=below:{$\scriptstyle{m^{\phantom{\prime}}_3}$}] {$\scriptscriptstyle{3}$};
	\node (g4) [gauge,right of=g3, label=below:{$\scriptstyle{m^{\phantom{\prime}}_4}$}] {$\scriptscriptstyle{4}$};
	\node (g5) [gauge,right of=g4, label=below:{$\scriptstyle{m^{\phantom{\prime}}_5}$}] {$\scriptscriptstyle{5}$};
	\node (g6) [gauge,right of=g5, label=below:{$\scriptstyle{m^{\phantom{\prime}}_6}$}] {$\scriptscriptstyle{6}$};
	\node (g7) [gauge,right of=g6, label=below:{$\scriptstyle{m^\prime_4}$}] {$\scriptscriptstyle{4}$};
	\node (g8) [gauge,right of=g7, label=below:{$\scriptstyle{m^\prime_2}$}] {$\scriptscriptstyle{2}$};
	\node (g9) [gauge,above of=g6, label=above:{$\scriptstyle{m^\prime_3}$}] {$\scriptscriptstyle{3}$};
	\draw (g1)--(g2) (g2)--(g3) (g3)--(g4) (g4)--(g5) (g5)--(g6) (g6)--(g7) 
(g7)--(g8) (g6)--(g9);
	\end{tikzpicture}
	} 
	\caption{}
	\label{fig:Kac}
    \end{subfigure}
    \begin{subfigure}{0.25\textwidth}
\begin{tabular}{c|cc}
\toprule
     $\ell$ & $\ess{G}{\ell}{r}$  & $\tee{G}{\ell}{r}$ \\ \midrule
     $2$ & $m^\prime_2=1$  &  $m_2=1$  \\
     $3$ & $m^\prime_3=1$ & $m_3=1$ \\
     $4$ & $m^\prime_4=1$ & $m_4=1$ \\ 
    \bottomrule
\end{tabular}
\caption{}
\label{fig:cases}
    \end{subfigure}
    \caption{(\subref{fig:Kac}) The Kac labels  $\{m_i,m^\prime_j\}$, which are non-negative integers, need to satisfy $\sum_{i=1}^6 a_i m_i + \sum_{j=2}^4 a^\prime_j m^\prime_j = \ell$ to define an embedding $\mathbb{Z}_\ell \hookrightarrow E_8$. The $\{a_i,a^\prime_j\}$ are the Dynkin labels of affine $E_8$, which are the numbers displayed inside the nodes. (\subref{fig:cases}) The relevant embeddings for the $\mathcal{S}$-fold theories considered here.}
    \label{fig:Kac_and_cases}
\end{figure}
\paragraph{6d origin of $\ess{G}{\ell}{r}$ theories.}
For the intents and purposes of this paper, there are three cases to consider for the $\ess{G}{\ell}{r}$ theories \cite{Giacomelli:2020jel}:
\begin{compactitem}
\item $\ell=2$: The boundary conditions preserve $\mathrm{SO}(16) \subset E_8$, see \cite[Sec.\ 3.3.3]{Cabrera:2019izd} as well as \cite[Sec.\ 5.1]{Mekareeya:2017jgc}. The 6d theory on the tensor branch is described by
\begin{align}
 \raisebox{-.5\height}{
 	\begin{tikzpicture}
	\tikzstyle{gauge} = [circle, draw,inner sep=3pt];
	\tikzstyle{flavour} = [regular polygon,regular polygon sides=4,inner 
sep=3pt, draw];
	\node (g1) [gauge,label={[rotate=-45]below right:{$\scriptstyle{\mathrm{SU}(2)}$}}] {};
	\node (g2) [right of=g1] {$\ldots$};
	\node (g3) [gauge,right of=g2,label={[rotate=-45]below right:{$\scriptstyle{\mathrm{SU}(2)}$}}] {};
	\node (g4) [gauge,right of=g3,label={[rotate=-45]below right:{$\scriptstyle{\mathrm{SU}(2)}$}}] {};
	\node (g5) [gauge,right of=g4,label={[rotate=-45]below right:{$\scriptstyle{\mathrm{USp}(2)}$}}] 
{};
	\node (f1) [flavour,left of=g1,label=below:{$\scriptstyle{\mathrm{SU}(2)}$}] {};
	\node (f5) [flavour,above of=g5,label=above:{$\scriptstyle{\mathrm{SO}(16)}$}] {};
	\draw (g1)--(g2) (g2)--(g3) (g3)--(g4) (g1)--(f1) (g4)--(g5) (g5)--(f5);
  \draw[decoration={brace,mirror,raise=20pt},decorate,thick]
  (-0.25,-0.6) -- node[below=20pt] {$r-1$} (3.75,-0.6);
	\end{tikzpicture}
	}  \quad  + \quad \text{$r$ tensors}\; ,
		\label{sukMQ_electric}
\end{align}
and the magnetic quiver of the Higgs branch at the origin of the tensor branch has been derived to be
\begin{align} 
  \raisebox{-.5\height}{
 	\begin{tikzpicture}
	\tikzstyle{gauge} = [circle, draw,inner sep=3pt];
	\tikzstyle{flavour} = [regular polygon,regular polygon sides=4,inner 
sep=3pt, draw];
	\node (g1) [gauge,label=below:{$\scriptstyle{1}$}] {};
	\node (g2) [gauge,right of=g1,label=below:{$\scriptstyle{2}$}] {};
	\node (g3) [gauge,right of=g2,label=below:{$\scriptstyle{r{+}2}$}] {};
	\node (g4) [gauge,right of=g3,label=below:{$\scriptstyle{2r{+}2}$}] {};
	\node (g5) [gauge,right of=g4,label=below:{$\scriptstyle{3r{+}2}$}] {};
	\node (g6) [gauge,right of=g5,label=below:{$\scriptstyle{4r{+}2}$}] {};
	\node (g7) [gauge,right of=g6,label=below:{$\scriptstyle{5r{+}2}$}] {};
	\node (g8) [gauge,right of=g7,label=below:{$\scriptstyle{6r{+}2}$}] {};
	\node (g9) [gauge,right of=g8,label=below:{$\scriptstyle{4r{+}1}$}] {};
	\node (g10) [gauge,above of=g8,label=above:{$\scriptstyle{3r{+}1}$}] {};
	\node (g11) [gauge,right of=g9,label=below:{$\scriptstyle{2r}$}] {};
	\draw (g1)--(g2) (g2)--(g3) (g3)--(g4) (g4)--(g5) (g5)--(g6) (g6)--(g7) 
(g7)--(g8) (g8)--(g9) (g8)--(g10) (g9)--(g11);
	\end{tikzpicture}
	} \,.
	\label{sukMQ}
\end{align}
\item $\ell=3$: The construction has a remaining $\mathrm{SU}(9) \subset E_8$ symmetry, see \cite[Sec.\ 3.4.5]{Cabrera:2019izd}. On a generic point of the tensor branch, the 6d theory is described by
\begin{align}
 \raisebox{-.5\height}{
 	\begin{tikzpicture}
	\tikzstyle{gauge} = [circle, draw,inner sep=3pt];
	\tikzstyle{flavour} = [regular polygon,regular polygon sides=4,inner 
sep=3pt, draw];
	\node (g1) [gauge,label={[rotate=-45]below right:{$\scriptstyle{\mathrm{SU}(3)}$}}] {};
	\node (g2) [right of=g1] {$\cdots$};
	\node (g3) [gauge,right of=g2,label={[rotate=-45]below right:{$\scriptstyle{\mathrm{SU}(3)}$}}] {};
	\node (g4) [gauge,right of=g3,label={[rotate=-45]below right:{$\scriptstyle{\mathrm{SU}(3)}$}}] {};
	\node (g5) [gauge,right of=g4,label={[rotate=-45]below right:{$\scriptstyle{\mathrm{SU}(3)}$}}] {};
	\node (f1) [flavour,left of=g1,label=below:{$\scriptstyle{\mathrm{SU}(3)}$}] {};
	\node (f5) [flavour,above of=g5,label=above:{$\scriptstyle{\mathrm{SU}(9)}$}] {};
	\draw (g1)--(g2) (g2)--(g3) (g3)--(g4) (g1)--(f1) (g4)--(g5) (g5)--(f5);
    \draw[decoration={brace,mirror,raise=20pt},decorate,thick]
  (-0.25,-0.6) -- node[below=20pt] {$r-1$} (3.75,-0.6);
	\end{tikzpicture}
	} \quad  + \quad  \text{$r$ tensors}\; ,
		\label{su3MQ_electric}
\end{align}
and the magnetic quiver for the Higgs branch at the origin of the tensor branch is
\begin{align}
  \raisebox{-.5\height}{
 	\begin{tikzpicture}
	\tikzstyle{gauge} = [circle, draw,inner sep=3pt];
	\tikzstyle{flavour} = [regular polygon,regular polygon sides=4,inner 
sep=3pt, draw];
	\node (g0) [gauge,label=below:{$\scriptstyle{1}$}] {};
	\node (g1) [gauge,right of=g0,label=below:{$\scriptstyle{2}$}] {};
	\node (g2) [gauge,right of=g1,label=below:{$\scriptstyle{3}$}] {};
	\node (g3) [gauge,right of=g2,label=below:{$\scriptstyle{r{+}3}$}] {};
	\node (g4) [gauge,right of=g3,label=below:{$\scriptstyle{2r{+}3}$}] {};
	\node (g5) [gauge,right of=g4,label=below:{$\scriptstyle{3r{+}3}$}] {};
	\node (g6) [gauge,right of=g5,label=below:{$\scriptstyle{4r{+}3}$}] {};
	\node (g7) [gauge,right of=g6,label=below:{$\scriptstyle{5r{+}3}$}] {};
	\node (g8) [gauge,right of=g7,label=below:{$\scriptstyle{6r{+}3}$}] {};
	\node (g9) [gauge,right of=g8,label=below:{$\scriptstyle{4r{+}2}$}] {};
	\node (g10) [gauge,right of=g9,label=below:{$\scriptstyle{2r{+}1}$}] {};
	\node (g11) [gauge,above of=g8,label=above:{$\scriptstyle{3r{+}1}$}] {};
	\draw (g0)--(g1) (g1)--(g2) (g2)--(g3) (g3)--(g4) (g4)--(g5) (g5)--(g6) 
(g6)--(g7) (g7)--(g8) (g8)--(g9) (g9)--(g10) (g8)--(g11);
	\end{tikzpicture}
	}  
	\label{su3MQ}
\end{align}
\item $\ell=4$: The preserved symmetry is $\mathrm{SU}(8) \times \mathrm{SU}(2) \subset E_8$, see \cite[Sec.\ 3.5.10]{Cabrera:2019izd} and also \cite[Sec.\ 5.3]{Mekareeya:2017jgc}. The 6d theory on the tensor branch is 
\begin{align}
 \raisebox{-.5\height}{
 	\begin{tikzpicture}
	\tikzstyle{gauge} = [circle, draw,inner sep=3pt];
	\tikzstyle{flavour} = [regular polygon,regular polygon sides=4,inner 
sep=3pt, draw];
	\node (g1) [gauge,label={[rotate=-45]below right:{$\scriptstyle{ \mathrm{SU}(4) }$}}] {};
	\node (g2) [right of=g1] {$\ldots$};
	\node (g3) [gauge,right of=g2,label={[rotate=-45]below right:{$\scriptstyle{ \mathrm{SU}(4) }$}}] {};
	\node (g4) [gauge,right of=g3,label={[rotate=-45]below right:{$\scriptstyle{ \mathrm{SU}(4) }$}}] {};
	\node (g5) [gauge,right of=g4,label={[rotate=-45]below right:{$\scriptstyle{ \mathrm{SU}(4) }$}}] {};
	\node (f1) [flavour,left of=g1,label=below:{$\scriptstyle{ \mathrm{SU}(4) }$}] {};
	\node (f5) [flavour,above of=g5,label=above:{$\scriptstyle{ \mathrm{SU}(8) }$}] {};
	\node (fAS) [flavour,right of=g5,label=below:{$\scriptstyle{1}$}] {};
	\draw (g1)--(g2) (g2)--(g3) (g3)--(g4) (g1)--(f1) (g4)--(g5) (g5)--(f5);
	\draw [line join=round,decorate, decoration={zigzag, segment length=4,amplitude=.9,post=lineto,post length=2pt}] (g5)--(fAS);
	\draw (4.5,0.25) node {$\scriptscriptstyle{\Lambda^2}$};
    \draw[decoration={brace,mirror,raise=20pt},decorate,thick]
  (-0.25,-0.6) -- node[below=20pt] {$r-1$} (3.75,-0.6);
	\end{tikzpicture}
	} \quad + \quad \text{$r$ tensors}\; ,
	\label{su3MQ2_electric}
\end{align}
and the magnetic quiver for the Higgs branch at the origin of the tensor branch is
\begin{align}
    \raisebox{-.5\height}{
 	\begin{tikzpicture}
	\tikzstyle{gauge} = [circle, draw,inner sep=3pt];
	\tikzstyle{flavour} = [regular polygon,regular polygon sides=4,inner 
sep=3pt, draw];
	\node (g1) [gauge,label=below:{$\scriptstyle{1}$}] {};
	\node (g2) [gauge,right of=g1,label=below:{$\scriptstyle{2}$}] {};
	\node (g3) [gauge,right of=g2,label=below:{$\scriptstyle{3}$}] {};
	\node (g4) [gauge,right of=g3,label=below:{$\scriptstyle{4}$}] {};
	\node (g5) [gauge,right of=g4,label=below:{$\scriptstyle{r{+}4}$}] {};
	\node (g6) [gauge,right of=g5,label=below:{$\scriptstyle{2r{+}4}$}] {};
	\node (g7) [gauge,right of=g6,label=below:{$\scriptstyle{3r{+}4}$}] {};
	\node (g8) [gauge,right of=g7,label=below:{$\scriptstyle{4r{+}4}$}] {};
	\node (g9) [gauge,right of=g8,label=below:{$\scriptstyle{5r{+}4}$}] {};
	\node (g10) [gauge,right of=g9,label=below:{$\scriptstyle{6r{+}4}$}] {};
	\node (g11) [gauge,right of=g10,label=below:{$\scriptstyle{4r{+}2}$}] {};
	\node (g12) [gauge,right of=g11,label=below:{$\scriptstyle{2r{+}1}$}] {};
	\node (g13) [gauge,above of=g10,label=above:{$\scriptstyle{3r{+}2}$}] {};
	\draw (g1)--(g2) (g2)--(g3) (g3)--(g4) (g4)--(g5) (g5)--(g6) 
(g6)--(g7) (g7)--(g8) (g8)--(g9) (g9)--(g10) (g10)--(g11) (g11)--(g12) 
(g10)--(g13);
	\end{tikzpicture}
	}  \,.
	\label{su3MQ2}
\end{align}
\end{compactitem}
In summary, equations \eqref{sukMQ}, \eqref{su3MQ},  and \eqref{su3MQ2} provide the magnetic quivers for the 6d theories that were used in \cite[Sec.\ 4.1]{Giacomelli:2020jel} to construct the $\ess{G}{\ell}{r}$ theories.
\paragraph{6d origin of $\tee{G}{\ell}{r}$ theories.}
As proposed in \cite{Giacomelli:2020jel}, the $\tee{G}{\ell}{r}$ theories originate from the following 6d $\mathcal{N}=(1,0)$ systems:
\begin{compactitem}
    \item $\ell=2$: The boundary conditions preserve $E_7 \times \mathrm{SU}(2) \subset E_8$, see \cite[Sec.\ 3.3.2]{Cabrera:2019izd} as well as \cite[Sec.\ 5.1]{Mekareeya:2017jgc}. The theory on the tensor branch is described by
    \begin{align}
 \raisebox{-.5\height}{
 	\begin{tikzpicture}
	\tikzstyle{gauge} = [circle, draw,inner sep=3pt];
	\tikzstyle{flavour} = [regular polygon,regular polygon sides=4,inner 
sep=3pt, draw];
	\node (g1) [gauge,label={[rotate=-45]below right:{$\scriptstyle{ \mathrm{SU}(2) }$}}] {};
	\node (g2) [right of=g1] {$\ldots$};
	\node (g3) [gauge,right of=g2,label={[rotate=-45]below right:{$\scriptstyle{ \mathrm{SU}(2) }$}}] {};
	\node (g4) [gauge,right of=g3,label={[rotate=-45]below right:{$\scriptstyle{ \mathrm{SU}(2) }$}}] {};
	\node (f1) [flavour,left of=g1,label=below:{$\scriptstyle{ \mathrm{SU}(2) }$}] {};
	\node (f5) [flavour,right of=g4,label=below:{$\scriptstyle{ \mathrm{SU}(2) }$}] {};
	\draw (g1)--(g2) (g2)--(g3) (g3)--(g4) (g1)--(f1) (g4)--(f5);
    \draw[decoration={brace,mirror,raise=20pt},decorate,thick]
  (-0.25,-0.6) -- node[below=20pt] {$r-1$} (3.75,-0.6);
	\end{tikzpicture}
	} 
	\quad + \quad
    \text{$r$ tensors} 
	\; ,
\end{align} 

and the magnetic quiver of the Higgs branch of the conformal fixed point is
\begin{align} 
  \raisebox{-.5\height}{
 	\begin{tikzpicture}
	\tikzstyle{gauge} = [circle, draw,inner sep=3pt];
	\tikzstyle{flavour} = [regular polygon,regular polygon sides=4,inner sep=3pt, draw];
	\node (g1) [gauge,label=below:{$\scriptstyle{1}$}] {};
	\node (g2) [gauge,right of=g1,label=below:{$\scriptstyle{2}$}] {};
	\node (g3) [gauge,right of=g2,label=below:{$\scriptstyle{r{+}1}$}] {};
	\node (g4) [gauge,right of=g3,label=below:{$\scriptstyle{2r}$}] {};
	\node (g5) [gauge,right of=g4,label=below:{$\scriptstyle{3r}$}] {};
	\node (g6) [gauge,right of=g5,label=below:{$\scriptstyle{4r}$}] {};
	\node (g7) [gauge,right of=g6,label=below:{$\scriptstyle{5r}$}] {};
	\node (g8) [gauge,right of=g7,label=below:{$\scriptstyle{6r}$}] {};
	\node (g9) [gauge,right of=g8,label=below:{$\scriptstyle{4r}$}] {};
	\node (g10) [gauge,above of=g8,label=above:{$\scriptstyle{3r}$}] {};
	\node (g11) [gauge,right of=g9,label=below:{$\scriptstyle{2r}$}] {};
	\draw (g1)--(g2) (g2)--(g3) (g3)--(g4) (g4)--(g5) (g5)--(g6) (g6)--(g7) 
(g7)--(g8) (g8)--(g9) (g8)--(g10) (g9)--(g11);
	\end{tikzpicture}
	} \,.
	\label{eq:magQuiv_Tee_l=2}
\end{align}
    \item $\ell=3$: The remaining symmetry is $E_6\times \mathrm{SU}(3) \subset E_8$, see \cite[Sec.\ 3.4.3]{Cabrera:2019izd}. The effective theory is summarised as follows:
    \begin{align}
 \raisebox{-.5\height}{
 	\begin{tikzpicture}
	\tikzstyle{gauge} = [circle, draw,inner sep=3pt];
	\tikzstyle{flavour} = [regular polygon,regular polygon sides=4,inner 
sep=3pt, draw];
	\node (g1) [gauge,label={[rotate=-45]below right:{$\scriptstyle{ \mathrm{SU}(3) }$}}] {};
	\node (g2) [right of=g1] {$\ldots$};
	\node (g3) [gauge,right of=g2,label={[rotate=-45]below right:{$\scriptstyle{ \mathrm{SU}(3) }$}}] {};
	\node (g4) [gauge,right of=g3,label={[rotate=-45]below right:{$\scriptstyle{ \mathrm{SU}(3) }$}}] {};
	\node (f1) [flavour,left of=g1,label=below:{$\scriptstyle{ \mathrm{SU}(3) }$}] {};
	\node (f5) [flavour,right of=g4,label=below:{$\scriptstyle{ \mathrm{SU}(3) }$}] {};
	\draw (g1)--(g2) (g2)--(g3) (g3)--(g4) (g1)--(f1) (g4)--(f5);
    \draw[decoration={brace,mirror,raise=20pt},decorate,thick]
  (-0.25,-0.6) -- node[below=20pt] {$r-1$} (3.75,-0.6);
	\end{tikzpicture}
	} 
	\quad + \quad
    \text{$r$ tensors} 
	\; ,
\end{align} 
such that the magnetic quiver of the Higgs branch at the origin of the tensor branch is given by
\begin{align}
  \raisebox{-.5\height}{
 	\begin{tikzpicture}
	\tikzstyle{gauge} = [circle, draw,inner sep=3pt];
	\tikzstyle{flavour} = [regular polygon,regular polygon sides=4,inner sep=3pt, draw];
	\node (g0) [gauge,label=below:{$\scriptstyle{1}$}] {};
	\node (g1) [gauge,right of=g0,label=below:{$\scriptstyle{2}$}] {};
	\node (g2) [gauge,right of=g1,label=below:{$\scriptstyle{3}$}] {};
	\node (g3) [gauge,right of=g2,label=below:{$\scriptstyle{r{+}2}$}] {};
	\node (g4) [gauge,right of=g3,label=below:{$\scriptstyle{2r{+}1}$}] {};
	\node (g5) [gauge,right of=g4,label=below:{$\scriptstyle{3r}$}] {};
	\node (g6) [gauge,right of=g5,label=below:{$\scriptstyle{4r}$}] {};
	\node (g7) [gauge,right of=g6,label=below:{$\scriptstyle{5r}$}] {};
	\node (g8) [gauge,right of=g7,label=below:{$\scriptstyle{6r}$}] {};
	\node (g9) [gauge,right of=g8,label=below:{$\scriptstyle{4r}$}] {};
	\node (g10) [gauge,right of=g9,label=below:{$\scriptstyle{2r}$}] {};
	\node (g11) [gauge,above of=g8,label=above:{$\scriptstyle{3r}$}] {};
	\draw (g0)--(g1) (g1)--(g2) (g2)--(g3) (g3)--(g4) (g4)--(g5) (g5)--(g6) 
(g6)--(g7) (g7)--(g8) (g8)--(g9) (g9)--(g10) (g8)--(g11);
	\end{tikzpicture}
	}  
	\,.
	\label{eq:magQuiv_Tee_l=3}
\end{align}
    \item $\ell=4$: The construction breaks the $E_8$ to $\mathrm{SO}(10)\times \mathrm{SU}(4) \subset E_8$, see \cite[Sec.\ 3.5.9]{Cabrera:2019izd} and \cite[Sec.\ 5.3]{Mekareeya:2017jgc}. On a generic point of the tensor branch, one finds
    \begin{align}
 \raisebox{-.5\height}{
 	\begin{tikzpicture}
	\tikzstyle{gauge} = [circle, draw,inner sep=3pt];
	\tikzstyle{flavour} = [regular polygon,regular polygon sides=4,inner 
sep=3pt, draw];
	\node (g1) [gauge,label={[rotate=-45]below right:{$\scriptstyle{ \mathrm{SU}(4) }$}}] {};
	\node (g2) [right of=g1] {$\ldots$};
	\node (g3) [gauge,right of=g2,label={[rotate=-45]below right:{$\scriptstyle{ \mathrm{SU}(4) }$}}] {};
	\node (g4) [gauge,right of=g3,label={[rotate=-45]below right:{$\scriptstyle{ \mathrm{SU}(4) }$}}] {};
	\node (f1) [flavour,left of=g1,label=below:{$\scriptstyle{ \mathrm{SU}(4) }$}] {};
	\node (f5) [flavour,right of=g4,label=below:{$\scriptstyle{ \mathrm{SU}(4) }$}] {};
	\draw (g1)--(g2) (g2)--(g3) (g3)--(g4) (g1)--(f1) (g4)--(f5);
    \draw[decoration={brace,mirror,raise=20pt},decorate,thick]
  (-0.25,-0.6) -- node[below=20pt] {$r-1$} (3.75,-0.6);
	\end{tikzpicture}
	} 
	\quad + \quad
    \text{$r$ tensors} 
	\; ,
\end{align} 
and the Higgs branch of the associated SCFT is described by the following magnetic quiver:
\begin{align}
    \raisebox{-.5\height}{
 	\begin{tikzpicture}
	\tikzstyle{gauge} = [circle, draw,inner sep=3pt];
	\tikzstyle{flavour} = [regular polygon,regular polygon sides=4,inner 
sep=3pt, draw];
	\node (g1) [gauge,label=below:{$\scriptstyle{1}$}] {};
	\node (g2) [gauge,right of=g1,label=below:{$\scriptstyle{2}$}] {};
	\node (g3) [gauge,right of=g2,label=below:{$\scriptstyle{3}$}] {};
	\node (g4) [gauge,right of=g3,label=below:{$\scriptstyle{4}$}] {};
	\node (g5) [gauge,right of=g4,label=below:{$\scriptstyle{r{+}3}$}] {};
	\node (g6) [gauge,right of=g5,label=below:{$\scriptstyle{2r{+}2}$}] {};
	\node (g7) [gauge,right of=g6,label=below:{$\scriptstyle{3r{+}1}$}] {};
	\node (g8) [gauge,right of=g7,label=below:{$\scriptstyle{4r}$}] {};
	\node (g9) [gauge,right of=g8,label=below:{$\scriptstyle{5r}$}] {};
	\node (g10) [gauge,right of=g9,label=below:{$\scriptstyle{6r}$}] {};
	\node (g11) [gauge,right of=g10,label=below:{$\scriptstyle{4r}$}] {};
	\node (g12) [gauge,right of=g11,label=below:{$\scriptstyle{2r}$}] {};
	\node (g13) [gauge,above of=g10,label=above:{$\scriptstyle{3r}$}] {};
	\draw (g1)--(g2) (g2)--(g3) (g3)--(g4) (g4)--(g5) (g5)--(g6) 
(g6)--(g7) (g7)--(g8) (g8)--(g9) (g9)--(g10) (g10)--(g11) (g11)--(g12) 
(g10)--(g13);
	\end{tikzpicture}
	}  \,.
	\label{eq:magQuiv_Tee_l=4}
\end{align}
\end{compactitem}
In summary, equations \eqref{eq:magQuiv_Tee_l=2}, \eqref{eq:magQuiv_Tee_l=3},  and \eqref{eq:magQuiv_Tee_l=4} provide the magnetic quivers for the 6d theories that were used in \cite[Sec.\ 4.2]{Giacomelli:2020jel} to construct the $\tee{G}{\ell}{r}$ theories.

\subsection{Analysis of FI deformations}
If we turn on a, say complex, FI parameter $\lambda$ at a $\mathrm{U}(N)$ gauge node in a quiver as in the following figure
\begin{align}
    \raisebox{-.5\height}{
\begin{tikzpicture}[thick, scale=0.4]
\node[](L3) at (-5,0) {$\Phi$};
\node[] (L2) at (-0.3,0.7) {$\widetilde{Q}_i,Q_i$};
\node[circle, draw, inner sep=2.5](L4) at (-2.5,0){$N$};
\node[rectangle, draw, inner sep=1.7,minimum height=.6cm,minimum width=.6cm](L5) at (2,0){$k$};
  \path[every node/.style={font=\sffamily\small,
  		fill=white,inner sep=1pt}]
(L4) edge [loop, out=145, in=215, looseness=4] (L4);
\draw[-] (L5) -- (L4);
\end{tikzpicture}
}
\end{align}
the superpotential becomes 
\begin{equation} 
\mathcal{W}= \widetilde{Q}_i\Phi Q^i - \lambda\Tr\Phi.
\end{equation}
Consequently, we have to solve the following F-term and D-term equations: 
\begin{subequations}
\begin{align}
Q^i\widetilde{Q}_i &= \lambda I_N \\ 
Q^iQ_i^{\dagger}-\widetilde{Q}^{\dagger i} \widetilde{Q}_i &=0 \,, 
\end{align}
\end{subequations}
 where the summation over flavor indices, indeed, includes a summation over all the bifundamental hypermultiplets charged under the $\mathrm{U}(N)$ gauge node. 

From F-terms we deduce that $\Tr(Q^i\widetilde{Q}_i)=N\lambda$ and from the cyclicity of the trace we have the identity $\Tr(Q^i\widetilde{Q}_i)=\Tr(\widetilde{Q}_iQ^i)$. Combining these relations we conclude that for any quiver with unitary gauge groups and bifundamental matter only (which is the only case considered below), the FI parameters obey the relation 
\be\label{constr} \sum_i N_i\lambda^i=0,\ee
where $N_i$ denotes the rank of the i-th node. We therefore see that we always need to turn on FI parameters at two nodes at least.

The easiest case to analyze is that of FI parameters turned on at two abelian nodes. Because of \eqref{constr}, the two FI parameters are $\lambda$ and $-\lambda$. In this case, the deformation induces a nontrivial expectation value for a chain of bifundamental multiplets connecting the two nodes. This follows from the constraint on the trace of bifundamental bi-linears discussed above. Finding explicitly a solution to F and D-term equations in this case is easy. First of all, we pick a subquiver starting and ending at the nodes at which we have turned on the FI parameter. All the bifundamental multiplets along the subquiver, which we denote as $B_i$ and $\widetilde{B}_i$, acquire a nontrivial VEV. Modulo gauge transformations, we can set 
\be\label{solab} \langle B_i\rangle=\sqrt{\lambda}(e_1,0,\dots,0); \quad \langle \widetilde{B}_i\rangle=\langle B_i\rangle^T, \ee 
where $e_1$ is the vector whose first entry is equal to one and all the others are trivial. Of course, the size of the matrix $\langle B_i\rangle$ is dictated by the rank of the gauge groups under which the bifundamental multiplet is charged. If in the original quiver two nodes are connected by more than one bifundamental hypermultiplet, modulo a flavor rotation we can turn on the VEV \eqref{solab} for one of them and set to zero the VEV of all the others. It is easy to check that this choice satisfies the equations of motion. 
All the unitary gauge groups along the subquiver are spontaneously broken as $\mathrm{U}(n_i)\rightarrow \mathrm{U}(n_i-1)$, whereas the other nodes in the quiver are unaffected. Among all the broken $\mathrm{U}(1)$ factors, the diagonal combination survives and gives rise to a new $\mathrm{U}(1)$ node which is coupled to all the nodes of the quiver connected to those of the subquiver. Overall, this is equivalent to subtracting from the original quiver an abelian quiver isomorphic to the subquiver mentioned before. 

The case of FI parameters turned on at nodes of the same rank, say $k$, is not harder to analyze: The VEV for the bifundamentals is obtained from \eqref{solab} by picking the tensor product with the $k\times k$ identity matrix $I_k$, all the groups along the subquiver are broken as $\mathrm{U}(n_i)\rightarrow \mathrm{U}(n_i-k)$ and finally we need to add a $\mathrm{U}(k)$ node associated with the surviving gauge symmetry. Overall, this corresponds to a modified quiver subtraction in which we rebalance with a $\mathrm{U}(k)$ node. Of course, the quiver we subtract has $\mathrm{U}(k)$ nodes only.

For our analysis we need a more elaborate variant of the above deformation which involves three nodes. Say, we turn on FI parameters at the nodes $\mathrm{U}(k)$, $\mathrm{U}(n)$ and $\mathrm{U}(n+k)$. The FI parameters satisfy the relation \eqref{constr} and we further impose the constraint $\lambda_k=\lambda_n$, so that we still have only one independent parameter. The equations of motion can be solved as follows: We set the VEV of all the bifundamental multiplets in the subquiver connecting the nodes $\mathrm{U}(k)$ and $\mathrm{U}(n+k)$ to be 
\be\label{solab1} \langle B_i\rangle=\sqrt{\lambda_k}(I_k,0,\dots,0)\,, \quad \langle \widetilde{B}_i\rangle=\langle B_i\rangle^T, \ee 
and the VEV of the bifundamentals in the subquiver connecting the nodes $\mathrm{U}(n+k)$ and $\mathrm{U}(n)$ to be 
\be\label{solab2} \langle B_i\rangle=\sqrt{\lambda_k}(I_n,0,\dots,0)\,, \quad \langle \widetilde{B}_i\rangle=\langle B_i\rangle^T. \ee 
Here, we are assuming that the two subquivers meet at the $\mathrm{U}(n+k)$ node only. 

Overall, the Higgsing of the theory can be described in terms of a sequence of two modified quiver subtraction: We first subtract a quiver of $\mathrm{U}(n)$ nodes going from node $\mathrm{U}(n)$ to node $\mathrm{U}(n+k)$ and as in the previous case we rebalance with a $\mathrm{U}(n)$ node. Then we subtract from the resulting quiver a quiver of $\mathrm{U}(k)$ nodes going from node $\mathrm{U}(k)$ to node $\mathrm{U}(n+k)$ and we rebalance with a $\mathrm{U}(k)$ node. A careful analysis of the Higgsing reveals that the $\mathrm{U}(n)$ node we introduced at the first step should not be rebalanced when we perform the second subtraction. 

Let us illustrate the procedure for $k=1$ and $n=2$ in the case of the $E_8$ quiver: 
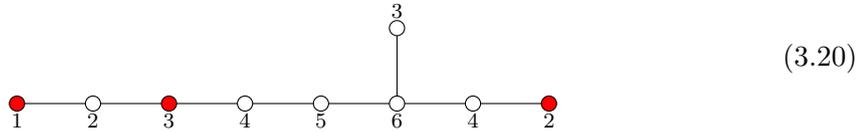
\be\label{E8MQ}
\raisebox{-.5\height}{
\begin{tikzpicture}
\filldraw[fill= red] (0,0) circle [radius=0.1] node[below] {\scriptsize 1};
\filldraw[fill= white] (1,0) circle [radius=0.1] node[below] {\scriptsize 2};
\filldraw[fill= red] (2,0) circle [radius=0.1] node[below] {\scriptsize 3};
\filldraw[fill= white] (3,0) circle [radius=0.1] node[below] {\scriptsize 4};
\filldraw[fill= white] (4,0) circle [radius=0.1] node[below] {\scriptsize 5};
\filldraw[fill= white] (5,0) circle [radius=0.1] node[below] {\scriptsize 6};
\filldraw[fill= white] (6,0) circle [radius=0.1] node[below] {\scriptsize 4};
\filldraw[fill= white] (5,1) circle [radius=0.1] node[above] {\scriptsize 3};
\filldraw[fill= red] (7,0) circle [radius=0.1] node[below] {\scriptsize 2};
\draw (0.1, 0) -- (0.9,0) ;
\draw (1.1, 0) -- (1.9,0) ;
\draw (2.1, 0) -- (2.9,0) ;
\draw (3.1, 0) -- (3.9,0) ;
\draw (4.1, 0) -- (4.9,0) ;
\draw (5.1, 0) -- (5.9,0) ;
\draw (5, 0.1) -- (5,0.9) ;
\draw (6.1, 0) -- (6.9,0) ;
\end{tikzpicture} 
}
\ee 
In the quiver \eqref{E8MQ}, we turn on FI parameters at the nodes colored in red. We first subtract an $A_6$ quiver with $\mathrm{U}(2)$ nodes, which results in 
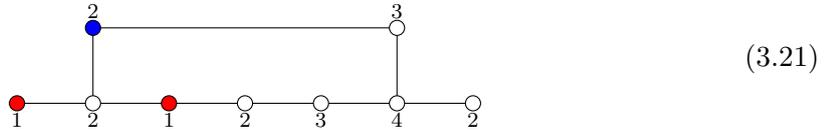
\be\label{E8MQ2}
\raisebox{-.5\height}{
\begin{tikzpicture}
\filldraw[fill= red] (0,0) circle [radius=0.1] node[below] {\scriptsize 1};
\filldraw[fill= white] (1,0) circle [radius=0.1] node[below] {\scriptsize 2};
\filldraw[fill= red] (2,0) circle [radius=0.1] node[below] {\scriptsize 1};
\filldraw[fill= white] (3,0) circle [radius=0.1] node[below] {\scriptsize 2};
\filldraw[fill= white] (4,0) circle [radius=0.1] node[below] {\scriptsize 3};
\filldraw[fill= white] (5,0) circle [radius=0.1] node[below] {\scriptsize 4};
\filldraw[fill= white] (6,0) circle [radius=0.1] node[below] {\scriptsize 2};
\filldraw[fill= white] (5,1) circle [radius=0.1] node[above] {\scriptsize 3};
\filldraw[fill= blue] (1,1) circle [radius=0.1] node[above] {\scriptsize 2};
\draw (0.1, 0) -- (0.9,0) ;
\draw (1.1, 0) -- (1.9,0) ;
\draw (2.1, 0) -- (2.9,0) ;
\draw (3.1, 0) -- (3.9,0) ;
\draw (4.1, 0) -- (4.9,0) ;
\draw (5.1, 0) -- (5.9,0) ;
\draw (5, 0.1) -- (5,0.9) ;
\draw (1.1, 1) -- (4.9,1) ;
\draw (1, 0.1) -- (1,0.9) ;
\end{tikzpicture} 
}
\ee 
The node colored in blue in \eqref{E8MQ2} is used to rebalance. Thereafter, we subtract an $A_3$ abelian quiver and obtain
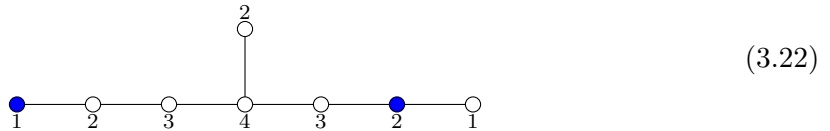
\be\label{E8MQ3}
\raisebox{-.5\height}{
\begin{tikzpicture}
\filldraw[fill= blue] (0,0) circle [radius=0.1] node[below] {\scriptsize 1};
\filldraw[fill= white] (1,0) circle [radius=0.1] node[below] {\scriptsize 2};
\filldraw[fill= white] (2,0) circle [radius=0.1] node[below] {\scriptsize 3};
\filldraw[fill= white] (3,0) circle [radius=0.1] node[below] {\scriptsize 4};
\filldraw[fill= white] (4,0) circle [radius=0.1] node[below] {\scriptsize 3};
\filldraw[fill= blue] (5,0) circle [radius=0.1] node[below] {\scriptsize 2};
\filldraw[fill= white] (6,0) circle [radius=0.1] node[below] {\scriptsize 1};
\filldraw[fill= white] (3,1) circle [radius=0.1] node[above] {\scriptsize 2};
\draw (0.1, 0) -- (0.9,0) ;
\draw (1.1, 0) -- (1.9,0) ;
\draw (2.1, 0) -- (2.9,0) ;
\draw (3.1, 0) -- (3.9,0) ;
\draw (4.1, 0) -- (4.9,0) ;
\draw (5.1, 0) -- (5.9,0) ;
\draw (3, 0.1) -- (3,0.9) ;
\end{tikzpicture} 
}
\ee 
The $\mathrm{U}(1)$ node colored in blue is, again, introduced to rebalance and, as we have explained before, is not connected to the other $\mathrm{U}(2)$ node. Overall, this FI deformation implements the mass deformation from the $E_8$ to the $E_7$ theory. We apply this procedure repeatedly below.

\subsection{Case \texorpdfstring{$\ell=2$}{l=2}} 
The Higgs branch of the 6d theory \eqref{sukMQ_electric} at the fixed point is described by the magnetic quiver \eqref{sukMQ}.
Upon dimensional reduction we get a 5d SCFT whose HB is described by the same magnetic quiver. We now turn on a mass deformation for the five dimensional theory, described by a FI deformation of the quiver.
\paragraph{$\ess{E_6}{2}{r}$ theories.}
Specifically, we furn on FI parameters at the nodes $\mathrm{U}(1)$, $\mathrm{U}(3r+1)$ and $\mathrm{U}(3r+2)$, subject to the constraint $\lambda_1=\lambda_{3r+1}$. 
\begin{figure}[t]
\begin{center}
\begin{tikzpicture}
\filldraw[fill= red] (0,0) circle [radius=0.1] node[below] {\scriptsize $1$};
\filldraw[fill= white] (1,0) circle [radius=0.1] node[below] {\scriptsize $2$};
\filldraw[fill= white] (2,0) circle [radius=0.1] node[below] {\scriptsize $r{+}2$};
\filldraw[fill= white] (3,0) circle [radius=0.1] node[below] {\scriptsize $2r{+}2$};
\filldraw[fill= red] (4,0) circle [radius=0.1] node[below] {\scriptsize $3r{+}2$};
\filldraw[fill= white] (5,0) circle [radius=0.1] node[below] {\scriptsize $4r{+}2$};
\filldraw[fill= white] (6,0) circle [radius=0.1] node[below] {\scriptsize $5r{+}2$};
\filldraw[fill= red] (7,1) circle [radius=0.1] node[above] {\scriptsize $3r{+}1$};
\filldraw[fill= white] (7,0) circle [radius=0.1] node[below] {\scriptsize $6r{+}2$};
\filldraw[fill= white] (8,0) circle [radius=0.1] node[below] {\scriptsize $4r{+}1$};
\filldraw[fill= white] (9,0) circle [radius=0.1] node[below] {\scriptsize $2r$};
\draw (0.1, 0) -- (0.9,0) ;
\draw (1.1, 0) -- (1.9,0) ;
\draw (2.1, 0) -- (2.9,0) ;
\draw (3.1, 0) -- (3.9,0) ;
\draw (4.1, 0) -- (4.9,0) ;
\draw (5.1, 0) -- (5.9,0) ;
\draw (7, 0.1) -- (7,0.9) ;
\draw (6.1, 0) -- (6.9,0) ;
\draw (7.1, 0) -- (7.9,0) ;
\draw (8.1, 0) -- (8.9,0) ;
\draw[->, thick] (4.5,-0.7)--(4.5,-2.2);
\filldraw[fill= white] (1,-3) circle [radius=0.1] node[below] {\scriptsize $1$};
\filldraw[fill= white] (2,-3) circle [radius=0.1] node[below] {\scriptsize $r{+}1$};
\filldraw[fill= white] (3,-3) circle [radius=0.1] node[below] {\scriptsize $2r{+}1$};
\filldraw[fill= white] (4,-3) circle [radius=0.1] node[below] {\scriptsize $3r{+}1$};
\filldraw[fill= white] (5,-3) circle [radius=0.1] node[below] {\scriptsize $4r{+}1$};
\filldraw[fill= white] (6,-3) circle [radius=0.1] node[below] {\scriptsize $3r{+}1$};
\filldraw[fill= white] (7,-3) circle [radius=0.1] node[below] {\scriptsize $2r{+}1$};
\filldraw[fill= white] (8,-3) circle [radius=0.1] node[below] {\scriptsize $r{+}1$};
\filldraw[fill= white] (9,-3) circle [radius=0.1] node[below] {\scriptsize $1$};
\filldraw[fill= white] (5,-2) circle [radius=0.1] node[above] {\scriptsize $2r$};
\draw (1.1, -3) -- (1.9,-3) ;
\draw (2.1, -3) -- (2.9,-3) ;
\draw (3.1, -3) -- (3.9,-3) ;
\draw (4.1, -3) -- (4.9,-3) ;
\draw (5.1, -3) -- (5.9,-3) ;
\draw (5, -2.9) -- (5,-2.1) ;
\draw (6.1, -3) -- (6.9,-3) ;
\draw (7.1, -3) -- (7.9,-3) ;
\draw (8.1, -3) -- (8.9,-3) ;
\end{tikzpicture} 
\end{center}
\caption{The FI deformation of the 6d magnetic quiver \eqref{sukMQ}. We turn on FI parameters at the nodes in red, resulting in the second quiver.}
\label{DeftoUSpUV}
\end{figure}
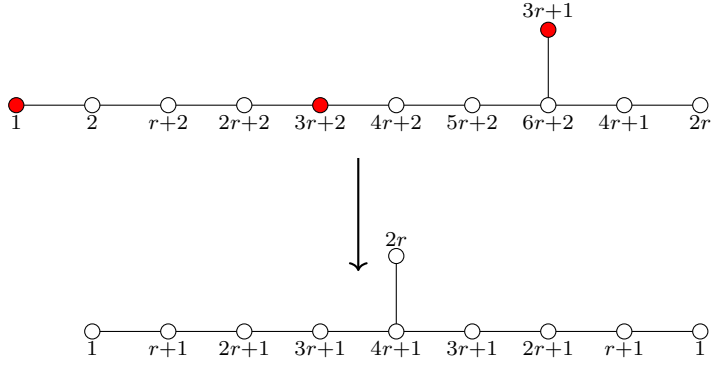
 This deformation, summarized in Figure \ref{DeftoUSpUV}, reduces the theory to the UV completion of the gauge theory 
\begin{align}
\label{lagrir}
\raisebox{-.5\height}{
\begin{tikzpicture}[x=1.5cm,y=.8cm]
\node (f1) at (0,0) [flavour0,label=below:{$\scriptstyle{3}$}] {};
\node (g1) at (1,0) [gauge,label=below:{$\scriptstyle{\mathrm{USp}(2r)}$}] {};
\node (g2) at (2,0) [gauge,label=below:{$\scriptstyle{\mathrm{USp}(2r)}$}] {};
\node (f2) at (3,0) [flavour0,label=below:{$\scriptstyle{3}$}] {};
\draw (f1)--(g1)--(g2)--(f2);
\end{tikzpicture}
} \,.
\end{align}
This can be seen by exhibiting the mass deformations leading to the 3d mirror of \eqref{lagrir}, see Figure \ref{DeftoUSp2} for details. 
\begin{figure}[t]
\begin{center}
\begin{tikzpicture}
\filldraw[fill= white] (1,-3) circle [radius=0.1] node[below] {\scriptsize $1$};
\filldraw[fill= white] (2,-3) circle [radius=0.1] node[below] {\scriptsize $r{+}1$};
\filldraw[fill= red] (3,-3) circle [radius=0.1] node[below] {\scriptsize $2r{+}1$};
\filldraw[fill= white] (4,-3) circle [radius=0.1] node[below] {\scriptsize $3r{+}1$};
\filldraw[fill= white] (5,-3) circle [radius=0.1] node[below] {\scriptsize $4r{+}1$};
\filldraw[fill= white] (6,-3) circle [radius=0.1] node[below] {\scriptsize $3r{+}1$};
\filldraw[fill= red] (7,-3) circle [radius=0.1] node[below] {\scriptsize $2r{+}1$};
\filldraw[fill= white] (8,-3) circle [radius=0.1] node[below] {\scriptsize $r{+}1$};
\filldraw[fill= white] (9,-3) circle [radius=0.1] node[below] {\scriptsize $1$};
\filldraw[fill= white] (5,-2) circle [radius=0.1] node[above] {\scriptsize $2r$};
\draw (1.1, -3) -- (1.9,-3) ;
\draw (2.1, -3) -- (2.9,-3) ;
\draw (3.1, -3) -- (3.9,-3) ;
\draw (4.1, -3) -- (4.9,-3) ;
\draw (5.1, -3) -- (5.9,-3) ;
\draw (5, -2.9) -- (5,-2.1) ;
\draw (6.1, -3) -- (6.9,-3) ;
\draw (7.1, -3) -- (7.9,-3) ;
\draw (8.1, -3) -- (8.9,-3) ;
\draw[->, thick] (5,-3.7)--(5,-5.2);
\filldraw[fill= red] (3,-5) circle [radius=0.1] node[above] {\scriptsize $1$};
\filldraw[fill= white] (4,-5) circle [radius=0.1] node[above] {\scriptsize $r{+}1$};
\filldraw[fill= white] (3,-6) circle [radius=0.1] node[below] {\scriptsize $r{+}1$};
\filldraw[fill= white] (4,-6) circle [radius=0.1] node[below] {\scriptsize $2r{+}1$};
\filldraw[fill= red] (2,-6) circle [radius=0.1] node[below] {\scriptsize $1$};
\filldraw[fill= white] (5,-6) circle [radius=0.1] node[below] {\scriptsize $2r$};
\filldraw[fill= white] (6,-6) circle [radius=0.1] node[below] {\scriptsize $2r$};
\filldraw[fill= white] (6,-5) circle [radius=0.1] node[above] {\scriptsize $r$};
\filldraw[fill= white] (7,-6) circle [radius=0.1] node[below] {\scriptsize $r$};
\draw (2.1, -6) -- (2.9,-6) ;
\draw (3.1, -6) -- (3.9,-6) ;
\draw (4.1, -6) -- (4.9,-6) ;
\draw (5.1, -6) -- (5.9,-6) ;
\draw (4, -5.9) -- (4,-5.1) ;
\draw (6.1, -6) -- (6.9,-6) ;
\draw (3.1, -5) -- (3.9,-5) ;
\draw (6, -5.9) -- (6,-5.1) ;
\draw[->, thick] (8,-6)--(9.5,-6);
\filldraw[fill= white] (11,-5) circle [radius=0.1] node[above] {\scriptsize $r$};
\filldraw[fill= white] (10,-6) circle [radius=0.1] node[below] {\scriptsize $r$};
\filldraw[fill= white] (11,-6) circle [radius=0.1] node[below] {\scriptsize $2r$};
\filldraw[fill= white] (12,-6) circle [radius=0.1] node[below] {\scriptsize $2r$};
\filldraw[fill= white] (12,-5) circle [radius=0.1] node[above] {\scriptsize $1$};
\filldraw[fill= white] (13,-6) circle [radius=0.1] node[below] {\scriptsize $2r$};
\filldraw[fill= white] (13,-5) circle [radius=0.1] node[above] {\scriptsize $r$};
\filldraw[fill= white] (14,-6) circle [radius=0.1] node[below] {\scriptsize $r$};
\draw (10.1, -6) -- (10.9,-6) ;
\draw (11.1, -6) -- (11.9,-6) ;
\draw (12.1, -6) -- (12.9,-6) ;
\draw (13.1, -6) -- (13.9,-6) ;
\draw (11, -5.9) -- (11,-5.1) ;
\draw (12, -5.9) -- (12,-5.1) ;
\draw (13, -5.9) -- (13,-5.1) ;
\end{tikzpicture} 
\end{center}
\caption{The FI deformations leading from the magnetic quiver of the 5d SCFT (the second quiver in Figure \ref{DeftoUSpUV}) to the 3d mirror of the $\mathrm{USp}(2r)^2$ gauge theory \eqref{lagrir}.}
\label{DeftoUSp2}
\end{figure}
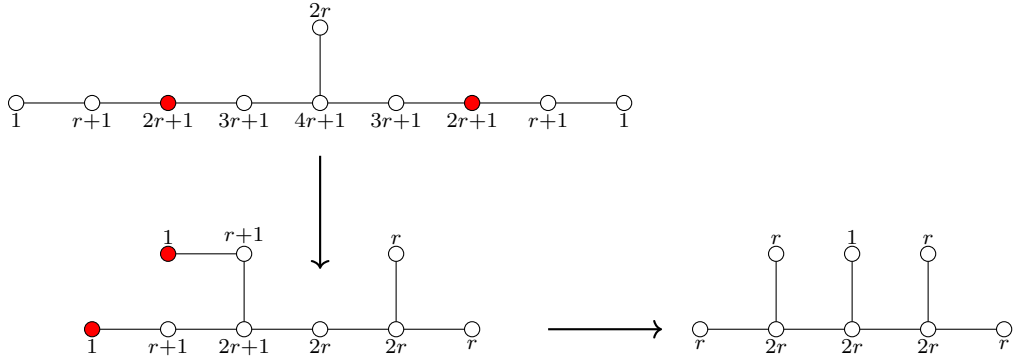
The reduction from 5d to 4d is now implemented by a $\mathbb{Z}_2$ folding of the second magnetic quiver in Figure \ref{DeftoUSpUV}, which leads to the magnetic quiver of the $\mathcal{S}^{(r)}_{E_6,2}$ series, see Table \ref{resulttable}.
\paragraph{$\ess{D_4}{2}{r}$ theories.}
Alternatively, we can turn on two mass deformations as displayed in Figure \ref{DeftoUSp22}, which lead to the UV completion of 
\begin{align}
 \raisebox{-.5\height}{
\begin{tikzpicture}[x=1.5cm,y=.8cm]
\node (f1) at (0,0) [flavour0,label=below:{$\scriptstyle{2}$}] {};
\node (g1) at (1,0) [gauge,label=below:{$\scriptstyle{\mathrm{USp}(2r)}$}] {};
\node (g2) at (2,0) [gauge,label=below:{$\scriptstyle{\mathrm{USp}(2r)}$}] {};
\node (f2) at (3,0) [flavour0,label=below:{$\scriptstyle{2}$}] {};
\draw (f1)--(g1)--(g2)--(f2);
\end{tikzpicture}
} \,.
	\label{eq:2_USp2r_USp2r_2}
\end{align}
The $\mathbb{Z}_2$ folding of the corresponding magnetic quiver (last quiver in Figure \ref{DeftoUSp22}) leads to the magnetic quiver of the $\mathcal{S}^{(r)}_{D_4,2}$ series, see Table \ref{resulttable}. 
\begin{figure}[t]
\begin{center}
\begin{tikzpicture}
\filldraw[fill= white] (1,-3) circle [radius=0.1] node[below] {\scriptsize $1$};
\filldraw[fill= white] (2,-3) circle [radius=0.1] node[below] {\scriptsize $r{+}1$};
\filldraw[fill= red] (3,-3) circle [radius=0.1] node[below] {\scriptsize $2r{+}1$};
\filldraw[fill= white] (4,-3) circle [radius=0.1] node[below] {\scriptsize $3r{+}1$};
\filldraw[fill= white] (5,-3) circle [radius=0.1] node[below] {\scriptsize $4r{+}1$};
\filldraw[fill= white] (6,-3) circle [radius=0.1] node[below] {\scriptsize $3r{+}1$};
\filldraw[fill= red] (7,-3) circle [radius=0.1] node[below] {\scriptsize $2r{+}1$};
\filldraw[fill= white] (8,-3) circle [radius=0.1] node[below] {\scriptsize $r{+}1$};
\filldraw[fill= white] (9,-3) circle [radius=0.1] node[below] {\scriptsize $1$};
\filldraw[fill= white] (5,-2) circle [radius=0.1] node[above] {\scriptsize $2r$};
\draw (1.1, -3) -- (1.9,-3) ;
\draw (2.1, -3) -- (2.9,-3) ;
\draw (3.1, -3) -- (3.9,-3) ;
\draw (4.1, -3) -- (4.9,-3) ;
\draw (5.1, -3) -- (5.9,-3) ;
\draw (5, -2.9) -- (5,-2.1) ;
\draw (6.1, -3) -- (6.9,-3) ;
\draw (7.1, -3) -- (7.9,-3) ;
\draw (8.1, -3) -- (8.9,-3) ;
\draw[->, thick] (5,-3.7)--(5,-5.2);
\filldraw[fill= white] (3,-5) circle [radius=0.1] node[above] {\scriptsize $1$};
\filldraw[fill= white] (4,-5) circle [radius=0.1] node[above] {\scriptsize $r{+}1$};
\filldraw[fill= white] (3,-6) circle [radius=0.1] node[below] {\scriptsize $r{+}1$};
\filldraw[fill= white] (4,-6) circle [radius=0.1] node[below] {\scriptsize $2r{+}1$};
\filldraw[fill= white] (2,-6) circle [radius=0.1] node[below] {\scriptsize $1$};
\filldraw[fill= white] (5,-6) circle [radius=0.1] node[below] {\scriptsize $2r$};
\filldraw[fill= white] (6,-6) circle [radius=0.1] node[below] {\scriptsize $2r$};
\filldraw[fill= red] (6,-5) circle [radius=0.1] node[above] {\scriptsize $r$};
\filldraw[fill= red] (7,-6) circle [radius=0.1] node[below] {\scriptsize $r$};
\draw (2.1, -6) -- (2.9,-6) ;
\draw (3.1, -6) -- (3.9,-6) ;
\draw (4.1, -6) -- (4.9,-6) ;
\draw (5.1, -6) -- (5.9,-6) ;
\draw (4, -5.9) -- (4,-5.1) ;
\draw (6.1, -6) -- (6.9,-6) ;
\draw (3.1, -5) -- (3.9,-5) ;
\draw (6, -5.9) -- (6,-5.1) ;
\draw[->, thick] (8,-6)--(9.5,-6);
\filldraw[fill= white] (11,-5) circle [radius=0.1] node[above] {\scriptsize $1$};
\filldraw[fill= white] (10,-6) circle [radius=0.1] node[below] {\scriptsize $1$};
\filldraw[fill= white] (11,-6) circle [radius=0.1] node[below] {\scriptsize $r{+}1$};
\filldraw[fill= white] (12,-6) circle [radius=0.1] node[below] {\scriptsize $2r{+}1$};
\filldraw[fill= white] (12,-5) circle [radius=0.1] node[above] {\scriptsize $r{+}1$};
\filldraw[fill= white] (13,-6) circle [radius=0.1] node[below] {\scriptsize $2r$};
\filldraw[fill= white] (13,-5) circle [radius=0.1] node[above] {\scriptsize $r$};
\filldraw[fill= white] (14,-6) circle [radius=0.1] node[below] {\scriptsize $r$};
\draw (10.1, -6) -- (10.9,-6) ;
\draw (11.1, -6) -- (11.9,-6) ;
\draw (12.1, -6) -- (12.9,-6) ;
\draw (13.1, -6) -- (13.9,-6) ;
\draw (11.1, -5) -- (11.9,-5) ;
\draw (12, -5.9) -- (12,-5.1) ;
\draw (13, -5.9) -- (13,-5.1) ;
\end{tikzpicture} 
\end{center}
\caption{The FI deformations of the last quiver in Figure \ref{DeftoUSpUV} leading to the 3d mirror of the $\mathrm{USp}(2r)^2$ gauge theory \eqref{eq:2_USp2r_USp2r_2} with 2 flavors on each side.}
\label{DeftoUSp22}
\end{figure}
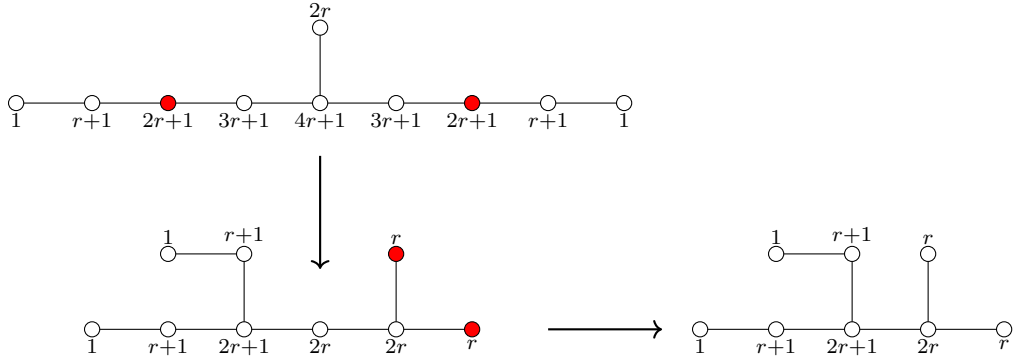
\paragraph{$\ess{A_2}{2}{r}$ theories.}
Finally, the magnetic quiver of the $\mathcal{S}^{(r)}_{A_2,2}$ series arises via a $\mathbb{Z}_2$ folding of the magnetic quiver which describes (a component of) the Higgs branch of the UV completion of 
\begin{align}
 \raisebox{-.5\height}{
\begin{tikzpicture}[x=1.5cm,y=.8cm]
\node (f1) at (0,0) [flavour0,label=below:{$\scriptstyle{1}$}] {};
\node (g1) at (1,0) [gauge,label=below:{$\scriptstyle{\mathrm{USp}(2r)}$}] {};
\node (g2) at (2,0) [gauge,label=below:{$\scriptstyle{\mathrm{USp}(2r)}$}] {};
\node (f2) at (3,0) [flavour0,label=below:{$\scriptstyle{1}$}] {};
\draw (f1)--(g1)--(g2)--(f2);
\end{tikzpicture}
} \,.
	\label{eq:1_USp2r_USp2r_1}
\end{align}
The relevant magnetic quiver for \eqref{eq:1_USp2r_USp2r_1} is given by 
\be\label{USpcft}
\raisebox{-.5\height}{
\begin{tikzpicture}
\filldraw[fill= white] (-1,1) circle [radius=0.1] node[left] {\scriptsize $1$};
\filldraw[fill= white] (0,0) circle [radius=0.1] node[left] {\scriptsize $r{+}1$};
\filldraw[fill= white] (-1,-1) circle [radius=0.1] node[left] {\scriptsize $1$};
\filldraw[fill= white] (1,-1) circle [radius=0.1] node[right] {\scriptsize $r$};
\filldraw[fill= white] (1,1) circle [radius=0.1] node[right] {\scriptsize $r$};
\draw (0.05, 0.05) -- (0.95,0.95) ;
\draw (0.05, -0.05) -- (0.95,-0.95) ;
\draw (-0.05, 0.05) -- (-0.95,0.95) ;
\draw (-0.05, -0.05) -- (-0.95,-0.95) ;
\draw (1,0.9) -- (1,-0.9);
\end{tikzpicture} 
}
\ee
and the $\mathbb{Z}_2$ folding produces the magnetic quiver for $\mathcal{S}^{(r)}_{A_2,2}$, see Table \ref{resulttable}.

\subsection{Case \texorpdfstring{$\ell=3$}{l=3}} 
The magnetic quiver for the Higgs branch at the RG-fixed point of the six-dimensional theory \eqref{su3MQ_electric} is given in \eqref{su3MQ}.
\paragraph{$\ess{D_4}{3}{r}$ theories.}
We now turn on two FI deformations starting from \eqref{su3MQ}. The first involves FI parameters at the nodes $\mathrm{U}(1)$, $\mathrm{U}(4r+2)$, and $\mathrm{U}(4r+3)$ with the constraint $\lambda_1=\lambda_{4r+2}$. In the resulting quiver, we then turn on FI parameters at the nodes $\mathrm{U}(1)$, $\mathrm{U}(3r+1)$, and $\mathrm{U}(3r+2)$, again subject to a constraint which in this case reads $\lambda_1=\lambda_{3r+1}$. Figure \ref{DeftoZ3} details the effects of these deformations.
\begin{figure}[t]
\begin{center}
\begin{tikzpicture}
\filldraw[fill= red] (-1,-1) circle [radius=0.1] node[below] {\scriptsize 1};
 \filldraw[fill= white] (0,-1) circle [radius=0.1] node[below] {\scriptsize 2};
 \filldraw[fill= white] (1,-1) circle [radius=0.1] node[below] {\scriptsize 3};
 \filldraw[fill= white] (2,-1) circle [radius=0.1] node[below] {\scriptsize r+3};
 \filldraw[fill= white] (3,-1) circle [radius=0.1] node[below] {\scriptsize 2r+3};
 \filldraw[fill= white] (4,-1) circle [radius=0.1] node[below] {\scriptsize 3r+3};
 \filldraw[fill= red] (5,-1) circle [radius=0.1] node[below] {\scriptsize 4r+3};
 \filldraw[fill= white] (6,-1) circle [radius=0.1] node[below] {\scriptsize 5r+3};
 \filldraw[fill= white] (7,0) circle [radius=0.1] node[above] {\scriptsize 3r+1};
 \filldraw[fill= white] (7,-1) circle [radius=0.1] node[below] {\scriptsize 6r+3};
 \filldraw[fill= red] (8,-1) circle [radius=0.1] node[below] {\scriptsize 4r+2};
 \filldraw[fill= white] (9,-1) circle [radius=0.1] node[below] {\scriptsize 2r+1};
 \draw (-0.1, -1) -- (-0.9,-1) ;
 \draw (0.1, -1) -- (0.9,-1) ;
 \draw (1.1, -1) -- (1.9,-1) ;
 \draw (2.1, -1) -- (2.9,-1) ;
 \draw (3.1, -1) -- (3.9,-1) ;
 \draw (4.1, -1) -- (4.9,-1) ;
 \draw (5.1, -1) -- (5.9,-1) ;
 \draw (7, -0.1) -- (7,-0.9) ;
 \draw (6.1, -1) -- (6.9,-1) ;
 \draw (7.1, -1) -- (7.9,-1) ;
 \draw (8.1, -1) -- (8.9,-1) ;
\draw[->, thick] (4,-1.5)--(4,-2.5);
\filldraw[fill= red] (0,-3) circle [radius=0.1] node[below] {\scriptsize 1};
\filldraw[fill= white] (1,-3) circle [radius=0.1] node[below] {\scriptsize 2};
\filldraw[fill= white] (2,-3) circle [radius=0.1] node[below] {\scriptsize r+2};
\filldraw[fill= white] (3,-3) circle [radius=0.1] node[below] {\scriptsize 2r+2};
\filldraw[fill= red] (4,-3) circle [radius=0.1] node[below] {\scriptsize 3r+2};
\filldraw[fill= white] (5,-3) circle [radius=0.1] node[below] {\scriptsize 4r+2};
\filldraw[fill= red] (6,-3) circle [radius=0.1] node[below] {\scriptsize 3r+1};
\filldraw[fill= white] (7,-3) circle [radius=0.1] node[below] {\scriptsize 2r+1};
\filldraw[fill= white] (8,-3) circle [radius=0.1] node[below] {\scriptsize r+1};
\filldraw[fill= white] (9,-3) circle [radius=0.1] node[below] {\scriptsize 1};
\filldraw[fill= white] (5,-2) circle [radius=0.1] node[above] {\scriptsize 2r+1};
\draw (0.1, -3) -- (0.9,-3) ;
\draw (1.1, -3) -- (1.9,-3) ;
\draw (2.1, -3) -- (2.9,-3) ;
\draw (3.1, -3) -- (3.9,-3) ;
\draw (4.1, -3) -- (4.9,-3) ;
\draw (5.1, -3) -- (5.9,-3) ;
\draw (5, -2.9) -- (5,-2.1) ;
\draw (6.1, -3) -- (6.9,-3) ;
\draw (7.1, -3) -- (7.9,-3) ;
\draw (8.1, -3) -- (8.9,-3) ;
\draw[->, thick] (5,-3.5)--(5,-4.5);
\filldraw[fill= white] (5,-8) circle [radius=0.1] node[right] {\scriptsize 1};
\filldraw[fill= white] (5,-7) circle [radius=0.1] node[right] {\scriptsize r+1};
\filldraw[fill= white] (3,-5) circle [radius=0.1] node[above] {\scriptsize r+1};
\filldraw[fill= white] (5,-6) circle [radius=0.1] node[right] {\scriptsize 2r+1};
\filldraw[fill= white] (4,-5) circle [radius=0.1] node[above] {\scriptsize 2r+1};
\filldraw[fill= white] (2,-5) circle [radius=0.1] node[above] {\scriptsize 1};
\filldraw[fill= white] (5,-5) circle [radius=0.1] node[above] {\scriptsize 3r+1};
\filldraw[fill= white] (6,-5) circle [radius=0.1] node[above] {\scriptsize 2r+1};
\filldraw[fill= white] (7,-5) circle [radius=0.1] node[above] {\scriptsize r+1};
\filldraw[fill= white] (8,-5) circle [radius=0.1] node[above] {\scriptsize 1};
\draw (2.1, -5) -- (2.9,-5) ;
\draw (3.1, -5) -- (3.9,-5) ;
\draw (4.1, -5) -- (4.9,-5) ;
\draw (5.1, -5) -- (5.9,-5) ;
\draw (5, -5.9) -- (5,-5.1) ;
\draw (6.1, -5) -- (6.9,-5) ;
\draw (7.1, -5) -- (7.9,-5) ;
\draw (5, -6.9) -- (5,-6.1) ;
\draw (5, -7.9) -- (5,-7.1) ;

\end{tikzpicture} 
\end{center}
\caption{The FI deformations leading to the magnetic quiver of the 5d SCFT which gives the quiver of  $\mathcal{S}^{(r)}_{D_4,3}$ theories upon $\mathbb{Z}_3$ folding.}\label{DeftoZ3}
\end{figure}

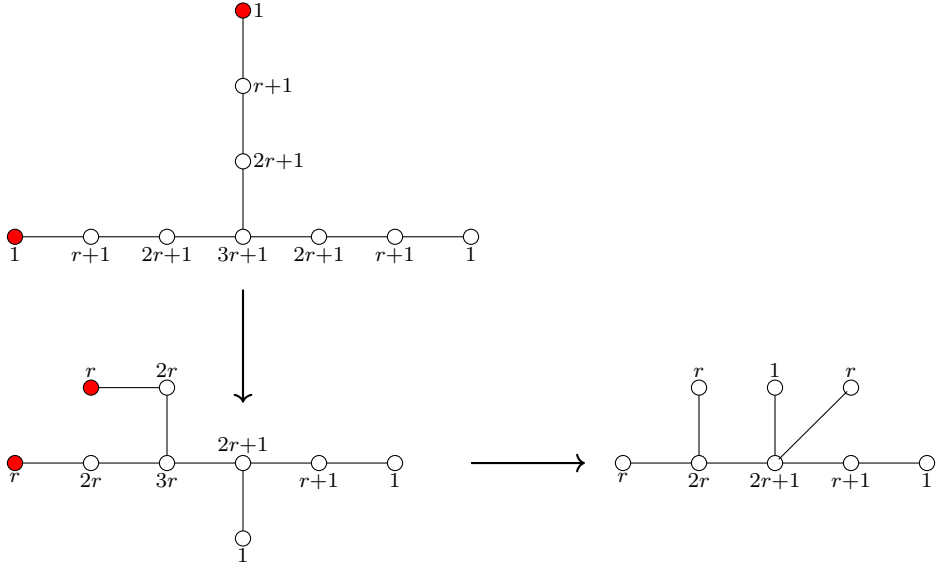
\begin{figure}[t]
\begin{center}
\begin{tikzpicture}
\filldraw[fill= red] (5,0) circle [radius=0.1] node[right] {\scriptsize $1$};
\filldraw[fill= white] (5,-1) circle [radius=0.1] node[right] {\scriptsize $r{+}1$};
\filldraw[fill= white] (3,-3) circle [radius=0.1] node[below] {\scriptsize $r{+}1$};
\filldraw[fill= white] (5,-2) circle [radius=0.1] node[right] {\scriptsize $2r{+}1$};
\filldraw[fill= white] (4,-3) circle [radius=0.1] node[below] {\scriptsize $2r{+}1$};
\filldraw[fill= red] (2,-3) circle [radius=0.1] node[below] {\scriptsize $1$};
\filldraw[fill= white] (5,-3) circle [radius=0.1] node[below] {\scriptsize $3r{+}1$};
\filldraw[fill= white] (6,-3) circle [radius=0.1] node[below] {\scriptsize $2r{+}1$};
\filldraw[fill= white] (7,-3) circle [radius=0.1] node[below] {\scriptsize $r{+}1$};
\filldraw[fill= white] (8,-3) circle [radius=0.1] node[below] {\scriptsize $1$};
\draw (2.1, -3) -- (2.9,-3) ;
\draw (3.1, -3) -- (3.9,-3) ;
\draw (4.1, -3) -- (4.9,-3) ;
\draw (5.1, -3) -- (5.9,-3) ;
\draw (5, -2.9) -- (5,-2.1) ;
\draw (6.1, -3) -- (6.9,-3) ;
\draw (7.1, -3) -- (7.9,-3) ;
\draw (5, -1.9) -- (5,-1.1) ;
\draw (5, -0.9) -- (5,-0.1) ;
\draw[->, thick] (5,-3.7)--(5,-5.2);
\filldraw[fill= red] (3,-5) circle [radius=0.1] node[above] {\scriptsize $r$};
\filldraw[fill= white] (4,-5) circle [radius=0.1] node[above] {\scriptsize $2r$};
\filldraw[fill= white] (3,-6) circle [radius=0.1] node[below] {\scriptsize $2r$};
\filldraw[fill= white] (4,-6) circle [radius=0.1] node[below] {\scriptsize $3r$};
\filldraw[fill= red] (2,-6) circle [radius=0.1] node[below] {\scriptsize $r$};
\filldraw[fill= white] (5,-6) circle [radius=0.1] node[above] {\scriptsize $2r{+}1$};
\filldraw[fill= white] (6,-6) circle [radius=0.1] node[below] {\scriptsize $r{+}1$};
\filldraw[fill= white] (5,-7) circle [radius=0.1] node[below] {\scriptsize $1$};
\filldraw[fill= white] (7,-6) circle [radius=0.1] node[below] {\scriptsize $1$};
\draw (2.1, -6) -- (2.9,-6) ;
\draw (3.1, -6) -- (3.9,-6) ;
\draw (4.1, -6) -- (4.9,-6) ;
\draw (5.1, -6) -- (5.9,-6) ;
\draw (4, -5.9) -- (4,-5.1) ;
\draw (6.1, -6) -- (6.9,-6) ;
\draw (3.1, -5) -- (3.9,-5) ;
\draw (5, -6.9) -- (5,-6.1) ;
\draw[->, thick] (8,-6)--(9.5,-6);
\filldraw[fill= white] (11,-5) circle [radius=0.1] node[above] {\scriptsize $r$};
\filldraw[fill= white] (10,-6) circle [radius=0.1] node[below] {\scriptsize $r$};
\filldraw[fill= white] (11,-6) circle [radius=0.1] node[below] {\scriptsize $2r$};
\filldraw[fill= white] (12,-6) circle [radius=0.1] node[below] {\scriptsize $2r{+}1$};
\filldraw[fill= white] (12,-5) circle [radius=0.1] node[above] {\scriptsize $1$};
\filldraw[fill= white] (13,-6) circle [radius=0.1] node[below] {\scriptsize $r{+}1$};
\filldraw[fill= white] (13,-5) circle [radius=0.1] node[above] {\scriptsize $r$};
\filldraw[fill= white] (14,-6) circle [radius=0.1] node[below] {\scriptsize $1$};
\draw (10.1, -6) -- (10.9,-6) ;
\draw (11.1, -6) -- (11.9,-6) ;
\draw (12.1, -6) -- (12.9,-6) ;
\draw (13.1, -6) -- (13.9,-6) ;
\draw (11, -5.9) -- (11,-5.1) ;
\draw (12, -5.9) -- (12,-5.1) ;
\draw (12.05, -5.95) -- (12.95,-5.05) ;
\end{tikzpicture} 
\end{center}
\caption{The FI deformations leading from the third quiver in Figure \ref{DeftoZ3} to the 3d mirror of the Lagrangian theory \eqref{eq:3_SU2r+1_USp2r_3}.}
\label{Deftolagr3}
\end{figure}

The third quiver in Figure \ref{DeftoZ3} gives the magnetic quiver of  $\mathcal{S}^{(r)}_{D_4,3}$ theories upon $\mathbb{Z}_3$ folding. Moreover, we also notice that the third quiver in Figure \ref{DeftoZ3} describes the UV completion of the Lagrangian theory 
\begin{equation}
\raisebox{-.5\height}{
\begin{tikzpicture}[x=1.5cm,y=.8cm]
\node (f1) at (0,0) [flavour0,label=below:{$\scriptstyle{3}$}] {};
\node (g1) at (1,0) [gauge,label=below:{$\scriptstyle{\mathrm{SU}(2r{+}1)}$}] {};
\node (g2) at (2,0) [gauge,label=below:{$\scriptstyle{\mathrm{USp}(2r)}$}] {};
\node (f2) at (3,0) [flavour0,label=below:{$\scriptstyle{2}$}] {};
\node (fAS) at (1,1) [flavour0,label=above:{$\scriptstyle{1}$}] {};
\draw (f1)--(g1)--(g2)--(f2);
\draw [line join=round,decorate, decoration={zigzag, segment length=4,amplitude=.9,post=lineto,post length=2pt}]  (g1)--(fAS);
\draw (1.15,0.5) node {$\scriptscriptstyle{\Lambda^2}$};
\end{tikzpicture}
}
\label{eq:3_SU2r+1_USp2r_3}
\end{equation}
where the wiggly line in \eqref{eq:3_SU2r+1_USp2r_3} denotes a hypermultiplet in the rank-2 anti-symmetric representation of $\mathrm{SU}(2r+1)$. This can be shown by turning on a sequence of FI deformations as displayed in Figure \ref{Deftolagr3}.

\begin{figure}[t]
\begin{center}
\begin{tikzpicture}
\filldraw[fill= white] (5,0) circle [radius=0.1] node[right] {\scriptsize $1$};
\filldraw[fill= red] (5,-1) circle [radius=0.1] node[right] {\scriptsize $r{+}1$};
\filldraw[fill= red] (3,-3) circle [radius=0.1] node[below] {\scriptsize $r{+}1$};
\filldraw[fill= white] (5,-2) circle [radius=0.1] node[right] {\scriptsize $2r{+}1$};
\filldraw[fill= white] (4,-3) circle [radius=0.1] node[below] {\scriptsize $2r{+}1$};
\filldraw[fill= white] (2,-3) circle [radius=0.1] node[below] {\scriptsize $1$};
\filldraw[fill= white] (5,-3) circle [radius=0.1] node[below] {\scriptsize $3r{+}1$};
\filldraw[fill= white] (6,-3) circle [radius=0.1] node[below] {\scriptsize $2r{+}1$};
\filldraw[fill= white] (7,-3) circle [radius=0.1] node[below] {\scriptsize $r{+}1$};
\filldraw[fill= white] (8,-3) circle [radius=0.1] node[below] {\scriptsize $1$};
\draw (2.1, -3) -- (2.9,-3) ;
\draw (3.1, -3) -- (3.9,-3) ;
\draw (4.1, -3) -- (4.9,-3) ;
\draw (5.1, -3) -- (5.9,-3) ;
\draw (5, -2.9) -- (5,-2.1) ;
\draw (6.1, -3) -- (6.9,-3) ;
\draw (7.1, -3) -- (7.9,-3) ;
\draw (5, -1.9) -- (5,-1.1) ;
\draw (5, -0.9) -- (5,-0.1) ;
\draw[->, thick] (8,-2)--(9.5,-2);
\filldraw[fill= red] (11,-2) circle [radius=0.1] node[above] {\scriptsize $r$};
\filldraw[fill= red] (10,-3) circle [radius=0.1] node[below] {\scriptsize $r$};
\filldraw[fill= white] (11,-3) circle [radius=0.1] node[below] {\scriptsize $2r$};
\filldraw[fill= white] (12,-3) circle [radius=0.1] node[below] {\scriptsize $2r{+}1$};
\filldraw[fill= white] (14,-3) circle [radius=0.1] node[below] {\scriptsize $1$};
\filldraw[fill= white] (13,-3) circle [radius=0.1] node[below] {\scriptsize $r{+}1$};
\filldraw[fill= white] (12,-2) circle [radius=0.1] node[left] {\scriptsize $r{+}1$};
\filldraw[fill= white] (12,-1) circle [radius=0.1] node[left] {\scriptsize $1$};
\filldraw[fill= white] (13,-2) circle [radius=0.1] node[below] {\scriptsize $1$};
\draw (10.1, -3) -- (10.9,-3) ;
\draw (11.1, -3) -- (11.9,-3) ;
\draw (12.1, -3) -- (12.9,-3) ;
\draw (13.1, -3) -- (13.9,-3) ;
\draw (11, -2.9) -- (11,-2.1) ;
\draw (12, -2.9) -- (12,-2.1) ;
\draw (12.1, -2) -- (12.9,-2) ;
\draw (12, -1.9) -- (12,-1.1) ;
\draw[->, thick] (12,-4)--(12,-5.5);
\filldraw[fill= red] (11,-5) circle [radius=0.1] node[above] {\scriptsize $r$};
\filldraw[fill= red] (13,-5) circle [radius=0.1] node[above] {\scriptsize $r$};
\filldraw[fill= white] (12,-6) circle [radius=0.1] node[below] {\scriptsize $2r{+}1$};
\filldraw[fill= white] (14,-6) circle [radius=0.1] node[below] {\scriptsize $1$};
\filldraw[fill= white] (13,-6) circle [radius=0.1] node[below] {\scriptsize $r{+}1$};
\filldraw[fill= white] (11,-6) circle [radius=0.1] node[above] {\scriptsize $r{+}1$};
\filldraw[fill= white] (11,-7) circle [radius=0.1] node[left] {\scriptsize $1$};
\filldraw[fill= white] (10,-6) circle [radius=0.1] node[above] {\scriptsize $1$};
\draw (11.1, -6) -- (11.9,-6) ;
\draw (12.1, -6) -- (12.9,-6) ;
\draw (13.1, -6) -- (13.9,-6) ;
\draw (11, -2.9) -- (11,-2.1) ;
\draw (11, -6.9) -- (11,-6.1) ;
\draw (10.1, -6) -- (10.9,-6) ;
\draw (11.05, -5.05) -- (11.95,-5.95) ;
\draw (12.05, -5.95) -- (12.95,-5.05) ;
\draw[->, thick] (9.5,-6)--(8,-6);
\filldraw[fill= white] (5,-5) circle [radius=0.1] node[above] {\scriptsize $r$};
\filldraw[fill= white] (5,-6) circle [radius=0.1] node[below] {\scriptsize $r{+}1$};
\filldraw[fill= white] (7,-6) circle [radius=0.1] node[below] {\scriptsize $1$};
\filldraw[fill= red] (6,-6) circle [radius=0.1] node[below] {\scriptsize $r{+}1$};
\filldraw[fill= red] (4,-6) circle [radius=0.1] node[below] {\scriptsize $r{+}1$};
\filldraw[fill= white] (4,-5) circle [radius=0.1] node[left] {\scriptsize $1$};
\filldraw[fill= white] (3,-6) circle [radius=0.1] node[below] {\scriptsize $1$};
\draw (4.1, -6) -- (4.9,-6) ;
\draw (5.1, -6) -- (5.9,-6) ;
\draw (6.1, -6) -- (6.9,-6) ;
\draw (4, -5.9) -- (4,-5.1) ;
\draw (3.1, -6) -- (3.9,-6) ;
\draw (5.05, -5.05) -- (5.95,-5.95) ;
\draw (4.05, -5.95) -- (4.95,-5.05) ;
\draw[->, thick] (5,-7)--(5,-8.5);
\filldraw[fill= white] (5,-9) circle [radius=0.1] node[left] {\scriptsize $r$};
\filldraw[fill= white] (6,-9) circle [radius=0.1] node[below] {\scriptsize $r{+}1$};
\filldraw[fill= white] (7,-9) circle [radius=0.1] node[right] {\scriptsize $1$};
\filldraw[fill= white] (7,-10) circle [radius=0.1] node[right] {\scriptsize $1$};
\filldraw[fill= white] (7,-8) circle [radius=0.1] node[right] {\scriptsize $1$};
\draw (6.1, -9) -- (6.9,-9) ;
\draw (5.1, -8.95) -- (5.9,-8.95) ;
\draw (5.1, -9.05) -- (5.9,-9.05) ;
\draw (6.05, -9.05) -- (6.95,-9.95) ;
\draw (6.05, -8.95) -- (6.95,-8.05) ;
\end{tikzpicture} 
\end{center}
\caption{The FI deformations leading from the third quiver in Figure \ref{DeftoZ3} to the magnetic quiver of the 5d SCFT which gives the quiver of  $\mathcal{S}^{(r)}_{A_1,3}$ theories upon $\mathbb{Z}_3$ folding.}
\label{Deftoh1}
\end{figure}
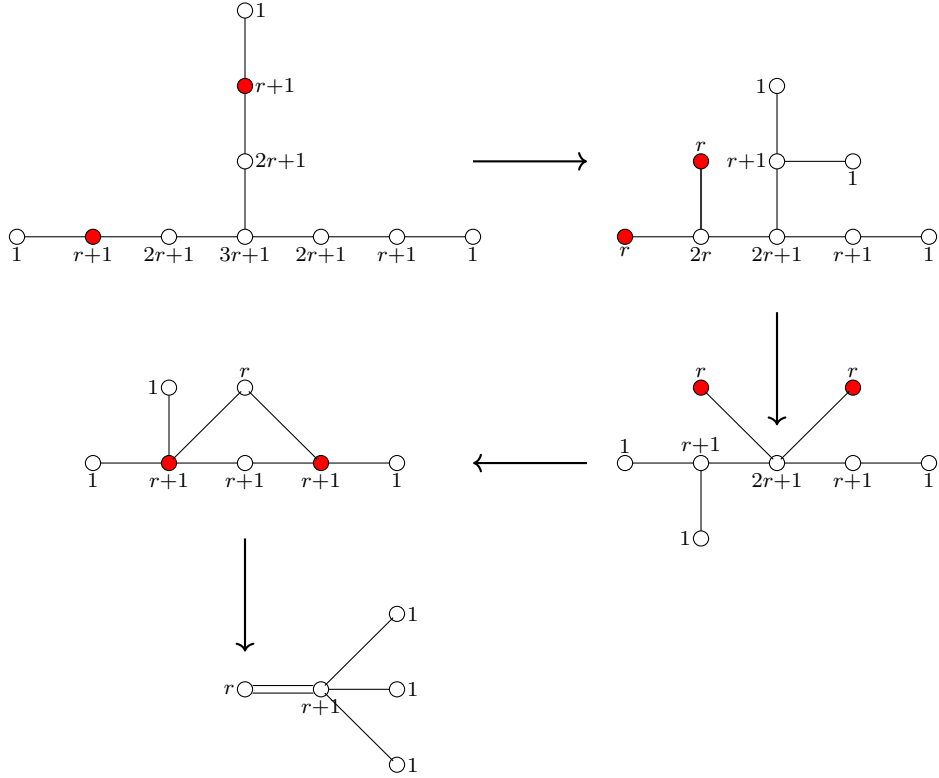

\paragraph{$\ess{A_1}{3}{r}$ theories.}
The $\mathcal{S}^{(r)}_{A_1,3}$ theories can be obtained by mass deforming the $\mathcal{S}^{(r)}_{D_4,3}$ theories and, similarly to the $\ell=2$ case, we can mass deform the 5d theory and then perform a $\mathbb{Z}_3$ folding of the corresponding magnetic quiver. Again, the relevant quiver can be found starting from the first quiver in Figure \ref{Deftolagr3} and activating FI deformations. The sequence of FI deformations is rather elaborate and is presented in Figure \ref{Deftoh1}. 
 We point out that the 5d SCFT described by the last quiver in Figure \ref{Deftoh1} flows in the infrared, upon mass deformation, to the Lagrangian theory 
\begin{equation}
\raisebox{-.5\height}{
\begin{tikzpicture}[x=1.5cm,y=.8cm]
\node (f1) at (0,0) [flavour0,label=below:{$\scriptstyle{1}$}] {};
\node (g1) at (1,0) [gauge,label=below:{$\scriptstyle{\mathrm{SU}(2r{+}1)}$}] {};
\node (g2) at (2,0) [gauge,label=below:{$\scriptstyle{\mathrm{USp}(2r)}$}] {};
\node (fAS) at (1,1) [flavour0,label=above:{$\scriptstyle{1}$}] {};
\draw (f1)--(g1)--(g2);
\draw [line join=round,decorate, decoration={zigzag, segment length=4,amplitude=.9,post=lineto,post length=2pt}]  (g1)--(fAS);
\draw (1.15,0.5) node {$\scriptscriptstyle{\Lambda^2}$};
\end{tikzpicture}
} \,.
\end{equation}

\subsection{Case \texorpdfstring{$\ell=4$}{l=4}}
The magnetic quiver associated to the Higgs branch at the UV fixed point of the six-dimensional theory \eqref{su3MQ2_electric} is given by \eqref{su3MQ2}.

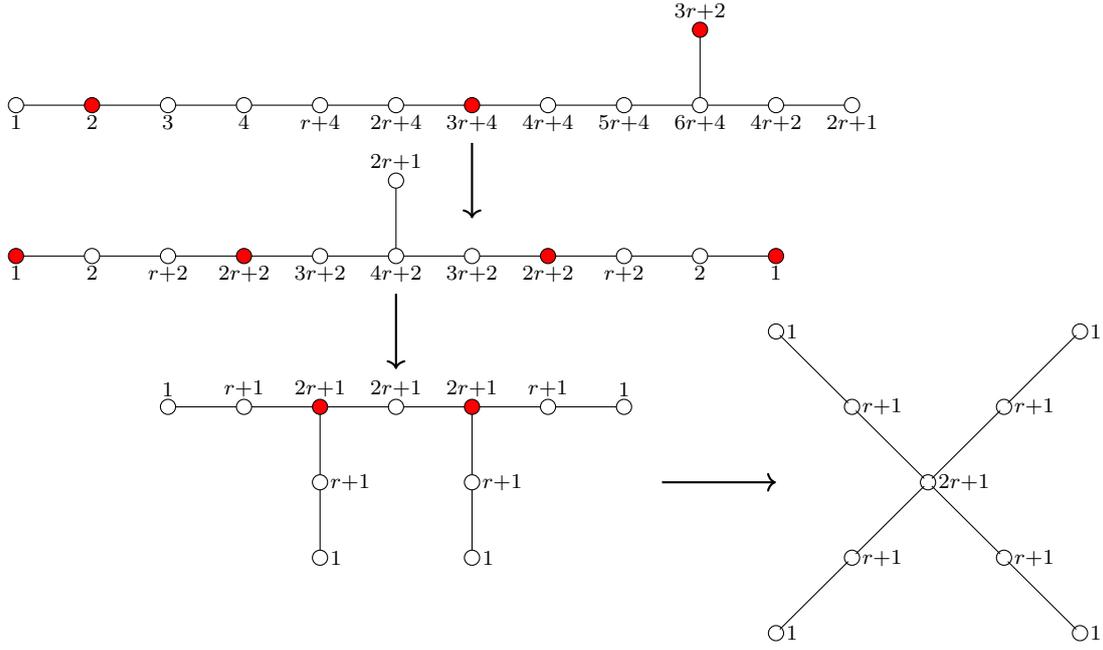
\begin{figure}[t]
\begin{center}
\begin{tikzpicture}
\filldraw[fill= white] (0,-2) circle [radius=0.1] node[below] {\scriptsize $1$};
 \filldraw[fill= red] (1,-2) circle [radius=0.1] node[below] {\scriptsize $2$};
 \filldraw[fill= white] (2,-2) circle [radius=0.1] node[below] {\scriptsize $3$};
 \filldraw[fill= white] (3,-2) circle [radius=0.1] node[below] {\scriptsize $4$};
 \filldraw[fill= white] (4,-2) circle [radius=0.1] node[below] {\scriptsize $r{+}4$};
 \filldraw[fill= white] (5,-2) circle [radius=0.1] node[below] {\scriptsize $2r{+}4$};
 \filldraw[fill= red] (6,-2) circle [radius=0.1] node[below] {\scriptsize $3r{+}4$};
 \filldraw[fill= white] (7,-2) circle [radius=0.1] node[below] {\scriptsize $4r{+}4$};
 \filldraw[fill= white] (8,-2) circle [radius=0.1] node[below] {\scriptsize $5r{+}4$};
 \filldraw[fill= red] (9,-1) circle [radius=0.1] node[above] {\scriptsize $3r{+}2$};
 \filldraw[fill= white] (9,-2) circle [radius=0.1] node[below] {\scriptsize $6r{+}4$};
 \filldraw[fill= white] (10,-2) circle [radius=0.1] node[below] {\scriptsize $4r{+}2$};
 \filldraw[fill= white] (11,-2) circle [radius=0.1] node[below] {\scriptsize $2r{+}1$};
 \draw (10.1, -2) -- (10.9,-2) ;
 \draw (9.1, -2) -- (9.9,-2) ;
 \draw (0.1, -2) -- (0.9,-2) ;
 \draw (1.1, -2) -- (1.9,-2) ;
 \draw (2.1, -2) -- (2.9,-2) ;
 \draw (3.1, -2) -- (3.9,-2) ;
 \draw (4.1, -2) -- (4.9,-2) ;
 \draw (5.1, -2) -- (5.9,-2) ;
 \draw (9, -1.1) -- (9,-1.9) ;
 \draw (6.1, -2) -- (6.9,-2) ;
 \draw (7.1, -2) -- (7.9,-2) ;
 \draw (8.1, -2) -- (8.9,-2) ;
\draw[->, thick] (6,-2.5)--(6,-3.5);
\filldraw[fill= red] (0,-4) circle [radius=0.1] node[below] {\scriptsize $1$};
\filldraw[fill= white] (1,-4) circle [radius=0.1] node[below] {\scriptsize $2$};
\filldraw[fill= white] (2,-4) circle [radius=0.1] node[below] {\scriptsize $r{+}2$};
\filldraw[fill= red] (3,-4) circle [radius=0.1] node[below] {\scriptsize $2r{+}2$};
\filldraw[fill= white] (4,-4) circle [radius=0.1] node[below] {\scriptsize $3r{+}2$};
\filldraw[fill= white] (5,-4) circle [radius=0.1] node[below] {\scriptsize $4r{+}2$};
\filldraw[fill= white] (6,-4) circle [radius=0.1] node[below] {\scriptsize $3r{+}2$};
\filldraw[fill= red] (7,-4) circle [radius=0.1] node[below] {\scriptsize $2r{+}2$};
\filldraw[fill= white] (8,-4) circle [radius=0.1] node[below] {\scriptsize $r{+}2$};
\filldraw[fill= white] (9,-4) circle [radius=0.1] node[below] {\scriptsize $2$};
\filldraw[fill= red] (10,-4) circle [radius=0.1] node[below] {\scriptsize $1$};
\filldraw[fill= white] (5,-3) circle [radius=0.1] node[above] {\scriptsize $2r{+}1$};
\draw (0.1, -4) -- (0.9,-4) ;
\draw (1.1, -4) -- (1.9,-4) ;
\draw (2.1, -4) -- (2.9,-4) ;
\draw (3.1, -4) -- (3.9,-4) ;
\draw (4.1, -4) -- (4.9,-4) ;
\draw (5.1, -4) -- (5.9,-4) ;
\draw (5, -3.9) -- (5,-3.1) ;
\draw (6.1, -4) -- (6.9,-4) ;
\draw (7.1, -4) -- (7.9,-4) ;
\draw (8.1, -4) -- (8.9,-4) ;
\draw (9.1, -4) -- (9.9,-4) ;
\draw[->, thick] (5,-4.5)--(5,-5.5);
\filldraw[fill= white] (4,-8) circle [radius=0.1] node[right] {\scriptsize $1$};
\filldraw[fill= white] (4,-7) circle [radius=0.1] node[right] {\scriptsize $r{+}1$};
\filldraw[fill= white] (3,-6) circle [radius=0.1] node[above] {\scriptsize $r{+}1$};
\filldraw[fill= white] (6,-8) circle [radius=0.1] node[right] {\scriptsize $1$};
\filldraw[fill= white] (6,-7) circle [radius=0.1] node[right] {\scriptsize $r{+}1$};
\filldraw[fill= red] (4,-6) circle [radius=0.1] node[above] {\scriptsize $2r{+}1$};
\filldraw[fill= white] (2,-6) circle [radius=0.1] node[above] {\scriptsize $1$};
\filldraw[fill= white] (5,-6) circle [radius=0.1] node[above] {\scriptsize $2r{+}1$};
\filldraw[fill= red] (6,-6) circle [radius=0.1] node[above] {\scriptsize $2r{+}1$};
\filldraw[fill= white] (7,-6) circle [radius=0.1] node[above] {\scriptsize $r{+}1$};
\filldraw[fill= white] (8,-6) circle [radius=0.1] node[above] {\scriptsize $1$};
\draw (2.1, -6) -- (2.9,-6) ;
\draw (3.1, -6) -- (3.9,-6) ;
\draw (4.1, -6) -- (4.9,-6) ;
\draw (5.1, -6) -- (5.9,-6) ;
\draw (4, -6.9) -- (4,-6.1) ;
\draw (6.1, -6) -- (6.9,-6) ;
\draw (7.1, -6) -- (7.9,-6) ;
\draw (4, -7.9) -- (4,-7.1) ;
\draw (6, -6.9) -- (6,-6.1) ;
\draw (6, -7.9) -- (6,-7.1) ;
\draw[->, thick] (8.5,-7)--(10,-7);
\filldraw[fill= white] (10,-5) circle [radius=0.1] node[right] {\scriptsize $1$};
\filldraw[fill= white] (11,-6) circle [radius=0.1] node[right] {\scriptsize $r{+}1$};
\filldraw[fill= white] (11,-8) circle [radius=0.1] node[right] {\scriptsize $r{+}1$};
\filldraw[fill= white] (10,-9) circle [radius=0.1] node[right] {\scriptsize $1$};
\filldraw[fill= white] (13,-8) circle [radius=0.1] node[right] {\scriptsize $r{+}1$};
\filldraw[fill= white] (12,-7) circle [radius=0.1] node[right] {\scriptsize $2r{+}1$};
\filldraw[fill= white] (14,-9) circle [radius=0.1] node[right] {\scriptsize $1$};
\filldraw[fill= white] (13,-6) circle [radius=0.1] node[right] {\scriptsize $r{+}1$};
\filldraw[fill= white] (14,-5) circle [radius=0.1] node[right] {\scriptsize $1$};
\draw (12.05, -6.95) -- (12.95,-6.05) ;
\draw (13.05, -5.95) -- (13.95,-5.05) ;
\draw (12.05, -7.05) -- (12.95,-7.95) ;
\draw (13.05, -8.05) -- (13.95,-8.95) ;
\draw (11.05, -7.95) -- (11.95,-7.05) ;
\draw (10.05, -5.05) -- (10.95,-5.95) ;
\draw (11.05, -6.05) -- (11.95,-6.95) ;
\draw (10.05, -8.95) -- (10.95,-8.05) ;
\end{tikzpicture} 
\end{center}
\caption{The FI deformations leading to a magnetic quiver which, upon $\mathbb{Z}_4$ folding, leads to the  magnetic quiver of the $\mathcal{S}^{(r)}_{A_2,4}$ theories, see Table \ref{resulttable}.}
\label{DeftoZ4}
\end{figure}

\paragraph{$\mathcal{S}^{(r)}_{A_2,4}$ theories.}
We turn on FI parameters at the nodes $\mathrm{U}(2)$, $\mathrm{U}(3r+4)$ and $\mathrm{U}(3r+2)$ subject to the constraint $\lambda_2=\lambda_{3r+2}$. We then turn on another FI deformation in the resulting quiver as depicted in Figure \ref{DeftoZ4}.
The FI deformation at the two abelian nodes and at the two $\mathrm{U}(2r+2)$ nodes satisfy the constraints $\lambda_1=\lambda_{2r+2}$ and also $\lambda'_{1}=\lambda'_{2r+2}$. Finally, a $\mathbb{Z}_4$ folding of the last quiver in Figure \ref{DeftoZ4} gives the
magnetic quiver of  $\mathcal{S}^{(r)}_{A_2,4}$ theories. 

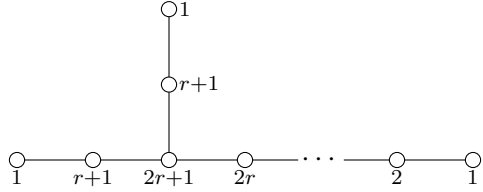
\begin{figure}[t]
\begin{center}
\begin{tikzpicture}
\filldraw[fill= white] (4,-4) circle [radius=0.1] node[right] {\scriptsize $1$};
\filldraw[fill= white] (4,-5) circle [radius=0.1] node[right] {\scriptsize $r{+}1$};
\filldraw[fill= white] (3,-6) circle [radius=0.1] node[below] {\scriptsize $r{+}1$};
\filldraw[fill= white] (4,-6) circle [radius=0.1] node[below] {\scriptsize $2r{+}1$};
\filldraw[fill= white] (2,-6) circle [radius=0.1] node[below] {\scriptsize $1$};
\filldraw[fill= white] (5,-6) circle [radius=0.1] node[below] {\scriptsize $2r$};
\node[](L) at (6,-6){$\cdots$};
\filldraw[fill= white] (7,-6) circle [radius=0.1] node[below] {\scriptsize $2$};
\filldraw[fill= white] (8,-6) circle [radius=0.1] node[below] {\scriptsize $1$};
\draw (2.1, -6) -- (2.9,-6) ;
\draw (3.1, -6) -- (3.9,-6) ;
\draw (4.1, -6) -- (4.9,-6) ;
\draw (5.1, -6) -- (5.7,-6) ;
\draw (4, -5.9) -- (4,-5.1) ;
\draw (6.3, -6) -- (6.9,-6) ;
\draw (7.1, -6) -- (7.9,-6) ;
\draw (4, -4.9) -- (4,-4.1) ;
\end{tikzpicture}  
\end{center}
\caption{The magnetic quiver for the 5d SCFT with $\mathrm{SO}(4r+6)\times \mathrm{U}(1)$ global symmetry. The last quiver in Figure \ref{DeftoZ4} describes an $\mathrm{SU}(2r+1)$ gauging of two copies of this theory.}
\label{So4r+6}
\end{figure}

The 5d theory described by the last quiver in Figure \ref{DeftoZ4} can be interpreted as an $\mathrm{SU}(2r+1)$ vector multiplet coupled to two copies of a SCFT with global symmetry $\mathrm{SO}(4r+6)\times \mathrm{U}(1)$, whose magnetic quiver is given in Figure \ref{So4r+6}. 

\begin{figure}[t]
\begin{center}
\begin{tikzpicture}
\filldraw[fill= red] (4,-4) circle [radius=0.1] node[right] {\scriptsize $1$};
\filldraw[fill= white] (4,-5) circle [radius=0.1] node[right] {\scriptsize $r{+}1$};
\filldraw[fill= white] (3,-6) circle [radius=0.1] node[below] {\scriptsize $r{+}1$};
\filldraw[fill= white] (4,-6) circle [radius=0.1] node[below] {\scriptsize $2r{+}1$};
\filldraw[fill= red] (2,-6) circle [radius=0.1] node[below] {\scriptsize $1$};
\filldraw[fill= white] (5,-6) circle [radius=0.1] node[below] {\scriptsize $2r$};
\node[](L) at (6,-6){$\cdots$};
\filldraw[fill= white] (7,-6) circle [radius=0.1] node[below] {\scriptsize $2$};
\filldraw[fill= white] (8,-6) circle [radius=0.1] node[below] {\scriptsize $1$};
\draw (2.1, -6) -- (2.9,-6) ;
\draw (3.1, -6) -- (3.9,-6) ;
\draw (4.1, -6) -- (4.9,-6) ;
\draw (5.1, -6) -- (5.7,-6) ;
\draw (4, -5.9) -- (4,-5.1) ;
\draw (6.3, -6) -- (6.9,-6) ;
\draw (7.1, -6) -- (7.9,-6) ;
\draw (4, -4.9) -- (4,-4.1) ;
\draw[->, thick] (8,-5.5)--(9.5,-5.5);
\filldraw[fill= white] (12,-5) circle [radius=0.1] node[right] {\scriptsize $1$};
\filldraw[fill= white] (11,-5) circle [radius=0.1] node[right] {\scriptsize $r$};
\filldraw[fill= white] (10,-6) circle [radius=0.1] node[below] {\scriptsize $r$};
\filldraw[fill= white] (11,-6) circle [radius=0.1] node[below] {\scriptsize $2r$};
\filldraw[fill= white] (12,-6) circle [radius=0.1] node[below] {\scriptsize $2r$};
\node[](L) at (13,-6){$\cdots$};
\filldraw[fill= white] (14,-6) circle [radius=0.1] node[below] {\scriptsize $2$};
\filldraw[fill= white] (15,-6) circle [radius=0.1] node[below] {\scriptsize $1$};
\draw (10.1, -6) -- (10.9,-6) ;
\draw (11.1, -6) -- (11.9,-6) ;
\draw (12.1, -6) -- (12.7,-6) ;
\draw (11, -5.9) -- (11,-5.1) ;
\draw (13.3, -6) -- (13.9,-6) ;
\draw (14.1, -6) -- (14.9,-6) ;
\draw (12, -5.9) -- (12,-5.1) ;
\end{tikzpicture}  
\end{center}
\caption{The FI deformation leading from the magnetic quiver of the $\mathrm{SO}(4r+6)\times \mathrm{U}(1)$ SCFT to that of the low-energy theory $\mathrm{USp}(2r)$ with $2r+3$ fundamentals.}
\label{Deftolagr4}
\end{figure}
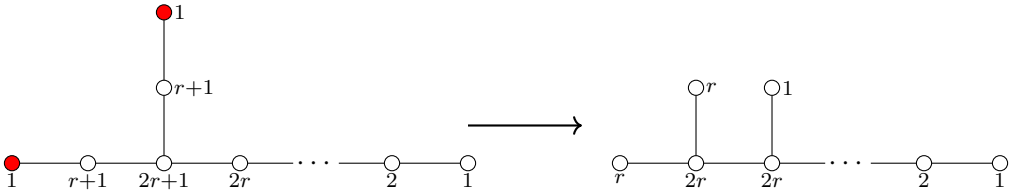

\noindent This theory is in turn known to describe the UV completion of $\mathrm{USp}(2r)$ SQCD with $2r+3$ flavors. The corresponding deformation leading to the Lagrangian theory can also be understood in terms of FI deformations, as depicted in Figure \ref{Deftolagr4}. We therefore conclude that by activating this deformation twice (one for each copy of the $\mathrm{SO}(4r+6)\times \mathrm{U}(1)$ SCFT) the 5d theory described by the last quiver in Figure \ref{DeftoZ4} flows in the infrared to the Lagrangian theory 
\begin{equation}
\label{eq:USp2r_SU2r+1_USp2r}
\raisebox{-.5\height}{
\begin{tikzpicture}[x=1.5cm,y=.8cm]
\node (f1) at (-1,0) [flavour0,label=below:{$\scriptstyle{2}$}] {};
\node (g0) at (0,0) [gauge,label=below:{$\scriptstyle{\mathrm{USp}(2r)}$}] {};
\node (g1) at (1,0) [gauge,label=below:{$\scriptstyle{\mathrm{SU}(2r{+}1)}$}] {};
\node (g2) at (2,0) [gauge,label=below:{$\scriptstyle{\mathrm{USp}(2r)}$}] {};
\node (f2) at (3,0) [flavour0,label=below:{$\scriptstyle{2}$}] {};
\draw (f1)--(g0)--(g1)--(g2)--(f2);
\end{tikzpicture}
} \,.
\end{equation}

\subsection{FI-deformations in brane systems}
In this subsection, we show that all the FI deformations admit a straightforward realisation in the Type IIA brane system. Hence, giving further evidence that the quiver subtraction approach to the deformation analysis is justified.

As specific example, we demonstrate the procedure on the $\ess{E_6}{2}{r}$ case. The brane configuration for the magnetic quiver \eqref{sukMQ} is given by
\begin{align}
\raisebox{-.5\height}{
\begin{tikzpicture}
\foreach \i in {0,...,9} {
\draw (\i,-1)--(\i,1); }
\draw[thick,dotted] (10,-1)--(10,1);
\draw (0,-0.5)--(1,-0.5) 
(1,0)--(2,0)
(2,-0.5)--(3,-0.5) 
(3,0)--(4,0)
(4,-0.5)--(5,-0.5) 
(5,0)--(6,0)
(6,-0.5)--(7,-0.5) 
(7,0)--(8,0)
(8,-0.5)--(9,-0.5);
\draw (8,0.5) .. controls (10.5,0.4) .. (9,0.3);
\node[circle, draw, fill=white] at (10,0.65){};
	\node[cross out, draw] at (10,0.65){};
\draw (0.5,-0.25) node {$\scriptstyle{1}$};
\draw (1.5,0.25) node {$\scriptstyle{2}$};
\draw (2.5,-0.25) node {$\scriptstyle{r{+}2}$};
\draw (3.5,0.25) node {$\scriptstyle{2r{+}2}$};
\draw (4.5,-0.25) node {$\scriptstyle{3r{+}2}$};
\draw (5.5,0.25) node {$\scriptstyle{4r{+}2}$};
\draw (6.5,-0.25) node {$\scriptstyle{5r{+}2}$};
\draw (7.5,0.25) node {$\scriptstyle{6r{+}2}$};
\draw (8.5,-0.25) node {$\scriptstyle{3r{+}1}$};
\draw (8.5,0.75) node {$\scriptstyle{4r{+}1}$};
\draw (10.25,1) node {$\scriptstyle{2r}$};
\end{tikzpicture}
}
\label{eq:6d_branes_fixed_pt}
\end{align}
where vertical solid lines represent D8 branes, horizontal solid lines are D6 branes, circles with crosses are NS5 branes, and the dashed vertical line is a O8${}^-$ orientifold. The numbers above solid vertical lines indicate the number of D6 branes which are suspended between D8 branes in each interval. The number $2r$ for the NS5 branes denotes how many half NS5 branes are stuck on the orientifold.
As detailed in \cite{Cabrera:2019izd}, the conformal fixed point is realized once the NS5 branes are on the orientifold, as in \eqref{eq:6d_branes_fixed_pt}.
Recalling that a FI deformation for an electric theory is given by NS5 branes moving away, the corresponding deformation in the magnetic theory is given by D6 branes that are moved away.
Thus, one can realize the FI deformations of Figure \ref{DeftoUSpUV} via
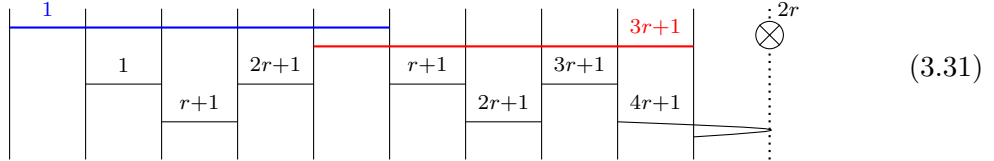
\begin{align}
\raisebox{-.5\height}{
\begin{tikzpicture}
\foreach \i in {0,...,9} {
\draw (\i,-1)--(\i,1); }
\draw[thick,dotted] (10,-1)--(10,1);
\draw 
(1,0)--(2,0)
(2,-0.5)--(3,-0.5) 
(3,0)--(4,0)
(5,0)--(6,0)
(6,-0.5)--(7,-0.5) 
(7,0)--(8,0)
;
\draw (8,-0.5) .. controls (10.5,-0.6) .. (9,-0.7);
\draw[red,thick] (4,0.5)--(9,0.5);
\draw[blue,thick] (0,0.75) -- (5,0.75);
\node[circle, draw, fill=white] at (10,0.65){};
	\node[cross out, draw] at (10,0.65){};
\draw[blue] (0.5,1) node {$\scriptstyle{1}$};
\draw (1.5,0.25) node {$\scriptstyle{1}$};
\draw (2.5,-0.25) node {$\scriptstyle{r{+}1}$};
\draw (3.5,0.25) node {$\scriptstyle{2r{+}1}$};
\draw (5.5,0.25) node {$\scriptstyle{r{+}1}$};
\draw (6.5,-0.25) node {$\scriptstyle{2r{+}1}$};
\draw (7.5,0.25) node {$\scriptstyle{3r{+}1}$};
\draw[red] (8.5,0.75) node {$\scriptstyle{3r{+}1}$};
\draw (8.5,-0.25) node {$\scriptstyle{4r{+}1}$};
\draw (10.25,1) node {$\scriptstyle{2r}$};
\end{tikzpicture}
}
\end{align}
where the alignment of D6 branes in blue corresponds to the FI deformation between the $\mathrm{U}(1)$ and $\mathrm{U}(3r+2)$ node, while the stack of $(3r+1)$ D6 branes in red give the deformation between the $\mathrm{U}(3r+2)$ and $\mathrm{U}(3r+1)$ magnetic gauge node. Then, reading the magnetic quiver from the Higgs branch degrees of freedom reproduces the correct magnetic quiver after FI deformation. 

Thereafter, the FI deformation of Figure \ref{DeftoUSp2} is realized in the brane configuration as follows:
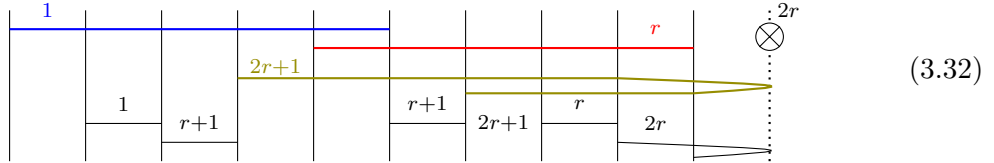
\begin{align}
\raisebox{-.5\height}{
\begin{tikzpicture}
\foreach \i in {0,...,9} {
\draw (\i,-1)--(\i,1); }
\draw[thick,dotted] (10,-1)--(10,1);
\draw 
(1,-0.5)--(2,-0.5)
(2,-0.75)--(3,-0.75) 
(5,-0.5)--(6,-0.5)
(7,-0.5)--(8,-0.5)
;
\draw (8,-0.75) .. controls (10.5,-0.85) .. (9,-0.95);
\draw[red,thick] (4,0.5)--(9,0.5);
\draw[blue,thick] (0,0.75) -- (5,0.75);
\draw[olive,thick] (8,0.1) .. controls (10.5,0) .. (9,-0.1);
\draw[olive,thick] (3,0.1) -- (8,0.1) (6,-0.1)--(9,-0.1);
\node[circle, draw, fill=white] at (10,0.65){};
	\node[cross out, draw] at (10,0.65){};
\draw[blue] (0.5,1) node {$\scriptstyle{1}$};
\draw (1.5,-0.25) node {$\scriptstyle{1}$};
\draw (2.5,-0.5) node {$\scriptstyle{r{+}1}$};
\draw[olive] (3.5,0.25) node {$\scriptstyle{2r{+}1}$};
\draw (5.5,-0.25) node {$\scriptstyle{r{+}1}$};
\draw (6.5,-0.5) node {$\scriptstyle{2r{+}1}$};
\draw (7.5,-0.25) node {$\scriptstyle{r}$};
\draw[red] (8.5,0.75) node {$\scriptstyle{r}$};
\draw (8.5,-0.5) node {$\scriptstyle{2r}$};
\draw (10.25,1) node {$\scriptstyle{2r}$};
\end{tikzpicture}
}
\end{align}
wherein the aligned stack of $(2r+1)$ D6 branes in olive are due to the chosen FI deformation between the two $\mathrm{U}(2r+1)$ magnetic gauge nodes. Note that for this alignment, the red stack of D6 branes had to be split. Reading off the magnetic quiver from the Higgs branch degrees of freedom, again, reproduces the magnetic quiver found in the analysis above.

For the other cases, a straightforward analogous analysis on the brane systems can be preformed.

\section{6d twisted \texorpdfstring{$S^1$}{S1} compactifications, brane webs and folding}
\label{sectionBW}
In this section, we present an alternative derivation of the magnetic quivers for the S-folds. It is similar in spirit to the route followed in Section \ref{SectionFI}, but is implemented using other string theory backgrounds, and in particular the formalism of five-brane webs. 

We first recall that the $\mathcal{S}^{(r)}_{G,\ell}$ theories can be constructed by torus compactification from certain 6d SCFTs. On their tensor branches, these theories can be described by the quiver \eqref{eq:6d_theory_ess}, which for convenience we recall here
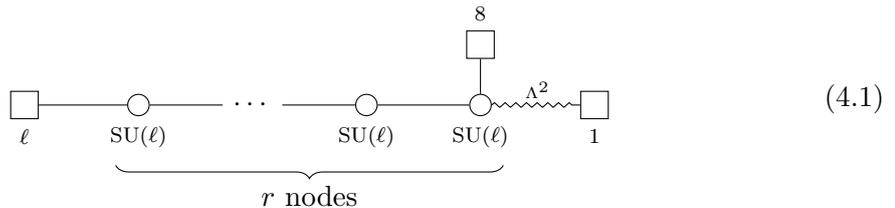
\begin{equation}
\label{6dquiver}
\raisebox{-.5\height}{
\begin{tikzpicture}[x=1.5cm,y=.8cm]
\node (g0) at (4,1) [flavour0,label=above:{$\scriptstyle{8}$}] {};
\node (g1) at (4,0) [gauge,label=below:{$\scriptstyle{\mathrm{SU}(\ell)}$}] {};
\node (g2) at (3,0) [gauge,label=below:{$\scriptstyle{\mathrm{SU}(\ell)}$}] {};
\node (g3) at (2,0) {$\cdots$};
\node (g4) at (1,0) [gauge,label=below:{$\scriptstyle{\mathrm{SU}(\ell)}$}] {};
\node (g5) at (0,0) [flavour0,label=below:{$\scriptstyle{\ell}$}] {};
\node (fAS) at (5,0) [flavour0,label=below:{$\scriptstyle{1}$}] {};
\draw (g0)--(g1)--(g2)--(g3)--(g4)--(g5);
	\draw [line join=round,decorate, decoration={zigzag, segment length=4,amplitude=.9,post=lineto,post length=2pt}]  (g1)--(fAS);
	\draw (4.5,0.25) node {$\scriptscriptstyle{\Lambda^2}$};
\draw [decorate,decoration={brace,amplitude=5pt,mirror,raise=0pt}]
(.8,-1) -- (4.2,-1) node [black,midway,yshift=-.4cm] {$r$ nodes};
\end{tikzpicture}
}
\end{equation}
for $\ell=2,3,4$. In order to obtain the magnetic quivers for the various $\mathcal{S}^{(r)}_{G,\ell}$ theories, we proceed in four steps \cite{Hayashi:2015fsa,Zafrir:2015rga,Ohmori:2018ona} : 
\begin{compactitem}
    \item First, one compactifies the 6d theory above on a circle and provide a description of the subsequent 5d theory, either in the form of a quiver, if available, or more generally in the form of a brane web. 
    \item The 5d theory obtained this way is, by definition, marginal (has a 6d UV completion). One then mass-deforms it in $\mathbb{Z}_{\ell}$ symmetric ways, in effect decoupling $s \ell$ hypermultiplets for some integer $s \geq 1$. 
    \item The resulting 5d theories can be brought to the SCFT point, where they develop a Higgs branch for which we can compute a magnetic quiver (see Appendix \ref{AppendixRules}), using the brane web realization or the corresponding generalized toric polygon (GTP) \cite{Aharony:1997bh,Benini:2009gi,vanBeest:2020kou}.  
    \item Finally, the 5d theories are compactified down to 4d theories on a circle with a $\mathbb{Z}_{\ell}$ twist. The corresponding magnetic quiver can also be derived from the brane web, using the rules recalled in Appendix \ref{AppendixRules}. 
\end{compactitem}

In the remainder of this section, we apply the above program to derive the magnetic quivers of Table \ref{resulttable}. We distinguish the cases of $\ell$ even, where a 5d Lagrangian gauge theory formulation is known for the 5d theory on its Coulomb branch, and the case $\ell=3$ where only a brane web is available. Note that the gauge theories discussed below are different from those presented in Section \ref{SectionFI} and we propose they are dual. This is consistent with the dualities for the corresponding marginal theories discussed in \cite{Hayashi:2015zka}.   

\begin{table}[t]
    \centering
    \begin{tabular}{c|c|c} \toprule 
   Weakly coupled theory & GTP & Magnetic quiver \\ \midrule 
\raisebox{-.5\height}{ \scalebox{0.76}{ \begin{tikzpicture}[x=2cm,y=.8cm]
\node (g0) at (0,0) [flavour0,label=below:{\small $3$}] {};
\node (g1) at (1,0) [gauge,label=below:{\small $\mathrm{SU}(2r{+}1)$}] {};
\node (g2) at (2,0) [flavour0,label=below:{\small $3$}] {};
\node (fAS) at (1,1) [flavour0,label=above:{\small $2$}] {};
\draw (g0)--(g1)--(g2);
	\draw [line join=round,decorate, decoration={zigzag, segment length=4,amplitude=.9,post=lineto,post length=2pt}]  (g1)--(fAS);
	\draw (1.15,0.5) node {$\scriptscriptstyle{\Lambda^2}$};
\end{tikzpicture}}} & \raisebox{-.5\height}{ \scalebox{0.76}{  
\begin{tikzpicture}[x=.5cm,y=.5cm] 
\draw[ligne] (0,3)--(0,0); 
\draw[ligne] (0,0)--(0,-3); 
\draw[ligne] (0,-3)--(6,-3); 
\draw[ligne] (6,-3)--(6,0); 
\draw[ligne] (6,0)--(6,3); 
\draw[ligne] (6,3)--(6,4); 
\draw[ligne] (6,4)--(6,7); 
\draw[ligne] (6,7)--(6,10); 
\draw[ligne] (6,10)--(0,10); 
\draw[ligne] (0,10)--(0,7); 
\draw[ligne] (0,7)--(0,4); 
\draw[ligne] (0,4)--(0,3); 
\node[bd] at (0,0) {}; 
\node[bd] at (0,-3) {}; 
\node[bd] at (6,-3) {}; 
\node[bd] at (6,0) {}; 
\node[bd] at (6,3) {}; 
\node[bd] at (6,4) {}; 
\node[bd] at (6,7) {}; 
\node[bd] at (6,10) {}; 
\node[bd] at (0,10) {}; 
\node[bd] at (0,7) {}; 
\node[bd] at (0,4) {}; 
\node[bd] at (0,3) {}; 
\node[wd] at (0,-1) {}; 
\node[wd] at (0,-2) {}; 
\node[wd] at (1,-3) {}; 
\node[wd] at (2,-3) {}; 
\node[wd] at (3,-3) {}; 
\node[wd] at (4,-3) {}; 
\node[wd] at (5,-3) {}; 
\node[wd] at (6,-2) {}; 
\node[wd] at (6,-1) {}; 
\node[wd] at (6,1) {}; 
\node[wd] at (6,2) {}; 
\node[wd] at (6,5) {}; 
\node[wd] at (6,6) {}; 
\node[wd] at (6,8) {}; 
\node[wd] at (6,9) {}; 
\node[wd] at (5,10) {}; 
\node[wd] at (4,10) {}; 
\node[wd] at (3,10) {}; 
\node[wd] at (2,10) {}; 
\node[wd] at (1,10) {}; 
\node[wd] at (0,9) {}; 
\node[wd] at (0,8) {}; 
\node[wd] at (0,6) {}; 
\node[wd] at (0,5) {}; 
\node[wd] at (0,2) {}; 
\node[wd] at (0,1) {}; 
\node at (0,-4) {};
\node at (0,11) {};
\end{tikzpicture}}}         &    
\raisebox{-.5\height}{\scalebox{0.76}{\begin{tikzpicture}[x=1.2cm,y=.8cm]
\node (g0) at (0,0) [gauge,label=below:{\small $1$}] {};
\node (g1) at (1,0) [gauge,label=below:{\small $1{+}r$}] {};
\node (g2) at (2,0) [gauge,label=below:{\small $1{+}2r$}] {};
\node (g3) at (3,0) [gauge,label=below:{\small $1{+}3r$}] {};
\node (g4) at (4,0) [gauge,label=below:{\small $1{+}2r$}] {};
\node (g5) at (5,0) [gauge,label=below:{\small $1{+}r$}] {};
\node (g6) at (6,0) [gauge,label=below:{\small $1$}] {};
\node (g12) at (3,1) [gauge,label=above:{\small $2r$}] {};
\draw (g0)--(g1)--(g2)--(g3)--(g4)--(g5)--(g6);
\draw (g3)--(g12);
\end{tikzpicture}}}    \\
\raisebox{-.5\height}{ \scalebox{0.76}{ \begin{tikzpicture}[x=2cm,y=.8cm]
\node (g0) at (0,0) [flavour0,label=below:{\small $2$}] {};
\node (g1) at (1,0) [gauge,label=below:{\small $\mathrm{SU}(2r{+}1)$}] {};
\node (g2) at (2,0) [flavour0,label=below:{\small $2$}] {};
\node (fAS) at (1,1) [flavour0,label=above:{\small $2$}] {};
\draw (g0)--(g1)--(g2);
	\draw [line join=round,decorate, decoration={zigzag, segment length=4,amplitude=.9,post=lineto,post length=2pt}]  (g1)--(fAS);
	\draw (1.15,0.5) node {$\scriptscriptstyle{\Lambda^2}$};
\end{tikzpicture}}} & 
\raisebox{-.5\height}{ \scalebox{0.76}{  
\begin{tikzpicture}[x=.5cm,y=.5cm] 
\draw[ligne] (0,3)--(0,0); 
\draw[ligne] (0,0)--(3,0); 
\draw[ligne] (3,0)--(6,0); 
\draw[ligne] (6,0)--(6,3); 
\draw[ligne] (6,3)--(6,4); 
\draw[ligne] (6,4)--(6,7); 
\draw[ligne] (6,7)--(3,7); 
\draw[ligne] (3,7)--(0,7); 
\draw[ligne] (0,7)--(0,4); 
\draw[ligne] (0,4)--(0,3); 
\node[bd] at (0,0) {}; 
\node[bd] at (3,0) {}; 
\node[bd] at (6,0) {}; 
\node[bd] at (6,3) {}; 
\node[bd] at (6,4) {}; 
\node[bd] at (6,7) {}; 
\node[bd] at (3,7) {}; 
\node[bd] at (0,7) {}; 
\node[bd] at (0,4) {}; 
\node[bd] at (0,3) {}; 
\node[wd] at (1,0) {}; 
\node[wd] at (2,0) {}; 
\node[wd] at (4,0) {}; 
\node[wd] at (5,0) {}; 
\node[wd] at (6,1) {}; 
\node[wd] at (6,2) {}; 
\node[wd] at (6,5) {}; 
\node[wd] at (6,6) {}; 
\node[wd] at (5,7) {}; 
\node[wd] at (4,7) {}; 
\node[wd] at (2,7) {}; 
\node[wd] at (1,7) {}; 
\node[wd] at (0,6) {}; 
\node[wd] at (0,5) {}; 
\node[wd] at (0,2) {}; 
\node[wd] at (0,1) {}; 
\node at (0,-1) {};
\node at (0,8) {};
\end{tikzpicture}}}& 
\raisebox{-.5\height}{\scalebox{0.76}{\begin{tikzpicture}[x=1.2cm,y=.8cm]
\node (g0) at (0,0) [gauge,label=below:{\small $1$}] {};
\node (g1) at (1,0) [gauge,label=below:{\small $1{+}r$}] {};
\node (g2) at (2,0) [gauge,label=below:{\small $1{+}2r$}] {};
\node (g3) at (3,0) [gauge,label=below:{\small $1{+}r$}] {};
\node (g4) at (4,0) [gauge,label=below:{\small $1$}] {};
\node (g11) at (1,1) [gauge,label=above:{\small $r$}] {};
\node (g12) at (2,1) [gauge,label=above:{\small $2r$}] {};
\node (g13) at (3,1) [gauge,label=above:{\small $r$}] {};
\draw (g0)--(g1)--(g2)--(g3)--(g4);
\draw (g11)--(g12)--(g13);
\draw (g2)--(g12);
\end{tikzpicture}}}     \\ 
\raisebox{-.5\height}{ \scalebox{0.76}{ \begin{tikzpicture}[x=2cm,y=.8cm]
\node (g0) at (0,0) [flavour0,label=below:{\small $1$}] {};
\node (g1) at (1,0) [gauge,label=below:{\small $\mathrm{SU}(2r{+}1)$}] {};
\node (g2) at (2,0) [flavour0,label=below:{\small $1$}] {};
\node (fAS) at (1,1) [flavour0,label=above:{\small $2$}] {};
\draw (g0)--(g1)--(g2);
\draw [line join=round,decorate, decoration={zigzag, segment length=4,amplitude=.9,post=lineto,post length=2pt}]  (g1)--(fAS);
	\draw (1.15,0.5) node {$\scriptscriptstyle{\Lambda^2}$};
\end{tikzpicture}}} &  \raisebox{-.5\height}{ \scalebox{0.76}{  
\begin{tikzpicture}[x=.5cm,y=.5cm] 
\draw[ligne] (0,3)--(0,0); 
\draw[ligne] (0,0)--(3,0); 
\draw[ligne] (3,0)--(6,3); 
\draw[ligne] (6,3)--(6,4); 
\draw[ligne] (6,4)--(6,7); 
\draw[ligne] (6,7)--(3,7); 
\draw[ligne] (3,7)--(0,4); 
\draw[ligne] (0,4)--(0,3); 
\node[bd] at (0,0) {}; 
\node[bd] at (3,0) {}; 
\node[bd] at (6,3) {}; 
\node[bd] at (6,4) {}; 
\node[bd] at (6,7) {}; 
\node[bd] at (3,7) {}; 
\node[bd] at (0,4) {}; 
\node[bd] at (0,3) {}; 
\node[wd] at (1,0) {}; 
\node[wd] at (2,0) {}; 
\node[wd] at (4,1) {}; 
\node[wd] at (5,2) {}; 
\node[wd] at (6,5) {}; 
\node[wd] at (6,6) {}; 
\node[wd] at (5,7) {}; 
\node[wd] at (4,7) {}; 
\node[wd] at (2,6) {}; 
\node[wd] at (1,5) {}; 
\node[wd] at (0,2) {}; 
\node[wd] at (0,1) {}; 
\node at (0,-1) {};
\node at (0,8) {};
\end{tikzpicture}}}   & 
\raisebox{-.5\height}{\scalebox{0.76}{\begin{tikzpicture}[x=1.2cm,y=.8cm]
\node (g0) at (0,0) [gauge,label=below:{\small $1$}] {};
\node (g1) at (1,0) [gauge,label=below:{\small $1{+}r$}] {};
\node (g4) at (2,0) [gauge,label=below:{\small $1$}] {};
\node (g10) at (0.5,1) [gauge,label=above:{\small $r$}] {};
\node (g11) at (1.5,1) [gauge,label=above:{\small $r$}] {};
\draw (g0)--(g1)--(g4);
\draw (g1)--(g10)--(g11)--(g1);
\end{tikzpicture} \hspace{.5cm} \begin{tikzpicture}[x=1.2cm,y=.8cm]
\draw (0.5,.95)--(1.5,.95);
\draw (0.5,1.05)--(1.5,1.05);
\node (g0) at (0,0) [gauge,label=below:{\small $1$}] {};
\node (g1) at (1,0) [gauge,label=below:{\small $1$}] {};
\node (g4) at (2,0) [gauge,label=below:{\small $1$}] {};
\node (g10) at (0.5,1) [gauge,label=above:{\small $r$}] {};
\node (g11) at (1.5,1) [gauge,label=above:{\small $r$}] {};
\draw (g10)--(g0)--(g1)--(g4)--(g11);
\end{tikzpicture}}}   \\ \midrule 
\raisebox{-.5\height}{ \scalebox{0.76}{ \begin{tikzpicture}[x=2cm,y=.8cm]
\node (g0) at (0,0) [flavour0,label=below:{\small $2$}] {};
\node (g1) at (1,0) [gauge,label=below:{\small $\mathrm{SU}(2r{+}1)$}] {};
\node (g2) at (2,0) [gauge,label=below:{\small $\mathrm{SU}(2r{+}1)$}] {};
\node (g3) at (3,0) [flavour0,label=below:{\small $2$}] {};
\node (fAS1) at (1,1) [flavour0,label=above:{\small $1$}] {};
\node (fAS2) at (2,1) [flavour0,label=above:{\small $1$}] {};
\draw (g0)--(g1)--(g2)--(g3);
\draw [line join=round,decorate, decoration={zigzag, segment length=4,amplitude=.9,post=lineto,post length=2pt}]  (g1)--(fAS1) (g2)--(fAS2);
	\draw (1.15,0.5) node {$\scriptscriptstyle{\Lambda^2}$};
	\draw (2.15,0.5) node {$\scriptscriptstyle{\Lambda^2}$};
\end{tikzpicture}}} & 
\raisebox{-.5\height}{ \scalebox{0.76}{  
\begin{tikzpicture}[x=.5cm,y=.5cm] 
\draw[ligne] (0,3)--(0,0); 
\draw[ligne] (0,0)--(3,0); 
\draw[ligne] (3,0)--(4,0); 
\draw[ligne] (4,0)--(7,0); 
\draw[ligne] (7,0)--(7,3); 
\draw[ligne] (7,3)--(7,4); 
\draw[ligne] (7,4)--(7,7); 
\draw[ligne] (7,7)--(4,7); 
\draw[ligne] (4,7)--(3,7); 
\draw[ligne] (3,7)--(0,7); 
\draw[ligne] (0,7)--(0,4); 
\draw[ligne] (0,4)--(0,3); 
\node[bd] at (0,0) {}; 
\node[bd] at (3,0) {}; 
\node[bd] at (4,0) {}; 
\node[bd] at (7,0) {}; 
\node[bd] at (7,3) {}; 
\node[bd] at (7,4) {}; 
\node[bd] at (7,7) {}; 
\node[bd] at (4,7) {}; 
\node[bd] at (3,7) {}; 
\node[bd] at (0,7) {}; 
\node[bd] at (0,4) {}; 
\node[bd] at (0,3) {}; 
\node[wd] at (1,0) {}; 
\node[wd] at (2,0) {}; 
\node[wd] at (5,0) {}; 
\node[wd] at (6,0) {}; 
\node[wd] at (7,1) {}; 
\node[wd] at (7,2) {}; 
\node[wd] at (7,5) {}; 
\node[wd] at (7,6) {}; 
\node[wd] at (6,7) {}; 
\node[wd] at (5,7) {}; 
\node[wd] at (2,7) {}; 
\node[wd] at (1,7) {}; 
\node[wd] at (0,6) {}; 
\node[wd] at (0,5) {}; 
\node[wd] at (0,2) {}; 
\node[wd] at (0,1) {}; 
\node at (0,-1) {};
\node at (0,8) {};
\end{tikzpicture}}} & 
\raisebox{-.5\height}{\scalebox{0.76}{
\begin{tikzpicture}[x=1.2cm,y=.8cm]
\node (g0) at (0,0) [gauge,label=below:{\small $1$}] {};
\node (g1) at (1,0) [gauge,label=below:{\small $1{+}r$}] {};
\node (g2) at (2,0) [gauge,label=below:{\small $1{+}2r$}] {};
\node (g3) at (3,0) [gauge,label=below:{\small $1{+}r$}] {};
\node (g4) at (4,0) [gauge,label=below:{\small $1$}] {};
\node (g10) at (0,1) [gauge,label=above:{\small $1$}] {};
\node (g11) at (1,1) [gauge,label=above:{\small $1{+}r$}] {};
\node (g12) at (2,1) [gauge,label=above:{\small $1{+}2r$}] {};
\node (g13) at (3,1) [gauge,label=above:{\small $1{+}r$}] {};
\node (g14) at (4,1) [gauge,label=above:{\small $1$}] {};
\draw (g0)--(g1)--(g2)--(g3)--(g4);
\draw (g10)--(g11)--(g12)--(g13)--(g14);
\draw (g12)--(g2);
\end{tikzpicture}}}
\\ \bottomrule
\end{tabular}
    \caption{For each 5d SCFT, which is the 5d $\mathcal{N}=1$ UV completion of the weakly coupled theory, one can construct the generalized toric polygon and deduce the corresponding magnetic quiver. This gives all the quivers for $\ell=2$ and $\ell=4$ upon $\mathbb{Z}_{\ell}$ folding. In the GTPs, every edge with white dots should contain exactly $r-1$ white dots ($r=3$ is drawn). Note that the last quiver differs from the one of Figure \ref{DeftoZ4}, illustrating the fact that a quotient on the brane web and on the quivers are different concepts.   }
    \label{tabGTPs}
\end{table}

\subsection{Cases \texorpdfstring{$\ell=2$ or $4$}{l= 2 or 4} }
This 6d theories corresponds after compactification on a circle to 5d marginal theories which are described in \cite{Zafrir:2015rga}, namely the UV fixed points of the 5d quiver theories 
\begin{equation}
\label{5dSCFT}
\raisebox{-.5\height}{
\begin{tikzpicture}[x=2cm,y=.8cm]
\node (fAS1) at (0,0) [flavour0,label=below:{$\scriptstyle{1}$}] {};
\node (g0) at (1,1) [flavour0,label=above:{$\scriptstyle{4}$}] {};
\node (g1) at (1,0) [gauge,label=below:{$\scriptstyle{\mathrm{SU}(2r{+}1)}$}] {};
\node (g2) at (2,0) [gauge,label=below:{$\scriptstyle{\mathrm{SU}(2r{+}1)}$}] {};
\node (g3) at (3,0) {$\cdots$};
\node (g4) at (4,0) [gauge,label=below:{$\scriptstyle{\mathrm{SU}(2r{+}1)}$}] {};
\node (g5) at (5,0) [gauge,label=below:{$\scriptstyle{\mathrm{SU}(2r{+}1)}$}] {};
\node (g6) at (5,1) [flavour0,label=above:{$\scriptstyle{4}$}] {};
\node (fAS2) at (6,0) [flavour0,label=below:{$\scriptstyle{1}$}] {};
\draw (g0)--(g1)--(g2)--(g3)--(g4)--(g5)--(g6);
	\draw [line join=round,decorate, decoration={zigzag, segment length=4,amplitude=.9,post=lineto,post length=2pt}]  (g1)--(fAS1) (g5)--(fAS2);
	\draw (5.5,0.25) node {$\scriptscriptstyle{\Lambda^2}$};
	\draw (0.5,0.25) node {$\scriptscriptstyle{\Lambda^2}$};
\draw [decorate,decoration={brace,amplitude=5pt,mirror,raise=0pt}]
(.6,-1) -- (5.4,-1) node [black,midway,yshift=-.4cm] {$\ell /2 $ nodes};
\end{tikzpicture}
}
\end{equation}
All the Chern-Simons levels are taken to be 0. The 4d theories are obtained by a further $\mathbb{Z}_{\ell}$ twisted compactification on a circle of 5d SCFT descendants of (\ref{5dSCFT}), where fundamental matter are decoupled. The twist is possible only if the fundamental hypermultiplets are decoupled on each side of (\ref{5dSCFT}) in a $\mathbb{Z}_{\ell}$ symmetric way (in particular the number of decoupled hypermultiplets has to be a multiple of $\ell$). This leaves us with four theories to study. These theories are 
\begin{compactitem}
    \item For $\ell=2$ with $s \ell$ decoupled hypermultiplets ($s=1,2,3$)
    \begin{equation}
    \label{quivell2}
    \raisebox{-.5\height}{
\begin{tikzpicture}[x=2cm,y=.8cm]
\node (g0) at (0,0) [flavour0,label=below:{\small $4{-}s$}] {};
\node (g1) at (1,0) [gauge,label=below:{\small $\mathrm{SU}(2r{+}1)$}] {};
\node (g2) at (2,0) [flavour0,label=below:{\small $4{-}s$}] {};
\node (fAS) at (1,1) [flavour0,label=above:{\small $2$}] {};
\draw (g0)--(g1)--(g2);
	\draw [line join=round,decorate, decoration={zigzag, segment length=4,amplitude=.9,post=lineto,post length=2pt}]  (g1)--(fAS);
	\draw (1.15,0.5) node {$\scriptscriptstyle{\Lambda^2}$};
\end{tikzpicture}
}
\end{equation}
\item For $\ell=4$ with $\ell$ decoupled hypermultiplets
\begin{equation}
\raisebox{-.5\height}{
\begin{tikzpicture}[x=2cm,y=.8cm]
\node (g0) at (0,0) [flavour0,label=below:{\small $2$}] {};
\node (g1) at (1,0) [gauge,label=below:{\small $\mathrm{SU}(2r{+}1)$}] {};
\node (g2) at (2,0) [gauge,label=below:{\small $\mathrm{SU}(2r{+}1)$}] {};
\node (g3) at (3,0) [flavour0,label=below:{\small $2$}] {};
\node (fAS1) at (1,1) [flavour0,label=above:{\small $1$}] {};
\node (fAS2) at (2,1) [flavour0,label=above:{\small $1$}] {};
\draw (g0)--(g1)--(g2)--(g3);
	\draw [line join=round,decorate, decoration={zigzag, segment length=4,amplitude=.9,post=lineto,post length=2pt}]  (g1)--(fAS1) (g2)--(fAS2);
	\draw (1.15,0.5) node {$\scriptscriptstyle{\Lambda^2}$};
	\draw (2.15,0.5) node {$\scriptscriptstyle{\Lambda^2}$};
\end{tikzpicture}
}
\end{equation}
\end{compactitem}
One can construct brane webs for these theories, or equivalently generalized toric polygons, and deduce the magnetic quivers, see Table \ref{tabGTPs}. In each case, the GTP has a manifest $\mathbb{Z}_{\ell}$ symmetry, and the folded magnetic quivers can be obtained from the rules outlined at the beginning of this section.  Upon $\mathbb{Z}_{\ell}$ folding, one indeed recovers the quivers of Table \ref{resulttable}. 

To illustrate the procedure, let us treat as an example the case $\ell=4$, $r=3$. The brane web is 
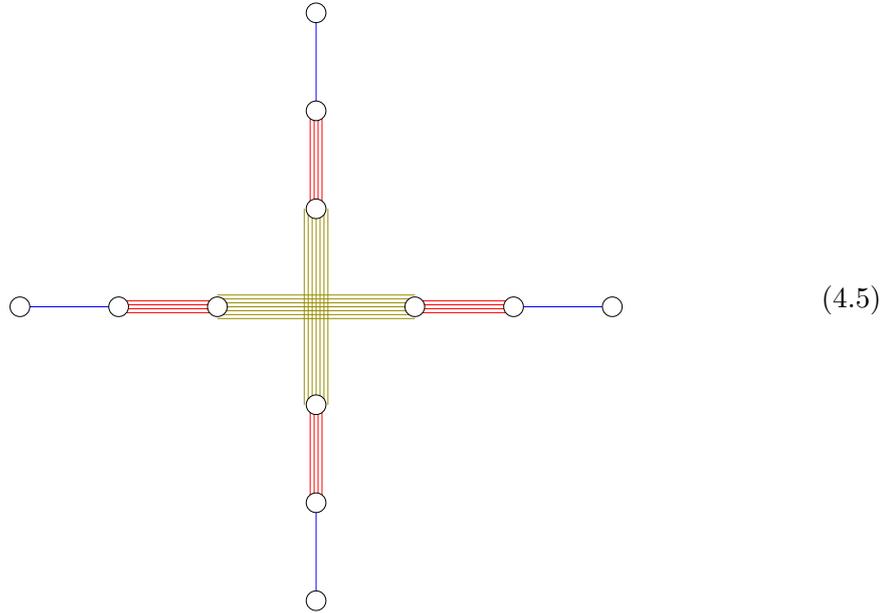
\begin{equation}
    \raisebox{-.5\height}{ \scalebox{.866}{  
\begin{tikzpicture}[x=1.5cm,y=1.5cm] 
\draw[blue,transform canvas={yshift=.00cm}] (0,3)--(1,3); 
\draw[red,transform canvas={yshift=.09cm}] (1,3)--(2,3); 
\draw[red,transform canvas={yshift=.03cm}] (1,3)--(2,3); 
\draw[red,transform canvas={yshift=-.03cm}] (1,3)--(2,3); 
\draw[red,transform canvas={yshift=-.09cm}] (1,3)--(2,3); 
\draw[olive,transform canvas={yshift=-.18cm}] (2,3)--(4,3); 
\draw[olive,transform canvas={yshift=-.12cm}] (2,3)--(4,3); 
\draw[olive,transform canvas={yshift=-.06cm}] (2,3)--(4,3); 
\draw[olive,transform canvas={yshift=.00cm}] (2,3)--(4,3); 
\draw[olive,transform canvas={yshift=.06cm}] (2,3)--(4,3); 
\draw[olive,transform canvas={yshift=.12cm}] (2,3)--(4,3); 
\draw[olive,transform canvas={yshift=.18cm}] (2,3)--(4,3); 
\draw[red,transform canvas={yshift=.09cm}] (4,3)--(5,3); 
\draw[red,transform canvas={yshift=.03cm}] (4,3)--(5,3); 
\draw[red,transform canvas={yshift=-.03cm}] (4,3)--(5,3); 
\draw[red,transform canvas={yshift=-.09cm}] (4,3)--(5,3); 
\draw[blue,transform canvas={yshift=.00cm}] (5,3)--(6,3); 
\node[D7] at (0,3) {}; 
\node[D7] at (1,3) {}; 
\node[D7] at (2,3) {}; 
\node[D7] at (4,3) {}; 
\node[D7] at (5,3) {}; 
\node[D7] at (6,3) {}; 
\draw[blue,transform canvas={xshift=.00cm}] (3,0)--(3,1); 
\draw[red,transform canvas={xshift=.09cm}] (3,1)--(3,2); 
\draw[red,transform canvas={xshift=.03cm}] (3,1)--(3,2); 
\draw[red,transform canvas={xshift=-.03cm}] (3,1)--(3,2); 
\draw[red,transform canvas={xshift=-.09cm}] (3,1)--(3,2); 
\draw[olive,transform canvas={xshift=-.18cm}] (3,2)--(3,4); 
\draw[olive,transform canvas={xshift=-.12cm}] (3,2)--(3,4); 
\draw[olive,transform canvas={xshift=-.06cm}] (3,2)--(3,4); 
\draw[olive,transform canvas={xshift=.00cm}] (3,2)--(3,4); 
\draw[olive,transform canvas={xshift=.06cm}] (3,2)--(3,4); 
\draw[olive,transform canvas={xshift=.12cm}] (3,2)--(3,4); 
\draw[olive,transform canvas={xshift=.18cm}] (3,2)--(3,4); 
\draw[red,transform canvas={xshift=.09cm}] (3,4)--(3,5); 
\draw[red,transform canvas={xshift=.03cm}] (3,4)--(3,5); 
\draw[red,transform canvas={xshift=-.03cm}] (3,4)--(3,5); 
\draw[red,transform canvas={xshift=-.09cm}] (3,4)--(3,5); 
\draw[blue,transform canvas={xshift=.00cm}] (3,5)--(3,6); 
\node[D7] at (3,0) {}; 
\node[D7] at (3,1) {}; 
\node[D7] at (3,2) {}; 
\node[D7] at (3,4) {}; 
\node[D7] at (3,5) {}; 
\node[D7] at (3,6) {}; 
\end{tikzpicture}}}
\end{equation}
The colors denote the $\mathbb{Z}_{4}$ orbits: blue and red are size 4 orbits and give short nodes, while olive has size 1 and gives a long node. The magnetic quiver is therefore 
\begin{equation}
\scalebox{.7}{\raisebox{-.5\height}{
    \begin{tikzpicture}
	\begin{pgfonlayer}{nodelayer}
		\node [style=gauge3,red] (2) at (-1, 0) {};
		\node [style=none] (6) at (-1, -0.5) {$1{+}r$};
		\node [style=gauge3,olive] (24) at (0.5, 0) {};
		\node [style=none] (29) at (0.5, -0.5) {$1{+}2r$};
		\node [style=none] (30) at (-1, 0.05) {};
		\node [style=none] (31) at (0.5, 0.05) {};
		\node [style=none] (32) at (-1, -0.075) {};
		\node [style=none] (33) at (0.5, -0.075) {};
		\node [style=none] (34) at (-0.5, 0) {};
		\node [style=none] (35) at (0, 0.5) {};
		\node [style=none] (36) at (0, -0.5) {};
		\node [style=none] (37) at (-1, 0.15) {};
		\node [style=none] (38) at (0.5, 0.15) {};
		\node [style=none] (39) at (-1, -0.175) {};
		\node [style=none] (40) at (0.5, -0.175) {};
		\node [style=gauge3,blue] (41) at (-2.5, 0) {};
		\node [style=none] (42) at (-2.5, -0.5) {$1$};
	\end{pgfonlayer}
	\begin{pgfonlayer}{edgelayer}
		\draw (30.center) to (31.center);
		\draw (33.center) to (32.center);
		\draw (35.center) to (34.center);
		\draw (34.center) to (36.center);
		\draw (37.center) to (38.center);
		\draw (39.center) to (40.center);
		\draw (41) to (2);
	\end{pgfonlayer}
\end{tikzpicture}
}}
\end{equation}
Note that by default the ungauging is located on the long side. 

A few comments are in order regarding Table \ref{tabGTPs}. First, note that for the first line of this table ($s=1$), the GTP read from quiver (\ref{quivell2}) is (drawn for $r=3$)
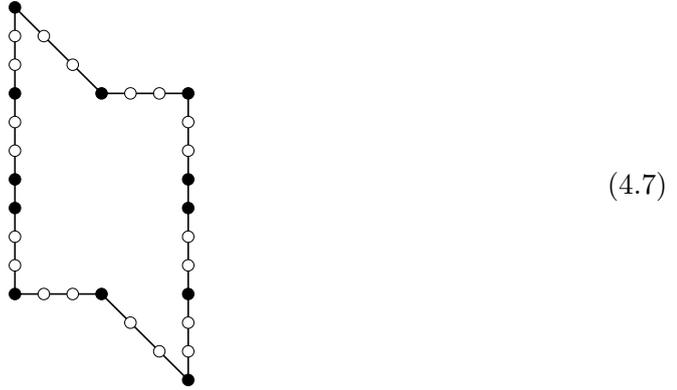
\begin{equation}
    \raisebox{-.5\height}{ \scalebox{.76}{  
\begin{tikzpicture}[x=.5cm,y=.5cm] 
\draw[ligne] (0,3)--(0,0); 
\draw[ligne] (0,0)--(3,0); 
\draw[ligne] (3,0)--(6,-3); 
\draw[ligne] (6,-3)--(6,0); 
\draw[ligne] (6,0)--(6,3); 
\draw[ligne] (6,3)--(6,4); 
\draw[ligne] (6,4)--(6,7); 
\draw[ligne] (6,7)--(3,7); 
\draw[ligne] (3,7)--(0,10); 
\draw[ligne] (0,10)--(0,7); 
\draw[ligne] (0,7)--(0,4); 
\draw[ligne] (0,4)--(0,3); 
\node[bd] at (0,0) {}; 
\node[bd] at (3,0) {}; 
\node[bd] at (6,-3) {}; 
\node[bd] at (6,0) {}; 
\node[bd] at (6,3) {}; 
\node[bd] at (6,4) {}; 
\node[bd] at (6,7) {}; 
\node[bd] at (3,7) {}; 
\node[bd] at (0,10) {}; 
\node[bd] at (0,7) {}; 
\node[bd] at (0,4) {}; 
\node[bd] at (0,3) {}; 
\node[wd] at (1,0) {}; 
\node[wd] at (2,0) {}; 
\node[wd] at (4,-1) {}; 
\node[wd] at (5,-2) {}; 
\node[wd] at (6,-2) {}; 
\node[wd] at (6,-1) {}; 
\node[wd] at (6,1) {}; 
\node[wd] at (6,2) {}; 
\node[wd] at (6,5) {}; 
\node[wd] at (6,6) {}; 
\node[wd] at (5,7) {}; 
\node[wd] at (4,7) {}; 
\node[wd] at (2,8) {}; 
\node[wd] at (1,9) {}; 
\node[wd] at (0,9) {}; 
\node[wd] at (0,8) {}; 
\node[wd] at (0,6) {}; 
\node[wd] at (0,5) {}; 
\node[wd] at (0,2) {}; 
\node[wd] at (0,1) {}; 
\end{tikzpicture}}}
\end{equation}
and one can perform a monodromy transformation to get a convex polygon, as shown in the table. Secondly, for the theory given in the third line of the table ($\ell=2$, $s=3$) there are two maximal decompositions of the brane web, giving a Higgs branch which is the union of two cones. This is why there are two magnetic quivers. Only the first one gives the magnetic quiver of the $\ess{A_2}{2}{r}$ theory after folding: the detailed explanation for that fact is presented in  Appendix \ref{AppendixCone}.

\subsection{Case \texorpdfstring{$\ell = 3$}{l=3}}
For this case, we need to study the brane web directly, as no Lagrangian 5d theory is known. We start from the 6d construction, and T-dualize along a direction common to the NS5 and D8 branes (the vertical direction in the brane web \eqref{bigWeb}). The $O8^-$ plane produces two $O7^-$ planes, which are resolved in two different ways: once with a pair of $\{(1,1),(1,-1)\}$ seven-branes (top), and once with a pair of $\{(0,1),(2,1)\}$ seven-branes (bottom). There are also $8+ \ell =11$ D7 branes to account for the fundamental hypermultiplets in \eqref{6dquiver} (5 to the left and 6 to the right). One then obtains the following brane web: 
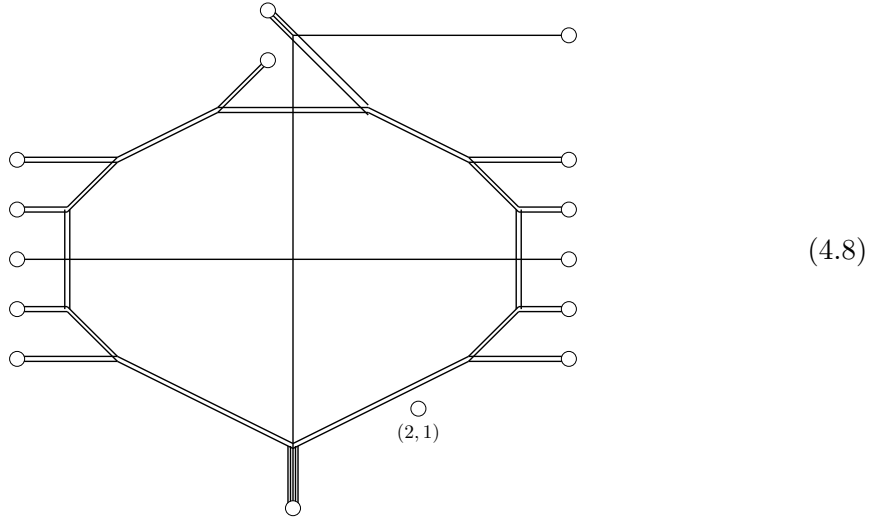
\begin{align}
\label{bigWeb}
   \raisebox{-.5\height}{
      \scalebox{0.66}{
   \begin{tikzpicture}
\draw[ligne,transform canvas={yshift=.05cm}] (0,2)--(2,2);
\draw[ligne,transform canvas={yshift=-.05cm}] (0,2)--(2,2);
\draw[ligne,transform canvas={yshift=.05cm}] (0,1)--(1,1);
\draw[ligne,transform canvas={yshift=-.05cm}] (0,1)--(1,1);
\draw[ligne,transform canvas={yshift=.05cm}] (0,-1)--(1,-1);
\draw[ligne,transform canvas={yshift=-.05cm}] (0,-1)--(1,-1);
\draw[ligne,transform canvas={yshift=.05cm}] (0,-2)--(2,-2);
\draw[ligne,transform canvas={yshift=-.05cm}] (0,-2)--(2,-2);
\draw[ligne,transform canvas={yshift=.05cm}] (2,2)--(4,3);
\draw[ligne,transform canvas={yshift=-.05cm}] (2,2)--(4,3);
\draw[ligne,transform canvas={yshift=.05cm}] (4,3)--(7,3);
\draw[ligne,transform canvas={yshift=-.05cm}] (4,3)--(7,3);
\draw[ligne,transform canvas={yshift=.05cm}] (7,3)--(9,2);
\draw[ligne,transform canvas={yshift=-.05cm}] (7,3)--(9,2);
\draw[ligne,transform canvas={yshift=.05cm}] (9,2)--(10,1);
\draw[ligne,transform canvas={yshift=-.05cm}] (9,2)--(10,1);
\draw[ligne,transform canvas={xshift=.05cm}] (10,1)--(10,-1);
\draw[ligne,transform canvas={xshift=-.05cm}] (10,1)--(10,-1);
\draw[ligne,transform canvas={yshift=.05cm}] (10,-1)--(9,-2);
\draw[ligne,transform canvas={yshift=-.05cm}] (10,-1)--(9,-2);
\draw[ligne,transform canvas={yshift=.05cm}] (9,-2)--(5.5,-3.75);
\draw[ligne,transform canvas={yshift=-.05cm}] (9,-2)--(5.5,-3.75);
\draw[ligne,transform canvas={yshift=.05cm}] (5.5,-3.75)--(2,-2);
\draw[ligne,transform canvas={yshift=-.05cm}] (5.5,-3.75)--(2,-2);
\draw[ligne,transform canvas={yshift=.05cm}] (2,-2)--(1,-1);
\draw[ligne,transform canvas={yshift=-.05cm}] (2,-2)--(1,-1);
\draw[ligne,transform canvas={xshift=.05cm}] (1,-1)--(1,1);
\draw[ligne,transform canvas={xshift=-.05cm}] (1,-1)--(1,1);
\draw[ligne,transform canvas={yshift=.05cm}] (1,1)--(2,2);
\draw[ligne,transform canvas={yshift=-.05cm}] (1,1)--(2,2);
\draw[ligne,transform canvas={yshift=.05cm}] (4,3)--(5,4);
\draw[ligne,transform canvas={yshift=-.05cm}] (4,3)--(5,4);
\draw[ligne,transform canvas={yshift=.1cm}] (7,3)--(5,5);
\draw[ligne,transform canvas={yshift=-.1cm}] (7,3)--(5,5);
\draw[ligne,transform canvas={yshift=.05cm}] (9,2)--(11,2);
\draw[ligne,transform canvas={yshift=-.05cm}] (9,2)--(11,2);
\draw[ligne,transform canvas={yshift=.05cm}] (10,1)--(11,1);
\draw[ligne,transform canvas={yshift=-.05cm}] (10,1)--(11,1);
\draw[ligne,transform canvas={yshift=.05cm}] (10,-1)--(11,-1);
\draw[ligne,transform canvas={yshift=-.05cm}] (10,-1)--(11,-1);
\draw[ligne,transform canvas={yshift=.05cm}] (9,-2)--(11,-2);
\draw[ligne,transform canvas={yshift=-.05cm}] (9,-2)--(11,-2);
\draw[ligne,transform canvas={xshift=.05cm}] (5.5,-3.75)--(5.5,-5);
\draw[ligne,transform canvas={xshift=.1cm}] (5.5,-3.75)--(5.5,-5);
\draw[ligne,transform canvas={xshift=-.05cm}] (5.5,-3.75)--(5.5,-5);
\draw[ligne,transform canvas={xshift=-.1cm}] (5.5,-3.75)--(5.5,-5);
\draw[ligne,] (5.5,-5)--(5.5,4.5);
\draw[ligne,] (11,4.5)--(5.5,4.5);
\draw[ligne,] (5,5)--(5.5,4.5);
\draw[ligne,] (0,0)--(11,0);
\node[D7] at (0,-2) {}; 
\node[D7] at (0,-1) {}; 
\node[D7] at (0,0) {}; 
\node[D7] at (0,1) {}; 
\node[D7] at (0,2) {}; 
\node[D7] at (11,-2) {}; 
\node[D7] at (11,-1) {}; 
\node[D7] at (11,0) {}; 
\node[D7] at (11,1) {}; 
\node[D7] at (11,2) {}; 
\node[D7] at (5.5,-5) {}; 
\node[D7] at (11,4.5) {}; 
\node[D7] at (5,4) {}; 
\node[D7] at (5,5) {}; 
\node[D7] at (8,-3) {}; 
\node at (8,-3.5) {$(2,1)$}; 
\end{tikzpicture}
  }
   } 
\end{align}
In the drawings, we take the rank to be $r=2$, but the drawings are similar for other ranks, each double line being replaced by $r$ lines. After some manipulations, similar to those performed in \cite[Fig. 14, 15]{Zafrir:2015rga}, the brane web can be brought to a $\mathbb{Z}_3$ symmetric form: 
\begin{align}
\label{finalWeb}
   \raisebox{-.5\height}{
  \scalebox{0.66}{
   \begin{tikzpicture}
\draw[ligne,transform canvas={yshift=.05cm}] (0,0)--(1,0);
\draw[ligne,transform canvas={yshift=-.05cm}] (0,0)--(1,0);
\draw[ligne,transform canvas={yshift=.05cm}] (1,0)--(2,-1);
\draw[ligne,transform canvas={yshift=-.05cm}] (1,0)--(2,-1);
\draw[ligne,transform canvas={xshift=.05cm}] (1,0)--(1,4);
\draw[ligne,transform canvas={xshift=-.05cm}] (1,0)--(1,4);
\draw[ligne,transform canvas={yshift=.05cm}] (0,1)--(2,1);
\draw[ligne,transform canvas={yshift=-.05cm}] (0,1)--(2,1);
\draw[ligne,transform canvas={yshift=.05cm}] (2,1)--(3,0);
\draw[ligne,transform canvas={yshift=-.05cm}] (2,1)--(3,0);
\draw[ligne,transform canvas={xshift=.05cm}] (2,1)--(2,4);
\draw[ligne,transform canvas={xshift=-.05cm}] (2,1)--(2,4);
\draw[ligne] (0,2)--(3,2);
\draw[ligne] (3,4)--(3,2);
\draw[ligne] (4,1)--(3,2);
\draw[ligne,transform canvas={yshift=.05cm}] (0,3)--(4,3);
\draw[ligne,transform canvas={yshift=-.05cm}] (0,3)--(4,3);
\draw[ligne,transform canvas={yshift=.05cm}] (4,3)--(5,2);
\draw[ligne,transform canvas={yshift=-.05cm}] (4,3)--(5,2);
\draw[ligne,transform canvas={xshift=.05cm}] (4,3)--(4,4);
\draw[ligne,transform canvas={xshift=-.05cm}] (4,3)--(4,4);
\node[D7] at (0,0) {}; 
\node[D7] at (0,1) {}; 
\node[D7] at (0,2) {}; 
\node[D7] at (0,3) {}; 
\node[D7] at (1,4) {};  
\node[D7] at (2,4) {}; 
\node[D7] at (3,4) {}; 
\node[D7] at (4,4) {};
\node[D7] at (2,-1) {}; 
\node[D7] at (3,0) {}; 
\node[D7] at (4,1) {}; 
\node[D7] at (5,2) {}; 
\node[D7] at (5,4) {}; 
\node[D7] at (0,4) {}; 
\node[D7] at (0,-1) {}; 
\node at (6,4) {$(1,0)$};
\node at (-1,4) {$(1,1)$};
\node at (-1,-1) {$(2,-1)$};
\end{tikzpicture}
  }
   } 
\end{align}
This represents a marginal 5d theory (with a 6d UV completion). As for the even $\ell$ cases, one needs to decouple $s \ell$ ``hypermultiplets" (for $s=1,2$) in order to get the magnetic quivers which, after $\mathbb{Z}_3$ folding, give the magnetic quivers for the $\ess{D_4}{3}{r}$ and $\ess{A_1}{3}{r}$ theories. An obvious guess for implementing the $s=1$ decoupling in the above web is to remove the three seven-branes which are disconnected from the rest. One can then go to the origin of the 5d Coulomb branch and compute the magnetic quiver, which is given in Table \ref{tabell3}. The case $s=2$ is treated in the standard way: in the generalized toric polygon, this corresponds to cutting a minimal triangle at every corner. The resulting polygon and associated quiver are given in Table \ref{tabell3}. Upon $\mathbb{Z}_3$ folding, following the rules given above, one recovers the quivers announced in Table \ref{resulttable}. 

\begin{table}[t]
    \centering
    \begin{tabular}{c|c} \toprule 
    GTP & Quiver \\ \midrule 
        \raisebox{-.5\height}{ \scalebox{0.76}{ 
\begin{tikzpicture}[x=.5cm,y=.5cm,rotate=-90] 
\draw[ligne] (0,2)--(0,0); 
\draw[ligne] (0,0)--(2,0); 
\draw[ligne] (2,0)--(3,0); 
\draw[ligne] (3,0)--(5,0); 
\draw[ligne] (5,0)--(7,0); 
\draw[ligne] (7,0)--(5,2); 
\draw[ligne] (5,2)--(3,4); 
\draw[ligne] (3,4)--(2,5); 
\draw[ligne] (2,5)--(0,7); 
\draw[ligne] (0,7)--(0,5); 
\draw[ligne] (0,5)--(0,3); 
\draw[ligne] (0,3)--(0,2); 
\node[bd] at (0,0) {}; 
\node[bd] at (2,0) {}; 
\node[bd] at (3,0) {}; 
\node[bd] at (5,0) {}; 
\node[bd] at (7,0) {}; 
\node[bd] at (5,2) {}; 
\node[bd] at (3,4) {}; 
\node[bd] at (2,5) {}; 
\node[bd] at (0,7) {}; 
\node[bd] at (0,5) {}; 
\node[bd] at (0,3) {}; 
\node[bd] at (0,2) {}; 
\node[wd] at (1,0) {}; 
\node[wd] at (4,0) {}; 
\node[wd] at (6,0) {}; 
\node[wd] at (6,1) {}; 
\node[wd] at (4,3) {}; 
\node[wd] at (1,6) {}; 
\node[wd] at (0,6) {}; 
\node[wd] at (0,4) {}; 
\node[wd] at (0,1) {}; 
\node at (-1,-1) {};
\node at (8,8) {};
\end{tikzpicture}}} & 
\raisebox{-.5\height}{\scalebox{0.76}{
\begin{tikzpicture}[x=1.2cm,y=.8cm]
\node (g0) at (0,0) [gauge,label=below:{\small $1$}] {};
\node (g1) at (1,0) [gauge,label=below:{\small $1{+}r$}] {};
\node (g2) at (2,0) [gauge,label=below:{\small $1{+}2r$}] {};
\node (g3) at (3,0) [gauge,label=below:{\small $1{+}3r$}] {};
\node (g41) at (4,-.5) [gauge,label=below:{\small $1{+}2r$}] {};
\node (g51) at (5,-.5) [gauge,label=below:{\small $1{+}r$}] {};
\node (g61) at (6,-.5) [gauge,label=below:{\small $1$}] {};
\node (g42) at (4,.5) [gauge,label=above:{\small $1{+}2r$}] {};
\node (g52) at (5,.5) [gauge,label=above:{\small $1{+}r$}] {};
\node (g62) at (6,.5) [gauge,label=above:{\small $1$}] {};
\draw (g0)--(g1)--(g2)--(g3)--(g41)--(g51)--(g61);
\draw (g3)--(g42)--(g52)--(g62);
\end{tikzpicture}}}
\\
       \raisebox{-.5\height}{ \scalebox{0.76}{ \begin{tikzpicture}[x=.5cm,y=.5cm,rotate=-90] 
\draw[ligne] (0,2)--(2,0); 
\draw[ligne] (2,0)--(3,0); 
\draw[ligne] (3,0)--(5,0); 
\draw[ligne] (5,0)--(5,2); 
\draw[ligne] (5,2)--(3,4); 
\draw[ligne] (3,4)--(2,5); 
\draw[ligne] (2,5)--(0,5); 
\draw[ligne] (0,5)--(0,3); 
\draw[ligne] (0,3)--(0,2); 
\node[bd] at (2,0) {}; 
\node[bd] at (3,0) {}; 
\node[bd] at (5,0) {}; 
\node[bd] at (5,2) {}; 
\node[bd] at (3,4) {}; 
\node[bd] at (2,5) {}; 
\node[bd] at (0,5) {}; 
\node[bd] at (0,3) {}; 
\node[bd] at (0,2) {}; 
\node[wd] at (4,0) {}; 
\node[wd] at (5,1) {}; 
\node[wd] at (4,3) {}; 
\node[wd] at (1,5) {}; 
\node[wd] at (0,4) {}; 
\node[wd] at (1,1) {}; 
\node at (-1,-1) {};
\node at (6,6) {};
\end{tikzpicture}}}  & 
\raisebox{-.5\height}{\scalebox{0.76}{
\begin{tikzpicture}[x=1.2cm,y=.8cm]
\node (g0) at (0,0) [gauge,label=below:{\small $r$}] {};
\node (g1) at (1,0) [gauge,label=below:{\small $1{+}r$}] {};
\node (g21) at (2,1) [gauge,label=right:{\small $1$}] {};
\node (g22) at (2,0) [gauge,label=right:{\small $1$}] {};
\node (g23) at (2,-1) [gauge,label=right:{\small $1$}] {};
\draw[transform canvas={yshift=.05cm}] (g0)--(g1);
\draw[transform canvas={yshift=-.05cm}] (g0)--(g1);
\draw (g1)--(g21);
\draw (g22)--(g1)--(g23);
\end{tikzpicture}}} \\ \bottomrule 
    \end{tabular}
    \caption{Generalized toric polygons and magnetic quivers for the 5d theories corresponding to $\ell=3$. Top for $s=1$, bottom for $s=2$. The GTPs are drawn for $r=2$, in order to match with the brane web (\ref{finalWeb}).  }
    \label{tabell3}
\end{table}

\section{Hasse diagrams}
\label{sec:Hasse}

\begin{figure}
    \centering
\vspace*{-1cm}\hspace*{-0.5cm}\scalebox{.92}{\begin{tikzpicture}[x=1.7cm,y=1.7cm]
\node (30) [hasse] at (3,0) {};
\node (21) [hasse,label=left:{$\mathcal{I}$}] at (2,-1) {};
\node (41) [hasse,label=right:{$\mathcal{I}$}] at (4,-1) {};
\node (12) [hasse,label=left:{$\mathcal{S}^{(1)}$}] at (1,-2) {};
\node (32) [hasse,label=right:{$\mathcal{I} \otimes \mathcal{I}$}] at (3,-2) {};
\node (03) [hasse,label=left:{$\mathcal{S}^{(1)} \otimes \mathcal{I}$}] at (0,-3) {};
\node (23) [hasse,label=right:{$\mathcal{T}^{(2)}$}] at (2,-3) {};
\node (14) [hasse,label=left:{$\mathcal{S}^{(2)}$}] at (1,-4) {};
\node (34) [hasse,label=right:{$\mathcal{T}^{(2)} \otimes \mathcal{I}$}] at (3,-4) {};
\node (05) [hasse,label=left:{$\mathcal{S}^{(2)} \otimes \mathcal{I}$}] at (0,-5) {};
\node (25) [hasse,label=right:{$\mathcal{T}^{(3)}$}] at (2,-5) {};
\node (16) [hasse,label=left:{$\mathcal{S}^{(3)}$}] at (1,-6) {};
\node (36) [hasse,label=right:{$\mathcal{T}^{(3)} \otimes \mathcal{I}$}] at (3,-6) {};
\node (07) [hasse,label=left:{$\mathcal{S}^{(3)} \otimes \mathcal{I}$}] at (0,-7) {};
\node (27) [hasse,label=right:{$\mathcal{T}^{(4)}$}] at (2,-7) {};
\node (18) [hasse,label=left:{$\mathcal{S}^{(4)}$}] at (1,-8) {};
\node (38) [hasse,label=right:{$\mathcal{T}^{(4)} \otimes \mathcal{I}$}] at (3,-8) {};
\draw (30) edge [] node[label={[label distance=-.2cm]135:$g$}] {} (21);
\draw (30) edge [] node[label={[label distance=-.2cm]45:$g$}] {} (41);
\draw (21) edge [] node[label={[label distance=-.2cm]135:$\color{red}{h_{m+1,\ell}}$}] {} (12);
\draw (21) edge [] node[label={[label distance=-.2cm]45:$g$}] {} (32);
\draw (41) edge [] node[label={[label distance=-.2cm]-45:$g$}] {} (32);
\draw (12) edge [] node[label={[label distance=-.2cm]135:$g$}] {} (03);
\draw (12) edge [] node[label={[label distance=-.2cm]45:$g'$}] {} (23);
\draw (32) edge [] node[label={[label distance=-.2cm]-45:$\color{red}{k_{\ell}}$}] {} (23);
\draw (03) edge [] node[label={[label distance=-.2cm]-135:$a_1$}] {} (14);
\draw (23) edge [] node[label={[label distance=-.2cm]-45:$h_{m,\ell}$}] {} (14);
\draw (23) edge [] node[label={[label distance=-.2cm]45:$g$}] {} (34);
\draw (14) edge [] node[label={[label distance=-.2cm]135:$g$}] {} (05);
\draw (14) edge [] node[label={[label distance=-.2cm]45:$g'$}] {} (25);
\draw (34) edge [] node[label={[label distance=-.2cm]-45:$a_1$}] {} (25);
\draw (05) edge [] node[label={[label distance=-.2cm]-135:$a_1$}] {} (16);
\draw (25) edge [] node[label={[label distance=-.2cm]-45:$h_{m,\ell}$}] {} (16);
\draw (25) edge [] node[label={[label distance=-.2cm]45:$g$}] {} (36);
\draw (16) edge [] node[label={[label distance=-.2cm]135:$g$}] {} (07);
\draw (16) edge [] node[label={[label distance=-.2cm]45:$g'$}] {} (27);
\draw (36) edge [] node[label={[label distance=-.2cm]-45:$a_1$}] {} (27);
\draw (07) edge [] node[label={[label distance=-.2cm]-135:$a_1$}] {} (18);
\draw (27) edge [] node[label={[label distance=-.2cm]-45:$h_{m,\ell}$}] {} (18);
\draw (27) edge [] node[label={[label distance=-.2cm]45:$g$}] {} (38);
\draw (1,-8)--(.5,-8.5) (1,-8)--(1.5,-8.5) (3,-8)--(2.5,-8.5);
\draw[dotted] (3,-8)--(3.5,-7.5) (3,-6)--(3.5,-5.5) (3,-4)--(3.5,-3.5);
\draw[dotted] (0,-7)--(-.5,-6.5) (0,-5)--(-.5,-4.5) (0,-3)--(-.5,-2.5);
\node (0000) at (0,-9) {};
\end{tikzpicture}
\begin{tikzpicture}[x=1.7cm,y=1.7cm]
\node (000) [hasse] at (2,1) {};
\node (10) [hasse,label=left:{$\mathring{\mathcal{S}}^{(0)}$}] at (1,0) {};
\node (30) [hasse,label=right:{$\mathcal{I}$}] at (3,0) {};
\node (01) [hasse,label=left:{$\mathring{\mathcal{S}}^{(0)} \otimes \mathcal{I}$}] at (0,-1) {};
\node (21) [hasse,label=right:{$\mathring{\mathcal{T}}^{(1)}$}] at (2,-1) {};
\node (12) [hasse,label=left:{$\mathring{\mathcal{S}}^{(1)}$}] at (1,-2) {};
\node (32) [hasse,label=right:{$\mathring{\mathcal{T}}^{(1)} \otimes \mathcal{I}$}] at (3,-2) {};
\node (03) [hasse,label=left:{$\mathring{\mathcal{S}}^{(1)} \otimes \mathcal{I}$}] at (0,-3) {};
\node (23) [hasse,label=right:{$\mathring{\mathcal{T}}^{(2)}$}] at (2,-3) {};
\node (14) [hasse,label=left:{$\mathring{\mathcal{S}}^{(2)}$}] at (1,-4) {};
\node (34) [hasse,label=right:{$\mathring{\mathcal{T}}^{(2)} \otimes \mathcal{I}$}] at (3,-4) {};
\node (05) [hasse,label=left:{$\mathring{\mathcal{S}}^{(2)} \otimes \mathcal{I}$}] at (0,-5) {};
\node (25) [hasse,label=right:{$\mathring{\mathcal{T}}^{(3)}$}] at (2,-5) {};
\node (16) [hasse,label=left:{$\mathring{\mathcal{S}}^{(3)}$}] at (1,-6) {};
\node (36) [hasse,label=right:{$\mathring{\mathcal{T}}^{(3)} \otimes \mathcal{I}$}] at (3,-6) {};
\node (07) [hasse,label=left:{$\mathring{\mathcal{S}}^{(3)} \otimes \mathcal{I}$}] at (0,-7) {};
\node (27) [hasse,label=right:{$\mathring{\mathcal{T}}^{(4)}$}] at (2,-7) {};
\node (18) [hasse,label=left:{$\mathring{\mathcal{S}}^{(4)}$}] at (1,-8) {};
\node (38) [hasse,label=right:{$\mathring{\mathcal{T}}^{(4)} \otimes \mathcal{I}$}] at (3,-8) {};\draw (000) edge [] node[label={[label distance=-.2cm]135:$h_{m,\ell}$}] {} (10);
\draw (000) edge [] node[label={[label distance=-.2cm]45:$g$}] {} (30);
\draw (10) edge [] node[label={[label distance=-.2cm]135:$g$}] {} (01);
\draw (10) edge [] node[label={[label distance=.2cm]90:$g'$}] {} (21);
\draw (30) edge [] node[label={[label distance=-.2cm]-45:$\color{red}{A_{\ell -1}}$}] {} (21);
\draw (30) edge [] node[label={[label distance=.2cm]-135:$h_{m,\ell}$}] {} (01);
\draw (01) edge [] node[label={[label distance=-.2cm]-135:$\color{red}{A_{\ell -1}}$}] {} (12);
\draw (21) edge [] node[label={[label distance=-.2cm]-45:$h_{m,\ell}$}] {} (12);
\draw (21) edge [] node[label={[label distance=-.2cm]45:$g$}] {} (32);
\draw (12) edge [] node[label={[label distance=-.2cm]135:$g$}] {} (03);
\draw (12) edge [] node[label={[label distance=-.2cm]45:$g'$}] {} (23);
\draw (32) edge [] node[label={[label distance=-.2cm]-45:$a_1$}] {} (23);
\draw (03) edge [] node[label={[label distance=-.2cm]-135:$a_1$}] {} (14);
\draw (23) edge [] node[label={[label distance=-.2cm]-45:$h_{m,\ell}$}] {} (14);
\draw (23) edge [] node[label={[label distance=-.2cm]45:$g$}] {} (34);
\draw (14) edge [] node[label={[label distance=-.2cm]135:$g$}] {} (05);
\draw (14) edge [] node[label={[label distance=-.2cm]45:$g'$}] {} (25);
\draw (34) edge [] node[label={[label distance=-.2cm]-45:$a_1$}] {} (25);
\draw (05) edge [] node[label={[label distance=-.2cm]-135:$a_1$}] {} (16);
\draw (25) edge [] node[label={[label distance=-.2cm]-45:$h_{m,\ell}$}] {} (16);
\draw (25) edge [] node[label={[label distance=-.2cm]45:$g$}] {} (36);
\draw (16) edge [] node[label={[label distance=-.2cm]135:$g$}] {} (07);
\draw (16) edge [] node[label={[label distance=-.2cm]45:$g'$}] {} (27);
\draw (36) edge [] node[label={[label distance=-.2cm]-45:$a_1$}] {} (27);
\draw (07) edge [] node[label={[label distance=-.2cm]-135:$a_1$}] {} (18);
\draw (27) edge [] node[label={[label distance=-.2cm]-45:$h_{m,\ell}$}] {} (18);
\draw (27) edge [] node[label={[label distance=-.2cm]45:$g$}] {} (38);
\draw (1,-8)--(.5,-8.5) (1,-8)--(1.5,-8.5) (3,-8)--(2.5,-8.5);
\draw[dotted] (3,-8)--(3.5,-7.5) (3,-6)--(3.5,-5.5) (3,-4)--(3.5,-3.5) (3,-2)--(3.5,-1.5);
\draw[dotted] (0,-7)--(-.5,-6.5) (0,-5)--(-.5,-4.5) (0,-3)--(-.5,-2.5);
\node (0000) at (0,-9) {};
\end{tikzpicture}} \\
	\scalebox{.8}{\begin{tabular}{ccccccc||ccccccc}
\toprule 
 $G$ & $\ell$ & $g$ & $g'$ & $h_{m+1,\ell}$ & $h_{m,\ell}$ & $k_{\ell}$ 
 & $G$ & $\ell$ & $g$ & $g'$ & $h_{m+1,\ell}$ & $h_{m,\ell}$ & $k_{\ell}$ \\ 
\midrule
      $E_6$ & $2$ & $e_6$ & $f_4$ & $c_5$ & $c_4$ & $a_1 \times a_1$& $D_4$ & $3$ & $d_4$ & $g_2$ & $h_{4,3}$ & $h_{3,3}$ & $k_3$		 \\  
      $D_4$ & $2$ & $d_4$ & $b_3$ & $c_3$ & $c_2$ & $a_1 \times a_1$ &  $A_1$ & $3$ & $a_1$ & $a_1$ & $h_{2,3}$ & $A_2$ & $k_3$		 \\ 
      $A_2$ & $2$ & $a_2$ & $a_2$ & $c_2$ & $c_1$ & $a_1 \times a_1$ & $A_2$ & $4$ & $a_2$ & $a_1$ & $h_{3,4}$ & $h_{2,4}$ & $k_4$	 \\ 
          \bottomrule
	\end{tabular}}
    \caption{Part of the Hasse diagrams of the $\ess{G}{\ell}{r}$ and $\tee{G}{\ell}{r}$ theories (left) and $\essdg{G}{\ell}{r}$ and $\teedg{G}{\ell}{r}$ theories (right). Each vertex $v$ is a symplectic leaf, and is labeled by the name of a 4d theory $T$, in such a way that the Hasse diagram for the Higgs branch of $T$ is the subdiagram made of all the leaves above $v$. Black lines show transitions as given by the table, these are elementary transitions except for $k_2$ which is an $a_1 \times a_1$ diamond (see Figure \ref{fig:Hasse_Tee_2_theories} for explicit drawing). The dotted lines show transitions to leaves which are not represented (the diagrams of $\ess{G}{\ell}{1}$, $\tee{G}{\ell}{2}$, $\essdg{G}{\ell}{0}$, $\essdg{G}{\ell}{1}$, and $\teedg{G}{\ell}{1}$ are therefore complete). The highest $r$-value displayed is $4$, but the diagrams are infinite towards the bottom. Elementary transitions other than $g$, $g'$, $a_1$ and $h_{m,\ell}$ are shown in red for ease of reading. }
    \label{fig:my_label}
\end{figure}

In order to understand the structure of the Higgs branch of the $\ess{G}{\ell}{r}$ and $\tee{G}{\ell}{r}$ theories of \cite{Giacomelli:2020jel} it is helpful to construct their Hasse diagram \cite{Bourget:2019aer}. As the magnetic quivers for these theories are provided in the previous sections, quiver subtraction can in principle be used to obtain the Hasse diagram of a given theory. This approach was already successfully used in \cite{Bourget:2020asf} for the $\ess{G}{\ell}{1}$ theories, there are however two caveats for $r>1$ and for the $\tee{G}{\ell}{r}$ theories:
\begin{compactenum}[(1)]
\item The magnetic quivers involved allow for subtraction of the same subdiagram representing an elementary slice more than once. This is not captured by the quiver subtraction algorithm presented in \cite{Bourget:2019aer} and a preliminary refinement of the algorithm is proposed in Appendix \ref{AppendixCartography}.
\item Despite not knowing the exact form, one can straightforwardly convince oneself that the full Hasse diagram for these theories quickly becomes messy and complicated as $r$ increases, similar to the moduli space of instantons of classical non A type, with complexity going like the sum of partitions $\sum_{i=1}^r p_i$, where $p_n$ is the number of partitions of $n$. See Equation (3.2) of \cite{Hanany:2018vph}.
\end{compactenum}
Nevertheless it is still possible to construct a subdiagram of the full Hasse diagram of these theories. As shown in \cite{Giacomelli:2020jel} there exists a partial Higgsing along the Higgs branch of $\ess{G}{\ell}{r}$ to $\tee{G}{\ell}{r}$, as well as a partial Higgsing along the Higgs branch of $\tee{G}{\ell}{r}$ to $\ess{G}{\ell}{r-1}$. Let us recall an important principle from \cite{Bourget:2019aer}. When a theory $T_1$ can be Higgsed along its Higgs branch to a theory $T_2$, then the Higgs branch of $T_2$ is the transverse slice to a symplectic leaf $L$ in the Higgs branch of $T_1$. If $T_1$ and $T_2$ have magnetic quivers $\mathsf{Q}_1$ and $\mathsf{Q}_2$ respectively, then one can compute the magnetic quiver for the closure of the leaf $L$ through a suitable subtraction of quivers, $\mathsf{Q}_1-\mathsf{Q}_2$\footnote{Up to possible decoration addressed in Appendix \ref{AppendixCartography}, which is not relevant here.}. It is not required that any of the magnetic quivers represent elementary slices, and one finds a subdiagram of the Hasse diagram of the Higgs branch of $T_1$.

Let us investigate this for the Higgsing of $\ess{E_6}{2}{r}$ to $\tee{E_6}{2}{r}$. Quiver subtraction of the corresponding magnetic quivers yields:
\begin{equation}
\vcenter{\hbox{\scalebox{0.70}{
\begin{tikzpicture}
\node[gauge3,label=below:{1}] (1) at (0,0) {};
\node[gauge3,label=below:{1+r}] (2) at (1,0) {};
\node[gauge3,label=below:{1+2r}] (3) at (2,0) {};
\node[gauge3,label=below:{1+3r}] (4) at (3,0) {};
\node[gauge3,label=below:{1+4r}] (5) at (4,0) {};
\node[gauge3,label=below:{2r}] (6) at (5,0) {};
\draw (1)--(2)--(3)--(4) (5)--(6);
\draw[transform canvas={yshift=-1.5pt}] (4)--(5);
\draw[transform canvas={yshift=1.5pt}] (4)--(5);
\draw (3.7,0.3)--(3.4,0)--(3.7,-0.3);
\end{tikzpicture}
}}}\quad
-\quad
\vcenter{\hbox{\scalebox{0.70}{
\begin{tikzpicture}
\node[gauge3,label=below:{1}] (1) at (0,0) {};
\node[gauge3,label=below:{r}] (2) at (1,0) {};
\node[gauge3,label=below:{2r}] (3) at (2,0) {};
\node[gauge3,label=below:{3r}] (4) at (3,0) {};
\node[gauge3,label=below:{4r}] (5) at (4,0) {};
\node[gauge3,label=below:{2r}] (6) at (5,0) {};
\draw (1)--(2)--(3)--(4) (5)--(6);
\draw[transform canvas={yshift=-1.5pt}] (4)--(5);
\draw[transform canvas={yshift=1.5pt}] (4)--(5);
\draw (3.7,0.3)--(3.4,0)--(3.7,-0.3);
\end{tikzpicture}
}}}\quad
=\quad
\vcenter{\hbox{\scalebox{0.70}{
\begin{tikzpicture}
{\color{red}\node[gauge3,label=left:{1}] (1) at (1,1) {};}
\node[gauge3,label=below:{1}] (2) at (1,0) {};
\node[gauge3,label=below:{1}] (3) at (2,0) {};
\node[gauge3,label=below:{1}] (4) at (3,0) {};
\node[gauge3,label=below:{1}] (5) at (4,0) {};
\draw (2)--(3)--(4);
\draw[transform canvas={yshift=-1.5pt}] (4)--(5);
\draw[transform canvas={yshift=1.5pt}] (4)--(5);
\draw (3.7,0.3)--(3.4,0)--(3.7,-0.3);
\draw[transform canvas={xshift=-1.5pt}] (1)--(2);
\draw[transform canvas={xshift=1.5pt}] (1)--(2);
\draw (0.7,0.7)--(1,0.4)--(1.3,0.7);
\end{tikzpicture}
}}}
\end{equation}
Where the $\mathrm{U}(1)$ node used to rebalance the resulting quiver is colored \textcolor{red}{red}, and we rebalance with a non-simply laced edge as described in Appendix \ref{AppendixCartography}. The result is an elementary slice, $c_4$, hence the corresponding Higgsing is minimal\footnote{One cannot Higgs to a third theory in between, without also unHiggsing.}.

Let us turn to the Higgsing of $\tee{E_6}{2}{r}$ to $\ess{E_6}{2}{r-1}$. Quiver subtraction of the corresponding magnetic quivers yields:
\begin{equation}
\vcenter{\hbox{\scalebox{0.70}{
\begin{tikzpicture}
\node[gauge3,label=below:{$1$}] (1) at (0,0) {};
\node[gauge3,label=below:{$r$}] (2) at (1,0) {};
\node[gauge3,label=below:{$2r$}] (3) at (2,0) {};
\node[gauge3,label=below:{$3r$}] (4) at (3,0) {};
\node[gauge3,label=below:{$4r$}] (5) at (4,0) {};
\node[gauge3,label=below:{$2r$}] (6) at (5,0) {};
\draw (1)--(2)--(3)--(4) (5)--(6);
\draw[transform canvas={yshift=-1.5pt}] (4)--(5);
\draw[transform canvas={yshift=1.5pt}] (4)--(5);
\draw (3.7,0.3)--(3.4,0)--(3.7,-0.3);
\end{tikzpicture}
}}}\quad
-\quad
\vcenter{\hbox{\scalebox{0.70}{
\begin{tikzpicture}
\node[gauge3,label=below:{$1$}] (1) at (0,0) {};
\node[gauge3,label=below:{$r$}] (2) at (1,0) {};
\node[gauge3,label=below:{$2r{-}1$}] (3) at (2,0) {};
\node[gauge3,label=below:{$3r{-}2$}] (4) at (3,0) {};
\node[gauge3,label=below:{$4r{-}3$}] (5) at (4,0) {};
\node[gauge3,label=below:{$2r{-}2$}] (6) at (5,0) {};
\draw (1)--(2)--(3)--(4) (5)--(6);
\draw[transform canvas={yshift=-1.5pt}] (4)--(5);
\draw[transform canvas={yshift=1.5pt}] (4)--(5);
\draw (3.7,0.3)--(3.4,0)--(3.7,-0.3);
\end{tikzpicture}
}}}\quad
=\quad
\vcenter{\hbox{\scalebox{0.70}{
\begin{tikzpicture}
\node[gauge3,label=below:{$1$}] (3) at (2,0) {};
\node[gauge3,label=below:{$2$}] (4) at (3,0) {};
\node[gauge3,label=below:{$3$}] (5) at (4,0) {};
\node[gauge3,label=below:{$2$}] (6) at (5,0) {};
{\color{red}\node[gauge3,label=below:{$1$}] (7) at (6,0) {};}
\draw (3)--(4) (5)--(6)--(7);
\draw[transform canvas={yshift=-1.5pt}] (4)--(5);
\draw[transform canvas={yshift=1.5pt}] (4)--(5);
\draw (3.7,0.3)--(3.4,0)--(3.7,-0.3);
\end{tikzpicture}
}}}
\label{eq:StoT}
\end{equation}
Again the result is an elementary slice, $f_4$, hence the corresponding Higgsing is minimal.

We can now turn to the Higgsing of $\ess{E_6}{2}{r}$ to $\ess{E_6}{2}{r-1}$. Quiver subtraction of the corresponding magnetic quivers yields:
\begin{equation}
\vcenter{\hbox{\scalebox{0.70}{
\begin{tikzpicture}
\node[gauge3,label=below:{$1$}] (1) at (0,0) {};
\node[gauge3,label=below:{$1{+}r$}] (2) at (1,0) {};
\node[gauge3,label=below:{$1{+}2r$}] (3) at (2,0) {};
\node[gauge3,label=below:{$1{+}3r$}] (4) at (3,0) {};
\node[gauge3,label=below:{$1{+}4r$}] (5) at (4,0) {};
\node[gauge3,label=below:{$2r$}] (6) at (5,0) {};
\draw (1)--(2)--(3)--(4) (5)--(6);
\draw[transform canvas={yshift=-1.5pt}] (4)--(5);
\draw[transform canvas={yshift=1.5pt}] (4)--(5);
\draw (3.7,0.3)--(3.4,0)--(3.7,-0.3);
\end{tikzpicture}
}}}\quad
-\quad
\vcenter{\hbox{\scalebox{0.70}{
\begin{tikzpicture}
\node[gauge3,label=below:{$1$}] (1) at (0,0) {};
\node[gauge3,label=below:{$r$}] (2) at (1,0) {};
\node[gauge3,label=below:{$2r{-}1$}] (3) at (2,0) {};
\node[gauge3,label=below:{$3r{-}2$}] (4) at (3,0) {};
\node[gauge3,label=below:{$4r{-}3$}] (5) at (4,0) {};
\node[gauge3,label=below:{$2r{-}2$}] (6) at (5,0) {};
\draw (1)--(2)--(3)--(4) (5)--(6);
\draw[transform canvas={yshift=-1.5pt}] (4)--(5);
\draw[transform canvas={yshift=1.5pt}] (4)--(5);
\draw (3.7,0.3)--(3.4,0)--(3.7,-0.3);
\end{tikzpicture}
}}}\quad
=\quad
\vcenter{\hbox{\scalebox{0.70}{
\begin{tikzpicture}
\node[gauge3,label=below:{$1$}] (2) at (1,0) {};
\node[gauge3,label=below:{$2$}] (3) at (2,0) {};
\node[gauge3,label=below:{$3$}] (4) at (3,0) {};
\node[gauge3,label=below:{$4$}] (5) at (4,0) {};
\node[gauge3,label=below:{$2$}] (6) at (5,0) {};
{\color{red}\node[gauge3,label=below:{$1$}] (7) at (6,0) {};}
\draw (2)--(3)--(4) (5)--(6)--(7);
\draw[transform canvas={yshift=-1.5pt}] (4)--(5);
\draw[transform canvas={yshift=1.5pt}] (4)--(5);
\draw (3.7,0.3)--(3.4,0)--(3.7,-0.3);
\end{tikzpicture}
}}}
\label{eq:TtoS}
\end{equation}
This is not a minimal slice, and we can ask for the Hasse diagram of its magnetic quiver. This is easily obtained from quiver subtraction:
\begin{equation}
\begin{tikzpicture}
\node (a) at (0,0) {$\scalebox{0.70}{
\begin{tikzpicture}
\node[gauge3,label=below:{1}] (2) at (1,0) {};
\node[gauge3,label=below:{2}] (3) at (2,0) {};
\node[gauge3,label=below:{3}] (4) at (3,0) {};
\node[gauge3,label=below:{4}] (5) at (4,0) {};
\node[gauge3,label=below:{2}] (6) at (5,0) {};
\node[gauge3,label=below:{1}] (7) at (6,0) {};
\draw (2)--(3)--(4) (5)--(6)--(7);
\draw[transform canvas={yshift=-1.5pt}] (4)--(5);
\draw[transform canvas={yshift=1.5pt}] (4)--(5);
\draw (3.7,0.3)--(3.4,0)--(3.7,-0.3);
\end{tikzpicture}
}$
};
\node (b) at (-3,-2) {$\scalebox{0.70}{
\begin{tikzpicture}
\node[gauge3,label=below:{1}] (7) at (6,0) {};
{\color{red}\node[gauge3,label=below:{1}] (8) at (7,0) {};}
\draw[transform canvas={yshift=-1.5pt}] (7)--(8);
\draw[transform canvas={yshift=1.5pt}] (7)--(8);
\end{tikzpicture}
}$
};
\node (c) at (3,-2) {$\scalebox{0.70}{
\begin{tikzpicture}
{\color{red}\node[gauge3,label=left:{1}] (1) at (1,1) {};}
\node[gauge3,label=below:{1}] (2) at (1,0) {};
\node[gauge3,label=below:{1}] (3) at (2,0) {};
\node[gauge3,label=below:{1}] (4) at (3,0) {};
\node[gauge3,label=below:{1}] (5) at (4,0) {};
\draw (2)--(3)--(4);
\draw[transform canvas={yshift=-1.5pt}] (4)--(5);
\draw[transform canvas={yshift=1.5pt}] (4)--(5);
\draw (3.7,0.3)--(3.4,0)--(3.7,-0.3);
\draw[transform canvas={xshift=-1.5pt}] (1)--(2);
\draw[transform canvas={xshift=1.5pt}] (1)--(2);
\draw (0.7,0.7)--(1,0.4)--(1.3,0.7);
\end{tikzpicture}
}$
};
\node (d) at (0,-4) {$\scalebox{0.70}{
\begin{tikzpicture}
{\color{red}\node[gauge3,label=below:{1}] at (0,0) {};}
\end{tikzpicture}
}$
};
\draw[->] (a)--(b);
\draw[->] (b)--(d);
\draw[->] (a)--(c);
\draw[->] (c)--(d);
\node at (-2.5,-0.75) {$-e_6$};
\node at (2.5,-0.75) {$-f_4$};
\node at (-2.5,-3.5) {$-A_1$};
\node at (2,-3.5) {$-c_4$};
\end{tikzpicture}
\label{eq:StoS}
\end{equation}
The $c_4$ slice in \eqref{eq:StoS} looks like the $c_4$ slice of \eqref{eq:StoT}, and the $f_4$ slice in \eqref{eq:StoS} looks like the $f_4$ slice of \eqref{eq:TtoS}. This is consistent with the minimal Higgsings $\ess{E_6}{2}{r}\rightarrow\tee{E_6}{2}{r}\rightarrow\ess{E_6}{2}{r-1}$. But what is the other theory we can obtain from $\ess{E_6}{2}{r}$ through minimal Higgsing, corresponding to the $A_1$ slice? Can we find a magnetic quiver for this theory, i.e.\ can we subtract a quiver from the magnetic quiver of $\ess{E_6}{2}{r}$ and obtain the $A_1$ of \eqref{eq:StoS}?

\begin{figure}[t]
    \centering
    \begin{tikzpicture}
    \node (a) at (0,0) {$\mathsf{Q}(\ess{E_6}{2}{r})=\vcenter{\hbox{\scalebox{0.70}{
    \begin{tikzpicture}
    \node[gauge3,label=below:{$1$}] (1) at (0,0) {};
    \node[gauge3,label=below:{$1{+}r$}] (2) at (1,0) {};
    \node[gauge3,label=below:{$1{+}2r$}] (3) at (2,0) {};
    \node[gauge3,label=below:{$1{+}3r$}] (4) at (3,0) {};
    \node[gauge3,label=below:{$1{+}4r$}] (5) at (4,0) {};
    \node[gauge3,label=below:{$2r$}] (6) at (5,0) {};
    \draw (1)--(2)--(3)--(4) (5)--(6);
    \draw[transform canvas={yshift=-1.5pt}] (4)--(5);
    \draw[transform canvas={yshift=1.5pt}] (4)--(5);
    \draw (3.7,0.3)--(3.4,0)--(3.7,-0.3);
    \end{tikzpicture}
    }}}$};
    \node (b) at (-3,-2) {$\scalebox{0.70}{
    \begin{tikzpicture}
    {\color{olive}\node[gauge3,label=left:{1}] (0) at (0,1) {};}
    \node[gauge3,label=below:{$1$}] (1) at (0,0) {};
    \node[gauge3,label=below:{$r$}] (2) at (1,0) {};
    \node[gauge3,label=below:{$2r{-}1$}] (3) at (2,0) {};
    \node[gauge3,label=below:{$3r{-}2$}] (4) at (3,0) {};
    \node[gauge3,label=below:{$4r{-}3$}] (5) at (4,0) {};
    \node[gauge3,label=below:{$2r{-}2$}] (6) at (5,0) {};
    \draw[transform canvas={xshift=-1.5pt}] (0)--(1);
    \draw[transform canvas={xshift=1.5pt}] (0)--(1);
    \draw (-0.3,0.7)--(0,0.4)--(0.3,0.7);
    \draw (1)--(2)--(3)--(4) (5)--(6);
    \draw[transform canvas={yshift=-1.5pt}] (4)--(5);
    \draw[transform canvas={yshift=1.5pt}] (4)--(5);
    \draw (3.7,0.3)--(3.4,0)--(3.7,-0.3);
    \end{tikzpicture}
    }$};
    \node (c) at (3,-4) {$\scalebox{0.70}{
    \begin{tikzpicture}
    \node[gauge3,label=below:{$1$}] (2) at (1,0) {};
    \node[gauge3,label=below:{$2$}] (3) at (2,0) {};
    \node[gauge3,label=below:{$3$}] (4) at (3,0) {};
    \node[gauge3,label=below:{$4$}] (5) at (4,0) {};
    \node[gauge3,label=below:{$2$}] (6) at (5,0) {};
    {\color{blue}\node[gauge3,label=right:{$1$}] (7) at (5,1) {};}
    \draw (2)--(3)--(4) (5)--(6)--(7);
    \draw[transform canvas={yshift=-1.5pt}] (4)--(5);
    \draw[transform canvas={yshift=1.5pt}] (4)--(5);
    \draw (3.7,0.3)--(3.4,0)--(3.7,-0.3);
    \end{tikzpicture}
    }$};
    \node (d) at (0,-6) {$\scalebox{0.70}{
    \begin{tikzpicture}
    {\color{olive}\node[gauge3,label=left:{$1$}] (1) at (1,0) {};}
    {\color{blue}\node[gauge3,label=right:{$1$}] (2) at (2,0) {};}
    \draw[transform canvas={yshift=-1.5pt}] (1)--(2);
    \draw[transform canvas={yshift=1.5pt}] (1)--(2);
    \end{tikzpicture}
    }$};
    \node (e) at (0,-8) {$\scalebox{0.70}{
    \begin{tikzpicture}
    {\color{red}\node[gauge3,label=below:{$1$}] (1) at (1,0) {};}
    \end{tikzpicture}
    }$};
    \draw[->] (a)--(b);
    \draw[->] (b)--(d);
    \draw[->] (a)--(c);
    \draw[->] (c)--(d);
    \draw[->] (d)--(e);
    \node at (-2,-0.8) {$-e_6$};
    \node at (3,-2) {$-\mathsf{Q}(\ess{E_6}{2}{r-1})$};
    \node at (2,-5.5) {$-e_6$};
    \node at (-3,-4) {$-\mathsf{Q}(\ess{E_6}{2}{r-1})$};
    \node at (-0.5,-7) {$-A_1$};
    \end{tikzpicture}
    \caption{Demonstration of the emergence of an $A_1$ slice in the Higgs branch of $\ess{E_6}{2}{r}$  via quiver subtraction on its magnetic quiver. $\mathsf{Q}(\ess{E_6}{2}{r-1})$ is a shorthand for the magnetic quiver of $\ess{E_6}{2}{r-1}$.  }
    \label{fig:SubtProd}
\end{figure}
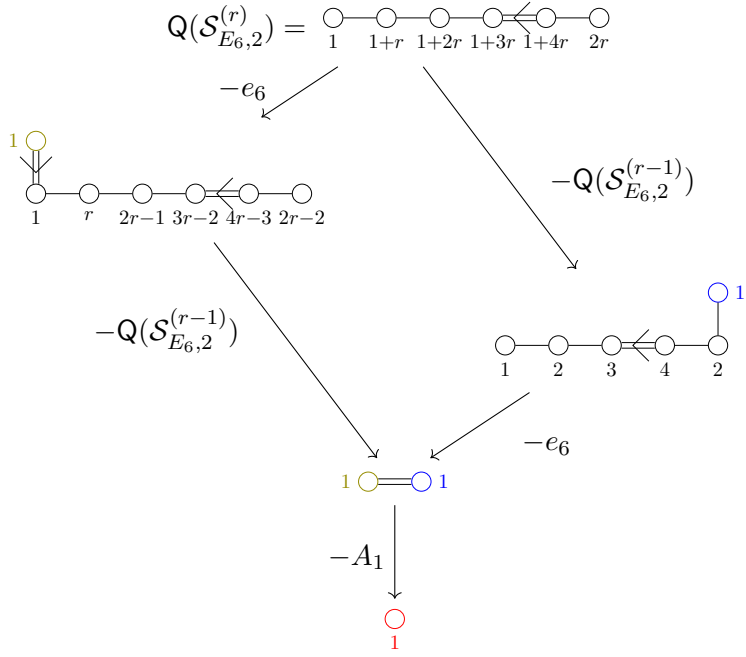

As shown in Figure \ref{fig:SubtProd}, when performing quiver subtraction on the magnetic quiver of $\ess{E_6}{2}{r}$, we can reach the $A_1$ slice of \eqref{eq:StoS} by subtracting the magnetic quiver of $\ess{E_6}{2}{r-1}$ and then subtracting $e_6$. However, the $\mathrm{U}(1)$ node used to rebalance after the first subtraction, colored {\color{blue}blue} in Figure \ref{fig:SubtProd}, is not affected by the following subtraction of the $e_6$ slice, and a second independent $\mathrm{U}(1)$ node, colored {\color{olive}olive} in Figure \ref{fig:SubtProd}, is added to rebalance. The $A_1$ slice at the bottom of the Hasse diagram of $\ess{E_6}{2}{r}$ is therefore made up of two `rebalancing $\mathrm{U}(1)$s', the blue and the olive. Therefore it is impossible to find a single quiver which can be subtracted from the magnetic quiver of $\ess{E_6}{2}{r}$ and obtain this $A_1$, as only one rebalancing $\mathrm{U}(1)$ would be added. Also shown in Figure \ref{fig:SubtProd} is that the order in which one subtracts the two slices does not matter. We can therefore expect the moduli space of the theory obtained through the Higgsing corresponding to the $A_1$ slice to be a product of $e_6$ and the Higgs branch of $\ess{E_6}{2}{r-1}$. In fact this leads to the conjecture, that the theory is really a product of $\ess{E_6}{2}{r-1}$ and $\mathcal{I}_{E_6}^{(1)}$\footnote{$\mathcal{I}_{G}^{(r)}$ denotes the theory whose Higgs branch is the reduced moduli space of $r$ $G$-instantons.}. In a brane picture the rebalancing of $\mathrm{U}(1)$s correspond to a collection of branes that align leading to the possibility of turning on Coulomb branch moduli. In the present case there are two independent such $\mathrm{U}(1)$s leading to the impression that there are two independent theories. This is consistent with the analysis of \cite{Giacomelli:2020jel}.

Similarly subtracting the magnetic quiver for $\tee{E_6}{2}{r-1}$ from the magnetic quiver for $\tee{E_6}{2}{r}$ yields a non-elementary slice:
\begin{equation}
    \vcenter{\hbox{\scalebox{0.70}{
    \begin{tikzpicture}
    \node[gauge3,label=below:{$1$}] (1) at (0,0) {};
    \node[gauge3,label=below:{$r$}] (2) at (1,0) {};
    \node[gauge3,label=below:{$2r$}] (3) at (2,0) {};
    \node[gauge3,label=below:{$3r$}] (4) at (3,0) {};
    \node[gauge3,label=below:{$4r$}] (5) at (4,0) {};
    \node[gauge3,label=below:{$2r$}] (6) at (5,0) {};
    \draw (1)--(2)--(3)--(4) (5)--(6);
    \draw[transform canvas={yshift=-1.5pt}] (4)--(5);
    \draw[transform canvas={yshift=1.5pt}] (4)--(5);
    \draw (3.7,0.3)--(3.4,0)--(3.7,-0.3);
    \end{tikzpicture}
    }}}\quad
    - \quad
    \vcenter{\hbox{\scalebox{0.70}{
    \begin{tikzpicture}
    \node[gauge3,label=below:{$1$}] (1) at (0,0) {};
    \node[gauge3,label=below:{$r{-}1$}] (2) at (1,0) {};
    \node[gauge3,label=below:{$2r{-}2$}] (3) at (2,0) {};
    \node[gauge3,label=below:{$3r{-}3$}] (4) at (3,0) {};
    \node[gauge3,label=below:{$4r{-}4$}] (5) at (4,0) {};
    \node[gauge3,label=below:{$2r{-}2$}] (6) at (5,0) {};
    \draw (1)--(2)--(3)--(4) (5)--(6);
    \draw[transform canvas={yshift=-1.5pt}] (4)--(5);
    \draw[transform canvas={yshift=1.5pt}] (4)--(5);
    \draw (3.7,0.3)--(3.4,0)--(3.7,-0.3);
    \end{tikzpicture}
    }}}\quad
    = \quad
    \vcenter{\hbox{\scalebox{0.70}{
    \begin{tikzpicture}
    {\color{red}\node[gauge3,label=left:{1}] (0) at (1,1) {};}
    \node[gauge3,label=below:{$1$}] (2) at (1,0) {};
    \node[gauge3,label=below:{$2$}] (3) at (2,0) {};
    \node[gauge3,label=below:{$3$}] (4) at (3,0) {};
    \node[gauge3,label=below:{$4$}] (5) at (4,0) {};
    \node[gauge3,label=below:{$2$}] (6) at (5,0) {};
    \draw[transform canvas={xshift=-1.5pt}] (0)--(2);
    \draw[transform canvas={xshift=1.5pt}] (0)--(2);
    \draw (0.7,0.7)--(1,0.4)--(1.3,0.7);
    \draw (2)--(3)--(4) (5)--(6);
    \draw[transform canvas={yshift=-1.5pt}] (4)--(5);
    \draw[transform canvas={yshift=1.5pt}] (4)--(5);
    \draw (3.7,0.3)--(3.4,0)--(3.7,-0.3);
    \end{tikzpicture}
    }}}
\end{equation}
And the same argument as in the previous paragraph can be used to argue the existence of the minimal Higgsing $\tee{E_6}{2}{r}\rightarrow\tee{E_6}{2}{r-1}\otimes\mathcal{I}_{E_6}^{(1)}$. 

Combining all ingredients in the above discussion into the general form of a subdiagram of the Hasse diagram of $\ess{G}{\ell}{r}$ yields the diagram in the left part of \ref{hasseFold}. Note that by doing so, we also obtain the subdiagrams for the $\tee{G}{\ell}{r}$ theories. Finally, similar computations allow to compute the analog diagram for the $\essdg{G}{\ell}{r}$ and $\teedg{G}{\ell}{r}$ theories, which is displayed on the right part of \ref{hasseFold}.

\section{Outlook}

In this article, we provide the magnetic quivers for the Higgs branches of the 4d $\mathcal{N}=2$ theories $\ess{G}{\ell}{r}$ and $\tee{G}{\ell}{r}$. The results of Tables \ref{resulttable} and \ref{resulttableT} are obtained via three independent approaches. 
\begin{compactitem}
\item First, based on recent progress in the understanding of magnetic quiver techniques and the underlying 3d $\mathcal{N}=4$ Coulomb branch moduli space, the candidate magnetic quivers for the $\mathcal{S}$-fold theories are derived in Section \ref{sectionMQ} from the known Higgs branch properties: dimension, global symmetry, and the order of the $\mathbb{Z}_{\ell}$ action.
\item Second, $\mathcal{S}$-fold theories are derived via $T^2$-compactification from certain 6d $\mathcal{N}=(1,0)$ theories. Starting from the magnetic quiver of the 6d SCFT, a sequence of FI-deformations results in an auxiliary magnetic quiver (corresponding to a 5d theory after mass deformation). A subsequent $\mathbb{Z}_{\ell}$ folding on this magnetic quiver produces the magnetic quiver for the 4d SCFT in Section \ref{SectionFI}.
\item Third, an $S^1$ compactification of 6d $\mathcal{N}=(1,0)$ theory, followed by a mass-deformation results in a 5d theory with a known brane-web realization. Building on the generalized toric polygons, the magnetic quiver for the 5d SCFT is derived and subsequently folded to the 4d magnetic quiver in Section \ref{sectionBW}.
\end{compactitem}
These independent derivations provide a non-trivial consistency check of the proposed magnetic quivers. Thereafter, we investigate the Higgs branch Hasse diagram of the $\mathcal{S}$-fold theories in Section \ref{sec:Hasse}. Here, our results show that a full analysis of the Hasse diagram requires further studies, beyond the comments in Appendix \ref{AppendixCartography}.
\paragraph{Discrete Gauging.} Since the magnetic quivers of Tables \ref{resulttable} and \ref{resulttableT} are non-simply laced one can change the ungauging scheme \cite{Hanany:2020jzl} and obtain new quivers, listed in Tables \ref{shortS} and \ref{shortR}. These quivers are proposed to be magnetic quivers for SCFTs obtained from discretely gauging the $\mathbb{Z}_{\ell}$ global symmetry of the $\ess{G}{\ell}{r}$ or $\tee{G}{\ell}{r}$ theories. We call these theories $\essdg{G}{\ell}{r}$ and $\teedg{G}{\ell}{r}$ respectively.
\paragraph{Open questions.} Before closing, we list open questions that are encountered:
\begin{compactenum}[(1)]
    \item The FI-deformations of Section \ref{SectionFI} require a more systematic analysis. In particular, it is desirable to understand the choice of FI-deformation and to determine the set of equivalent FI-deformations.
    \item In view of the Hasse diagrams in Figure \ref{fig:my_label}, the Higgsing $\tee{G}{\ell}{2} \to \mathcal{I}_{G}^{(2)}$ predicted in \cite{Giacomelli:2020jel} is beyond the reach of current magnetic quiver techniques. More generally, the quiver subtraction algorithm, used to derive the Hasse diagram from magnetic quivers, is not complete yet, as discussed in Appendix \ref{AppendixCartography}.
    \item As seen above, by ungauging on the short node we can describe moduli spaces of instantons with outer automorphisms twists and it would be interesting to confirm our results with an explicit construction of these spaces. It would also be important to clarify the interpretation of this ungauging scheme from the viewpoint of four-dimensional SCFTs and its connection with discrete gaugings.
    \item The $\essdg{G}{\ell}{1}$ theories are not part of the classification of $4d$ $\mathcal{N}=2$ SCFTs, which leaves the question whether the set of rank 1 SCFTs is larger than previously expected.
\end{compactenum}

\section*{Acknowledgements}
We are grateful to Marieke Van Beest, Julius Eckhard, Mario Martone, Sakura Sch\"afer-Nameki, Yuji Tachikawa and Gabi Zafrir for helpful discussions. The work of AB, JFG, AH and ZZ is supported by the STFC Consolidated Grant  ST/P000762/1 and ST/T000791/1. The work of SG is supported by the ERC Consolidator Grant 682608 ``Higgs bundles: Supersymmetric Gauge Theories and Geometry (HIGGSBNDL)''. The work of MS is supported by the National Thousand-Young-Talents Program of China, the National Natural Science Foundation of China (grant no.\ 11950410497), and the China Postdoctoral Science Foundation (grant no.\ 2019M650616).

\appendix

\section{Two cones for \texorpdfstring{$\ess{A_2}{2}{r}$}{S A2 2 r}}
\label{AppendixCone}
As mentioned in Section \ref{sectionBW}, as well as in \cite{Bourget:2020asf} for the rank $1$ case, the 5d SCFT, whose twisted compactification yields the $\ess{A_2}{2}{r}$ theory, has two magnetic quivers. These two magnetic quivers correspond to two maximal decompositions of the brane web, namely 
\begin{alignat}{2}
\label{firstConeMQ}
     &\raisebox{-.5\height}{\scalebox{1}{
   \begin{tikzpicture}
\draw[green] (-1,0)--(1,0);
\draw[green] (-1,.05)--(1,0.05);
\draw[green] (-1,.00)--(1,0.00);
\draw[green] (-1,-.05)--(1,-0.05);
\draw[red] (-.025,-1)--(-0.025,1);
\draw[red] (.025,-1)--(0.025,1);
\draw[blue] (-1.025,.975)--(.975,-1.025);
\draw[blue] (-.975,1.025)--(1.025,-.975);
\draw[orange] (-2,0)--(-1,0);
\draw[purple] (2,0)--(1,0);
\node[D7] at (-2,0) {};  
\node[D7] at (-1,0) {};  
\node[D7] at (1,0) {};  
\node[D7] at (2,0) {};  
\node[D7] at (-1,1) {};  
\node[D7] at (1,-1) {};  
\node[D7] at (0,1) {};  
\node[D7] at (0,-1) {};  
\end{tikzpicture}
   }} & \qquad \qquad  &\raisebox{-.5\height}{  \scalebox{0.96}{ \begin{tikzpicture}[x=1.2cm,y=.8cm]
    \node (g0) at (0,0) [gauge,label=below:{\small $1$},orange] {};
    \node (g1) at (1,0) [gauge,label=below:{\small $1{+}r$},green] {};
    \node (g4) at (2,0) [gauge,label=below:{\small $1$},purple] {};
    \node (g10) at (0.5,1) [gauge,label=above:{\small $r$},red] {};
    \node (g11) at (1.5,1) [gauge,label=above:{\small $r$},blue] {};
    \draw (g0)--(g1)--(g4);
    \draw (g1)--(g10)--(g11)--(g1);
    \end{tikzpicture}}}
    \,,\\
\label{secondConeMQ}
     &\raisebox{-.5\height}{\scalebox{1}{
   \begin{tikzpicture}
\draw[green] (-1,0)--(1,0);
\draw[blue] (-1,.05)--(0,0.05);
\draw[blue] (-1,.1)--(0,.1);
\draw[blue] (0,.05)--(0,1);
\draw[blue] (.05,.05)--(0.05,1);
\draw[blue] (.05,.05)--(1.05,-.95);
\draw[blue] (.00,.05)--(1.00,-.95);
\draw[red] (0,-.05)--(1,-0.05);
\draw[red] (0,-.1)--(1,-.1);
\draw[red] (0,-.05)--(0,-1);
\draw[red] (-.05,-.05)--(-0.05,-1);
\draw[red] (-.05,-.05)--(-1.05,.95);
\draw[red] (.00,-.05)--(-1.00,.95);
\draw[orange] (-2,0)--(-1,0);
\draw[purple] (2,0)--(1,0);
\node[D7] at (-2,0) {};  
\node[D7] at (-1,0) {};  
\node[D7] at (1,0) {};  
\node[D7] at (2,0) {};  
\node[D7] at (-1,1) {};  
\node[D7] at (1,-1) {};  
\node[D7] at (0,1) {};  
\node[D7] at (0,-1) {};  
\end{tikzpicture}
   }}  &\qquad \qquad   &\raisebox{-.5\height}{  \scalebox{0.96}{\begin{tikzpicture}[x=1.2cm,y=.8cm]
\draw (0.5,.95)--(1.5,.95);
\draw (0.5,1.05)--(1.5,1.05);
\node (g0) at (0,0) [gauge,label=below:{\small $1$},orange] {};
\node (g1) at (1,0) [gauge,label=below:{\small $1$},green] {};
\node (g4) at (2,0) [gauge,label=below:{\small $1$},purple] {};
\node (g10) at (0.5,1) [gauge,label=above:{\small $r$},red] {};
\node (g11) at (1.5,1) [gauge,label=above:{\small $r$},blue] {};
\draw (g10)--(g0)--(g1)--(g4)--(g11);
\end{tikzpicture}}}
\,.
\end{alignat}
However, the $\ess{A_2}{2}{r}$ theory has only one magnetic quiver, which is obtained from (\ref{firstConeMQ}) by $\mathbb{Z}_2$ folding. It is natural to ask why (\ref{firstConeMQ}) should be chosen for folding rather than (\ref{secondConeMQ}). The goal of this Appendix is to answer this question. 

First of all, a complete description of the 5d Higgs branch requires not only the two magnetic quivers \eqref{firstConeMQ} and \eqref{secondConeMQ}, each giving one cone, but also a magnetic quiver for their intersection. The brane web decomposition is straightforward, and we conjecture the following magnetic quiver: 
\begin{equation}
\label{quiverIntersection}
  \raisebox{-.5\height}{ \begin{tikzpicture}
\draw[green] (-1,0)--(1,0);
\draw[blue] (-1,.05)--(0,0.05);
\draw[blue] (-1,.1)--(0,.1);
\draw[blue] (0,.05)--(0,1);
\draw[blue] (.05,.05)--(0.05,1);
\draw[blue] (.05,.05)--(1.05,-.95);
\draw[blue] (.00,.05)--(1.00,-.95);
\draw[blue] (0,-.05)--(1,-0.05);
\draw[blue] (0,-.1)--(1,-.1);
\draw[blue] (0,-.05)--(0,-1);
\draw[blue] (-.05,-.05)--(-0.05,-1);
\draw[blue] (-.05,-.05)--(-1.05,.95);
\draw[blue] (.00,-.05)--(-1.00,.95);
\draw[orange] (-2,0)--(-1,0);
\draw[purple] (2,0)--(1,0);
\node[D7] at (-2,0) {};  
\node[D7] at (-1,0) {};  
\node[D7] at (1,0) {};  
\node[D7] at (2,0) {};  
\node[D7] at (-1,1) {};  
\node[D7] at (1,-1) {};  
\node[D7] at (0,1) {};  
\node[D7] at (0,-1) {};  
\end{tikzpicture}}
\qquad \qquad \qquad 
        \raisebox{-.5\height}{\scalebox{0.96}{\begin{tikzpicture}[x=1.2cm,y=.8cm]
    \node (g0) at (0,0) [gauge,label=below:{\small $1$},orange] {};
    \node (g1) at (1,0) [gauge,label=below:{\small $1$},green] {};
    \node (g4) at (2,0) [gauge,label=below:{\small $1$},purple] {};
    \node (g10) at (1,1) [gauge,label=left:{\small $r$},blue] {};
    \draw (g10)--(g0)--(g1)--(g4)--(g10);
    \draw (g10) to [out=45,in=135,looseness=10] (g10);
    \end{tikzpicture}}}
    \, .
\end{equation}
The identification of the adjoint hypermultiplet in the quiver, represented by a loop, is not covered by the rules in \cite{Cabrera:2018jxt}, but does show up in simpler brane systems like in \cite[Sec.\ 2.3]{Cabrera:2019izd}. We give arguments that there is an analogous phenomenon in brane webs. It comes from the fact that the blue subweb is a stack of two identical subwebs, which have a self-intersection number of 0, while all the other colored subwebs in \eqref{firstConeMQ}, \eqref{secondConeMQ}, and \eqref{quiverIntersection} are stacks of subwebs with self-intersection number $-2$. This is explained in more detail in Appendix \ref{AppendixRules}. From the point of view of quiver subtraction, subtracting an $a_2$ quiver $r$ times from \eqref{firstConeMQ} or an $a_1$ quiver $r$ times from \eqref{secondConeMQ}, and taking into account that one needs to rebalance with a $\mathrm{U}(r)$ with an adjoint as detailed in Appendix \ref{AppendixCartography}, one obtains the quiver \eqref{quiverIntersection}. 

Next, we consider the effect of the $\mathbb{Z}_2$ folding on those brane webs and quivers. On the brane web, the $\mathbb{Z}_2$ acts by a $180$ degrees rotation around the center. Crucially, note that the brane web decompositions in \eqref{firstConeMQ} and \eqref{quiverIntersection} are $\mathbb{Z}_2$ invariant \emph{up to identification of the degrees of freedom of two disconnected subwebs}, namely the orange and purple segments. On the other hand, the brane web decomposition \eqref{secondConeMQ} does not share this property: in order to make the decomposition $\mathbb{Z}_2$ invariant, one needs to identify the degrees of freedom of the blue and red subwebs, effectively reducing to the decomposition \eqref{quiverIntersection}. Consequently, after folding we end up with the two following quivers: 
\begin{alignat}{2}
\label{quiverAfterFoldingHigh}
     &\raisebox{-.5\height}{\scalebox{1}{
   \begin{tikzpicture}
\draw[green] (-1,0)--(1,0);
\draw[green] (-1,.05)--(1,0.05);
\draw[green] (-1,.00)--(1,0.00);
\draw[green] (-1,-.05)--(1,-0.05);
\draw[red] (-.025,-1)--(-0.025,1);
\draw[red] (.025,-1)--(0.025,1);
\draw[blue] (-1.025,.975)--(.975,-1.025);
\draw[blue] (-.975,1.025)--(1.025,-.975);
\draw[orange] (-2,0)--(-1,0);
\draw[orange] (2,0)--(1,0);
\node[D7] at (-2,0) {};  
\node[D7] at (-1,0) {};  
\node[D7] at (1,0) {};  
\node[D7] at (2,0) {};  
\node[D7] at (-1,1) {};  
\node[D7] at (1,-1) {};  
\node[D7] at (0,1) {};  
\node[D7] at (0,-1) {};  
\node at (2,-1) {$/ \mathbb{Z}_2$};
\end{tikzpicture}
   }}  &\qquad \qquad    &\raisebox{-.5\height}{  \scalebox{0.96}{ \begin{tikzpicture}[x=1.2cm,y=.8cm]
   \draw (0,.05)--(1,.05);
\draw (0,-.05)--(1,-.05);
\draw (.4,0)--(.6,.2);
\draw (.4,0)--(.6,-.2);
    \node (g0) at (0,0) [gauge,label=below:{\small $1$},orange] {};
    \node (g1) at (1,0) [gauge,label=below:{\small $1{+}r$},green] {};
    \node (g10) at (2,1) [gauge,label=right:{\small $r$},red] {};
    \node (g11) at (2,-1) [gauge,label=right:{\small $r$},blue] {};
    \draw (g1)--(g10)--(g11)--(g1);
    \end{tikzpicture}}}
    \,, \\
\label{quiverAfterFoldingLow}
  &\raisebox{-.5\height}{ \begin{tikzpicture}
\draw[green] (-1,0)--(1,0);
\draw[blue] (-1,.05)--(0,0.05);
\draw[blue] (-1,.1)--(0,.1);
\draw[blue] (0,.05)--(0,1);
\draw[blue] (.05,.05)--(0.05,1);
\draw[blue] (.05,.05)--(1.05,-.95);
\draw[blue] (.00,.05)--(1.00,-.95);
\draw[blue] (0,-.05)--(1,-0.05);
\draw[blue] (0,-.1)--(1,-.1);
\draw[blue] (0,-.05)--(0,-1);
\draw[blue] (-.05,-.05)--(-0.05,-1);
\draw[blue] (-.05,-.05)--(-1.05,.95);
\draw[blue] (.00,-.05)--(-1.00,.95);
\draw[orange] (-2,0)--(-1,0);
\draw[orange] (2,0)--(1,0);
\node[D7] at (-2,0) {};  
\node[D7] at (-1,0) {};  
\node[D7] at (1,0) {};  
\node[D7] at (2,0) {};  
\node[D7] at (-1,1) {};  
\node[D7] at (1,-1) {};  
\node[D7] at (0,1) {};  
\node[D7] at (0,-1) {};  
\node at (2,-1) {$/ \mathbb{Z}_2$};
\end{tikzpicture}}
&\qquad \qquad 
        &\raisebox{-.5\height}{\scalebox{0.96}{\begin{tikzpicture}[x=1.2cm,y=.8cm]
\draw (0,.05)--(1,.05);
\draw (0,-.05)--(1,-.05);
\draw (.4,0)--(.6,.2);
\draw (.4,0)--(.6,-.2);
\draw (0,.05)--(-1,.05);
\draw (0,-.05)--(-1,-.05);
\draw (-.4,0)--(-.6,.2);
\draw (-.4,0)--(-.6,-.2);
\node (g0) at (-1,0) [gauge,label=above:{\small $r$},blue] {};
\node (g1) at (0,0) [gauge,label=above:{\small $1$},orange] {};
\node (g2) at (1,0) [gauge,label=above:{\small $1$},green] {};
\draw (g0) to [out=135,in=225,looseness=10] (g0);
\end{tikzpicture}}}   
\,.
\end{alignat}
Again, the adjoint loop in the last quiver above comes from the self-intersection 0 of the blue subweb. 
In terms of Hasse diagrams, one can represent the operation as follows: 
\begin{equation}
\label{hasseFold}
   \raisebox{-.5\height}{ \begin{tikzpicture}
\node[bd,label=above:{(\ref{firstConeMQ})}] at (0,4) {};
\node[bd,label=above:{(\ref{secondConeMQ})}] at (1,3) {};
\node[bd,label=left:{(\ref{quiverIntersection})}] at (0,2) {};
\node[bd] at (0,0) {};
\draw[dashed] (0,0)--(0,4) (0,2)--(1,3);
\node[bd,label=above:{(\ref{quiverAfterFoldingHigh})}] at (4,3.7) {};
\node[bd,label=right:{(\ref{quiverAfterFoldingLow})}] at (4,1.7) {};
\node[bd] at (4,0) {};
\draw[dashed] (4,0)--(4,3.7);
\draw[->] (.3,4)--(3.7,3.7);
\draw[->] (1.3,3)--(3.7,1.8);
\draw[->] (0.3,2)--(3.7,1.7);
\end{tikzpicture} }
\,.
\end{equation}
The dots represent symplectic leaves whose closures are given by the magnetic quivers labeled next to them, the dashed lines represent (non necessarily elementary) transitions, and the black arrows show the effect of the $\mathbb{Z}_2$ folding. 

In a nutshell, the diagram (\ref{hasseFold}) shows that the full Higgs branch of the $\ess{A_2}{2}{r}$ is described by only one quiver, (\ref{quiverAfterFoldingHigh}), which appears in Table \ref{resulttable}, whereas the folding of the second quiver appears as a set of leaves in the same Hasse diagram. Both cones are accounted for.

\section{From brane web decompositions to magnetic quivers}
\label{AppendixRules}

In this section, we recall the rules of \cite{Cabrera:2018jxt} to read the magnetic quiver associated to a brane web decomposition, and extend them in order to include adjoint matter in the magnetic quiver, and to include the possibility of a discrete quotient on the brane web. 

\paragraph{Brane web with no discrete quotient.}
Consider a brane web $W$, which is decomposed into stacks of subwebs $m_i W_i$ ($i=1,\dots ,n$), where the $W_i$ are pairwise distinct and a given $W_i$ can not be written as $m>1$ times another subweb. Note that we do not require that the $W_i$ are irreducible. 

For each pair of indices $(i,j)$, not necessarily distinct, one can compute the intersection number $W_i \cdot W_j$ between the two subwebs $W_i$ and $W_j$. This intersection number is the sum of three contributions: 
\begin{equation}
\label{formulaIntersection}
    W_i \cdot W_j = \mathrm{SI}_{i,j} + \mathrm{X}_{i,j} - \mathrm{Y}_{i,j}
\end{equation}
where 
\begin{compactitem}
    \item $\mathrm{SI}_{i,j}$ is the stable intersection between the tropical curves of $W_i$ and $W_j$;
    \item $\mathrm{X}_{i,j}$ is the number of pairs of five-branes (one from $W_i$ and one from $W_j$) ending on the same seven-brane from opposite sides;
    \item $\mathrm{Y}_{i,j}$ is the number of pairs of five-branes (one from $W_i$ and one from $W_j$) ending on the same seven-brane from the same side. 
\end{compactitem}
The associated magnetic quiver contains $n$ unitary nodes $\mathrm{U}(m_i)$, and is simply laced, with links as follows: 
\begin{compactitem}
    \item For each pair of distinct indices $i \neq j$, the nodes $\mathrm{U}(m_i)$ and $\mathrm{U}(m_j)$ are connected by $W_i \cdot W_j$ lines (standing for hypermultiplets in the bifundamental representation); 
    \item For each index $i$, the node $\mathrm{U}(m_i)$ has $\frac{1}{2}W_i \cdot W_i +1$ loops (standing for hypermultiplets in the adjoint representation). 
\end{compactitem}

\paragraph{Example. }
As an example, we can check the computation of the quiver \eqref{quiverIntersection}. There are four nodes, corresponding to the four colors in the web. The computation of the number of bifundamental hypermultiplets is straightforward, so we consider only the computation of the number of adjoint loops. For the orange node, \eqref{formulaIntersection} gives
\begin{equation}
    W_{\textrm{orange}} \cdot W_{\textrm{orange}} = 0 + 0 - 2 = -2
\end{equation}
so there are $\frac{1}{2}(-2)+1=0$ adjoint loops. The same applies to the green and purple nodes. The blue node is more interesting; the relevant subweb is 
\begin{equation}
     W_{\textrm{blue}} =  \raisebox{-.5\height}{ \begin{tikzpicture}
\draw[blue] (-1,0)--(1,0);
\draw[blue] (0,-1)--(0,1);
\draw[blue] (-1,1)--(1,-1);
\node[D7] at (-1,0) {};  
\node[D7] at (1,0) {};  
\node[D7] at (-1,1) {};  
\node[D7] at (1,-1) {};  
\node[D7] at (0,1) {};  
\node[D7] at (0,-1) {};  
\end{tikzpicture}} \, . 
\end{equation}
Note that it is \emph{not} irreducible. 
The stable intersection can be computed by taking a copy of the web (dashed below), and computing the intersection with the initial web: 
\begin{equation}
\raisebox{-.5\height}{ \begin{tikzpicture}
\draw[blue] (-1,0)--(1,0);
\draw[blue] (0,-1)--(0,1);
\draw[blue] (-1,1)--(1,-1);
\draw[blue,dashed] (-1,.1)--(1,.1);
\draw[blue,dashed] (.1,-1)--(.1,1);
\draw[blue,dashed] (-.9,1.1)--(1.1,-.9);
\node[D7] at (-1,0) {};  
\node[D7] at (1,0) {};  
\node[D7] at (-1,1) {};  
\node[D7] at (1,-1) {};  
\node[D7] at (0,1) {};  
\node[D7] at (0,-1) {};  
\end{tikzpicture}} \, , 
\end{equation}
which gives 
\begin{equation}
    W_{\textrm{blue}} \cdot W_{\textrm{blue}} = 6 + 0 - 6 = 0 \, , 
\end{equation}
so there is one adjoint loop on that node.\\

Note that the rule for adding adjoint matter, which does not appear in \cite{Cabrera:2018jxt}, doesn't affect the magnetic quiver results of \cite{Cabrera:2018jxt}, as the nodes which would obtain an adjoint node are all $\mathrm{U}(1)$'s.

\paragraph{Brane web with discrete quotient.}
Assume now that a finite group $\mathbb{Z}_{\ell}$ acts on the brane web. Pick an \emph{adapted} decomposition into stacks of subwebs $m_i W_i$, which means that in addition to the requirements made above, the webs $W_i$ are invariant under the action of $G$ and are in one of the following situations: 
\begin{compactenum}[(1)]
    \item $W_i$ is made up of exactly $\ell$ disconnected components; 
    \item $W_i$ is made up of exactly 1 connected component. 
\end{compactenum}

One can then compute intersection numbers using \eqref{formulaIntersection}, and the magnetic quiver is a \emph{non simply laced} quiver. To each $W_i$ one associates a unitary node $\mathrm{U}(m_i)$, which is \emph{short} if $W_i$ is in case (1), and \emph{long} if it is in case (2). The rules for connectivity in the quiver are then exactly as in the simply laced case, except that a link connecting a long node and a short node is an $\ell$-fold oriented link. 

\section{State of the Art Cartography of the Hasse diagram}
\label{AppendixCartography}
The goal of quiver subtraction is to understand the Hasse diagram of the Coulomb branch of a quiver\footnote{The Hasse diagram of the Higgs branch of a theory is studied by computing the Hasse diagram of the Coulomb branch of the magnetic quiver.}. The rules of quiver subtraction have been explored for some years, however they are still elusive in full generality. So far these rules, and the interpretation of the quivers involved, were derived from Kraft-Procesi transitions in brane constructions, based on the classical Higgs mechanism, or through other empirical methods \cite{Cabrera:2018ann,Bourget:2019aer,Hanany:2019tji,Grimminger:2020dmg,Bourget:2020gzi}. There are various difficulties in coming up with a full set of rules for quiver subtraction:
\begin{compactenum}[(1)]
    \item The complete catalog of all elementary slices, and all their Coulomb branch representations, is still unknown.
    \item Quivers involving non-unitary nodes are notoriously difficult to deal with.
    \item For quivers involving non-simply laced edges rebalancing is tricky. 
    \item Non-standard matter is poorly studied.
    \item When the same slice can be subtracted more than once the naive rules give a wrong result when comparing with other methods, a refinement of the rules is needed.
\end{compactenum}
For the quivers at hand we have to address the third and the fifth point for unitary quivers. We provide our current conjecture on how to deal with these issues, using what we already know about the theories we study. As an example, we construct the Hasse diagram for the $\tee{E_6}{2}{2}$ theory.

\paragraph{Rebalancing non-simply laced.}
As already noted in \cite{Bourget:2019aer}, where possible it is convenient to work with unframed quivers in order to identify all subtractable pieces. In the following all quivers are ungauged on a long node. When subtracting from a non-simply laced quiver, however, the $\mathrm{U}(1)$ node used to rebalance sometimes has to be attached to the short end of the quiver. In this case, we \emph{propose} that the $\mathrm{U}(1)$ node is added with a non-simply laced edge of the same kind. In other words, one has to rebalance with a long $\mathrm{U}(1)$ node as follows:
\begin{equation}
\raisebox{-.5\height}{
    \begin{tikzpicture}
    \node[gauge3,label=below:{1}] (1) at (0,0) {};
    \node[gauge3,label=below:{2}] (2) at (1,0) {};
    \node[gauge3,label=below:{4}] (3) at (2,0) {};
    \node[gauge3,label=below:{6}] (4) at (3,0) {};
    \node[gauge3,label=below:{8}] (5) at (4,0) {};
    \node[gauge3,label=below:{4}] (6) at (5,0) {};
    \draw (1)--(2)--(3)--(4) (5)--(6);
    \draw[transform canvas={yshift=-1.5pt}] (4)--(5);
    \draw[transform canvas={yshift=1.5pt}] (4)--(5);
    \draw (3.7,0.3)--(3.4,0)--(3.7,-0.3);
    \begin{scope}[shift={(0,-2)}]
    \node at (0,0) {$-$};
    \node[gauge3,label=below:{1}] (22) at (1,0) {};
    \node[gauge3,label=below:{2}] (23) at (2,0) {};
    \node[gauge3,label=below:{3}] (24) at (3,0) {};
    \node[gauge3,label=below:{4}] (25) at (4,0) {};
    \node[gauge3,label=below:{2}] (26) at (5,0) {};
    \draw (22)--(23)--(24) (25)--(26);
    \draw[transform canvas={yshift=-1.5pt}] (24)--(25);
    \draw[transform canvas={yshift=1.5pt}] (24)--(25);
    \draw (3.7,0.3)--(3.4,0)--(3.7,-0.3);
    \end{scope}
    \begin{scope}[shift={(0,-4)}]
    {\color{red}\node[gauge3,label=left:{1}] (10) at (0,1) {};}
    \node[gauge3,label=below:{1}] (11) at (0,0) {};
    \node[gauge3,label=below:{1}] (12) at (1,0) {};
    \node[gauge3,label=below:{2}] (13) at (2,0) {};
    \node[gauge3,label=below:{3}] (14) at (3,0) {};
    \node[gauge3,label=below:{4}] (15) at (4,0) {};
    \node[gauge3,label=below:{2}] (16) at (5,0) {};
    \draw[transform canvas={xshift=-1.5pt}] (10)--(11);
    \draw[transform canvas={xshift=1.5pt}] (10)--(11);
    \draw (0.3,0.7)--(0,0.4)--(-0.3,0.7);
    \draw (11)--(12)--(13)--(14) (15)--(16);
    \draw[transform canvas={yshift=-1.5pt}] (14)--(15);
    \draw[transform canvas={yshift=1.5pt}] (14)--(15);
    \draw (3.7,0.3)--(3.4,0)--(3.7,-0.3);
    \end{scope}
    \draw[->] (5.5,0) to [out=300,in=60,looseness=1] (5.5,-3.8);
    \node at (8,-4) {Not yet decorated.};
    \end{tikzpicture}
    }
    \label{eq:rebalance}
\end{equation}
It is important not to rebalance, for instance, with a flavour node, as only now we do see the possibility of subtracting either a $c_5$ or a $e_6$ slice. In the next paragraph we argue, that one needs to `decorate' the resulting quiver in order to proceed with quiver subtraction correctly.
\paragraph{Same slice subtraction.}
Consider the result after the $e_6$ quiver subtraction in \eqref{eq:rebalance} and then subtract the $c_5$ slice, then there seems to be no issue. However, when we subtract the second $e_6$ slice naively we obtain the following:
\begin{equation}
\raisebox{-.5\height}{
    \begin{tikzpicture}
    {\color{red}\node[gauge3,label=left:{1}] (10) at (0,1) {};}
    \node[gauge3,label=below:{1}] (11) at (0,0) {};
    \node[gauge3,label=below:{1}] (12) at (1,0) {};
    \node[gauge3,label=below:{2}] (13) at (2,0) {};
    \node[gauge3,label=below:{3}] (14) at (3,0) {};
    \node[gauge3,label=below:{4}] (15) at (4,0) {};
    \node[gauge3,label=below:{2}] (16) at (5,0) {};
    \draw[transform canvas={xshift=-1.5pt}] (10)--(11);
    \draw[transform canvas={xshift=1.5pt}] (10)--(11);
    \draw (0.3,0.7)--(0,0.4)--(-0.3,0.7);
    \draw (11)--(12)--(13)--(14) (15)--(16);
    \draw[transform canvas={yshift=-1.5pt}] (14)--(15);
    \draw[transform canvas={yshift=1.5pt}] (14)--(15);
    \draw (3.7,0.3)--(3.4,0)--(3.7,-0.3);
        \begin{scope}[shift={(0,-2)}]
    \node at (0,0) {$-$};
    \node[gauge3,label=below:{1}] (22) at (1,0) {};
    \node[gauge3,label=below:{2}] (23) at (2,0) {};
    \node[gauge3,label=below:{3}] (24) at (3,0) {};
    \node[gauge3,label=below:{4}] (25) at (4,0) {};
    \node[gauge3,label=below:{2}] (26) at (5,0) {};
    \draw (22)--(23)--(24) (25)--(26);
    \draw[transform canvas={yshift=-1.5pt}] (24)--(25);
    \draw[transform canvas={yshift=1.5pt}] (24)--(25);
    \draw (3.7,0.3)--(3.4,0)--(3.7,-0.3);
    \end{scope}
        \begin{scope}[shift={(0,-4)}]
        \begin{scope}[rotate=45]
    {\color{red}\node[gauge3,label=left:{1}] (30) at (0,1) {};}
    \node[gauge3,label=below:{1}] (31) at (0,0) {};
    \draw[transform canvas={xshift=-1.5pt}] (30)--(31);
    \draw[transform canvas={xshift=1.5pt}] (30)--(31);
    \draw (0.3,0.7)--(0,0.4)--(-0.3,0.7);
    \end{scope}\begin{scope}[rotate=-45]
    {\color{blue}\node[gauge3,label=right:{1}] (32) at (0,1) {};}
    \draw[transform canvas={xshift=-1.5pt}] (32)--(31);
    \draw[transform canvas={xshift=1.5pt}] (32)--(31);
    \draw (0.3,0.7)--(0,0.4)--(-0.3,0.7);
    \end{scope}
    \end{scope}
    \node at (3,-4) {wrong!};
    \draw[->] (5.5,0) to [out=300,in=60,looseness=1] (5.5,-3.8);
    \end{tikzpicture}
    }
    \label{eq:wrong}
\end{equation}
The result of \eqref{eq:wrong}, however, is not consistent with the known global symmetry of the $\tee{E_6}{2}{2}$ theory. The global symmetry of the slice with the {\color{red}red} and {\color{blue}blue} rebalancing $\mathrm{U}(1)$s is $C_2$, not $A_1A_1$. In fact, already the quiver obtained from subtracting the first $e_6$ in \eqref{eq:rebalance} has an incorrect global symmetry\footnote{The quiver resulting after subtraction in \eqref{eq:rebalance} has Coulomb branch global symmetry $C_5C_2$ rather than $C_5A_1A_1$ as the top quiver in \eqref{eq:rebalance}}. It is instructive to think about what the rebalancing $\mathrm{U}(1)$s mean in terms of branes. Subtracting the same slice twice corresponds to aligning two previously indistinguishable sets of branes. When aligning only one such set, i.e.\ performing only one subtraction, there is no surprise, and we rebalance with a $\mathrm{U}(1)$. When aligning both sets, i.e.\ performing both subtractions, however, we should really rebalance by a single node. We propose that this node should have an adjoint loop, which is in agreement with what is read from the brane web in Appendix \ref{AppendixRules}. When subtracting the same quiver multiple times, we should therefore remember that the rebalancing $\mathrm{U}(1)$ nodes and the same slice, which can still be subtracted, are linked. One way to achieve this is via \emph{decorating} the quiver. The full subtraction would now become:
\begin{equation}
\raisebox{-.5\height}{
    \begin{tikzpicture}
    \node[gauge3,label=below:{1}] (1) at (0,0) {};
    \node[gauge3,label=below:{2}] (2) at (1,0) {};
    \node[gauge3,label=below:{4}] (3) at (2,0) {};
    \node[gauge3,label=below:{6}] (4) at (3,0) {};
    \node[gauge3,label=below:{8}] (5) at (4,0) {};
    \node[gauge3,label=below:{4}] (6) at (5,0) {};
    \draw (1)--(2)--(3)--(4) (5)--(6);
    \draw[transform canvas={yshift=-1.5pt}] (4)--(5);
    \draw[transform canvas={yshift=1.5pt}] (4)--(5);
    \draw (3.7,0.3)--(3.4,0)--(3.7,-0.3);
    \begin{scope}[shift={(0,-2)}]
    \node at (0,0) {$-$};
    \node[gauge3,label=below:{1}] (22) at (1,0) {};
    \node[gauge3,label=below:{2}] (23) at (2,0) {};
    \node[gauge3,label=below:{3}] (24) at (3,0) {};
    \node[gauge3,label=below:{4}] (25) at (4,0) {};
    \node[gauge3,label=below:{2}] (26) at (5,0) {};
    \draw (22)--(23)--(24) (25)--(26);
    \draw[transform canvas={yshift=-1.5pt}] (24)--(25);
    \draw[transform canvas={yshift=1.5pt}] (24)--(25);
    \draw (3.7,0.3)--(3.4,0)--(3.7,-0.3);
    \end{scope}
        \begin{scope}[shift={(0,-4)}]
    {\color{red}\node[gauge3,label=left:{1}] (10) at (0,1) {};}
    \node[gauge3,label=below:{1}] (11) at (0,0) {};
    {\color{red}\node[gauge3,label=below:{1}] (12) at (1,0) {};
    \node[gauge3,label=below:{2}] (13) at (2,0) {};
    \node[gauge3,label=below:{3}] (14) at (3,0) {};
    \node[gauge3,label=below:{4}] (15) at (4,0) {};
    \node[gauge3,label=below:{2}] (16) at (5,0) {};}
    \draw[transform canvas={xshift=-1.5pt}] (10)--(11);
    \draw[transform canvas={xshift=1.5pt}] (10)--(11);
    \draw (0.3,0.7)--(0,0.4)--(-0.3,0.7);
    \draw (11)--(12)--(13)--(14) (15)--(16);
    \draw[transform canvas={yshift=-1.5pt}] (14)--(15);
    \draw[transform canvas={yshift=1.5pt}] (14)--(15);
    \draw (3.7,0.3)--(3.4,0)--(3.7,-0.3);
    \end{scope}
        \begin{scope}[shift={(0,-6)}]
    \node at (0,0) {$-$};
    \node[gauge3,label=below:{1}] (42) at (1,0) {};
    \node[gauge3,label=below:{2}] (43) at (2,0) {};
    \node[gauge3,label=below:{3}] (44) at (3,0) {};
    \node[gauge3,label=below:{4}] (45) at (4,0) {};
    \node[gauge3,label=below:{2}] (46) at (5,0) {};
    \draw (42)--(43)--(44) (45)--(46);
    \draw[transform canvas={yshift=-1.5pt}] (44)--(45);
    \draw[transform canvas={yshift=1.5pt}] (44)--(45);
    \draw (3.7,0.3)--(3.4,0)--(3.7,-0.3);
    \end{scope}
        \begin{scope}[shift={(0,-8)}]
    {\color{red}\node[gauge3,label=left:{2}] (50) at (0,1) {};}
    \node[gauge3,label=below:{1}] (51) at (0,0) {};
    \draw[transform canvas={xshift=-1.5pt}] (50)--(51);
    \draw[transform canvas={xshift=1.5pt}] (50)--(51);
    \draw (0.3,0.7)--(0,0.4)--(-0.3,0.7);
    \draw (50) to [out=45,in=135,looseness=10] (50);
    \end{scope}
    \draw[->] (5.5,0) to [out=300,in=60,looseness=1] (5.5,-3.8);
    \draw[->] (5.5,-4.2) to [out=300,in=60,looseness=1] (5.5,-7.8);
    \end{tikzpicture}
    }
    \label{eq:right}
\end{equation}
The quiver in the middle has a \emph{decoration} (the red colour) to denote that the remaining $e_6$ slice and the $\mathrm{U}(1)$ node are linked. One can compute that the quiver obtained from subtracting two $e_6$ slices in \eqref{eq:right} is the product $a_1\times a_1$. When we subtract a $c_5$ from the decorated quiver in \eqref{eq:right} the decoration can be disregarded. We can now attempt to produce the full Hasse diagram of the $\tee{E_6}{2}{2}$ theory from quiver subtraction:
\begin{equation}
\raisebox{-.5\height}{
    \scalebox{0.8}{
    \begin{tikzpicture}
    \node[hasse] (1) at (0,0) {};
    \node[hasse] (2) at (-1,-1) {};
    \node[hasse] (4) at (0,-2) {};
    \node[hasse] (5) at (-2,-2) {};
    \node[hasse] (6) at (-1,-3) {};
    \node[hasse] (7) at (0,-3) {};
    \node[hasse] (8) at (-1,-4) {};
    \draw (1)--(2)--(4)--(6)--(8)--(7)--(4) (2)--(5)--(8);
    \node at (-0.7,-0.3) {$e_6$};
    \node at (-0.7,-1.7) {$e_6$};
    \node at (-1.7,-1.3) {$c_5$};
    \node at (-0.7,-2.3) {$A_1$};
    \node at (-0.8,-3.5) {$A_1$};
    \node at (0.3,-2.5) {$A_1$};
    \node at (-0.2,-3.7) {$A_1$};
    \node at (-1.8,-3) {$f_4$};
    \end{tikzpicture}
    }
    }
\end{equation}
According to \cite{Giacomelli:2020jel}, the theory obtained from Higgsing along the $A_1\times A_1$ slice is the $\mathcal{I}_{E_6}^{(1)}\otimes\mathcal{I}_{E_6}^{(1)}$ theory. Hence, there should be an additional leaf in the Hasse diagram
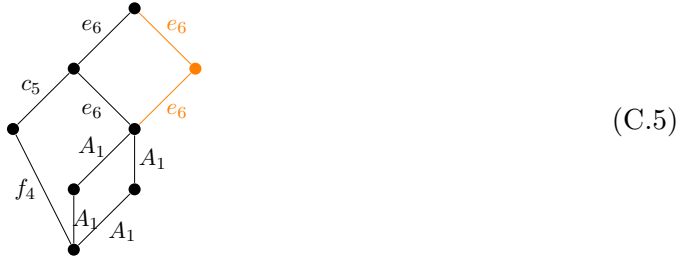
\begin{equation}
\raisebox{-.5\height}{
    \scalebox{0.8}{
    \begin{tikzpicture}
    \node[hasse] (1) at (0,0) {};
    \node[hasse] (2) at (-1,-1) {};
    {\color{orange}\node[hasse] (3) at (1,-1) {};}
    \node[hasse] (4) at (0,-2) {};
    \node[hasse] (5) at (-2,-2) {};
    \node[hasse] (6) at (-1,-3) {};
    \node[hasse] (7) at (0,-3) {};
    \node[hasse] (8) at (-1,-4) {};
    \draw (1)--(2)--(4)--(6)--(8)--(7)--(4) (2)--(5)--(8);
    {\color{orange}\draw (1)--(3)--(4);}
    \node at (-0.7,-0.3) {$e_6$};
    \node at (-0.7,-1.7) {$e_6$};
    \node at (-1.7,-1.3) {$c_5$};
    \node at (-0.7,-2.3) {$A_1$};
    \node at (-0.8,-3.5) {$A_1$};
    \node at (0.3,-2.5) {$A_1$};
    \node at (-0.2,-3.7) {$A_1$};
    \node at (-1.8,-3) {$f_4$};
    {\color{orange}\node at (0.7,-0.3) {$e_6$};
    \node at (0.7,-1.7) {$e_6$};}
    \end{tikzpicture}
    }
    }
\end{equation}
which we added in \textcolor{orange}{orange}.
It is so far unclear, how to obtain this Hasse diagram from quiver subtraction alone. Furthermore, we are unable to see the Higgsing of $\tee{E_6}{2}{2}$ to $\mathcal{I}_{E_6}^{(2)}$ of \cite{Giacomelli:2020jel} in this Hasse diagram. One question that remains is, what are the theories associated to the two leaves reached after a $A_1$ transition. Again, one cannot obtain a magnetic quiver in a straightforward manner. However, we \emph{conjecture} that both theories correspond to a $\mathbb{Z}_2$ discrete gauging of $\tee{E_6}{2}{1}\otimes\mathcal{I}_{E_6}^{(1)}=\mathcal{I}_{E_6}^{(1)}\otimes\mathcal{I}_{E_6}^{(1)}\otimes\mathbb{H}$, where $\mathbb{H}$ denotes a free hypermultiplet. In fact, this theory can be Higgsed to the product $\mathcal{I}_{E_6}^{(1)}\otimes\mathcal{I}_{E_6}^{(1)}$, via an $A_1$ transition.

The identification of adjoint matter and the decoration of a magnetic quiver - as in the quiver subtraction presented above - from a brane system is under active investigation. Decorated quivers and their Coulomb branches need to be understood in a systematic way.

\begin{figure}[t]
\centering
\begin{subfigure}{0.3\textwidth}
\centering
\raisebox{-.5\height}{
    \scalebox{0.8}{
    \begin{tikzpicture}
    \node[hasse] (1) at (0,0) {};
    \node[hasse] (2) at (-1,-1) {};
    {\color{orange}\node[hasse] (3) at (1,-1) {};}
    \node[hasse] (4) at (0,-2) {};
    \node[hasse] (5) at (-2,-2) {};
    \node[hasse] (6) at (-1,-3) {};
    \node[hasse] (7) at (0,-3) {};
    \node[hasse] (8) at (-1,-4) {};
    \draw (1)--(2)--(4)--(6)--(8)--(7)--(4) (2)--(5)--(8);
    {\color{orange}\draw (1)--(3)--(4);}
    \node at (-0.7,-0.3) {$e_6$};
    \node at (-0.7,-1.7) {$e_6$};
    \node at (-1.7,-1.3) {$c_5$};
    \node at (-0.7,-2.3) {$A_1$};
    \node at (-0.8,-3.5) {$A_1$};
    \node at (0.3,-2.5) {$A_1$};
    \node at (-0.2,-3.7) {$A_1$};
    \node at (-1.8,-3) {$f_4$};
    {\color{orange}\node at (0.7,-0.3) {$e_6$};
    \node at (0.7,-1.7) {$e_6$};}
    \end{tikzpicture}
    }
    }
\caption{$\tee{E_6}{2}{2}$}
\end{subfigure}
\begin{subfigure}{0.30\textwidth}
\centering
\raisebox{-.5\height}{
    \scalebox{0.8}{
    \begin{tikzpicture}
    \node[hasse] (1) at (0,0) {};
    \node[hasse] (2) at (-1,-1) {};
    {\color{orange}\node[hasse] (3) at (1,-1) {};}
    \node[hasse] (4) at (0,-2) {};
    \node[hasse] (5) at (-2,-2) {};
    \node[hasse] (6) at (-1,-3) {};
    \node[hasse] (7) at (0,-3) {};
    \node[hasse] (8) at (-1,-4) {};
    \draw (1)--(2)--(4)--(6)--(8)--(7)--(4) (2)--(5)--(8);
    {\color{orange}\draw (1)--(3)--(4);}
    \node at (-0.7,-0.3) {$d_4$};
    \node at (-0.7,-1.7) {$d_4$};
    \node at (-1.7,-1.3) {$c_3$};
    \node at (-0.7,-2.3) {$A_1$};
    \node at (-0.8,-3.5) {$A_1$};
    \node at (0.3,-2.5) {$A_1$};
    \node at (-0.2,-3.7) {$A_1$};
    \node at (-1.8,-3) {$b_3$};
    {\color{orange}\node at (0.7,-0.3) {$d_4$};
    \node at (0.7,-1.7) {$d_4$};}
    \end{tikzpicture}
    }
    }
\caption{$\tee{D_4}{2}{2}$}
\end{subfigure}
\begin{subfigure}{0.30\textwidth}
\centering
\raisebox{-.5\height}{
    \scalebox{0.8}{
    \begin{tikzpicture}
    \node[hasse] (1) at (0,0) {};
    \node[hasse] (2) at (-1,-1) {};
    {\color{orange}\node[hasse] (3) at (1,-1) {};}
    \node[hasse] (4) at (0,-2) {};
    \node[hasse] (5) at (-2,-2) {};
    \node[hasse] (6) at (-1,-3) {};
    \node[hasse] (7) at (0,-3) {};
    \node[hasse] (8) at (-1,-4) {};
    \draw (1)--(2)--(4)--(6)--(8)--(7)--(4) (2)--(5)--(8);
    {\color{orange}\draw (1)--(3)--(4);}
    \node at (-0.7,-0.3) {$a_2$};
    \node at (-0.7,-1.7) {$a_2$};
    \node at (-1.7,-1.3) {$c_2$};
    \node at (-0.7,-2.3) {$A_1$};
    \node at (-0.8,-3.5) {$A_1$};
    \node at (0.3,-2.5) {$A_1$};
    \node at (-0.2,-3.7) {$A_1$};
    \node at (-1.8,-3) {$a_2$};
    {\color{orange}\node at (0.7,-0.3) {$a_2$};
    \node at (0.7,-1.7) {$a_2$};}
    \end{tikzpicture}
    }
    }
\caption{$\tee{A_2}{2}{2}$}
\end{subfigure}
\begin{subfigure}{0.30\textwidth}
\centering
\raisebox{-.5\height}{
    \scalebox{0.8}{
    \begin{tikzpicture}
    \node at (0,1) {};
    \node[hasse] (1) at (0,0) {};
    \node[hasse] (2) at (-1,-1) {};
    {\color{orange}\node[hasse] (3) at (1,-1) {};}
    \node[hasse] (4) at (0,-2) {};
    \node[hasse] (5) at (-2,-2) {};
    \node[hasse] (8) at (-1,-4) {};
    \draw (1)--(2)--(4)--(8) (2)--(5)--(8);
    {\color{orange}\draw (1)--(3)--(4);}
    \node at (-0.7,-0.3) {$d_4$};
    \node at (-0.7,-1.7) {$d_4$};
    \node at (-1.7,-1.3) {$h_{4,3}$};
    \node at (-0.2,-3.3) {$k_3$};
    \node at (-1.8,-3) {$g_2$};
    {\color{orange}\node at (0.7,-0.3) {$d_4$};
    \node at (0.7,-1.7) {$d_4$};}
    \end{tikzpicture}
    }
    }
\caption{$\tee{D_4}{3}{2}$}
\end{subfigure}
\begin{subfigure}{0.30\textwidth}
\centering
\raisebox{-.5\height}{
    \scalebox{0.8}{
    \begin{tikzpicture}
    \node at (0,1) {};
    \node[hasse] (1) at (0,0) {};
    \node[hasse] (2) at (-1,-1) {};
    {\color{orange}\node[hasse] (3) at (1,-1) {};}
    \node[hasse] (4) at (0,-2) {};
    \node[hasse] (5) at (-2,-2) {};
    \node[hasse] (8) at (-1,-4) {};
    \draw (1)--(2)--(4)--(8) (2)--(5)--(8);
    {\color{orange}\draw (1)--(3)--(4);}
    \node at (-0.7,-0.3) {$A_1$};
    \node at (-0.7,-1.7) {$A_1$};
    \node at (-1.7,-1.3) {$h_{2,3}$};
    \node at (-0.2,-3.3) {$k_3$};
    \node at (-1.8,-3) {$A_1$};
    {\color{orange}\node at (0.7,-0.3) {$A_1$};
    \node at (0.7,-1.7) {$A_1$};}
    \end{tikzpicture}
    }
    }
\caption{$\tee{A_1}{3}{2}$}
\end{subfigure}
\begin{subfigure}{0.30\textwidth}
\centering
\raisebox{-.5\height}{
    \scalebox{0.8}{
    \begin{tikzpicture}
    \node at (0,1) {};
    \node[hasse] (1) at (0,0) {};
    \node[hasse] (2) at (-1,-1) {};
    {\color{orange}\node[hasse] (3) at (1,-1) {};}
    \node[hasse] (4) at (0,-2) {};
    \node[hasse] (5) at (-2,-2) {};
    \node[hasse] (8) at (-1,-4) {};
    \draw (1)--(2)--(4)--(8) (2)--(5)--(8);
    {\color{orange}\draw (1)--(3)--(4);}
    \node at (-0.7,-0.3) {$a_2$};
    \node at (-0.7,-1.7) {$a_2$};
    \node at (-1.7,-1.3) {$h_{3,4}$};
    \node at (-0.2,-3.3) {$k_4$};
    \node at (-1.8,-3) {$A_1$};
    {\color{orange}\node at (0.7,-0.3) {$a_2$};
    \node at (0.7,-1.7) {$a_2$};}
    \end{tikzpicture}
    }
    }
\caption{$\tee{A_2}{4}{2}$}
\end{subfigure}
\caption{Proposed Hasse diagram for $\tee{G}{\ell}{2}$ theories. The slice $k_n$ is detailed in \eqref{newslice}. These are subdiagrams of the left part of Figure \ref{hasseFold}. }
\label{fig:Hasse_Tee_2_theories}
\end{figure}
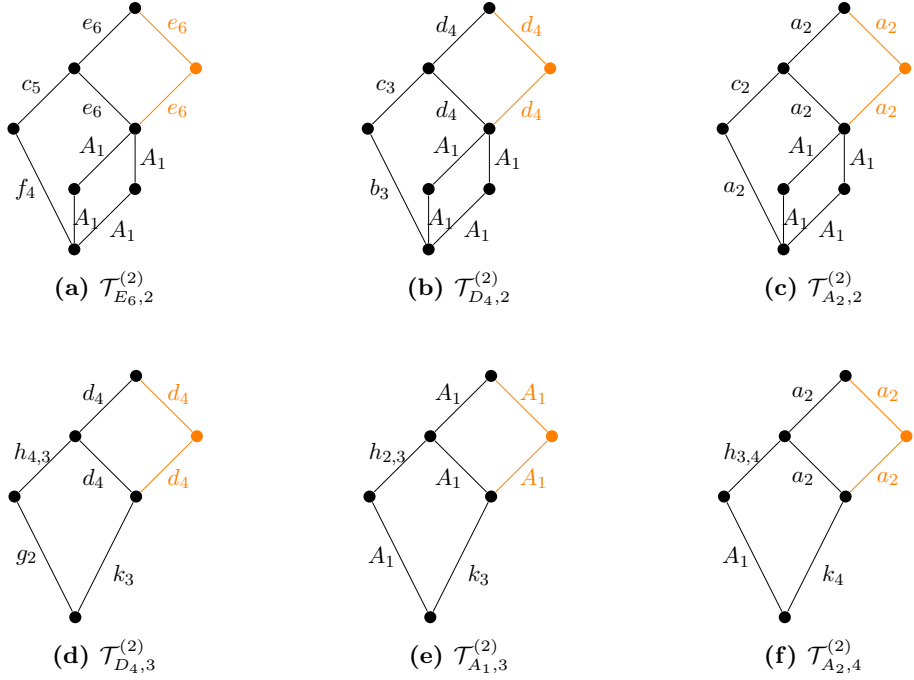

\paragraph{$\tee{G}{\ell}{2}$ Hasse diagrams.}
For completeness, the proposed Hasse diagram for all the $\tee{G}{\ell}{2}$ theories are provided in Figure \ref{fig:Hasse_Tee_2_theories}. 
The appearing slice $k_n$ is defined as:
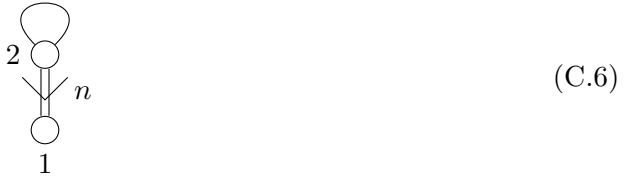
\begin{equation}
\raisebox{-.5\height}{
    \begin{tikzpicture}

        \begin{scope}[shift={(0,-8)}]
   \node[gauge3,label=left:{2}] (50) at (0,1) {};
    \node[gauge3,label=below:{1}] (51) at (0,0) {};
        \node at (0.5,0.5) {$n$};
    \draw[transform canvas={xshift=-1.5pt}] (50)--(51);
    \draw[transform canvas={xshift=1.5pt}] (50)--(51);
    \draw (0.3,0.7)--(0,0.4)--(-0.3,0.7);
    \draw (50) to [out=45,in=135,looseness=10] (50);
    \end{scope}
    \end{tikzpicture}
    }
    \label{newslice}
\end{equation}
where $n$ is the multiplicity of the non-simply laced edge. For $n=2$, this slice is studied above where the Coulomb branch is $a_1 \times a_1$. For $n>2$, the Coulomb branch global symmetry is $A_1$. It is not yet clear what the Coulomb branch Hasse diagram of (\ref{newslice}) is. We will leave a more thorough investigation of this family of quivers for the future. However, the fact that the slice has $A_1$ global symmetry ensures the global symmetry read from the Hasse diagram is consistent with expectation in Table \ref{resulttableT}. The highest weight generating function (HWG) of \eqref{newslice} for any $n$ is:
\begin{equation}
    \mathrm{HWG}(\nu,t)=\mathrm{PE}\big[(\nu^2t^2 + t^4 + \nu^nt^n+\nu^nt^{n+2}-\nu^{2n}t^{2n+4}\big]
\end{equation}
where $\nu$ is the highest weight fugacity of $A_1$ and $\mathrm{PE}$ is the plethystic exponential. See \cite{Hanany:2014dia} for a more detailed introduction of HWGs.

\begin{figure}[t]
\centering
\begin{subfigure}{0.3\textwidth}
\centering
\raisebox{-.5\height}{
    \scalebox{0.8}{
 \begin{tikzpicture}
	\begin{pgfonlayer}{nodelayer}
		\node [style=hasse] (0) at (0, 0) {};
		\node [style=hasse] (1) at (-0.5, -1) {};
		\node [style=hasse] (2) at (2.25, -2.25) {};
		\node [style=hasse] (3) at (1.75, -3.25) {};
		\node [style=hasse] (4) at (-0.5, -3) {};
		\node [style=hasse] (5) at (0.5, -4.75) {};
		\node [style=none] (6) at (1.5, -1) {$e_6$};
		\node [style=none] (7) at (-0.75, -0.25) {$c_4$};
		\node [style=none] (8) at (2.5, -3) {$c_4$};
		\node [style=none] (9) at (0.5, -1.5) {$e_6$};
		\node [style=none] (10) at (-1, -2) {$f_4$};
		\node [style=none] (11) at (-0.25, -4.25) {$c_4$};
		\node [style=none] (12) at (1.5, -4.25) {$A_1$};
		\node [style=none] (13) at (0, -2.5) {$A_1$};
	\end{pgfonlayer}
	\begin{pgfonlayer}{edgelayer}
		\draw (0) to (2);
		\draw (1) to (3);
		\draw (3) to (2);
		\draw (0) to (1);
		\draw (1) to (4);
		\draw (4) to (2);
		\draw (4) to (5);
		\draw (5) to (3);
	\end{pgfonlayer}
\end{tikzpicture}}}
\caption{$\essdg{E_6}{2}{1}$}
\end{subfigure}
\begin{subfigure}{0.30\textwidth}
\centering
\raisebox{-.5\height}{
    \scalebox{0.8}{\begin{tikzpicture}
	\begin{pgfonlayer}{nodelayer}
		\node [style=hasse] (0) at (0, 0) {};
		\node [style=hasse] (1) at (-0.5, -1) {};
		\node [style=hasse] (2) at (2.25, -2.25) {};
		\node [style=hasse] (3) at (1.75, -3.25) {};
		\node [style=hasse] (4) at (-0.5, -3) {};
		\node [style=hasse] (5) at (0.5, -4.75) {};
		\node [style=none] (6) at (1.5, -1) {$d_4$};
		\node [style=none] (7) at (-0.75, -0.25) {$c_2$};
		\node [style=none] (8) at (2.5, -3) {$c_2$};
		\node [style=none] (9) at (0.5, -1.5) {$d_4$};
		\node [style=none] (10) at (-1, -2) {$b_3$};
		\node [style=none] (11) at (-0.25, -4.25) {$c_2$};
		\node [style=none] (12) at (1.5, -4.25) {$A_1$};
		\node [style=none] (13) at (0, -2.5) {$A_1$};
	\end{pgfonlayer}
	\begin{pgfonlayer}{edgelayer}
		\draw (0) to (2);
		\draw (1) to (3);
		\draw (3) to (2);
		\draw (0) to (1);
		\draw (1) to (4);
		\draw (4) to (2);
		\draw (4) to (5);
		\draw (5) to (3);
	\end{pgfonlayer}
\end{tikzpicture}
 }
    }
\caption{$\essdg{D_4}{2}{1}$}
\end{subfigure}
\begin{subfigure}{0.30\textwidth}
\centering
\raisebox{-.5\height}{
    \scalebox{0.8}{
  \begin{tikzpicture}
	\begin{pgfonlayer}{nodelayer}
		\node [style=hasse] (0) at (0, 0) {};
		\node [style=hasse] (1) at (-0.5, -1) {};
		\node [style=hasse] (2) at (2.25, -2.25) {};
		\node [style=hasse] (3) at (1.75, -3.25) {};
		\node [style=hasse] (4) at (-0.5, -3) {};
		\node [style=hasse] (5) at (0.5, -4.75) {};
		\node [style=none] (6) at (1.5, -1) {$a_2$};
		\node [style=none] (7) at (-0.75, -0.25) {$A_1$};
		\node [style=none] (8) at (2.5, -3) {$A_1$};
		\node [style=none] (9) at (0.5, -1.5) {$a_2$};
		\node [style=none] (10) at (-1, -2) {$a_2$};
		\node [style=none] (11) at (-0.25, -4.25) {$A_1$};
		\node [style=none] (12) at (1.5, -4.25) {$A_1$};
		\node [style=none] (13) at (0, -2.5) {$A_1$};
	\end{pgfonlayer}
	\begin{pgfonlayer}{edgelayer}
		\draw (0) to (2);
		\draw (1) to (3);
		\draw (3) to (2);
		\draw (0) to (1);
		\draw (1) to (4);
		\draw (4) to (2);
		\draw (4) to (5);
		\draw (5) to (3);
	\end{pgfonlayer}
\end{tikzpicture}
    }
    }
\caption{$\essdg{A_2}{2}{1}$}
\end{subfigure}
%
%
\begin{subfigure}{0.30\textwidth}
\centering
\raisebox{-.5\height}{
    \scalebox{0.8}{
\begin{tikzpicture}
	\begin{pgfonlayer}{nodelayer}
		\node [style=hasse] (0) at (0, 0) {};
		\node [style=hasse] (1) at (-0.5, -1) {};
		\node [style=hasse] (2) at (2.25, -2.25) {};
		\node [style=hasse] (3) at (1.75, -3.25) {};
		\node [style=hasse] (4) at (-0.5, -3) {};
		\node [style=hasse] (5) at (0.5, -4.75) {};
		\node [style=none] (6) at (1.5, -1) {$d_4$};
		\node [style=none] (7) at (-0.75, -0.25) {$h_{3,3}$};
		\node [style=none] (8) at (2.5, -3) {$h_{3,3}$};
		\node [style=none] (9) at (0.5, -1.5) {$d_4$};
		\node [style=none] (10) at (-1, -2) {$g_2$};
		\node [style=none] (11) at (-0.25, -4.25) {$h_{3,3}$};
		\node [style=none] (12) at (1.5, -4.25) {$A_2$};
		\node [style=none] (13) at (0, -2.5) {$A_2$};
		\node [style=none] (144) at (0,1) {};
	\end{pgfonlayer}
	\begin{pgfonlayer}{edgelayer}
		\draw (0) to (2);
		\draw (1) to (3);
		\draw (3) to (2);
		\draw (0) to (1);
		\draw (1) to (4);
		\draw (4) to (2);
		\draw (4) to (5);
		\draw (5) to (3);
	\end{pgfonlayer}
\end{tikzpicture}
    }
    }
\caption{$\essdg{D_4}{3}{1}$}
\end{subfigure}
\begin{subfigure}{0.30\textwidth}
\centering
\raisebox{-.5\height}{
    \scalebox{0.8}{
\begin{tikzpicture}
	\begin{pgfonlayer}{nodelayer}
		\node [style=hasse] (0) at (0, 0) {};
		\node [style=hasse] (1) at (-0.5, -1) {};
		\node [style=hasse] (2) at (2.25, -2.25) {};
		\node [style=hasse] (3) at (1.75, -3.25) {};
		\node [style=hasse] (4) at (-0.5, -3) {};
		\node [style=hasse] (5) at (0.5, -4.75) {};
		\node [style=none] (6) at (2.25, -3) {$A_2$};
		\node [style=none] (7) at (0.75, -1.75) {$A_1$};
		\node [style=none] (8) at (1.5, -1) {$A_1$};
		\node [style=none] (9) at (-0.75, -0.25) {$A_2$};
		\node [style=none] (10) at (0, -2.5) {$A_2$};
		\node [style=none] (11) at (-0.25, -4.25) {$A_2$};
		\node [style=none] (12) at (1.5, -4.25) {$A_2$};
		\node [style=none] (13) at (-1, -2) {$A_1$};
		\node [style=none] (144) at (0,1) {};
	\end{pgfonlayer}
	\begin{pgfonlayer}{edgelayer}
		\draw (0) to (2);
		\draw (1) to (3);
		\draw (3) to (2);
		\draw (0) to (1);
		\draw (1) to (4);
		\draw (4) to (2);
		\draw (4) to (5);
		\draw (5) to (3);
	\end{pgfonlayer}
\end{tikzpicture}
    }
    }
\caption{$\essdg{A_1}{3}{1}$}
\end{subfigure}
\begin{subfigure}{0.30\textwidth}
\centering
\raisebox{-.5\height}{
    \scalebox{0.8}{
\begin{tikzpicture}
	\begin{pgfonlayer}{nodelayer}
		\node [style=hasse] (0) at (0, 0) {};
		\node [style=hasse] (1) at (-0.5, -1) {};
		\node [style=hasse] (2) at (2.25, -2.25) {};
		\node [style=hasse] (3) at (1.75, -3.25) {};
		\node [style=hasse] (4) at (-0.5, -3) {};
		\node [style=hasse] (5) at (0.5, -4.75) {};
		\node [style=none] (6) at (-0.75, -0.5) {$h_{2,4}$};
		\node [style=none] (7) at (1.25, -0.75) {$a_2$};
		\node [style=none] (8) at (0.5, -1.75) {$a_2$};
		\node [style=none] (9) at (2.5, -2.75) {$h_{2,4}$};
		\node [style=none] (10) at (0, -2.5) {$A_3$};
		\node [style=none] (11) at (-0.25, -4.25) {$h_{2,4}$};
		\node [style=none] (12) at (1.5, -4.25) {$A_3$};
		\node [style=none] (13) at (-0.75, -2) {$A_1$};
		\node [style=none] (144) at (0,1) {};
	\end{pgfonlayer}
	\begin{pgfonlayer}{edgelayer}
		\draw (0) to (2);
		\draw (1) to (3);
		\draw (3) to (2);
		\draw (0) to (1);
		\draw (1) to (4);
		\draw (4) to (2);
		\draw (4) to (5);
		\draw (5) to (3);
	\end{pgfonlayer}
\end{tikzpicture}
    }
    }
\caption{$\essdg{A_2}{4}{1}$}
\end{subfigure}
\caption{Proposed Hasse diagram for $\essdg{G}{\ell}{1}$ theories. These are subdiagrams of the right part of Figure \ref{hasseFold}. }
\label{fig:Hasse_essdg_theories}
\end{figure}
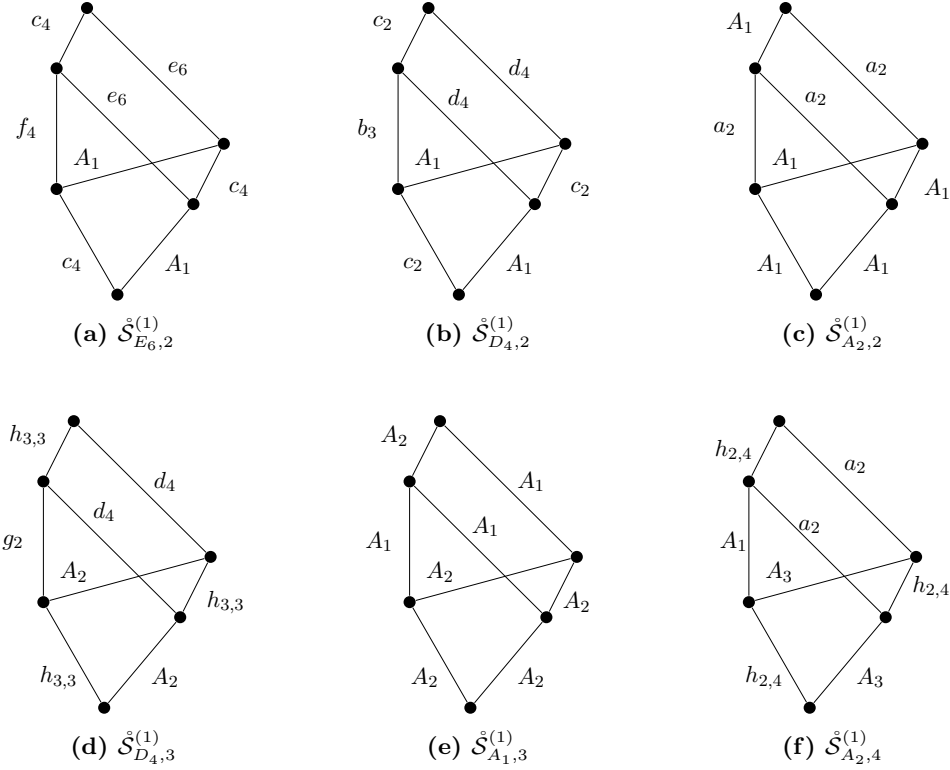

\paragraph{$\essdg{G}{\ell}{1}$ Hasse diagrams.}
For completeness, the proposed Hasse diagram for all the $\essdg{G}{\ell}{1}$ theories are provided in Figure \ref{fig:Hasse_essdg_theories}. 

\section{\texorpdfstring{Hilbert series of $\essdg{G}{\ell}{r}$ and  $\teedg{G}{\ell}{r}$}{Hilbert series of S and  T}}
\label{shortHS}
In this section, we tabulate the Hilbert series of the $\essdg{G}{\ell}{r}$ and  $\teedg{G}{\ell}{r}$ theories as detailed in Section \ref{shortside}. The global symmetry for general $r$ is given in Tables \ref{shortS} and \ref{shortR}. The Hilbert series and plethystic logarithm (PL) for $r=1,2$ are given in Tables \ref{shortS1}, \ref{shortS2}, \ref{shortT1}, \ref{shortT2} where an exact Hilbert series is given if possible. Let us recall the behavior of the global symmetry for long ungauging. For $\ess{G}{\ell}{r}$ theories there are symmetry enhancements for $r=1$ and for $\tee{G}{\ell}{r}$ theories the symmetry enhancement happens at $r=1,2$. In comparison, we notice that for  $\essdg{G}{\ell}{r}$ and  $\teedg{G}{\ell}{r}$ theories, the global symmetry remains the same for any $r$. This can be seen at order $t^2$ in the PL, where the coefficient gives the dimension of the global symmetry.

\begin{table}[t]
\small
\centering
\begin{adjustbox}{center}
\scalebox{0.8}{
	\begin{tabular}{ccc}
\toprule
		SCFT  & Hilbert Series & PL[HS] \\ 
\midrule
       $\essdg{E_6}{2}{1}$ &$1+39 t^2+84 t^3+888 t^4+3276 t^5+17251 t^6+68756 t^7+O\left(t^{8}\right)$	 &$39 t^2+84 t^3+108 t^4-1191 t^6-5836 t^7+O\left(t^8\right)$	\\ \hline
          $\essdg{D_4}{2}{1}$ &$\dfrac
	{\scriptsize \left(\begin{array}{c}1+2 t+14 t^2+48 t^3+172 t^4+498 t^5+1364 t^6+3294 t^7+7396 t^8\\+14930 t^9+27966 t^{10}+47994 t^{11}+76553 t^{12}+112762
   t^{13}\\+154822 t^{14}+197182 t^{15}+234766 t^{16}+260090 t^{17}+269508 t^{18} +\dots \mathrm{palindrome} +\dots + t^{36}
 \end{array}\right)} {(-1+t)^{16} (1+t)^{10} \left(1+t^2\right)^5 \left(1+t+t^2\right)^8}$	 & $\begin{array}{c} 16 t^2+30 t^3+29 t^4-40 t^5-263 t^6-476 t^7\\+277 t^8+3930 t^9+9560 t^{10} +O\left(t^{11}\right)\end{array}$ 
 \\ 
    \midrule
    $\essdg{A_2}{2}{1}$ &$\dfrac
	{\scriptsize \left(\begin{array}{c}1+2 t+7 t^2+16 t^3+40 t^4+72 t^5+123 t^6+170 t^7+231 t^8+268 t^9+292 t^{10}\\+268 t^{11}+231 t^{12}+170 t^{13}+123 t^{14}+72
   t^{15}+40 t^{16}+16 t^{17}+7 t^{18}+2 t^{19}+t^{20}
 \end{array}\right)} {(-1+t)^8 (1+t)^6 \left(1+t^2\right)^3 \left(1+t+t^2\right)^4}$		 	& $\begin{array}{c}7 t^2+8 t^3+8 t^4-8 t^5-31 t^6-32 t^7\\+30 t^8+168 t^9+194 t^{10}+O\left(t^{11}\right)\end{array}$ 
 \\ 
    \midrule
     $\essdg{D_4}{3}{1}$ &$\dfrac
	{\scriptsize \left(\begin{array}{c}1+4 t^2+14 t^3+34 t^4+94 t^5+254 t^6+574 t^7+1254 t^8+2418 t^9+4380 t^{10}\\+7102 t^{11}+10422 t^{12}+13786 t^{13}+15825
   t^{14}+14648 t^{15}+7785 t^{16}\\-6480 t^{17}-28546 t^{18}-56628 t^{19}-86567 t^{20}-109784 t^{21}-117179 t^{22}-98692 t^{23}\\-50401
   t^{24}+25892 t^{25}+121671 t^{26}+217074 t^{27}+290997 t^{28}+318062 t^{29}\\+285380 t^{30}+190202 t^{31}+44449 t^{32}-124370
   t^{33}-282653 t^{34}-393912 t^{35}-434220 t^{36}\\ +\dots \mathrm{palindrome} +\dots + t^{72}
 \end{array}\right)} {\left(1-t^2\right)^5 \left(1-t^3\right)^8 \left(1-t^4\right)^5 \left(1-t^6\right)^6}$		 	&$\begin{array}{c}9 t^2+22 t^3+29 t^4+38 t^5+39 t^6-54 t^7\\-405 t^8-1304 t^9-2541 t^{10}+O\left(t^{11}\right)\end{array}$ \\ 
    \midrule
      $\essdg{A_1}{3}{1}$&$\dfrac
	{\scriptsize \left(\begin{array}{c}1-2 t+3 t^2-2 t^3+3 t^4+3 t^8-2 t^9+3 t^{10}-2 t^{11}+t^{12}
 \end{array}\right)} {(1-t)^2 \left(1-t^3\right)^2 \left(1-t^4\right) \left(1-t^6\right)}$	
   	 &$\begin{array}{c} 2 t^2+4 t^3+3 t^4+4 t^5+t^6-4 t^7\\-8 t^8-8 t^9-6 t^{10}+O\left(t^{11}\right)+O\left(t^{11}\right)\end{array}$  \\  
    \midrule
        $\essdg{A_2}{4}{1}$ &$\dfrac
	{\scriptsize \left(\begin{array}{c}1-t^2+13 t^4+5 t^6+71 t^8+31 t^{10}+171 t^{12}\\+42 t^{14}+171 t^{16}+31 t^{18}+71 t^{20}+5 t^{22}+13
   t^{24}-t^{26}+t^{28}
 \end{array}\right)} {(-1+t)^{10} (1+t)^{10} \left(1+t^2\right)^5 \left(1+t^4\right)^2}$	 &$\begin{array}{c}4 t^2+16 t^4+18 t^6-114 t^{10}-309 t^{12}+213 t^{14}\\+2711 t^{16}+4612 t^{18}-11245 t^{20}+O\left(t^{21}\right)+O\left(t^{21}\right) \end{array}$ \\ 
    \bottomrule
	\end{tabular}}
\end{adjustbox}
\caption{Coulomb branch Hilbert series and plethystic logarithm for the magnetic quivers of $\essdg{G}{\ell}{r=1}$ theories in Table \ref{shortS}. The unrefined PL confirms the dimension of the global symmetry. }
\label{shortS1}
\end{table}

\begin{table}[t]
\small
\centering
\begin{adjustbox}{center}
\scalebox{0.8}{
	\begin{tabular}{ccc}
\toprule
		SCFT  & Hilbert Series & PL[HS] \\ 
\midrule
       $\essdg{E_6}{2}{2}$ &$1+39 t^2+84 t^3+893 t^4+3444 t^5+O\left(t^{6}\right)$	 &$\begin{array}{c} 39 t^2+84 t^3+113 t^4+168 t^5+O\left(t^6\right) \end{array}$ 	\\ \hline
          $\essdg{D_4}{2}{2}$ &$1+16t^2+30t^3+180t^4+540t^5+2029t^6+6240t^7++O\left(t^{8}\right)$	 & $\begin{array}{c}16 t^2+30 t^3+44 t^4+60 t^5+44 t^6-120 t^7+O\left(t^8\right)\end{array}$ 
 \\ 
    \midrule
    $\essdg{A_2}{2}{2}$ &	$\dfrac
	{\scriptsize (1-t)\left(\begin{array}{c}1+t+4 t^2+10 t^3+30 t^4+61 t^5+142 t^6+277 t^7+545 t^8+965 t^9+1695 t^{10}\\+2710 t^{11}+4262 t^{12}+6252 t^{13}+8892
   t^{14}+11964 t^{15}+15617 t^{16}\\+19303 t^{17}+23178 t^{18}+26476 t^{19}+29302 t^{20}+30943 t^{21}+31682 t^{22}\\ +\dots \mathrm{palindrome} +\dots + t^{44}
 \end{array}\right)} { \left(1-t^2\right)^4 \left(1-t^3\right)^2 \left(1-t^4\right)^3 \left(1-t^5\right)^3 \left(1-t^6\right)^3}$	&$\begin{array}{c}7 t^2+8 t^3+17 t^4+16 t^5+11 t^6\\-24 t^7-101 t^8-184 t^9-204 t^{10} +O\left(t^{11}\right)\end{array}$ 
 \\ 
    \midrule
     $\essdg{D_4}{3}{2}$ &$1+9 t^2+22 t^3+74 t^4+236 t^5+O\left(t^{6}\right)$		 	&$\begin{array}{c} 9 t^2+22 t^3+29 t^4+38 t^5+O\left(t^{6}\right)\end{array}$ \\ 
    \midrule
      $\essdg{A_1}{3}{2}$&	$\dfrac
	{\scriptsize \left(\begin{array}{c}1+t^2+t^3+3 t^4+6 t^5+10 t^6+13 t^7+21 t^8+25 t^9+37 t^{10}+48 t^{11}+62 t^{12}\\+72 t^{13}+88 t^{14}+92 t^{15}+105 t^{16}+107
   t^{17}+109 t^{18}+\dots \mathrm{palindrome} +\dots + t^{36}
 \end{array}\right)} {\left(1-t^2\right) \left(1-t^3\right)^3 \left(1-t^4\right)^2 \left(1-t^5\right) \left(1-t^6\right) \left(1-t^7\right)
   \left(1-t^9\right)}$	  &$\begin{array}{c}2 t^2+4 t^3+4 t^4+6 t^5+7 t^6\\+6 t^7+3 t^8-4 t^9-18 t^{10} +O\left(t^{11}\right)\end{array}$  \\  
    \midrule
        $\essdg{A_2}{4}{2}$ &	$\dfrac
	{\scriptsize \left(\begin{array}{c}1-t^2+13 t^4+6 t^6+99 t^8+141 t^{10}+665 t^{12}+1080 t^{14}+3278 t^{16}\\+5163 t^{18}+11492 t^{20}+16505
   t^{22}+29419 t^{24}\\+37560 t^{26}+56423 t^{28}+63691 t^{30}+82499 t^{32}+82607 t^{34}+93470 t^{36}\\ +\dots \mathrm{palindrome} +\dots + t^{72}
 \end{array}\right)} {\left(1-t^2\right)^5 \left(1-t^4\right)^3 \left(1-t^6\right)^3 \left(1-t^8\right)^3 \left(1-t^{12}\right)^2}$ &$\begin{array}{c} 4 t^2+16 t^4+22 t^6+30 t^8+12 t^{10}-70 t^{12}\\-394 t^{14}-1135 t^{16}-1125 t^{18}+3825
   t^{20}+O\left(t^{21}\right)  \end{array}$ \\ 
    \bottomrule
	\end{tabular}}
\end{adjustbox}
\caption{Coulomb branch Hilbert series and plethystic logarithm for the magnetic quivers of $\essdg{G}{\ell}{r=2}$ theories in Table \ref{shortS}. The unrefined PL confirms the dimension of the global symmetry. }
\label{shortS2}
\end{table}

\begin{table}[t]
\small
\centering
\begin{adjustbox}{center}
\scalebox{0.8}{
	\begin{tabular}{ccc}
\toprule
		SCFT  & Hilbert Series & PL[HS] \\ 
\midrule
       $\teedg{E_6}{2}{1}$ &$\dfrac
	{\scriptsize \left(\begin{array}{c} (1+t^2)(1-2 t+33 t^2-12 t^3+455 t^4+12 t^5+2904 t^6+56 t^7+8930 t^8-54 t^9+13145 t^{10}\\-54 t^{11}+8930
   t^{12}+56 t^{13}+2904 t^{14}+12 t^{15}+455 t^{16}-12 t^{17}+33 t^{18}-2 t^{19}+t^{20})
 \end{array}\right)} {(-1+t)^{24} (1+t)^{22}}$	 &$\begin{array}{c} 55 t^2+52 t^3-2 t^4-650 t^5-2925 t^6+1248 t^7\\+51701 t^8+152958 t^9-346372 t^{10} +O\left(t^{11}\right)\end{array}$ 	\\ \hline
          $\teedg{D_4}{2}{1}$ &$\dfrac{\left(1+t^2\right) \left(1-2 t+14 t^2-12 t^3+40 t^4-12 t^5+14 t^6-2 t^7+t^8\right)}{(-1+t)^{12} (1+t)^{10}}$	 & $\begin{array}{c} 24 t^2+14 t^3-37 t^4-98 t^5+35 t^6\\+672 t^7+588 t^8-3962 t^9-9373 t^{10}+O\left(t^{11}\right)\end{array}$ 
 \\  
    \midrule
    $\teedg{A_2}{2}{1}$ &$\dfrac{\left(1+t^2\right) \left(1+4 t^2+t^4\right)}{(-1+t)^6 (1+t)^6}$	&$\begin{array}{c} 11 t^2-10 t^4+16 t^6-45 t^8+144 t^{10}+O\left(t^{11}\right)\end{array}$ 
 \\  
    \midrule
     $\teedg{D_4}{3}{1}$ &$\dfrac{\left(1+t^2\right) \left(1-t+4 t^2+11 t^3+23 t^4+8 t^5+23 t^6+11 t^7+4 t^8-t^9+t^{10}\right)}{(-1+t)^{12} (1+t)^{10}
   \left(1+t+t^2\right)}$		 	&$\begin{array}{c} 15 t^2+16 t^3+27 t^4-14 t^5-183 t^6-434 t^7\\-175 t^8+2436 t^9+8197 t^{10}+O\left(t^{11}\right)\end{array}$ \\  
    \midrule
      $\teedg{A_1}{3}{1}$& $\dfrac{\left(1+t^2\right) \left(1-t+t^2\right)}{(-1+t)^4 (1+t)^2 \left(1+t+t^2\right)} $&$\begin{array}{c} 4 t^2+2 t^3-t^4-t^6\end{array}$  \\   
    \midrule
        $\teedg{A_2}{4}{1}$ &$\dfrac{1-2 t+3 t^2+2 t^4+3 t^6-2 t^7+t^8}{(-1+t)^6 (1+t)^4 \left(1+t^2\right)^2}$ &$\begin{array}{c} 4 t^2+4 t^3+7 t^4+4 t^5-5 t^6-16 t^7\\-31 t^8-20 t^9+30 t^{10}+O\left(t^{11}\right)\end{array}$ \\ 
    \bottomrule
	\end{tabular}}
\end{adjustbox}
\caption{Coulomb branch Hilbert series and plethystic logarithm for the magnetic quivers of $\teedg{G}{\ell}{r=1}$ theories in Table \ref{shortR}. The unrefined PL confirms the dimension of the global symmetry. }
\label{shortT1}
\end{table}

\begin{table}[t]
\small
\centering
\begin{adjustbox}{center}
\scalebox{0.8}{
	\begin{tabular}{ccc}
\toprule
		SCFT  & Hilbert Series & PL[HS] \\ 
\midrule
       $\teedg{E_6}{2}{2}$ &$1+55 t^2+52 t^3+1701 t^4+2964 t^5+O\left(t^6\right)$	 &$\begin{array}{c} 55 t^2+52 t^3+161 t^4+104 t^5+O\left(t^6\right)\end{array}$ 	\\ \hline
          $\teedg{D_4}{2}{2}$ &$1+24 t^2+14 t^3+368 t^4+364 t^5+4198 t^6+5488 t^7+O\left(t^8\right)$	 & $\begin{array}{c}24 t^2+14 t^3+68 t^4+28 t^5-139 t^6-336 t^7+O\left(t^8\right)\end{array}$ 
 \\ 
    \midrule
    $\teedg{A_2}{2}{2}$ &$\dfrac
	{\scriptsize \left(\begin{array}{c}1+5 t^2+38 t^4+111 t^6+297 t^8+442 t^{10}+576 t^{12}\\+442 t^{14}+297 t^{16}+111 t^{18}+38 t^{20}+5 t^{22}+t^{24}
 \end{array}\right)} {(-1+t)^{12}
   (1+t)^{12} \left(1+t^2\right)^6}$	&$\begin{array}{c} 11 t^2+29 t^4-39 t^6\\-199 t^8+608 t^{10}+O\left(t^{11}\right)\end{array}$ 
 \\ 
    \midrule
     $\teedg{D_4}{3}{2}$ &$1+15 t^2+16 t^3+149 t^4+284 t^5+O\left(t^6\right)$		 	& $\begin{array}{c} 15 t^2+16 t^3+29 t^4+44 t^5+O\left(t^6\right)\end{array}$ \\ 
    \midrule
      $\teedg{A_1}{3}{2}$&	$\dfrac
	{\scriptsize \left(\begin{array}{c}(1+t^4)(1+3 t^2+t^3+7 t^4+8 t^5+14 t^6+20 t^7+26 t^8+27 t^9+37 t^{10}+29 t^{11}\\+37 t^{12}+27 t^{13}+26
   t^{14}+20 t^{15}+14 t^{16}+8 t^{17}+7 t^{18}+t^{19}+3 t^{20}+t^{22})
 \end{array}\right)} {\left(1-t^2\right) \left(1-t^3\right) \left(1-t^4\right)^2 \left(1-t^5\right)^3 \left(1-t^6\right)}$ &$\begin{array}{c}4 t^2+2 t^3+4 t^4+8 t^5+t^6-2 t^7\\-5 t^8-16 t^9-20 t^{10} +O\left(t^{11}\right)\end{array}$  \\  
    \midrule
        $\teedg{A_2}{4}{2}$ &$\dfrac{\scriptsize  \left(\begin{array}{c} 1-2 t+4 t^2-4 t^3+10 t^4-6 t^5+21 t^6-8 t^7+53 t^8-10 t^9+104 t^{10}\\+174 t^{12}+22 t^{13}+253 t^{14}+52
   t^{15}+310 t^{16}+76 t^{17}+324 t^{18}\\+\dots \mathrm{palindrome} +\dots + t^{36}\end{array}\right)}{(1-t)^2 \left(1-t^2\right) \left(1-t^3\right)^2 \left(1-t^4\right)^3 \left(1-t^6\right)^3 \left(1-t^8\right)}$ & $\begin{array}{c} 4 t^2+4 t^3+8 t^4+8 t^5+13 t^6+8 t^7\\+9 t^8-4 t^9-38 t^{10}+O\left(t^{11}\right)\end{array}$\\ 
    \bottomrule
	\end{tabular}}
\end{adjustbox}
\caption{Coulomb branch Hilbert series and plethystic logarithm for the magnetic quivers of $\teedg{G}{\ell}{r=2}$ theories in Table \ref{shortR}. The unrefined PL confirms the dimension of the global symmetry. }
\label{shortT2}
\end{table}

\clearpage

\bibliographystyle{JHEP2}
\bibliography{bibli.bib}

\end{document}